\documentclass[11pt]{cernrep} 
\usepackage{graphicx} 
\usepackage{here} 
\usepackage{epsfig} 
\include{rlfig} 
\begin{document} 
\def    \nn             {\nonumber} 
\def    \=              {\;=\;} 
\def    \frac           #1#2{{#1 \over #2}} 
\def    \ret            {\\[\eqskip]} 
\def    \ie             {{\em i.e.\/} } 
\def    \eg             {{\em e.g.\/} } 
\def    \lsim           {\raisebox{-3pt}{$\>\stackrel{<}{\scriptstyle\sim}\>$}} 
\def    \gsim           {\raisebox{-3pt}{$\>\stackrel{>}{\scriptstyle\sim}\>$}} 
\def    \gtrsim         {\raisebox{-3pt}{$\>\stackrel{>}{\scriptstyle\sim}\>$}} 
\def    \esim           {\raisebox{-3pt}{$\>\stackrel{-}{\scriptstyle\sim}\>$}} 
\newcommand     \be     {\begin{equation}} 
\newcommand     \ee     {\end{equation}} 
\newcommand     \ba     {\begin{eqnarray}} 
\newcommand     \ea     {\end{eqnarray}} 
\newcommand     \sst            {\scriptstyle} 
\newcommand     \sss            {\scriptscriptstyle} 
\newcommand     \auno{a^{(1)}} 
\newcommand     \avg[1]         {\left\langle #1 \right\rangle} 
\newcommand     \Ca             {{C_{\rm A}}} 
\newcommand     \Cf             {{C_{\rm f}}} 
\newcommand     \lambdamsb     {\ifmmode 
          \Lambda_4^{\rm \scriptscriptstyle \overline{MS}} \else 
         $\Lambda_4^{\rm \scriptscriptstyle \overline{MS}}$ \fi} 
\newcommand     \MSB            {\ifmmode {\overline{\rm MS}} \else 
                                 $\overline{\rm MS}$  \fi} 
\newcommand     \nf             {n_{\rm f}} 
\newcommand     \nlf            {n_{\rm lf}} 
\newcommand     \ptmin     {\ifmmode p_{\scriptscriptstyle T}^{\sss min} \else 
                           $p_{\scriptscriptstyle T}^{\sss min}$ \fi} 
\def     \muf           {\mbox{$\mu_{\sss F}$}} 
\def     \mur            {\mbox{$\mu_{\sss R}$}} 
\def    \muo            {\mbox{$\mu_0$}} 
\newcommand\as{\alpha_{\sss S}} 
\newcommand\astwo{\alpha_{\sss S}^2} 
\newcommand\asthree{\alpha_{\sss S}^3} 
\newcommand\asfour{\alpha_{\sss S}^4} 
\newcommand\epb{\overline{\epsilon}} 
\newcommand\aem{\alpha_{\rm em}} 
\newcommand\QQb{{Q\overline{Q}}} 
\newcommand\qqb{{q\overline{q}}} 
\newcommand\cb{\overline{c}} 
\newcommand\bb{\overline{b}} 
\newcommand\tb{\overline{t}} 
\newcommand\Qb{\overline{Q}} 
\newcommand\qq{{\scriptscriptstyle Q\overline{Q}}} 
\def \asopi{\mbox{$\frac{\as}{\pi}$}} 
\def \oacube {\mbox{${\cal O}(\asthree)$}} 
\def \oatwo {\mbox{${\cal O}(\astwo)$}} 
\def \oas   {\mbox{${\cal O}(\as)$}} 
\def \ppbar {\mbox{$p \bar p$}} 
\def \ttbar {\mbox{$t \bar t$}} 
\def \bbbar {\mbox{$b \bar b$}} 
\def \ccbar {\mbox{$c \bar c$}} 
\def \mtt   {\mbox{$M_{\scriptscriptstyle t\bar t}$}} 
\def \pt   {\mbox{$p_{\scriptscriptstyle T}$}} 
\def \ptpair   {\mbox{$p_{\scriptscriptstyle T}^{\scriptscriptstyle 
                t\bar t}$}} 
\def \et   {\mbox{$E_{\scriptscriptstyle T}$}} 
\def \etsq {\mbox{$E_{\scriptscriptstyle T}^2$}} 
\def \rap   {\mbox{$\eta$}} 
\def \deltar {\mbox{$\Delta R$}} 
\def \dphi {\mbox{$\Delta \phi$}} 
\def \to   {\mbox{$\rightarrow$}} 
\def    \mb             {\mbox{$m_b$}} 
\def    \mc             {\mbox{$m_c$}} 
\def    \mt             {\mbox{$m_t$}} 
\newcommand \jpsi{\ifmmode{J/\psi 
    }\else{$J/\psi$}\fi} 
\def\calF{{\cal F}} 
\def\calP{{\cal P}} 
\def\calM{{\cal M}} 
\def\calO{{\cal O}} 
\newcommand{\lj}{$\ell J/\psi$} 
\newcommand{\mmax}{$M^{max}$} 
\newcommand{\mmaxl}{$M^{max}_{\ell J/\psi}$} 
\newcommand{\pyth}{P{\footnotesize YTHIA}} 
\newcommand{\herw}{{\small HERWIG}} 
\newcommand{\isajet}{{\small ISAJET}} 
\def        \ATLAS      {\mbox{ATLAS}} 
\def        \CMS        {\mbox{CMS}} 
\def        \abseta     {\mbox{$|\eta|$}} 
\newcommand \etmiss     {\ifmmode E_{\sss T}^{\sss miss} \else 
                            $E_{\sss T}^{\sss miss}$ \fi} 
\newcommand \ptl        {\ifmmode p_{\sss T}^{\sss \ell} \else 
                            $p_{\sss T}^{\sss\ell}$  \fi} 
\def        \eb         {\mbox{$\epsilon_b$}} 
\def        \detadphi   {\mbox{$\delta\eta\times\delta\phi$}} 
\def        \dr         {\mbox{$\Delta R$}} 
\def        \dmt        {\mbox{$\delta m_t$}} 
\def        \Dmt        {\mbox{$\Delta m_t$}} 
\def        \ADmt       {\mbox{$|\Delta m_t|$}} 
\def        \infb       {\mbox{fb$^{-1}$}} 
\def        \jj         {\mbox{$jj$}} 
\def        \jjb        {\mbox{$jjb$}} 
\def        \mjj        {\mbox{$m_{jj}$}} 
\def        \mjjb       {\mbox{$m_{jjb}$}} 
\def        \mlnb       {\mbox{$m_{\ell\nu b}$}} 
\def        \mW         {\mbox{$m_W$}} 
\def        \mcone      {\mbox{$m_{cone}$}} 
\def        \tjjb       {\mbox{$t\rightarrow jjb$}} 
\def        \tWb        {\mbox{$t\rightarrow Wb$}} 
\def        \Wjj        {\mbox{$W\rightarrow jj$}} 
\def        \Wlnu       {\mbox{$W\rightarrow \ell\nu$}} 
\def        \Wenu       {\mbox{$W\rightarrow e\nu$}} 
\def        \Wqqbar     {\mbox{$W\rightarrow q \bar{q}$}} 
\def        \Wbbbar     {\mbox{$Wb\bar{b}$}} 
\def        \Zll        {\mbox{$Z\rightarrow \ell^+ \ell^-$}} 
\def        \Ztt        {\mbox{$Z\rightarrow \tau\tau$}} 
\def        \Zqqbar     {\mbox{$Z\rightarrow q\bar{q}$}} 
\def        \tljpsi     {\mbox{$t \rightarrow \ell+\jpsi+X$}} 
\def        \ljpsi     {\mbox{$\ell+\jpsi$}} 
\def        \mlb        {\mbox{$m_{\ell b}$}} 
\def        \mll        {\mbox{$m_{\ell\ell}$}} 
\def        \tlnub      {\mbox{$t\rightarrow \ell\nu b$}} 
\def        \blnuc      {\mbox{$b\rightarrow \ell\nu c$}} 
 
\newcommand \avmlb {\ifmmode \langle m_{\ell b}^{\sss 2}\rangle \else  
  $\langle m_{\ell 
    b}^{\sss 2} \rangle $\fi} 
\newcommand \avthetalb {\ifmmode 
  \langle \cos\theta_{\ell b}\rangle \else $\langle \cos\theta_{\ell 
  b}\rangle $\fi} 
\def\lta{\;\raisebox{-.5ex}{\rlap{$\sim$}} \raisebox{.5ex}{$<$}\;} 
\def\gta{\;\raisebox{-.5ex}{\rlap{$\sim$}} \raisebox{.5ex}{$>$}\;} 
\newcommand{\permille}{$^0 \!\!\!\: / \! _{00}\;$} 
\newcommand{\GeV}{GeV}

\newcommand{\mtsq}{m_{t}^2} 
\newcommand{\mw}{M_{W}} 
\newcommand{\mww}{M_{W}^{2}} 
 
\newcommand{\md}{m_{d}} 
\newcommand{\ms}{m_{s}} 
\newcommand{\mbb}{m_{b}^2} 
\newcommand{\mh}{m_{H}} 
\newcommand{\mhh}{m_{H}^2} 
\newcommand{\mz}{M_{Z}} 
\newcommand{\mzz}{M_{Z}^{2}} 
 
\newcommand{\lra}{\leftrightarrow} 
 
\def\Ww{{\mbox{\boldmath $W$}}} 
\def\B{{\mbox{\boldmath $B$}}} 
\def\nn{\noindent} 
 
\newcommand{\sinsq}{\sin^2\theta} 
\newcommand{\cossq}{\cos^2\theta} 
\newcommand{\nl}{\nonumber \\} 
\newcommand{\eqn}[1]{Eq.({#1})} 
\newcommand{\ibidem}{{\it ibidem\/},} 
\newcommand{\into}{\;\;\to\;\;} 
\newcommand{\wws}[2]{\langle #1 #2\rangle^{\star}} 
\newcommand{\p}[1]{{\scriptstyle{\,(#1)}}} 
\newcommand{\ru}[1]{\raisebox{-.2ex}{#1}} 
\newcommand{\epem}{$e^{+} e^{-}\;$} 
\newcommand{\tcht}{$t\to c H\;$} 
\newcommand{\tczt}{$t\to c Z\;$} 
\newcommand{\tcgt}{$t\to c g\;$} 
\newcommand{\tcft}{$t\to c \gamma\;$} 
\newcommand{\tchm}{$t\to c H$} 
\newcommand{\tczm}{$t\to c Z$} 
\newcommand{\tcgm}{$t\to c g$} 
\newcommand{\tcfm}{$t\to c \gamma$} 
\newcommand{\tch}{t\to c H} 
\newcommand{\tcz}{t\to c Z} 
\newcommand{\tcg}{t\to c g} 
\newcommand{\tcf}{t\to c \gamma} 
 
\newcommand{\tbwh}{t\to b W H} 
\newcommand{\tbwz}{t\to b W Z} 
\newcommand{\tbwht}{$t\to b W H\;$} 
\newcommand{\tbwzt}{$t\to b W Z\;$} 
\newcommand{\tbwhm}{$t\to b W H$} 
\newcommand{\tbwzm}{$t\to b W Z$} 
 
\newcommand{\tbwf}{t\to b W \gamma} 
\newcommand{\tbwg}{t\to b W g} 
\newcommand{\tbwft}{$t\to b W \gamma\;$} 
\newcommand{\tbwgt}{$t\to b W g\;$} 
\newcommand{\tbwfm}{$t\to b W \gamma$} 
\newcommand{\tbwgm}{$t\to b W g$}

\newcommand{\tcww}{t\to c W W} 
\newcommand{\tcwwt}{$t\to c W W\;$} 
\newcommand{\tcwwm}{$t\to c W W$} 
 
\newcommand{\tuww}{t\to u W W} 
\newcommand{\tuwwt}{$t\to u W W\;$} 
\newcommand{\tuwwm}{$t\to u W W$} 
 
\newcommand{\tqw}{t\to q W} 
 
\newcommand{\tbw}{t\to b W} 
\newcommand{\tbwt}{$t\to b W\;$} 
\newcommand{\tbwm}{$t\to b W$} 
 
\newcommand{\tsw}{t\to s W} 
\newcommand{\tswt}{$t\to s W\;$} 
\newcommand{\tswm}{$t\to s W$} 
\newcommand{\tdw}{t\to d W} 
\newcommand{\tdwt}{$t\to d W\;$} 
\newcommand{\tdwm}{$t\to d W$} 
 
\newcommand{\Gt}{\Gamma(t\to b W)} 
 
\newcommand{\lolumi}{\mbox{$\rm {10^{33}~cm^{-2}\,s^{-1}}$}} 
\newcommand{\hilumi}{\mbox{$\rm {10^{34}~cm^{-2}\,s^{-1}}$}} 
\newcommand{\pT}{\mbox{$p_{T}$}} 
\newcommand{\invfb}{\mbox{fb$^{-1}$}} 
\newcommand{\ra}{\mbox{$\rightarrow$}} 
\newcommand{\mZ}{\mbox{$m_{Z}$}} 
\newcommand{\ET}{\mbox{$E_{T}$}} 
\newcommand{\lplm}{\mbox{$l^{+}l^{-}$}} 
\newcommand{\modeta}{\mbox{$\mid \eta \mid$}} 
\newcommand{\pTmiss}{\ifmmode p_{\sss T}^{\sss miss} \else 
                            $p_{\sss T}^{\sss miss}$ \fi} 
\newcommand{\mqZ}{\mbox{$m_{qZ}$}} 
\newcommand{\Ldt}{\mbox{$\int{\cal L} \cdot {\rm dt}$}} 
\newcommand{\Wpl}{\mbox{$W^{+}$}} 
\newcommand{\csq}{\mbox{$c^{2}$}} 
\newcommand{\ETmiss}{\etmiss} 
 
\title{TOP QUARK PHYSICS\footnote{~~To appear in the Report of the ``1999
  CERN Workshop on SM physics (and more) at the LHC''.}} 
\author{{\bf Conveners}: M. Beneke, I.
  Efthymiopoulos, M.L. Mangano, J.~Womersley \\ {\bf Contributing
    authors}: A.~Ahmadov, G.~Azuelos, U.~Baur, A.~Belyaev,
  E.L.~Berger, W.~Bernreuther, E.E.~Boos, M.~Bosman, A.~Brandenburg,
  R.~Brock, M.~Buice, N.~Cartiglia, F.~Cerutti, A.~Cheplakov,
  L.~Chikovani, M.~Cobal-Grassmann, G.~Corcella, F.~del Aguila,
  T.~Djobava, J.~Dodd, V.~Drollinger, A.~Dubak, S.~Frixione,
  D.~Froidevaux, B.~Gonz\'{a}lez Pi\~{n}eiro, Y.P.~Gouz, D.~Green,
  P.~Grenier, S.~Heinemeyer, W.~Hollik, V.~Ilyin, C.~Kao,
  A.~Kharchilava, R. Kinnunen, V.V.~Kukhtin, S.~Kunori, L.~La~Rotonda,
  A.~Lagatta, M.~Lefebvre, K.~Maeshima, G.~Mahlon, S.~Mc Grath, 
  G.~Medin, R.~Mehdiyev, B.~Mele, Z.~Metreveli, D.~O'Neil, L.H.~Orr,
  D.~Pallin, S.~Parke, J.~Parsons, D.~Popovic, L.~Reina,
  E.~Richter-Was, T.G.~Rizzo, D.~Salihagic, M.~Sapinski, M.H.~Seymour,
  V.~Simak, L.~Simic, G.~Skoro, S.R.~Slabospitsky, J.~Smolik,
  L.~Sonnenschein, T.~Stelzer, N.~Stepanov, Z.~Sullivan, T.~Tait,
  I.~Vichou, R.~Vidal, D.~Wackeroth, G.~Weiglein, S.~Willenbrock,
  W.~Wu} \institute{~}
 
\maketitle 
 
\section{INTRODUCTION} 
The top quark, when it was finally discovered at Fermilab in  
1995~\cite{Abe:1994xt,Abe:1995hr,Abe:1995eh}, completed the  
three-generation structure of the Standard Model (SM) and opened  
up the new field of top quark physics. Viewed as just another  
SM quark, the top quark appears to be a rather uninteresting  
species. Produced predominantly, in hadron-hadron collisions, through  
strong interactions, it decays rapidly without forming hadrons,  
and almost exclusively through the single mode $t\to W b$. The  
relevant CKM coupling $V_{tb}$ is already determined by the  
(three-generation) 
unitarity of the CKM matrix. Rare decays and CP violation are  
unmeasurably small in the SM. 
 
Yet the top quark is distinguished by its large mass, about 
35 times larger than the mass of the next heavy quark, and  
intriguingly close to the scale of electroweak (EW) symmetry breaking.  
This unique property raises a number of interesting questions. 
Is the top quark mass generated by the Higgs mechanism as the  
SM predicts and is its mass related to the top-Higgs-Yukawa  
coupling? Or does it play an even more fundamental role in the  
EW symmetry breaking mechanism? If there are new  
particles lighter than the top quark, does the top quark decay  
into them? Could non-SM physics first manifest itself in non-standard 
couplings of the top quark which show up as anomalies in top  
quark production and decays? Top quark physics tries to answer  
these questions. 
 
Several properties of the top quark have already been examined at the 
Tevatron. These include studies of the kinematical properties of top 
production~\cite{Abe:1999pm}, the measurements of the top 
mass~\cite{Abachi:1997jv,Abe:1998vq}, of the top production 
cross-section~\cite{Abachi:1997re,Abe:1998ue}, the reconstruction of 
$t\bar t$ pairs in the fully hadronic final 
states~\cite{Abe:1997rh,Abbott:1998nn}, the study of $\tau$ decays of 
the top quark~\cite{Abe:1997uk}, the reconstruction of hadronic decays 
of the $W$ boson from top decays~\cite{Abe:1998ev}, the search for 
flavour changing neutral current decays~\cite{Abe:1998fz}, the 
measurement of the $W$ helicity in top 
decays~\cite{Affolder:2000mp}, and bounds on $t\bar{t}$ spin 
correlations~\cite{Abbott:2000dt}. Most of these measurements are limited 
by the small sample of top quarks collected at the Tevatron up to 
now. The LHC is, in comparison, a top factory, producing about 8 
million $t\bar{t}$ pairs per experiment per year at low luminosity 
(10$\,\mbox{fb}^{-1}/$year), and another few million (anti-)tops in 
EW single (anti-)top quark production. We therefore expect 
that top quark properties can be examined with significant precision 
at the LHC. Entirely new measurements can be contemplated on the basis 
of the large available statistics. 
 
In this chapter we summarize the top physics potential of the LHC 
experiments. An important aspect of this chapter is to document SM 
model properties of the top quark against which anomalous behaviour 
has to be compared. In each section (with the exception of the one 
devoted to anomalous couplings) we begin by summarizing SM 
expectations and review the current theoretical status on a particular 
topic.  This is followed by a detailed description of experimental 
analysis strategies in the context of the ATLAS and CMS 
experiments. Particular emphasis is given to new simulations carried 
out in the course of this workshop. In detail, the outline of this 
chapter is as follows: 
 
In Section~\ref{sec:ewk} we summarize {\em SM precision calculations} 
of the {\em top quark mass relations} and of the {\em  
total top quark width}. We 
then recall  the importance of the top quark mass in 
{EW precision measurements}. We discuss, in particular, the 
role of EW precision measurements under the assumption that a 
SM Higgs boson has been discovered. 
 
Section~\ref{sec:ttprod} deals with the $t\bar{t}$ {\em production  
process}: expectations for and measurements of the total cross  
section, the transverse momentum and $t\bar{t}$ invariant mass  
distribution are discussed. A separate subsection is devoted to  
EW radiative corrections  
to $t\bar{t}$ production, and to radiative corrections in the  
Minimal Supersymmetric SM (MSSM).  
 
The prospects for an accurate {\em top quark mass measurement } 
are detailed in Section~\ref{sec:mass} Next to ``standard''  
measurements in the lepton+jets and di-lepton channels, two mass   
measurements are discussed that make use of the large number  
of top quarks available at the LHC: the selection of top quarks  
with large transverse momentum in the lepton+jets channel and  
the measurement of $\ell J/\psi$ correlations in $t\to \ell J/\psi X$  
decays. This decay mode appears to be particularly promising and  
the systematic uncertainties are analyzed in considerable detail. 
 
{\em Single top quark production} through EW interactions  
provides the only known way to directly measure the CKM matrix element  
$V_{tb}$ at hadron colliders. It also probes the nature of the  
top quark charged current. In Section~\ref{sec:onetop} the SM  
expectations for the three basic single top production mechanisms  
and their detection are documented, including the possibility to  
measure the high degree of polarisation in the SM.  
 
The issue of top quark spin is pursued in Section~\ref{TTSPIN} Here  
we summarize expectations on {\em spin correlations in $t\bar{t}$ 
production},  
the construction of observables sensitive to such correlations and  
the results of a simulation study of di-lepton angular correlations  
sensitive to spin correlations. Possible non-SM CP violating  
couplings of the top quark can be revealed through anomalous  
spin-momentum correlations and are also discussed here. 
 
As mentioned above, the search for {\em anomalous (i.e. non-SM) interactions} 
is one of the main motivations for top quark physics. In  
Section~\ref{ANOM} the sensitivity of the LHC experiments to the  
following couplings is investigated:  
$gt\bar{t}$ couplings and anomalous $W tb$ couplings in top production,  
flavour-changing neutral currents (FCNCs) in top production and  
decay. 
 
Section~\ref{RARE} is devoted to {\em rare top decays}.  
The SM expectations for radiative top decays  
and FCNC decays are documented. Decay rates large enough to be of  
interest require physics beyond the SM. The two Higgs Doublet Models,  
the MSSM and generic anomalous couplings are considered explicitly  
followed by ATLAS and CMS studies on the expected sensitivity in  
particular decay channels. 
 
Finally, the measurement of the {\em top quark Yukawa coupling in 
$t\bar{t}H$ production} is considered (Section~\ref{sec:ttH}). The SM 
cross sections are tabulated in the various production channels at the 
LHC. For the case of a low mass Higgs boson, the results of a 
realistic study using a simulation of the ATLAS detector are discussed. 
 
The following topics are collected in the appendices: $b$-quark  
tagging and the calibration of the jet energy scale in top  
events; the direct measurement of the top quark spin (as opposed to  
that of a top squark) and and of top quark electric charge; the total  
cross section for production of a fourth generation heavy quark;  
a compendium of Monte Carlo event generators available for  
top production and its backgrounds. 
 
The internal ATLAS and CMS notes quoted in the bibliography can 
be obtained from the collaborations' web pages \cite{Atlasnotes,CMSnotes}. 
Updated versions of this document, as well as a list of addenda and 
errata, will be available on the web page of the LHC Workshop top  
working group~\cite{Topgrouppage}.  
 
\section{TOP QUARK PROPERTIES AND ELECTROWEAK PRECISION  
MEASUREMENTS\protect\footnote{Section coordinators: M.~Beneke, G.~Weiglein.}} 
\label{sec:ewk} 
The top quark is, according to the Standard Model (SM), a spin-1/2 and 
charge-$2/3$ fermion, transforming as a colour triplet under the group 
$SU(3)$ of the strong interactions and as the weak-isospin partner of 
the bottom quark. None of these quantum numbers has been directly 
measured so far, although a large amount of indirect evidence supports 
these assignments. The analysis of EW observables in $Z^0$ 
decays~\cite{lepewwg:1999} requires the 
existence of a $T_3=1/2$, charge-2/3 fermion, with a mass in the range 
of 170~GeV, consistent with the direct Tevatron measurements.  The 
measurement of the total cross section at the Tevatron, and its 
comparison with the theoretical estimates, are consistent with the 
production of a spin-1/2 and colour-triplet particle.   
The LHC should provide a 
direct measurement of the top quantum numbers.  We present the results 
of some studies in this direction in Appendix~\ref{app:app1}. 
 
\subsection{Top quark mass and width} \label{sec:topprop} 
In addition to its quantum numbers, the two most fundamental properties 
of the top quark are its mass $m_t$ and width $\Gamma_t$, defined 
through the position of the single particle pole 
$m_t^\star=m_t-i\Gamma_t/2$ in the perturbative top quark propagator. In 
the SM $m_t$ is related to the top Yukawa coupling: 
\begin{equation} 
y_t(\mu) = 2^{3/4} G_F^{1/2} m_t \left(1+\delta_t(\mu)\right) , 
\end{equation} 
where $\delta_t(\mu)$ accounts for radiative corrections. 
Besides the top quark pole mass, the top quark $\overline{\rm MS}$ mass 
$\overline{m}_t(\mu)$ is often used. The definition of 
$\overline{m}_t(\mu)$ including EW corrections is subtle 
(see the discussion in \cite{Hempfling:1995ar}). 
As usually done in the literature, 
we define the $\overline{\rm MS}$ mass by including only pure 
QCD corrections: 
\begin{equation} 
\overline{m}_t(\mu) = m_t \left(1+\delta_{\rm QCD}(\mu)\right)^{-1}. 
\end{equation} 
The conversion 
factor $\delta_{\rm QCD}(\mu)$ is very well known 
\cite{Melnikov:2000qh}. Defining 
$\overline{m}_t=\overline{m}_t(\overline{m}_t)$ and 
$a_s=\alpha_s^{\rm \overline{MS}}(\overline{m}_t)/\pi$, we have 
\begin{eqnarray} 
\delta_{\rm QCD}(\overline{m}_t) &=& \frac{4}{3}\,a_s+ 
8.2366 \,a_s^2+ 73.638\,a_s^3+\ldots 
\nonumber\\ 
&=& (4.63+0.99+0.31+0.11^{+0.11}_{-0.11})\% = 
(6.05^{+0.11}_{-0.11})\%. 
\end{eqnarray} 
This assumes five massless flavours besides the top quark and we use 
$a_s=0.03475$ which 
corresponds to $\alpha_s^{\rm \overline{MS}}(m_Z)=0.119$ 
and $\overline{m}_t=165\,$GeV. 
The error estimate translates into an absolute uncertainty of 
$\pm 180\,$MeV in $m_t-\overline{m}_t$ and uses an estimate of the 
four-loop contribution. Note that the difference between the 
two mass definitions, $m_t-\overline{m}_t$, 
is about $10\,$GeV. This means that any observable that is supposed to 
measure a top quark mass with an accuracy of $1$--$2\,$GeV and which 
is known only at leading order (LO) must come with an explanation for why 
higher order corrections are small when the observable is expressed in 
terms 
of that top quark mass definition that it is supposed to determine 
accurately. We will return to this point in Section~\ref{sec:mass} 
 
The on-shell decay width $\Gamma_t$ is less well known, but the 
theoretical accuracy ($<1\%$) is more than sufficient compared to the accuracy 
of foreseeable measurements. The decay through $t\to bW$ is by far 
dominant and we restrict the discussion to this decay mode. It is  
useful to quantify the decay width in units of the lowest order decay  
width with $M_W$ and $m_b$ set to zero and $|V_{tb}|$ set to 1:  
\be 
\label{mwzero} 
\Gamma_0 = \frac{G_F m_t^3}{8\pi \sqrt{2}} = 1.76\,\mbox{GeV}. 
\ee 
Incorporating $M_W$ the leading order result reads 
\be 
\label{topLO} 
\Gamma_{\rm LO}(t \to bW)/|V_{tb}|^2 = \Gamma_0 
\left(1-3\frac{M_W^4}{m_t^4}+ 
2\frac{M_W^6}{m_t^6}\right) = 0.885\,\Gamma_0 = 1.56\,\mbox{GeV}. 
\ee 
The correction for non-vanishing bottom quark mass is about $-0.2\%$  
in units of $\Gamma_0$. Likewise corrections to treating the $W$ boson 
as a stable particle are negligible. Radiative corrections are known  
to second order in QCD and to first order in the EW  
theory. Table~\ref{mbtab1} summarises the known corrections to the  
limiting case (\ref{mwzero}). Putting all effects together we obtain: 
\begin{equation} 
\Gamma(t \to bW)/|V_{tb}|^2 \approx 0.807\,\Gamma_0 
=1.42\,\mbox{GeV}. 
\end{equation} 
The top quark lifetime is small compared to the time scale  
for hadronisation~\cite{Bigi:1986jk}. For this reason, top-hadron  
spectroscopy is not expected to be the subject of LHC measurements. 
   
\begin{table} 
\renewcommand{\arraystretch}{1.3} 
\begin{center} 
\caption{\label{mbtab1} Corrections to the top quark width $\Gamma_0$  
($M_W=0$, lowest order) in units of $\Gamma_0$. The best estimate of  
$\Gamma(t \to bW)/|V_{tb}|^2$ is obtained by adding all corrections  
together. Parameters: $a_s=0.03475$, $M_W=80.4\,$GeV 
and $m_t=175\,$GeV.} \vspace*{0.1cm} 
\begin{tabular}{|c||c|} 
\hline 
$M_W\not=0$ correction at lowest order, see (\ref{topLO}) & $-11.5\%$ \\ 
$\alpha_s$ correction, $M_W=0$ & $-9.5\%$ \\ 
$\alpha_s$ correction, $M_W\not=0$ correction & $+1.8\%$ \\ 
$\alpha_s^2$ correction, $M_W=0$ 
\cite{Czarnecki:1999qc,Chetyrkin:1999ju} 
& $-2.0\%$ \\ 
$\alpha_s^2$ correction, $M_W\not=0$ correction 
\cite{Chetyrkin:1999ju} 
& $+0.1\%$ \\ 
EW correction \cite{Denner:1991ns} & $+1.7\%$\\ 
\hline 
\end{tabular} 
\end{center} 
\renewcommand{\arraystretch}{1} 
\end{table} 
 
\subsection{Role of ${\bf m_t}$ in EW precision 
  physics} 
 \label{sec:weiglein} 
 The 
 EW precision observables serve as an important tool for 
 testing the theory, as they provide an important consistency test for 
 every model under consideration. By comparing the EW 
 precision data with the predictions (incorporating quantum 
 corrections) within the SM or its extensions, most notably the 
 minimal supersymmetric extension of the Standard Model 
 (MSSM)~\cite{susy}, it is in principle possible to derive 
 indirect constraints on all parameters of the model. The information 
 obtained in this way, for instance, on the mass of the Higgs boson in 
 the SM or on the masses of supersymmetric particles is complementary 
 to the information gained from the direct production of these 
 particles. 
 
In order to derive precise theoretical predictions, two kinds of 
theoretical uncertainties have to be kept under control: the 
uncertainties from unknown higher-order corrections, as the 
predictions are derived only up to a finite order in perturbation 
theory, and the parametric uncertainties caused by the experimental 
errors of the input parameters.  The top quark mass enters the 
EW precision observables as an input parameter via quantum 
effects, i.e.\ loop corrections. As a distinctive feature, the large 
numerical value of $m_t$ gives rise to sizable corrections that behave 
as powers of $m_t$. This is in contrast to the corrections associated 
with all other particles of the SM. 
In particular, the dependence on the mass of the Higgs boson is only 
logarithmic in leading order and therefore much weaker than the 
dependence on $m_t$. In the MSSM large corrections from SUSY particles 
are only possible for large splittings in the SUSY spectrum, while the 
SUSY particles in general decouple for large masses. 
 
The most important $m_t$-dependent contribution to the EW 
precision observables in the SM and the MSSM enters via the universal 
parameter $\Delta\rho$ which is proportional to 
$m_t^2$~\cite{Veltman:1977kh}, 
\begin{equation} 
\Delta\rho = \left(\frac{\Sigma^{Z}(0)}{M_Z^2} - 
\frac{\Sigma^{W}(0)}{M_W^2} \right)_{t,b} = N_C \frac{\alpha}{16 \pi 
s_W^2 c_W^2} \frac{m_t^2}{M_Z^2} , 
\end{equation} 
where the limit $m_b \to 0$ has been taken, $s_W$ ($c_W$) is the  
sin (cos) of the weak mixing angle, and $\Sigma^{Z}(0)$ and 
$\Sigma^{W}(0)$ indicate the transverse parts of the gauge-boson 
self-energies at zero momentum transfer. 
 
The theoretical prediction for $M_W$ is obtained from the relation 
between the vector-boson masses and the Fermi constant, 
\begin{equation} 
M_W^2 \left(1 - \frac{M_W^2}{M_Z^2}\right) = 
\frac{\pi \alpha}{\sqrt{2} G_F} \left(1 + \Delta r\right), 
\end{equation} 
where the quantity $\Delta r$~\cite{Sirlin:1980nh} 
is derived from muon decay and contains the radiative corrections. 
At one-loop order, $\Delta r$ can be written as 
$\Delta r = \Delta \alpha - \frac{c_W^2}{s_W^2} \Delta\rho + (\Delta 
r)_{\mathrm{nl}} $, 
where $\Delta \alpha$ contains the large logarithmic contributions from 
the light fermions, and the non-leading terms are collected in 
$(\Delta r)_{\mathrm{nl}}$. 
 
The leptonic effective weak mixing angle is determined from the 
effective couplings of the neutral current at the Z-boson resonance to 
charged leptons, 
$J_{\mu}^{\mathrm{NC}} = \left( \sqrt{2} G_F M_Z^2 \right)^{1/2} 
\left[g_V \gamma_{\mu} - g_A \gamma_{\mu} \gamma_5 \right]$, according to 
\begin{equation} 
\sin^2 \theta^{\mathrm{lept}}_{\mathrm{eff}} = 
\frac{1}{4} \left(1 - \frac{\mbox{Re}\,(g_V)}{\mbox{Re}\,(g_A)} 
\right) . 
\end{equation} 
In $\sin^2 \theta^{\mathrm{lept}}_{\mathrm{eff}}$ the leading $m_t$-dependent 
contributions enter via 
$\delta \sin^2 \theta^{\mathrm{lept}}_{\mathrm{eff}} = 
- (c_W^2 s_W^2)/(c_W^2 - s_W^2) \Delta\rho$. 
 
The precision observables $M_W$ and 
$\sin^2 \theta^{\mathrm{lept}}_{\mathrm{eff}}$ are 
currently known with experimental accuracies of $0.05\%$ and $0.07\%$, 
respectively~\cite{lepewwg:1999}. 
The accuracy in $M_W$ will be further improved 
at the LHC by about a factor of three (see the EW chapter of this 
Yellow Report). 
Besides the universal correction $\Delta \rho$, there is also a 
non-universal correction proportional to $m_t^2$ in the $Zb \bar b$ 
coupling, which however is less accurately measured experimentally 
compared to $M_W$ and $\sin^2 \theta^{\mathrm{lept}}_{\mathrm{eff}}$. 
The strong dependence of the SM radiative corrections to the precision 
observables on the input value of $m_t$ made it possible to predict the 
value of $m_t$ from the precision measurements prior to its actual 
experimental discovery, and the predicted value turned out to be in 
remarkable agreement with the experimental 
result~\cite{Abachi:1997jv,Abe:1998vq}. 
 
Within the MSSM, the mass of the lightest {\cal CP}-even Higgs boson, 
$m_h$, is a further observable whose theoretical prediction strongly 
depends on $m_t$. While in the SM the Higgs-boson mass is a free 
parameter, $m_h$ is calculable from the other SUSY parameters in the 
MSSM and is bounded to be lighter than $M_Z$ at the tree level. 
The dominant one-loop corrections arise from the top and scalar-top 
sector via terms of the form 
$G_F m_t^4 \ln (m_{\tilde{t}_1} 
m_{\tilde{t}_2}/m_t^2)$~\cite{Haber:1991aw}. 
As a rule of thumb, a 
variation of $m_t$ by 1~GeV, keeping all other parameters fixed,  
roughly translates into a shift of the 
predicted value of $m_h$ by 1~GeV. If the lightest {\cal CP}-even Higgs 
boson of the MSSM will be detected at the LHC, its mass will be measurable 
with an accuracy of about $\Delta m_h = 0.2$~GeV~\cite{atlasphystdr}. 
 
Due to the sensitive dependence of the EW precision 
observables on the numerical value of $m_t$, a high accuracy in the input  
value of $m_t$ is very important for stringent consistency 
tests of a model, for constraints on the model's parameters (e.g.\ the 
Higgs boson mass within the SM), and for a high sensitivity to 
possible effects of new physics. It should be noted that this calls not only 
for a high precision in the experimental measurement of the top quark 
mass, but also for a detailed investigation of how the quantity that 
is actually determined experimentally is related to the parameter 
$m_t$ used as input in higher-order calculations. While these 
quantities are the same in the simplest approximation, their relation 
is non-trivial in general due to higher-order contributions and 
hadronisation effects. A further discussion of this problem, which can 
be regarded as a systematic uncertainty in the experimental 
determination of $m_t$, is given in Section~\ref{sec:mass} 
 
\subsection{Physics gain from improving $\bf \Delta m_t$ from 
$\bf \Delta m_t = 2$~GeV to $\bf\Delta m_t = 1$~GeV} 
 
During this workshop the question was investigated of how much 
information one could gain from the EW precision observables by 
improving the experimental precision in $m_t$ from $\Delta m_t = 
2$~GeV, reachable within the first year of LHC running (see 
Section~\ref{sec:masslj}), to $\Delta m_t = 1$~GeV, possibly 
attainable on a longer time scale (see Section~\ref{sec:tjpsi}). 
 
In order to analyse this question quantitatively, we have considered the 
case of the SM and the MSSM and assumed that the Higgs boson has been 
found at the LHC. For the uncertainty in $\Delta \alpha_{\mathrm{had}}$  
(the hadronic contribution to the electromagnetic coupling at the  
scale $M_Z$) we 
have adopted $\delta (\Delta \alpha_{\mathrm{had}}) = 
0.00016$, which corresponds to the ``theory driven'' analyses  
of~\cite{Kuhn:1998ze}. 
 
Concerning the current theoretical prediction for $M_W$ and 
$\sin^2 \theta^{\mathrm{lept}}_{\mathrm{eff}}$ in the SM, the theoretical 
uncertainty from unknown higher-order corrections has been estimated to be 
about $\Delta M_W = 6$~MeV and 
$\Delta \sin^2 \theta^{\mathrm{lept}}_{\mathrm{eff}} = 
4 \times 10^{-5}$~\cite{Bardin:1999gt}. 
In Table~\ref{tab:precobsuncert} the theoretical uncertainties for 
$M_W$ and $\sin^2 \theta^{\mathrm{lept}}_{\mathrm{eff}}$ from unknown 
higher-order corrections are compared with the parametric uncertainty 
from the input parameters $\Delta \alpha_{\mathrm{had}}$ and $m_t$ for 
$\Delta m_t = 2$~GeV as well as $\Delta m_t = 1$~GeV. The parametric 
uncertainties from the other parameters, supposing that the SM Higgs 
boson has been found at the LHC in the currently preferred range, are 
negligible compared to the uncertainties from $\Delta 
\alpha_{\mathrm{had}}$ and $m_t$. The resulting uncertainties in 
$M_W$ and $\sin^2 \theta^{\mathrm{lept}}_{\mathrm{eff}}$ have been 
obtained using the parameterisation of the results for these quantities 
given in~\cite{Degrassi:1998iy}. As can be seen in the table, for 
$\Delta m_t = 2$~GeV the parametric uncertainty in $m_t$ gives rise to 
the largest theoretical uncertainty in both precision observables. While for 
$\sin^2 \theta^{\mathrm{lept}}_{\mathrm{eff}}$ the uncertainty induced 
from the error in $m_t$ is comparable to the one from the error in 
$\Delta \alpha_{\mathrm{had}}$, for $M_W$ the uncertainty from the error 
in $m_t$ is twice as big as the one from unknown higher-order 
corrections and four times as big as the one from the error in $\Delta 
\alpha_{\mathrm{had}}$. A reduction of the error from $\Delta m_t = 
2$~GeV to $\Delta m_t = 1$~GeV will thus mainly improve the precision in 
the prediction for $M_W$. The uncertainty induced in $M_W$ by $\Delta 
m_t = 1$~GeV is about the same as the current uncertainty from unknown 
higher-order corrections. The latter uncertainty can of course be 
improved by going beyond the present level in the perturbative evaluation 
of $\Delta r$.

\begin{table} 
\renewcommand{\arraystretch}{1.3} 
\begin{center} 
\caption{Comparison of the current theoretical uncertainty from 
unknown higher-order corrections ($\Delta_{\mathrm{theo}}$) in 
$M_W$ and $\sin^2 \theta^{\mathrm{lept}}_{\mathrm{eff}}$ with the 
parametric uncertainties from the error in $\Delta 
\alpha_{\mathrm{had}}$ and $m_t$.} \label{tab:precobsuncert} 
\vspace*{0.1cm} 
\begin{tabular}{|c||c|c|c|c|} 
\cline{2-5} \multicolumn{1}{c|}{} 
  & $\Delta_{\mathrm{theo}}$ 
  & $\delta (\Delta \alpha_{\mathrm{had}}) = 0.00016$ 
  & $\Delta m_t = 2$ GeV & $\Delta m_t = 1$ GeV \\ 
    \hline \hline 
$ \Delta M_W/\mathrm{MeV} $ & 6 & 3.0 & 12 & 6.1 \\ \hline 
$ \Delta \sin^2 \theta^{\mathrm{lept}}_{\mathrm{eff}} \times 10^5 $ & 4 & 
  5.6 & 6.1 & 3.1 \\ \hline 
\end{tabular} 
\end{center} 
\renewcommand{\arraystretch}{1} 
\end{table} 

\begin{figure} 
\begin{center} 
\includegraphics[width=0.6\textwidth,clip]{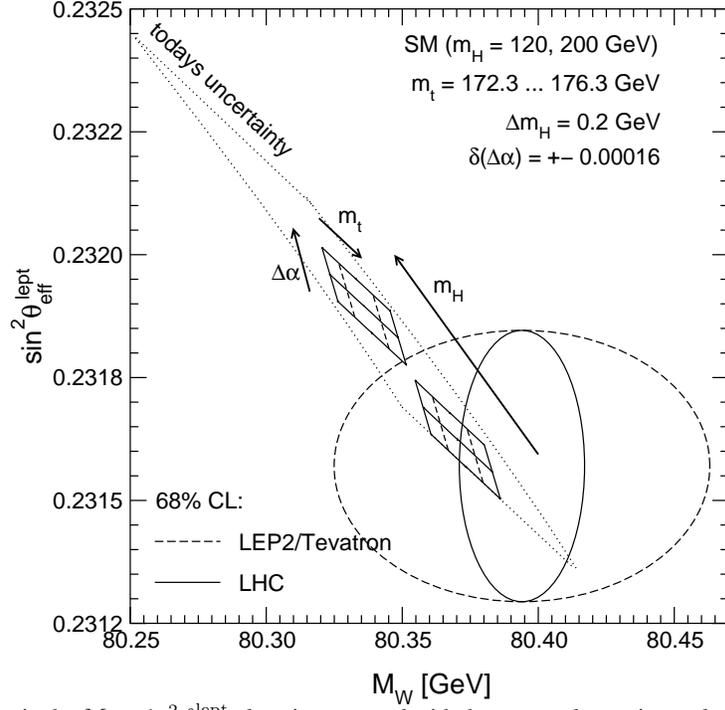} 
\end{center} 
\vskip -1cm 
\caption{The SM prediction in the 
$M_W$--$\sin^2 \theta^{\mathrm{lept}}_{\mathrm{eff}}$ plane is compared 
with the expected experimental accuracy at LEP2/Tevatron 
($\Delta M_W = 30$~MeV, 
$\sin^2 \theta^{\mathrm{lept}}_{\mathrm{eff}} = 1.7 \times 10^{-4}$) 
and at the LHC ($\Delta M_W = 15$~MeV, 
$\sin^2 \theta^{\mathrm{lept}}_{\mathrm{eff}} = 1.7 \times 10^{-4}$). 
The theoretical uncertainties induced by 
$\delta (\Delta \alpha_{\mathrm{had}}) = 0.00016$ and $\Delta m_t = 2$~GeV 
(full line) as well as $\Delta m_t = 1$~GeV (dashed line) are shown for two 
values of the Higgs boson mass $m_H$. 
\label{fig:precobs}} 
\end{figure}

In Fig.~\ref{fig:precobs} the theoretical predictions for $M_W$ and 
$\sin^2 \theta^{\mathrm{lept}}_{\mathrm{eff}}$ (see~\cite{Heinemeyer:1999zd}  
and references therein) are compared with the 
expected accuracies for these observables at LEP2/Tevatron and at the 
LHC (for the central values, the current experimental values are 
taken).  The parametric uncertainties corresponding to 
$\delta (\Delta \alpha_{\mathrm{had}}) = 0.00016$ and $\Delta m_t = 2$~GeV, 
$\Delta m_t = 1$~GeV are shown for two values of the Higgs boson 
mass, $m_H = 120$~GeV and $m_H = 200$~GeV, and the present theoretical 
uncertainty is also indicated (here $m_H$ is varied within 
100~GeV$ \leq m_H \leq 400$~GeV and $\Delta m_t = 5.1$~GeV). The figure 
shows that, assuming that the Higgs boson will be discovered at the LHC, 
the improved accuracy in $m_t$ and $M_W$ at the LHC will allow a stringent 
consistency test of the theory. A reduction of the experimental error in 
$m_t$ from $\Delta m_t = 2$~GeV to $\Delta m_t = 1$~GeV leads to a 
sizable improvement in the accuracy of the theoretical 
prediction. In view of the precision tests of the theory a further 
reduction of the experimental error in $M_W$ and 
$\sin^2 \theta^{\mathrm{lept}}_{\mathrm{eff}}$ 
would clearly be very desirable.

While within the MSSM the improved accuracy in $m_t$ and $M_W$ at the LHC 
will have a similar impact on the analysis of the precision observables 
as in the SM, the detection of the mass of the lightest {\cal CP}-even 
Higgs boson will provide a further stringent test of the model. 
The prediction for $m_h$ within the MSSM is particularly sensitive to 
the parameters in the $t$--$\tilde{t}$~sector, while in the region 
of large $M_A$ and large $\tan\beta$ (giving rise to Higgs masses beyond 
the reach of LEP2) the dependence on the latter two parameters is 
relatively mild. A precise measurement of $m_h$ can thus be used to 
constrain the parameters in the $t$--$\tilde{t}$~sector of the MSSM.

\begin{figure} 
\begin{center} 
\includegraphics[width=0.6\textwidth,clip]{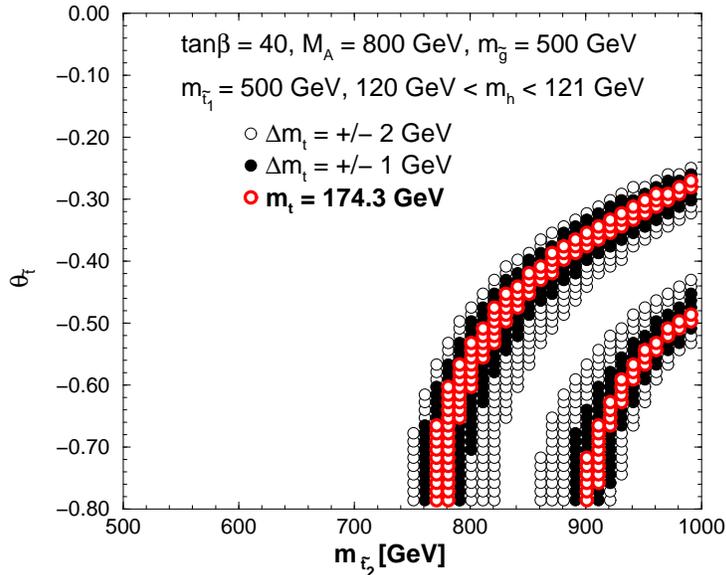} 
\end{center} 
\vskip -1cm 
\caption{Indirect constraints on the parameters of the scalar top sector 
of the MSSM from the measurement of $m_h$ at the LHC. The effect of the 
experimental error in $m_t$ is shown for $\Delta m_t = 2$~GeV and 
$\Delta m_t = 1$~GeV. 
\label{fig:mhconstrlhc}} 
\end{figure} 
 
In Fig.~\ref{fig:mhconstrlhc} it is assumed that the mass of the lightest 
scalar top quark, $m_{\tilde{t}_1}$, is known with high precision, while 
the mass of the heavier scalar top quark, $m_{\tilde{t}_2}$, and the mixing 
angle $\theta_{\tilde{t}}$ are treated as free parameters. The Higgs boson 
mass is assumed to be known with an experimental precision of $\pm 0.5$~GeV 
and the impact of $\Delta m_t = 2$~GeV and $\Delta m_t = 1$~GeV is shown 
(the theoretical uncertainty in the Higgs-mass prediction from unknown 
higher-order contributions and the parametric uncertainties besides the 
ones induced by $m_{\tilde{t}_2}$, $\theta_{\tilde{t}}$ and $m_t$ have 
been neglected here). The two bands represent the values of 
$m_{\tilde{t}_2}$, $\theta_{\tilde{t}}$ which are compatible with a 
Higgs-mass prediction of $m_h = 120.5 \pm 0.5$~GeV, where the two-loop 
result of~\cite{Heinemeyer:1999np} has been used (the bands 
corresponding to smaller and larger values of $m_{\tilde{t}_2}$ are 
related to smaller and larger values of the off-diagonal entry in the 
scalar top mixing matrix, respectively). 
Combining the constraints on the parameters in the scalar top sector obtained 
in this way with the results of the direct search for the scalar top 
quarks will allow a sensitive test of the MSSM. As can be seen in the 
figure, a reduction of $\Delta m_t$ from $\Delta m_t = 2$~GeV to 
$\Delta m_t = 1$~GeV will lead to a considerable reduction of the allowed 
parameter space in the $m_{\tilde{t}_2}$--$\theta_{\tilde{t}}$ plane. 
 
\section{$\bf t\bar t$ PRODUCTION AT THE LHC\protect\footnote{Section 
    coordinators: M.L.~Mangano, D.~Wackeroth, M.~Cobal (ATLAS), 
    J.~Parsons (ATLAS).}} 
\label{sec:ttprod} 
The determination of the top production characteristics will be one of the 
first measurements to be carried out with the large statistics 
available at the LHC. The large top quark mass ensures that top 
production is a short-distance process, and that the perturbative 
expansion, given by a series in powers of the small parameter 
$\as(\mt)\sim 0.1$, converges rapidly.  Because of the large 
statistics (of the order of $10^7$ top quark pairs produced per year), 
the measurements and their interpretation will be dominated by 
experimental and theoretical systematic errors.  Statistical 
uncertainties will be below the percent level for most observables.  
It will therefore be a severe challenge to reduce experimental and  
theoretical systematic uncertainties to a comparable level. 
In addition to providing interesting tests of QCD, accurate studies of 
the top production and decay mechanisms will be the basis for the 
evaluation of the intrinsic properties of the top quark and of its EW 
interactions. An accurate determination of the production cross 
section, for example, provides an independent indirect determination 
of $\mt$.  Asymmetries in the rapidity distributions of top and 
antitop quarks~\cite{Kuhn:1999kw} are sensitive to the light-quark parton  
distribution functions of the proton.  Anomalies in the total $\ttbar$ 
rate would indicate the presence of non-QCD production channels, to be 
confirmed by precise studies of the top quark distributions ({\em e.g.}   
$\pt$ and $\ttbar$ invariant mass spectra). These would be distorted 
by the presence of anomalous couplings or $s$-channel resonances  
expected in several beyond-the-SM (BSM) scenarios. 
Parity-violating asymmetries (for example in the rapidity 
distributions of right and left handed top quarks) are sensitive to 
the top EW couplings, and can be affected by the presence of BSM 
processes, such as the exchange of supersymmetric particles. As 
already observed at the Tevatron~\cite{Abachi:1997jv,Abe:1998vq}, the 
structure of the $\ttbar$ final state affects the direct determination 
of \mt. Initial and final-state gluon radiation do in fact contribute 
to the amount of energy carried by the jets produced in the decay of 
top quarks, and therefore need to be taken into proper account when 
jets are combined to extract \mt.  The details of the structure of 
these jets ({\em e.g.} their fragmentation function and their shapes), will 
also influence the experimental determination of the jet energy scales 
(important for the extraction of \mt), as well as the determination of 
the efficiency with which $b$-jets will be tagged (important for the 
measurement of the production cross section). 
 
It is therefore clear that an accurate understanding of the QCD 
dynamics is required to make full use of the rich 
statistics of $\ttbar$ final states in the study of the SM properties 
of top quarks, as well as to explore the presence of possible 
deviations from the SM.  In this section we review the current state 
of the art in predicting the production 
properties for top quark pairs (for a more detailed review of the 
theory of heavy quark production, see~\cite{Nason92}).  
The study of single top production will 
be presented in Section~\ref{sec:onetop} 
 
\subsection{Tools for QCD calculations} 
\label{sec:qcdtools} 
Full next-to-leading-order (NLO, ${\cal O}(\asthree)$) calculations 
are available for the following quantities: 
\begin{enumerate} 
\item Total cross sections~\cite{Nason:1988xz} 
\item Single-inclusive $p_T$ and $y$ spectra~\cite{Nason:1989zy} 
\item Double-differential spectra ($m_{t\bar t}$, azimuthal 
correlations $\Delta\Phi$, 
etc.)~\cite{Mangano:1992jk} 
\end{enumerate} 
All of the above calculations are available in the form of Fortran 
programs~\cite{Mangano:1992jk,MNRcode}, so that  
kinematical distributions can be evaluated  
at NLO~\cite{Frixione:1995fj} even in 
the presence of analysis cuts. 
 
Theoretical progress over the last few years has led to the 
resummation of Sudakov-type logarithms~\cite{Sterman:1987aj} which 
appear at all orders in the perturbative expansion for the total cross 
sections~\cite{Catani:1996dj,Berger:1998gz}. More recently, the 
accuracy of these resummations has been extended to the 
next-to-leading logarithmic (NLL) 
level~\cite{Kidonakis:1997gm,Bonciani:1998vc}. For a review of the 
theoretical aspects of Sudakov resummation, see the QCD chapter of 
this report.  As will be shown later, while the inclusion of these 
higher-order terms does not affect significantly the total production 
rate, it stabilises the theoretical predictions 
under changes in the renormalisation and factorisation scales, 
hence improving the predictive power. 
 
Unfortunately, the results of these resummed calculations are not available 
in a form suitable to implement selection cuts, as they 
only provide results for total cross-sections, fully integrated over 
all of phase space. The formalism has been generalised to 
the case of one-particle inclusive distributions  
in~\cite{Laenen:1998qw}, although no complete numerical analyses 
have been performed yet. 
 
The corrections of $\calO(\asthree)$ to the full production and decay 
should include the effect of gluon radiation off the quarks produced 
in the top decay. Interference effects are expected to take place 
between soft gluons emitted before and after the decay, at least for 
gluon energies not much larger than the top decay width.  While these 
correlations are not expected to affect the measurement of generic 
distributions, even small soft-gluon corrections can have an impact on 
the determination of the top mass.  Matrix elements for hard-gluon 
emission in $t\bar t$ production and decay ($p\bar p \to W^+ b W^-\bar 
b g$, with $t$ and $\bar t$ intermediate states) are implemented in a 
parton-level generator~\cite{Orr:1997pe}. The one-gluon emission off 
the light quarks from the $W$ decays was implemented, in the 
soft-gluon approximation, in the parton-level calculation  
of~\cite{Masuda:1996vs}. 
 
The above results refer to the production of top quarks treated as 
free, stable partons. Parton-shower Monte Carlo 
programs are available (\herw~\cite{Marchesini:1992ch}, 
\pyth~\cite{Sjostrand:1994yb}, \isajet~\cite{Paige:1981fb})  
for a complete description of the final state, 
including the full development of the perturbative gluon shower from 
both initial and final states, the decay of the top quarks, and the 
hadronisation of the final-state partons.   
These will be reviewed in 
Appendix~\ref{app:mc}. Recently, 
$\calO(\as)$ matrix element corrections to the decay of the top quark 
($t\to W b g$) have been included in the \herw\  Monte 
Carlo~\cite{Corcella:1998rs}. The impact of these corrections will be 
reviewed in Sections~\ref{sec:topdist} and~\ref{sec:ljpsisyst}. 
 
\subsection{Total $\bf t\bar{t}$ production rates} 
\label{sec:rates} 
In this section we collect the current theoretical predictions for 
cross sections and distributions, providing our best estimates of the 
systematic uncertainties. The theoretical uncertainties we shall 
consider include renormalisation (\mur) and factorisation (\muf) scale 
variations, and the choice of parton distribution functions (PDF's); 
 
We shall explore the first two by varying the scales over the range 
$\muo /2 < \mu < 2\muo$, where $\mu=\mur=\muf$ and 
  \begin{itemize} 
    \item $\muo=m_t$ for the total cross sections 
    \item $\muo=\sqrt{m_t^2+p_T^2}$ for single inclusive distributions 
    \item $\muo=\sqrt{m_t^2+(p_{T,t}^2+p_{T,\bar t}^2)/2}$ for double 
      inclusive distributions 
   \end{itemize} 
In the case of PDF's, we shall consider the latest 
fits of the CTEQ~\cite{Lai:1999wy} and of the 
MRST~\cite{Martin:1998sq,Martin:1999ww} groups: 
\begin{itemize} 
    \item MRST ($\as(M_Z)=0.1175$, $\langle k_T 
      \rangle=0.4$~GeV)  (default) 
    \item MRST$(g\downarrow)$  ($\as=0.1175$, $\langle k_T 
      \rangle=0.64$~GeV) 
    \item MRST$(g\uparrow)$  ($\as=0.1175$, $\langle k_T 
      \rangle=0$) 
    \item MRST$(\as\downarrow\downarrow)$ ($\as=0.1125$, $\langle k_T 
      \rangle=0.4$~GeV) 
    \item MRST$(\as\uparrow\uparrow)$ ($\as=0.1225$, $\langle k_T 
      \rangle=0.4$~GeV) 
    \item CTEQ5M ($\as=0.118$) 
    \item CTEQ5HJ ($\as=0.118$, enhanced weight for Tevatron high-$E_T$ jets) 
    \item CTEQ5HQ ($\as=0.118$, using the ACOT heavy flavour 
      scheme~\cite{Aivazis:1994pi}.) 
\end{itemize} 
All our numerical results relative to the MRST sets refer to the 
updated fits provided in~\cite{Martin:1999ww}. These give total 
rates which are on average 5\% larger than the fits in~\cite{Martin:1998sq}. 
\begin{figure}
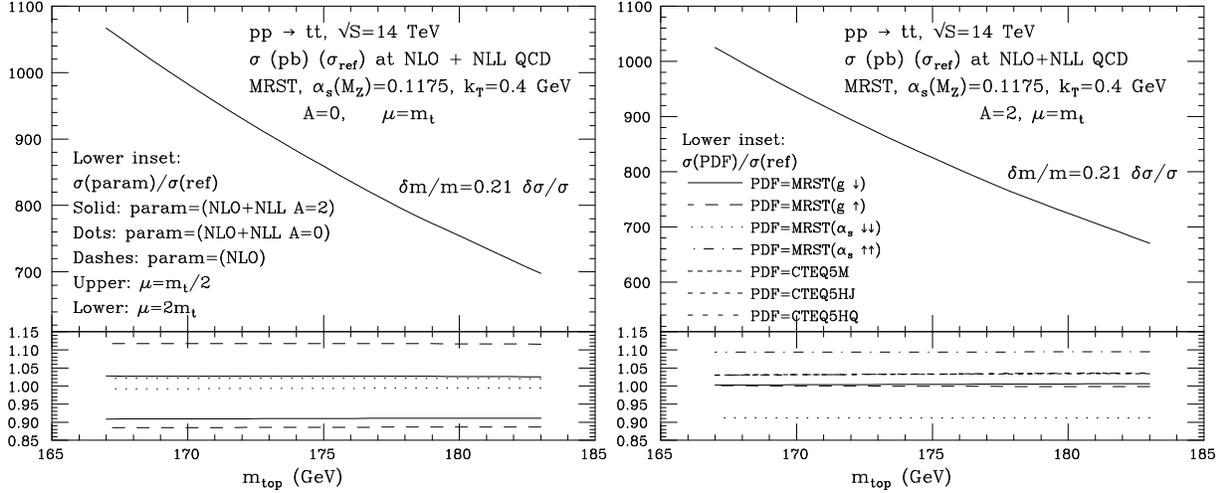
 
  \begin{center} 
    \centerline{ 
      \includegraphics[width=0.5\textwidth,clip]{sigtot_res.eps} \hfil 
      \includegraphics[width=0.5\textwidth,clip]{sigtot.eps}} 
\vskip -0.5cm 
    \caption{$\ttbar$ production rates. Left: scale 
      dependence at fixed order (NLO, dashed lines in the lower inset), and at 
      NLO+NLL (solid lines). Right: PDF dependence. See the text 
      for details.} 
    \label{fig:sigtot} 
  \end{center} 
\end{figure} 
The total $\ttbar$ production cross section is given in 
Fig.~\ref{fig:sigtot}, as a function of the top mass. As a reference 
set of parameters, we adopt $\muo=\mt$ and MRST. Full NLO+NLL 
corrections are included.  The upper inset shows the dependence of the 
cross section on the top mass. A fit to the distribution shows that 
$\Delta\sigma/\sigma \sim 5 \Delta \mt/\mt$. As a result, a 5\% 
measurement of the total cross section is equivalent to a 1\% 
determination of $\mt$ (approximately 2~GeV). As will be shown later 
on, 2~GeV is a rather safe estimate of the expected experimental 
accuracy in the determination of $\mt$ (1~GeV being the optimistic 
ultimate limit). It follows that 5\% should be a minimal goal in the 
overall precision for the measurement of $\sigma(\ttbar)$. The scale 
uncertainty of the theoretical predictions is shown in the lower inset 
of Fig.~\ref{fig:sigtot}.  The dashed lines refer to the NLO scale 
dependence, which is of the order of $\pm 12\%$. The dotted lines 
refer to the inclusion of the NLL corrections, according to the 
results of~\cite{Bonciani:1998vc}. The solid lines include the 
resummation of NLL effects, but assume a different structure of yet 
higher order (NNLL) corrections, relative to those contained in the 
reference NLL results (this is indicated by the value of the $A$ 
parameter equal to 2, see~\cite{Bonciani:1998vc} for the details). 
The scale uncertainty, after inclusion of NLL corrections, is 
significantly reduced. In the most conservative case of $A=2$, we have 
a $\pm 6\%$ variation. 
A detailed breakdown of the NLO  $\calO(\astwo+\asthree)$ 
and higher-order $\calO(\as^{\sss \ge 4})$ contributions, as a function of 
the scale and of the value of the parameter $A$, is given in 
Table~\ref{tab:sigtot}. 
A recent study~\cite{Cacciari:1999sy} of resummation effects on 
the total cross section for photo- and hadro-production of quarkonium 
states indicates that allowing $\mur\ne\muf$ increases the scale 
dependence of the NLL resummed cross-sections to almost match the scale 
dependence of the NLO results~\cite{Petrelli:1998ge}. Preliminary results of 
this study also suggest a similar increase of scale dependence in the 
case of $\ttbar$ production, if $\mur$ and $\muf$ are varied 
independently. This dependence can however be reduced by replacing  
$\mur$ with $\muf$ as the argument of $\as$ in the  
sub-leading coefficients of the resummed exponent~\cite{Cacciaripc}. 
{\renewcommand{\arraystretch}{1.2} 
\begin{table} 
\begin{center} 
\caption{Resummation contributions to the total $t\bar t$ 
  cross-sections ($m_t=175$~GeV) in pb. PDF set MRST.} 
\label{tab:sigtot}\vspace*{0.1cm} 
\begin{tabular}{|c|c|cc|cc|} \hline 
& & \multicolumn{2}{c|}{NLL resummed, A=2} 
& \multicolumn{2}{c|}{NLL resummed, A=0}\\ 
\cline{3-6} 
 $\mur=\muf$ & NLO  & ${\cal O}(\as^{\sss \ge 4})$ & NLO+NLL & ${\cal 
   O}(\as^{\sss \ge 4})$ & NLO+NLL \\ \hline\hline 
\mt/2 & 890 & $-7$ & 883 & $-12$ & 878 
\\ \hline 
\mt   & 796 & 29 & 825 & 63 & 859 
\\ \hline 
2\mt  & 705 & 77 & 782 & 148 & 853 
\\ \hline 
\end{tabular} 
\end{center} 
\end{table} } 
 
{\renewcommand{\arraystretch}{1.2} 
\begin{table} 
\begin{center} 
\caption{Total $t\bar t$ 
  cross-sections ($m_t=175$~GeV) in pb. NLO+NLL ($A=0$).} 
\label{tab:sigtotpdf} \vspace*{0.1cm} 
\begin{tabular}{|l|lll|} \hline 
  {PDF} & $ \mu=m_{t}/2 $ & $ \mu=m_{t} $ & $ \mu=2m_{t} $ \\ 
 \hline \hline 
  {MRST}                   &   877 &          859 &          853      \\ 
  {MRST} $g \uparrow $     &   881 &          862 &          857\\ 
  {MRST} $g \downarrow $   &   876 &          858 &          852   \\ 
  {MRST} $ \as\downarrow $ &   796 &          781 &          777 \\ 
  {MRST} $ \as\uparrow $   &   964 &          942 &          934\\ 
  {CTEQ5M}                 &   904 &          886 &          881\\ 
  {CTEQ5HJ}                &   905 &          886 &          881 
\\ \hline 
\end{tabular} 
\end{center} 
\end{table} }

The PDF dependence is shown on the right hand side of 
Fig.~\ref{fig:sigtot}, and given in detail for $\mt=175$~GeV in 
Table~\ref{tab:sigtotpdf}. The current uncertainty is at the level 
of $\pm 10$\%. Notice that the largest deviations from the default 
set occur for sets using different input values of $\as(M_Z)$. The 
difference between the reference sets of the two groups (MRST and 
CTEQ5M) is at the level of 3\%. It is interesting to explore 
potential correlations between the PDF dependence of top 
production, and the PDF dependence of other processes induced by 
initial states with similar parton composition and range in $x$. 
One such example is given by inclusive jet production. 
\begin{figure}
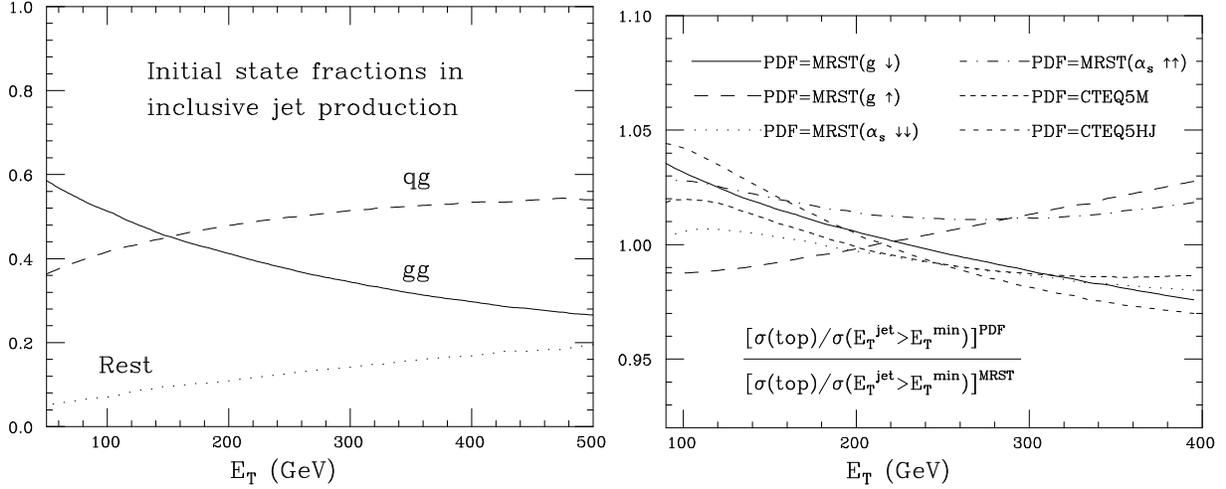
 
  \begin{center} 
    \centerline{ 
      \includegraphics[width=0.5\textwidth,clip]{incjet.eps} \hfil 
      \includegraphics[width=0.5\textwidth ,clip]{jettop_rat.eps}} 
\vskip -0.5cm 
    \caption{Left: initial state composition in inclusive jet 
      events, as a function of the jet $\et$ ($\vert \eta \vert 
      <2.5$). Right: PDF dependence of the top-to-jet cross-section 
      ratio, as a function of the minimum jet $\et$.} 
    \label{fig:incjet} 
  \end{center} 
\end{figure} 
Fig.~\ref{fig:incjet} shows the initial-state fraction of inclusive 
jet final states (with $\vert \eta \vert<2.5$) as a function of the 
jet-$\et$ threshold. For values of $\et\sim 200$~GeV, 90\% of the jets 
come from processes with at least one gluon in the initial state. 
This fraction is similar to that present in $\ttbar$ production, where 
90\% of the rate is due to $gg$ collisions.  On the right side of 
Fig.~\ref{fig:incjet} we show the double ratios: 
\be \frac{\left[ \sigma(\ttbar)/\sigma(\mbox{jet},\et>E_T^{\rm 
       min})\right]_{\rm PDF} } {\left[ 
     \sigma(\ttbar)/\sigma(\mbox{jet},\et>E_T^{\rm min})\right]_{\rm MRST} 
   }  
\ee  
As the plot shows, there is a strong correlation between the PDF 
dependences of the two processes. The correlation is maximal for 
$E_T^{\rm min}\sim 200$~GeV, as expected, since for this value the flavour 
composition of the initial states and the range of partonic momentum 
fractions probed in the two production processes are similar.  In the 
range $180 \lsim E_T^{\rm min} \lsim 260$~GeV the PDF dependence of the 
ratio $\sigma(\ttbar)/\sigma(\mbox{jet},\et>E_T^{\rm min})$ is reduced 
to a level of $\pm 1$\%, even for those sets for which the absolute 
top cross-section varies by $\pm 10$\%. The jet cross-sections were 
calculated~\cite{Frixione:1997np} using a scale 
$\mu^{jet}=\et\equiv\mu_0^{jet}$. If we vary the scales for $\ttbar$ 
and jet production in a correlated way (\ie\ selecting 
$\mu^{jet}/\mu_0^{jet}=\mu^{\sss t\bar t}/\mu_0^{\sss t\bar t}$), no 
significant scale dependence is observed. There is however no a-priori 
guarantee that the scales should be correlated. Unless this 
correlation can be proved to exist, use of the inclusive-jet cross 
section to normalise the $\ttbar$ cross section will therefore leave a 
residual systematic uncertainty which is no smaller than the scale 
dependence of the jet cross section. We do not expect this to become 
any smaller than the PDF dependence in the near future. 
 
Combining in quadrature the scale and PDF dependence of the total 
$\ttbar$ cross section, we are left with an overall 12\% 
theoretical systematic uncertainty, corresponding to a 4~GeV 
uncertainty on the determination of the top mass from the total cross 
section.

\subsection{Kinematical properties of $\bf t\bar{t}$ production} 
\label{sec:topdist} 
\begin{figure}[t]
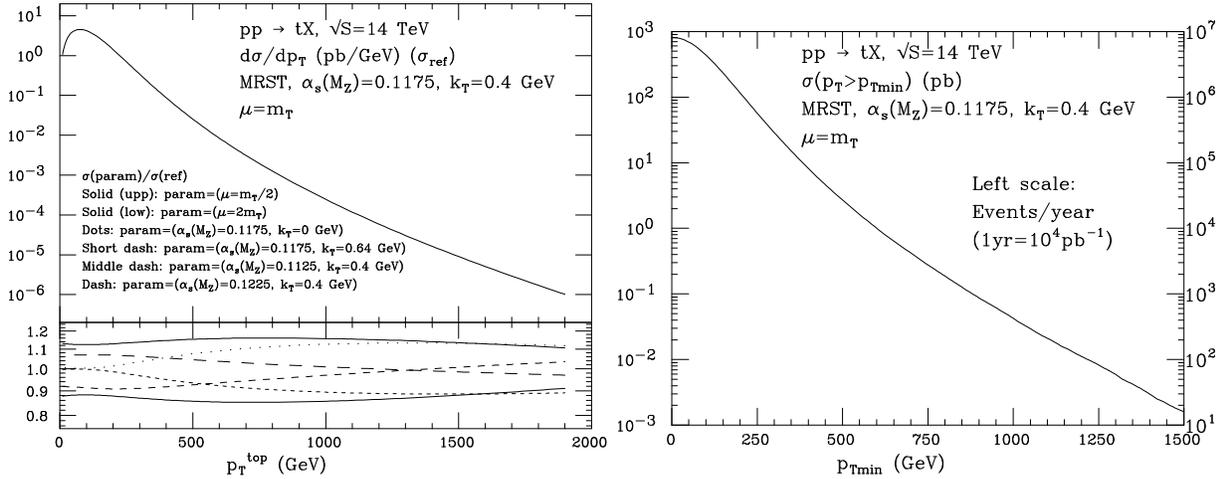
 
  \begin{center} 
   \centerline{ 
    \includegraphics[width=0.5\textwidth,clip]{ptrat.eps} \hfil 
    \includegraphics[width=0.5\textwidth,clip]{ptint.eps}} 
  \vskip -0.5cm 
    \caption{Inclusive top $p_T$ spectrum. Left: scale and PDF 
   dependence at NLO. Right: event rates above a given $p_T$ threshold.} 
    \label{fig:ptrat} 
  \end{center} 
\end{figure}

We start from the most inclusive quantity, the top $\pt$ 
spectrum. The NLO predictions are shown in Fig.~\ref{fig:ptrat}. Here 
we also explore the dependence on scale variations and on the choice of PDF. 
The uncertainties are $\pm 15$\% and $\pm 10$\%, 
respectively. 
\begin{figure}
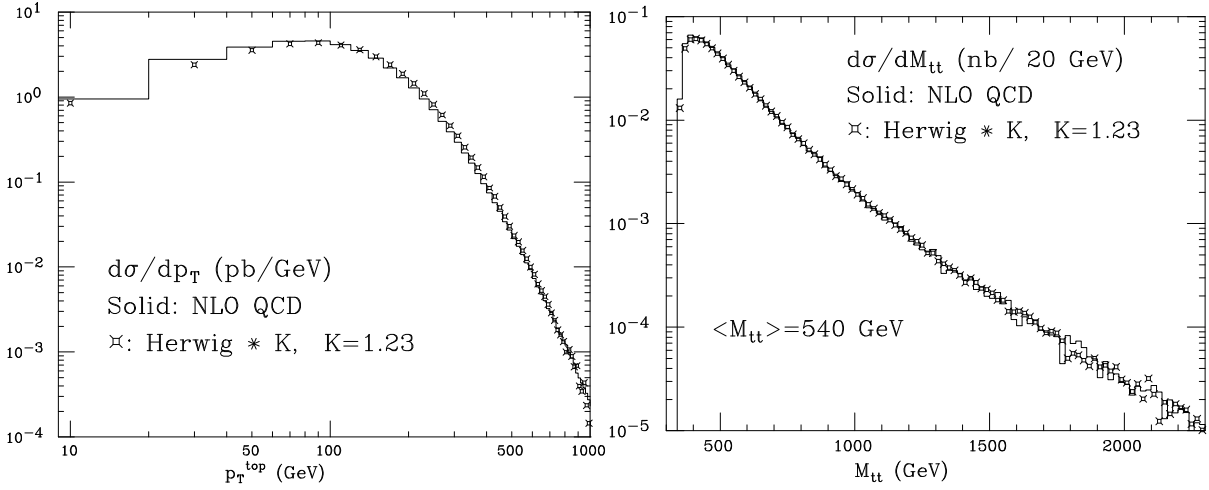
 
  \begin{center} 
   \centerline{ 
    \includegraphics[width=0.5\textwidth,clip]{hwpt.eps}\hfil 
    \includegraphics[width=0.5\textwidth,clip]{hwmtt.eps}} 
  \vskip -0.5cm 
    \caption{Comparison of NLO and 
    (rescaled)  \herw\  spectra. Left: inclusive top $p_T$ spectrum. 
      Right: inclusive $\mtt$ spectrum.} 
    \label{fig:hwpt} 
  \end{center} 
\end{figure} 
The reconstruction of top quarks and their momenta, as well as the 
determination of the reconstruction efficiencies and of the possible 
biases induced by the experimental selection cuts, depend on the 
detailed structure of the final state. It is important to verify that 
inclusive distributions as predicted by the most accurate NLO 
calculations are faithfully reproduced by the shower Monte Carlo 
calculations, used for all experimental studies.  This is done in 
Fig.~\ref{fig:hwpt}, where the NLO calculation is 
compared to the result of the \herw\  Monte Carlo, after a proper 
rescaling by an overall constant $K$-factor. The bin-by-bin agreement 
between the two calculations is at the level of 10\%, which should be 
adequate for a determination of acceptances and efficiencies at the 
percent level. 
 
Similar results are obtained for the invariant mass distribution of 
top quark pairs, shown in the plot of Fig.~\ref{fig:hwpt}. 
The scale and PDF dependence of the NLO calculation are similar to those 
found for the inclusive $p_T$ spectrum, and are not reported in the 
figure. 
 
\begin{figure}
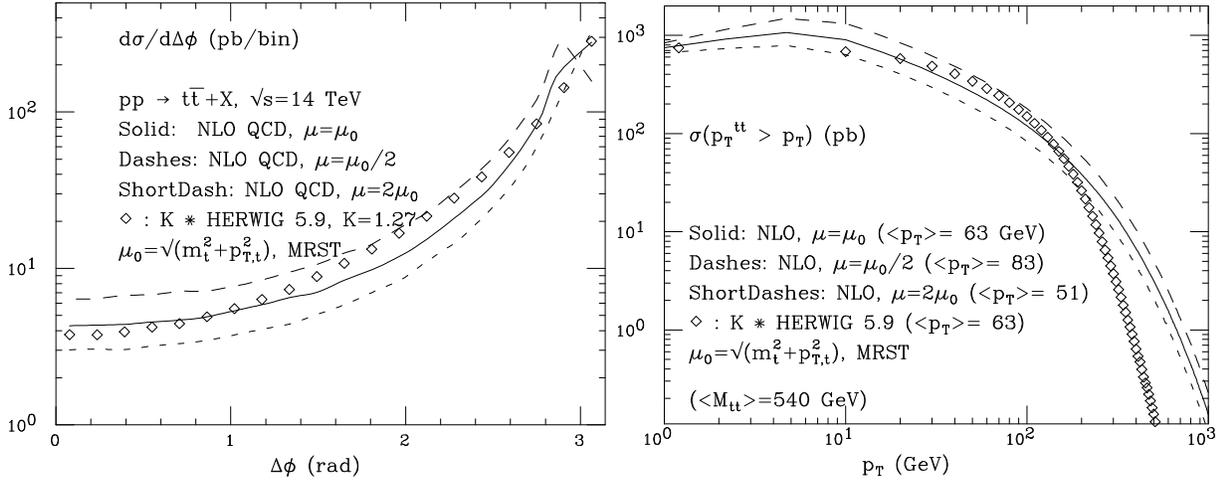
 
  \begin{center} 
   \centerline{ 
    \includegraphics[width=0.5\textwidth,clip]{phicomp.eps} \hfil 
    \includegraphics[width=0.5\textwidth,clip]{jptcomp.eps} } 
  \vskip -0.5cm 
    \caption{Left: azimuthal correlation between the $t$ and $\bar 
   t$. Right: integrated transverse momentum spectrum of the top quark 
   pair. Continuous lines correspond to the parton-level NLO 
   calculation, for different scale choices; the plots correspond to 
   the result of the \herw\  Monte Carlo.} 
    \label{fig:phicomp} 
  \end{center} 
\end{figure}

Contrary to the case of inclusive $\pt$ and $\mtt$ spectra, other 
kinematical distributions show large differences when comparing NLO 
and Monte Carlo results~\cite{Frixione:1995fj}. This is the case of 
distributions which are trivial at LO, and which are sensitive to 
Sudakov-like effects, such as the azimuthal correlations or the 
spectrum of the $\ttbar$ pair transverse momentum $\ptpair$. These two 
distributions are shown in the two plots of Fig.~\ref{fig:phicomp}. 
Notice that the scale uncertainty at NLO is larger for these 
distributions than for previous inclusive quantities. These 
kinematical quantities are in fact trivial at ${\cal O}(\astwo)$ 
(proportional to $\delta$-functions), and their evaluation at ${\cal 
O}(\asthree)$ is therefore not a true NLO prediction. 
The regions $\ptpair \to 0$ and $\dphi \to \pi$ are sensitive to 
multiple soft-gluon emission, and the differences between the NLO 
calculation (which only accounts for the emission of one gluon) and 
the Monte Carlo prediction (which includes the multi-gluon emission) 
is large. The region $\ptpair \gg \mt$ is vice-versa sensitive to the 
emission of individual hard gluons, a process which is more accurately 
accounted for by the full $\calO(\asthree)$ matrix elements included 
in the NLO calculation than by the Monte Carlo approach. 
Notice that the average value of $\ptpair$ is quite 
large, above 50~GeV. This is reasonable, as it is of the order of 
$\as$ times the average value of the hardness of the process ($\langle 
\mtt \rangle \sim 540$~GeV). It is found that this large transverse 
momentum is compensated by the emission of a jet recoiling 
against the top pair, with a smaller fraction of events where the 
$\ptpair$ comes from  emission of hard gluons from 
the final state top quarks. 
The large-$\ptpair$ discrepancy observed in Fig.~\ref{fig:phicomp} 
should be eliminated once the matrix element corrections to top 
production will be incorporated in \herw\ $\!$, along the lines of the work 
done for Drell-Yan production in~\cite{Corcella:1999tb}. 
 
\begin{figure}
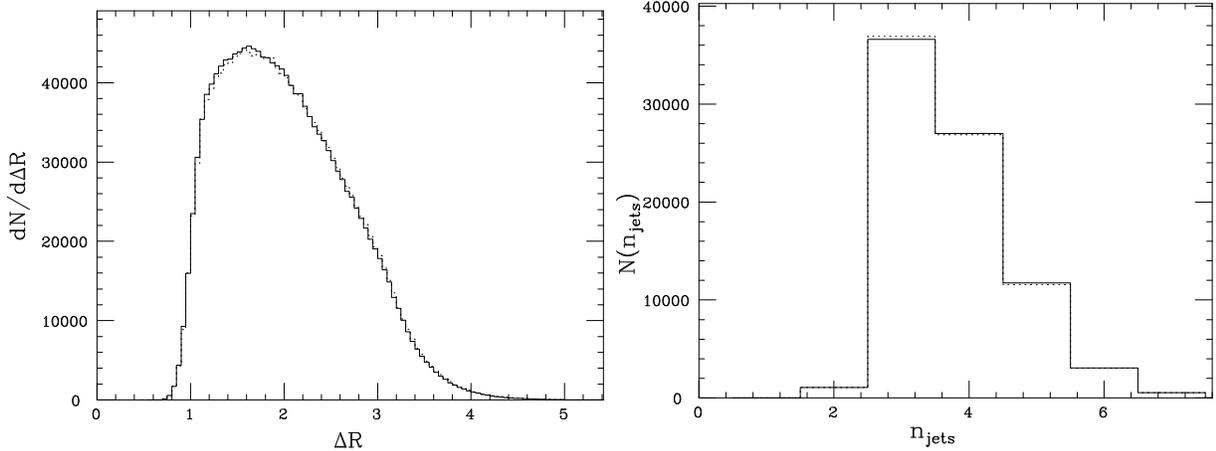
 \centerline{ 
    \includegraphics[width=0.5\textwidth,clip]{hw61dr.eps} \hfil 
    \includegraphics[width=0.5\textwidth,clip]{hw61nj.eps} } 
\vskip -0.5cm 
   \caption{Left: Distributions of the minimum invariant opening angle 
   $\Delta R$ among the three hardest jets, with (\herw\  6.1, 
   solid line) and without (6.0, dotted) matrix element corrections in 
   the $t\to bW$ decay. Right: jet multiplicity distributions.} 
   \label{fig:hw61dr} 
\end{figure}

Emission of extra jets is also expected from the evolution of the 
decay products of the top quarks ($b$'s, as well as the jets from the 
hadronic $W$ decays). Gluon radiation off the decay products is 
included in the shower Monte Carlo calculations. In the case of the 
latest version of \herw\ (v6.1)~\cite{Marchesini:1992ch}, the emission 
of the hardest gluon from the $b$ quarks is evaluated using the exact 
matrix elements~\cite{Corcella:1998rs}. This improvement, in addition 
to a few bug fixes, resolve the discrepancies uncovered  
in~\cite{Orr:1997pe} between an exact parton level calculation and 
previous versions of \herw.  The matrix-element corrections do not 
alter significantly most of the inclusive jet observables. As 
examples, we show in Fig.~\ref{fig:hw61dr} the $\deltar$ and the jet 
multiplicity distributions for events where both $W$'s decay 
leptonically. More details can be found in~\cite{Corcella00}. 
Jets are defined using the $k_T$ algorithm~\cite{Catani:1993hr}, with 
radius parameter $R=1$.  As can be seen, the impact of the exact 
matrix element corrections is limited, mostly because the extra-jet 
emission is dominated by initial-state radiation. 
 
The impact on quantities which more directly affect the determination 
of the top mass remains to be fully evaluated. Given the large rate of 
high-$\et$ jet emissions, their proper description will be a 
fundamental ingredient in the accurate reconstruction of the top 
quarks from the final state jets, and in the determination of the top 
quark mass.  A complete analysis will only be possible once the matrix 
element corrections to the $\ttbar$ production will be incorporated in 
the Monte Carlos. Work in this direction is in progress (G.~Corcella 
and M.H.~Seymour). 
 
\subsection{Non-QCD radiative corrections to $\bf t\bar{t}$  
production} 

The production and decay of top quarks at hadron colliders is a   
promising environment for the detection and study of loop induced SUSY 
effects:~at the parton level there is a large center of mass energy 
$\hat s$ available and owing to its large mass, the top quark strongly 
couples to the (virtual) Higgs bosons, a coupling  
which is additionally enhanced 
in SUSY models. Moreover, it might turn out that SUSY loop effects in 
connection with top and Higgs boson interactions less rapidly decouple 
than the ones to gauge boson observables. 
 
To fully explore the potential of precision top physics at the LHC and 
at the Tevatron~\cite{runtwoworkshop} to detect, discriminate and 
constrain new physics, the theoretical predictions for top quark 
observables need to be calculated beyond leading order (LO) in 
perturbation theory.  Here we will concentrate on the effects of 
non-QCD radiative corrections to the production processes $gg 
\to t\bar t$ and $q \bar q \to t \bar t$, including supersymmetric  
corrections.  When searching for quantum 
signatures of new physics also the SM loop effects have to be under 
control.  The present SM prediction for $t \bar t$ observables includes 
the QCD corrections as discussed above and the EW 
one-loop contributions to the QCD $t \bar t$ production 
processes~\cite{Beenakker:1994yr,Kao:1997bs,Stange:1993td}. The latter 
modify the $g t \bar t (q\bar q g)$ vertex by the virtual presence of 
the EW gauge bosons and the SM Higgs boson.  At the parton level, the 
EW radiative corrections can enhance the LO cross sections by up to 
$\approx 30 \%$ close to the threshold $\sqrt{\hat s} 
\stackrel{>}{\sim} 2 m_t$ when the SM Higgs boson is light and 
reduce the LO cross sections with increasing $\hat s$ by up to the 
same order of magnitude.  After convoluting with the parton 
distribution functions (PDF's), however, they only reduce the LO 
production cross section $\sigma(pp \to t \bar tX)$ at the LHC by a few 
percent~\cite{Beenakker:1994yr}: up to $2.5 (1.8) \%$ for the following 
cuts on the transverse momentum $p_T$ and the pseudo rapidity 
$\eta$ of the top quark:~$p_T>100(20)$ GeV and $|\eta|<2.5$. 
 
So far, the studies of loop induced effects of BSM physics in 
$t \bar t$ production at hadron colliders include the following 
calculations: 
 
{\bf The $\bf {\cal O}(\alpha)$ \bf corrections} 
within a general two Higgs doublet 
model (G2HDM) (=SM with two Higgs doublets but without imposing SUSY 
constraints) to $q \bar q \to t \bar 
t$~\cite{Zhou:1997dx,Hollik:1998hm} and $gg \to t \bar 
t$~\cite{Hollik:1998hm}. 
In addition to the contribution of the $W$ and $Z$, 
the $gt \bar t (q\bar qg)$ vertex is modified by the virtual presence 
of five physical Higgs bosons which appear in any G2HDM after 
spontaneous symmetry breaking:~$H^0,h^0,A^0,H^{\pm}$.  Thus, the G2HDM 
predictions for $t \bar t$ observables depend on their masses and on 
two mixing angles, $\beta$ and $\alpha$.  The G2HDM radiative 
corrections are especially large for light Higgs bosons and for very 
small ($<1$) and very large values of $\tan\beta$ due to the enhanced 
Yukawa-like couplings of the top quark to the (virtual) Higgs bosons. 
Moreover, there is a possible source for large corrections due to a 
threshold effect in the renormalised top quark self-energy, i.e. when 
$m_t \approx M_{H^{\pm}} +m_b$.  In~\cite{Hollik:1998hm} the 
$s$-channel Higgs exchange diagrams in the gluon fusion subprocess, 
$gg \to h^0,H^0\to t \bar t$, had been included. For this workshop 
we also considered the $gg \to A^0 \to t \bar t$ 
contribution~\cite{Wackeroth:mc2000}.  A study of the $s$-channel  
Higgs exchange diagrams alone, 
can be found in~\cite{Gaemers:1984} ($H^0$) 
and~\cite{Dicus:1994bm,Bernreuther:1998gs} ($H^0$ and $A^0$). 
They are of particular 
interest, since they can cause a peak-dip structure in the invariant 
$t \bar t$ mass distribution for heavy Higgs bosons, $M_{H^0,A^0} > 2 
m_t$, when interfered with the LO QCD $t\bar t$ production 
processes. 
 
{\bf The SUSY EW ${\cal O}(\alpha)$ corrections} within the MSSM to $q \bar 
q \to t \bar 
t$~\cite{Hollik:1998hm,Yang:1995hq,Li:1997ae,Kim:1996nz} 
and $gg \to t \bar t$~\cite{Hollik:1998hm,Zhou:1998dc}.   
In~\cite{Hollik:1998hm} also the squark loop contribution to the $gg 
\to h^0,H^0$ production process in the $s$ channel Higgs exchange 
diagrams has been taken into account. 
The SUSY EW corrections comprise the contributions of the 
supersymmetric Higgs sector, and the genuine SUSY contributions due to 
the virtual presence of two charginos $\tilde \chi^{\pm}$, four 
neutralinos $\tilde \chi^0$, two top squarks $\tilde t_{L,R}$ and two bottom 
squarks $\tilde b_{L,R}$.  The MSSM input parameters can be fixed in 
such a way that the $t \bar t$ observables including MSSM loop 
corrections depend on a relatively small set of 
parameters~\cite{Hollik:1998hm}:~$\tan\beta,M_{A^0},m_{\tilde 
t_1},m_{\tilde b_L},\Phi_{\tilde t},\mu,M_2$, where LR mixing is 
considered only in the top squark sector, parametrized by the mixing 
angle $\Phi_{\tilde t}$. $m_{\tilde t_1}$ and $m_{\tilde 
b_L}=m_{\tilde b_R}$ denote the mass of the lighter top squark and the 
bottom squark, respectively.  The effects of the supersymmetric Higgs 
sector tend to be less pronounced than the ones of the G2HDM:~since 
supersymmetry tightly correlates the parameters of the Higgs 
potential, the freedom to choose that set of parameters which yield 
the maximum effect is rather limited.  On the other hand, they can be 
enhanced by the genuine SUSY contribution depending on the choice of 
the MSSM input parameters. The SUSY EW corrections can become large 
close to the threshold for the top quark decay $t \to \tilde t+\tilde 
\chi^0$.  They are enhanced for very small ($<1$) and very large 
values of $\tan\beta$ and when there exists a light top squark 
($m_{\tilde t_1}\approx 100$ GeV). 
 
{\bf The SUSY QCD $\bf {\cal O}(\alpha_s)$ \bf corrections}  
to $q \bar q \to t \bar 
t$~\cite{Kim:1996nz,Li:1996jf, 
Sullivan:1997ry,Li:1998gh,Wackeroth:1998wm} and 
$gg \to t \bar t$~\cite{Zhou:1997fw}.  So far, there are 
only results available separately for the $q \bar q \to t \bar t$ 
(Tevatron) and the $gg \to t \bar t$ (LHC) production processes. The 
combination of both is work in progress and will be presented 
in~\cite{Wackeroth:2000sqcd}. 
The SUSY QCD contribution describes the modification of the $gt \bar t 
(q \bar qg)$ vertex and the gluon vacuum polarisation due to the 
virtual presence of gluinos and squarks.  Thus, additionally to the 
dependence on squarks masses (and on mixing angles if LR mixing is 
considered) the SUSY QCD corrections introduce a sensitivity of $t 
\bar t$ observables on the gluino mass $m_{\tilde g}$. As expected, 
the effects are the largest the lighter the gluino and/or the squarks. 
Again, there are possible enhancements due to threshold effects, for 
instance close to the anomalous threshold $m_t^2= m_{\tilde 
g}^2+m_{\tilde t_1}^2$. 
 
The $t\bar t$ observables under investigation so far comprise the 
total $t \bar t$ production cross section $\sigma_{t \bar t}$, the 
invariant $t \bar t$ mass distribution $d\sigma/ d M_{t \bar t}$ and 
parity violating asymmetries ${\cal A}_{LR}$ in the production of left 
and right handed top quark pairs. At present, the numerical discussion 
is concentrated on the impact of BSM quantum effects on $t \bar t$ 
observables in $p \,p\hskip-7pt\hbox{$^{^{(\!-\!)}}$} \to t \bar t X$. 
A parton level Monte Carlo program for $p 
\,p\hskip-7pt\hbox{$^{^{(\!-\!)}}$} \to t \bar t \to W^+ W^- b \bar b 
\to (f_i \bar f_i')(f_j' \bar f_j) b \bar b$ is presently under 
construction~\cite{Wackeroth:mc2000}. This will allow a more 
realistic study of the sensitivity of a variety of kinematical 
distributions to SUSY quantum signatures in the $t \bar t$ production 
processes, for instance by taking into account detector effects. 
 
In the following we give an overview of the present status 
of BSM quantum effects in $t \bar t$ observables at the LHC: 
 
$\underline{\bf{\sigma_{t \bar t}:}}$  
In Table~\ref{tab:totrel} 
we provide the relative corrections for $\sigma_{t\bar t}$ at the LHC 
for different BSM physics scenarios. They reflect the 
typical maximum size of the radiative corrections within the 
models under consideration.  As already mentioned there are 
possible enhancements due to threshold effects, which can yield 
much larger relative corrections. However, they only arise for 
very specific choices of the MSSM input parameters. The SUSY EW 
one-loop corrections always reduce the LO production cross 
sections and range from SM values, to up to $\approx -5 \%$ for 
heavy squarks and up to $\approx -20\%$ close to $m_t=m_{\tilde 
t_1}+m_{\tilde \chi^0}$.  The SUSY QCD one-loop corrections, 
however, can either reduce or enhance $\sigma_{t\bar t}$.  The 
relative corrections are negative for small $m_{\tilde g}$ and 
increase with decreasing gluino and/or squark masses. They change 
sign when approaching the threshold for real sparticle production 
and reach a maximum at $m_{\tilde g} \approx 200$ GeV of about $+2 
\%$~\cite{Zhou:1997fw}.  Again, very large corrections 
arise in the vicinity of a threshold for real sparticle 
production, $m_t=m_{\tilde g}+m_{\tilde t_1}$.  The SUSY EW and 
QCD one-loop corrections, so far, have only been combined for the 
$q\bar q\to t \bar t$ production process and numerical results are 
provided for the Tevatron $p\bar p$ collider 
in~\cite{Kim:1996nz,Wackeroth:1998wm}. 
 To summarise, apart from exceptional regions in the MSSM 
parameter space, it will be difficult to detect SUSY through loop 
contributions to the $t \bar t$ production rate.  If light sparticles exist, 
they are most likely directly observed first. Then, the comparison 
of the precisely measured top production rate with the MSSM 
predictions will test the consistency of the model under consideration 
at quantum level and might yield additional information on the 
parameter space, for instance constraints on $\tan\beta$ and 
$\Phi_{\tilde t}$. 
 
\begin{table} 
\begin{center} 
\caption{The relative corrections to $pp \to t \bar t X$ at the 
LHC when only including SUSY QCD one-loop 
corrections~\cite{Zhou:1997fw} (with $p_T>$20 GeV, 
$|\eta|<2.5$) or only the EW one-loop corrections within the 
G2HDM and the MSSM~\cite{Hollik:1998hm} ($p_T>100$ GeV). For 
comparison the SM prediction is also listed.} \label{tab:totrel} 
\vspace*{0.1cm} 
\begin{tabular}{|c|c|c|c|c|}\hline 
               & SM ($M_H=100$ GeV)   & G2HDM & SUSY EW   & SUSY QCD \\ 
               \hline\hline 
$|\sigma^{NLO}_{t \bar t}-\sigma^{LO}_{t \bar t}|/\sigma^{LO}_{t \bar t}$ 
               & $2.5\%$        & $\le 4 \%$  & $ \le 10 \%$ & $ \le 4 \%$    \\ \hline 
\end{tabular} 
\end{center} 
\end{table} 
 
\begin{figure} 
\centerline{ 
  \includegraphics[width=0.48\textwidth,clip]{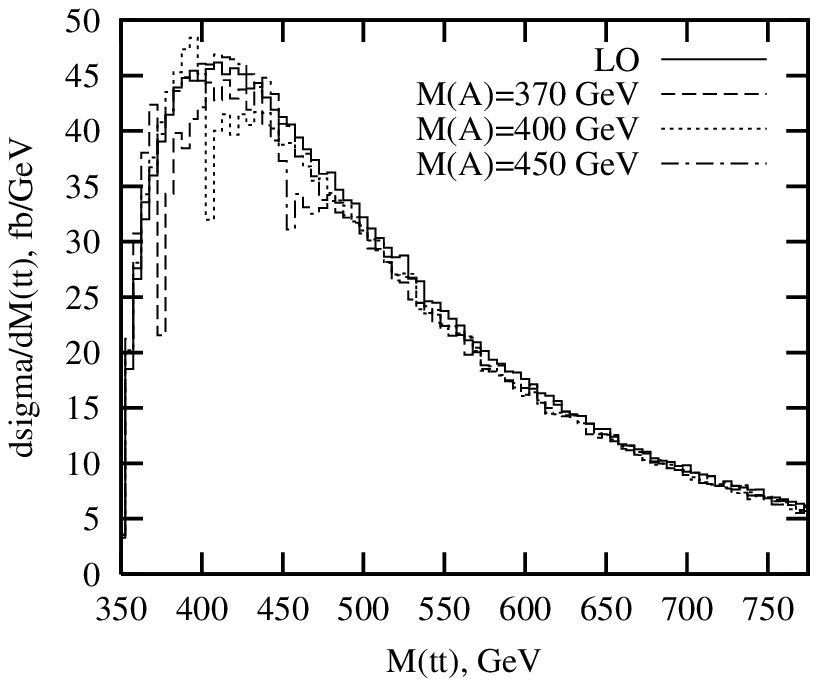} \hfill 
  \includegraphics[width=0.48\textwidth,clip]{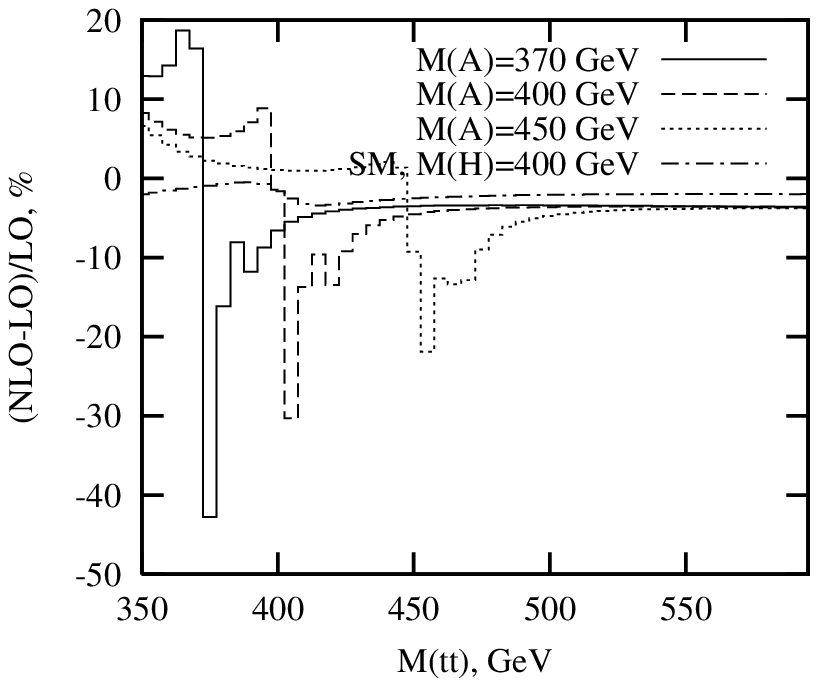}} 
\vskip -0.5cm 
\caption{The LO and NLO invariant mass distributions 
  $d\sigma/dM_{t\bar t}$ (left) and the relative corrections (right) 
  to the reaction 
  $pp\to t \bar t \to (\nu_e e^+) (d \bar u) b \bar b$ at the LHC 
  with $p_T(e,q)>20$ GeV, $E_T(\nu)>20$ GeV and $|\eta(e,q)|<3.2$ for 
  different values of $M_{A^0}$ ($\tan\beta=1.6$, $\Phi_{\tilde 
    t}=\pi/8$, $m_{\tilde t_1}=160$ GeV, $m_{\tilde b_L}=500$ GeV, 
  $\mu=120$ GeV and $M_2=3|\mu|$.).  For comparison the relative 
  correction when only taking into account the EW SM one-loop 
  contribution is also shown.  The CTEQ3LO set of PDF's is used and 
  $m_t=174$ GeV.} 
\label{fig:mtt} 
\end{figure}

$\underline{\bf{ d \sigma/dM_{t \bar t}:}}$  
More promising are the distributions of kinematic variables.  Here we 
will concentrate on the impact of SUSY quantum signatures on the 
invariant $t \bar t$ mass distribution. Results for the effects of EW 
one-loop corrections within the G2HDM and the MSSM on $d \sigma/d M_{t 
  \bar t}$ at the LHC are provided in~\cite{Hollik:1998hm}. So far, 
the impact of the SUSY QCD one-loop contribution on $d\sigma/M_{t \bar 
  t}$ has only been discussed for the Tevatron $p \bar p$ 
collider~\cite{Sullivan:1997ry}, where it turned out that they can 
significantly change the normalisation and distort the shape of 
$d\sigma/d M_{t\bar t}$. 
As already mentioned, there is the possibility for an interesting 
peak-dip structure due to a heavy neutral Higgs resonance in $gg \to t 
\bar t$ within two Higgs doublet models.  The potential of the LHC for 
the observation of such resonances has been studied 
in~\cite{Dicus:1994bm,Bernreuther:1998qv}.  In Section~\ref{sec:prodmeas} the 
results of an ATLAS analysis of the observability of the $H/A \to t 
\bar t$ channel for different luminosities are presented.  In 
Fig.~\ref{fig:mtt} we show preliminary results for the invariant $t 
\bar t$ mass distribution to $pp\to t \bar t \to W^+W^- b\bar b \to 
(\nu_e e^+)(d\bar u)b \bar b$ at the LHC when including MSSM EW 
one-loop corrections~\cite{Wackeroth:mc2000}.   
When $M_{A^0}>2 m_t$ the $gg\to H^0,A^0 \to t \bar t$ 
contributions can cause an excess of $t \bar t$ events at 
$M_{t\bar t}$ slightly below $M_{A^0}$, when the Higgs bosons are 
not too heavy, and a dip in the distribution slightly above $M_{t 
\bar t}=M_{A^0}$. For the choice of MSSM parameters used in 
Fig.~\ref{fig:mtt} the peak vanishes for $M_{A^0}> 400$ GeV and 
only a deficiency of events survives which decreases rapidly for 
increasing $M_{A^0}$. These effects can be enhanced when the SUSY 
QCD contributions are taken into account. 
 
\begin{figure} 
\centerline{ 
  \includegraphics[width=0.48\textwidth,clip]{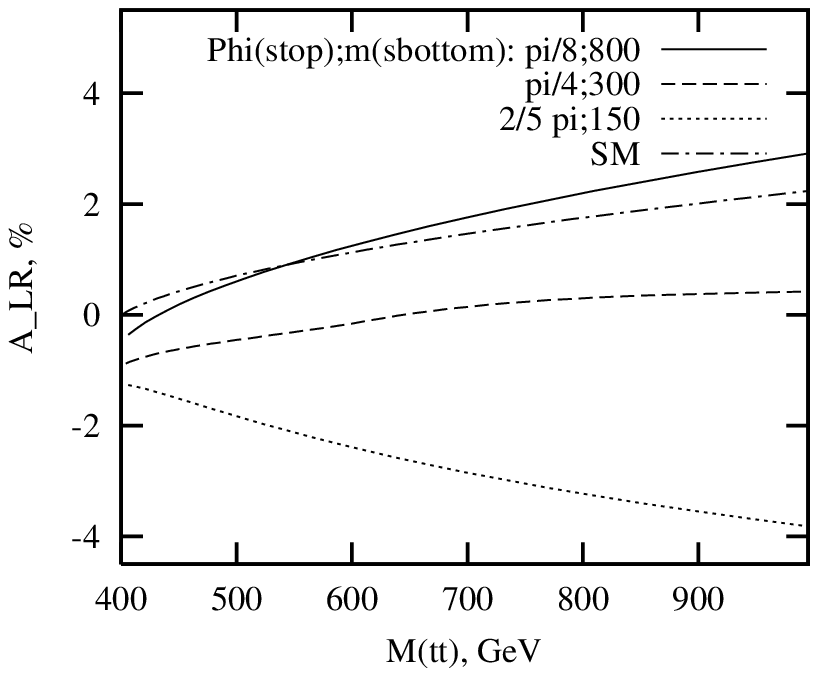} \hfill 
  \includegraphics[width=0.48\textwidth,clip]{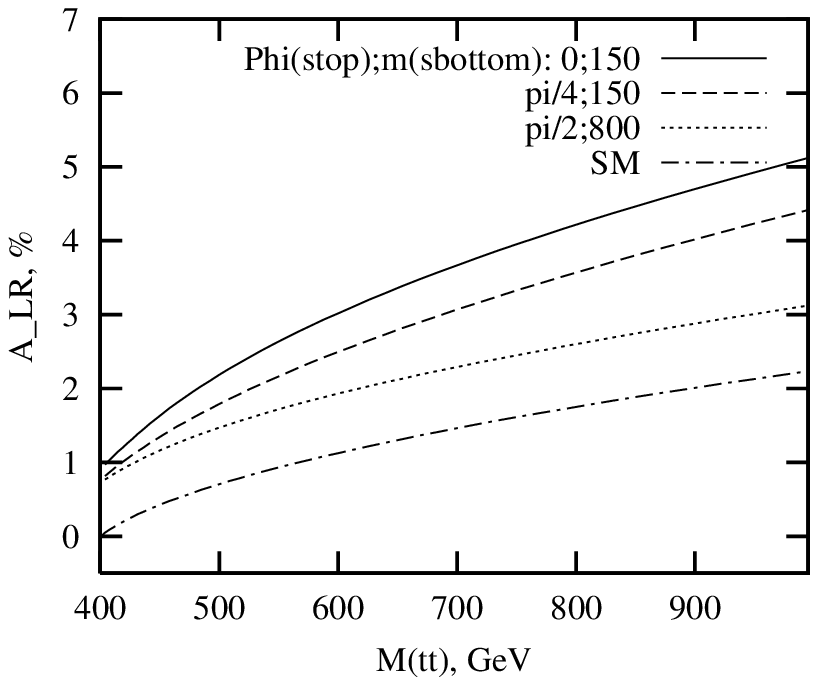}} 
\vskip -0.5cm 
\caption{The left-right asymmetry ${\cal A}_{LR}$ 
  in the invariant $t \bar t$ mass distribution to $pp \to t \bar t X$ 
  at the LHC with $p_T>100$ GeV for different values of 
  $\Phi_{\tilde t}$ and $m_{\tilde b_L}$ and for two extreme choices 
  of $\tan\beta$: $\tan\beta=0.7$ (left) and $\tan\beta=50$ (right) 
  (with $m_{\tilde t_1}=90$ GeV, $M_{H^{\pm}}$=110 GeV, $\mu=120$ GeV 
  and $M_2=3|\mu|$).} 
\label{fig:alrmtt} 
\end{figure}

$\underline{\bf{{\cal A}_{LR}:}}$ 
Parity violating asymmetries in the distribution of left and 
right-handed top quark pairs at hadron colliders directly probe 
the parity non-conserving parts of the non-QCD one-loop corrections to the 
$t \bar t$ production processes within the model under 
consideration and have been studied at the 
Tevatron~\cite{Li:1997ae, 
Li:1998gh,Kao:1995rn,Kao:1997bs,Sullivan:1997ry,Wackeroth:1998wm} 
and at the LHC~\cite{Kao:1999kj}.  In Fig.~\ref{fig:alrmtt} we 
show the left-right asymmetries ${\cal A}_{LR}$ in the invariant 
mass distribution of (longitudinally) polarised top quark pairs in 
$pp \to t_{L,R}\bar t_{L,R} X$, induced by SM and MSSM EW one-loop 
corrections~\cite{Kao:1999kj}.  The parity violating asymmetry 
within the MSSM results from the interplay of the supersymmetric 
Higgs sector ($M_{H^\pm}$) and the genuine SUSY contributions 
($\tilde \chi^{\pm},\tilde \chi^0$).  The contribution from the 
charged Higgs boson can either be enhanced or diminished depending 
on the values of $m_{\tilde t_1}$ and $\Phi_{\tilde t}$.  Within 
the G2HDM the loop-induced asymmetries are most pronounced for a 
light charged Higgs boson and very small and very large values of 
$\tan\beta$.  At the LHC, the G2HDM and MSSM EW one-loop 
corrections induce asymmetries in the total production rate of 
left and right-handed top quark pairs of up to about $2.5 \%$ and 
$3.2 \%$, respectively, and thus can be considerably larger than 
the SM expectation (SM: $1.2 \%$).  When the squarks are 
non-degenerate in mass also the SUSY QCD one-loop corrections 
induce parity violating asymmetries in strong $t \bar t$ 
production.  So far, there exist only studies for the 
Tevatron~\cite{Li:1998gh,Sullivan:1997ry,Wackeroth:1998wm}. 
 
\subsection{Measurement of $\bf t\bar{t}$ production 
  properties}\label{sec:prodmeas} 
According to the SM, the top quark decays almost 
exclusively via $t \rightarrow  Wb$. The final state topology 
of $t \bar t$ events then depends on the decay modes 
of the $W$ bosons. In approximately 65.5\% of $t\bar t$ events, 
both $W$ bosons decay hadronically via $W \rightarrow jj$, 
or at least one $W$ decays via $W \rightarrow \tau \nu$. 
These events are difficult to extract cleanly above the large QCD 
multi-jet background, and are for the most part not considered further. 
Instead, the analyses presented here concentrate on leptonic $t \bar t$ 
events, where at least one of the $W$ bosons decays 
via $W \rightarrow \ell \nu$ 
($\ell=e,\mu$). The lepton plus 
large $\etmiss$, due to the escaping neutrino(s), provide a large 
suppression against multi-jet backgrounds. The leptonic events, which 
account for approximately 34.5\% of all $t \bar t$ events, can be subdivided 
into a ``single lepton plus jets'' sample and a ``di-lepton'' 
sample, depending on whether one or both W bosons decay leptonically. 
As discussed below, the selection cuts and background issues are 
quite different for the various final state topologies. 
 
An important experimental tool for selecting clean top quark samples 
is the ability to identify $b$-jets. Techniques for $b$-tagging, 
using secondary vertices, semi-leptonic $b$-decays, and other 
characteristics of $b$-jets, have been extensively studied. 
Both ATLAS and CMS expect to achieve, for a $b$-tagging 
efficiency of 60\%, a rejection of at least 100 against prompt 
jets (i.e. jets containing no long-lived 
particles) at low luminosity. At high luminosity, a rejection factor 
of around 100 can be obtained with a somewhat reduced 
$b$-tagging efficiency of typically 50\%. 
 
 
All the results presented in this section are obtained using for the 
signal the \pyth~ Monte Carlo program. Most background processes have 
also been generated with \pyth, with the exception of $Wb \bar b$, 
which has been produced using the \herw~ 
implementation~\cite{Mangano:1993kp} of the exact massive 
matrix-element calculation. 
 
\subsubsection{Single lepton plus jets sample} 
The single lepton plus jets topology, 
$t \bar t \rightarrow WWb\bar b \rightarrow (\ell\nu)(jj)b\bar b$ arises in 
$2 \times 2/9 \times 6/9 \approx 29.6$\% of all $t \bar t$ events. 
One expects, therefore, production of almost 2.5 million single 
lepton plus jet events for an integrated luminosity of 10 fb$^{-1}$, 
corresponding to one year of LHC running at $10^{33}$ cm$^{-2}$ s$^{-1}$. 
The presence of a high $p_T$ isolated lepton provides an 
efficient trigger. The lepton and the high value of $\etmiss$ give a 
large suppression of backgrounds from QCD multi-jets and $b \bar b$ 
production. 
 
For the single lepton plus jets sample, it is possible to fully 
reconstruct the final state. The four-momentum 
of the missing neutrino can be reconstructed by setting $M^\nu$ = 0, 
assigning $E_T^\nu$ = $\etmiss$, and calculating $p_z^\nu$, 
with a quadratic ambiguity, by applying the constraint that 
$M^{\ell\nu}$ = $M_W$. 
 
An analysis by ATLAS~\cite{atlasphystdr} examined a 
typical set of selection cuts.  First, the presence of an isolated electron 
or muon with $p_T >$ 20 GeV and $|\eta | < 2.5$ was required, along with 
a value of $\etmiss >$  20 GeV. 
At least four jets with $p_T >$ 20 GeV were required, where 
the jets were reconstructed using a 
fixed cone algorithm with cone size of $\Delta R$ = 0.7. 
After cuts, the major sources of backgrounds were $W$+jet 
production with $W \rightarrow \ell \nu$ decay, 
and $Z$+jet events with $Z \rightarrow \ell^+\ell^-$. 
Potential backgrounds from $WW$, $WZ$, and 
$ZZ$ gauge boson pair production have also been considered, but are 
reduced to a negligible level after cuts. 
 
A clean sample of $t \bar t$ events was obtained using 
$b$-tagging. Requiring that at least one of the jets be tagged 
as a $b$-jet yielded a selection efficiency (not counting 
branching ratios) of 33.3\%. For 
an integrated luminosity of 10 fb$^{-1}$, this would correspond to 
a signal of 820,000 $t \bar t$ events.  The total background, 
dominated by $W$+jet production, leads to a 
signal-to-background ratio (S/B) of 18.6. 
Tighter cuts can be used 
to select a particularly clean sample. Examples of this will be given 
in Section~\ref{sec:mass} 
 
\subsubsection{Di-lepton sample} 
Di-lepton events, where each W decays leptonically, provide a 
particularly clean sample of $t \bar t$ events, although the product of 
branching ratios is small, $2/9 \times 2/9 \approx$ 4.9\%. 
With this branching ratio, one expects the production of over 
400,000 di-lepton events for an integrated luminosity of 10 fb$^{-1}$. 
 
The presence of two high $p_T$ isolated leptons allows these events 
to be triggered efficiently. 
Backgrounds arise from Drell-Yan processes associated with jets, 
$Z \rightarrow \tau^+\tau^-$ associated with jets, 
$WW$+jets, and $b \bar b$ production. Typical 
selection criteria~\cite{atlasphystdr,atlasphysnote99044} 
require two opposite-sign leptons within 
$|\eta | < 2.5$, with $p_T >$ 35 and 25 GeV respectively, and with 
$\etmiss >$ 40 GeV. For the case of like-flavour leptons ($e^+e^-$ and 
$\mu^+\mu^-$), an additional cut $|M^{\ell\ell} - M^{Z}| > 10$ 
GeV was made on the 
di-lepton mass to remove $Z$ candidates. 
Requiring, in addition, at least 
two jets with $p_T >$ 25 GeV produced a signal of 80,000 events 
for 10 fb$^{-1}$, with S/B around 10.  Introducing the requirement 
that at least one jet be tagged as a $b$-jet reduced the signal to 
about 58,000 events while improving the purity to S/B $\approx $ 50. 
 
\subsubsection{Multi-jet sample} 
 
The largest sample of $t \bar t$ events consists of the topology 
$t \bar t \rightarrow WWb\bar b \rightarrow 
 (jj)(jj)b\bar b$. The product of branching 
ratios of $6/9 \times 6/9 \approx$ 44.4\% implies production of 3.7 
million multi-jet events for an integrated luminosity of 10 fb$^{-1}$. 
However, these events suffer from a very large background from QCD 
multi-jet events. In addition, the all-jet final state poses 
difficulties for triggering. For example, the trigger menus 
examined so far by ATLAS~\cite{atlasphystdr} 
consider multi-jet trigger thresholds only up to 
four jets, for which a jet $E_T$ threshold of 55 GeV is applied at 
low luminosity. Further study is required to determine appropriate 
thresholds for a six-jet topology. 
 
 
At the Fermilab Tevatron Collider, both the CDF and D0 collaborations have 
 shown that it is possible to isolate a $t \bar t$ signal in this channel. 
 The CDF collaboration  has obtained a signal significance over background 
  of better than three standard deviations~\cite{Abe:1997rh} by applying 
simple 
   selection cuts and relying on the high $b$-tagging efficiency ($\simeq$ 
   46\%). To compensate for the less efficient $b$-tagging, the D0 
collaboration has  developed a more sophisticated event selection 
technique~\cite{Abbott:1998nn}. 
   Ten kinematic variables to separate signal and background were used in a 
    neural network, and the output was combined in a second network together 
     with three additional variables designed to best characterise 
     the $t \bar t$ 
     events.

ATLAS has made a very preliminary 
investigation~\cite{atlasphystdr,atlasphysnote99057} 
of a simple selection and reconstruction algorithm for attempting to extract 
the multi-jet $t \bar t$ signal from the background. 
Events were selected by requiring six or more jets with $p_T >$ 15 GeV, 
and with at least two of them tagged as $b$-jets. 
Jets were required to satisfy $|\eta | < 3$ ($|\eta | 
< 2.5$ for $b$-jet candidates). 
In addition, the scalar sum of the transverse momenta of the jets 
was required to be greater than 200 GeV. The $t \bar t$ signal efficiency 
for these cuts was 19.3\%, while only 0.29\% of the QCD 
multi-jet events survived. With this selection, and assuming a 
QCD multi-jet cross-section of 1.4 $\times 10^{-3}$ mb for 
$p_T$(hard process) $>$ 100 GeV, one obtains a signal-to-background ratio 
S/B $\approx $ 1/57. 
 
Reconstruction of the $t \bar t$ final state proceeded by first selecting 
di-jet pairs, from among those jets not tagged as $b$-jets, to 
form $W \rightarrow jj$ candidates. A $\chi^2_W$ was 
calculated from the deviations 
of the two $M_{jj}$ values from the known value of $M_W$. The combination 
which minimised the value of $\chi^2_W$ was  selected, and events with 
$\chi^2_W >$ 3.5 were rejected. For accepted events, the two $W$ candidates 
were then combined with $b$-tagged jets to form top and anti-top 
quark candidates, and a $\chi^2_t$ was calculated as the deviation from the 
condition that the top and anti-top masses are equal. Again, the 
combination with the lowest $\chi^2_t$  was selected, and events with 
$\chi^2_t >$ 7 were rejected. After this reconstruction procedure and 
cuts, the value of S/B improved to 1/8 within the mass window 
130-200 GeV.  Increasing the $p_T$ threshold for jets led to some 
further improvement; for example, requiring $p_T^j >$ 25 GeV yielded 
S/B = 1/6. 
 
The isolation of a top signal can be further improved in a number 
of ways, such as using a multivariate discriminant based on 
kinematic variables like aplanarity, sphericity or $\Delta R$(jet-jet), 
or restricting the analysis to a sample of high $p_T^{top}$ events. 
These techniques are undergoing further investigation, but it will 
be very difficult to reliably extract the signal from the background 
in this channel. In particular, the multi-jet rates and topologies 
suffer from very large uncertainties. 
 
\subsubsection{Measurement of the $t \bar t$ invariant mass spectrum} 
As discussed previously, properties of $t \bar t$ events provide 
important probes of both SM and BSM physics. For example, a 
heavy resonance decaying to $t \bar t$ might enhance the cross-section, and 
might produce a peak in the $M_{t\bar t}$ invariant mass spectrum. 
Deviations from the 
SM top quark branching ratios, due for example to a large rate of 
$t \rightarrow H^+b$, could lead to an apparent deficit in the 
$t \bar t$ cross-section measured with the assumption that 
BR($t \rightarrow Wb$) $\approx$ 1. 
 
Due to the very large samples of top quarks which will be produced 
at the LHC, measurements of the total cross-section 
$\sigma(t\bar t)$ will be limited by the uncertainty of the integrated 
luminosity determination, which is currently estimated to be 5\%-10\%. 
The cross-section relative to some other hard 
process, such as $Z$ production, should be measured more precisely. 
 
Concerning differential cross-sections, particular attention has 
thus far been paid by ATLAS~\cite{atlasphystdr} 
to measurement of the $M_{t \bar t}$ invariant mass spectrum. 
A number of theoretical models predict the existence of heavy resonances 
which decay to $t\bar t$. An example within the SM is the Higgs 
boson, which will decay to $t\bar t$ provided the decay is kinematically 
allowed. However, the strong coupling of the SM Higgs boson to the 
$W$ and $Z$ implies that the branching ratio to $t \bar t$ 
is never very large.  For example, for $M_H$ = 500 GeV, the SM Higgs 
natural width would be 63 GeV, and BR($H \rightarrow t \bar t$) 
$\approx 17$\%. The resulting value of $\sigma \times $BR for 
$H \rightarrow t\bar t$ 
in the SM is not sufficiently large to see a Higgs peak above the 
large background from continuum $t\bar t$ production. 
In the case of MSSM, however, if $M_{H,A} > 2 m_t$, then 
BR($H/A \rightarrow t\bar t$) $\approx $  100\% for $\tan \beta \approx 1$. 
For the case of scalar or pseudo-scalar Higgs resonances, it has been 
pointed out~\cite{Gaemers:1984,Dicus:1994bm} 
that interference can occur between the amplitude for the production 
of the resonance via $gg \rightarrow H/A \rightarrow t \bar t$ and the usual 
gluon fusion process $gg \rightarrow t \bar t$. The interference 
effects become stronger as the Higgs' mass and width increase, 
severely complicating attempts to extract a resonance signal. 
 
The possible existence of heavy resonances decaying to $t \bar t$ arises in 
technicolor models~\cite{Lane:1995} as well as other models 
of strong EW symmetry breaking~\cite{Hill:1994hs}. 
Recent variants of technicolor theories, such as 
Topcolor~\cite{Hill:1991}, 
posit new interactions which are specifically associated with the 
top quark, and could give rise to heavy particles decaying to 
$t \bar t$. Since $t\bar t$ production at the LHC is dominated by $gg$ 
fusion, colour octet resonances (``colourons'') could also be 
produced~\cite{Simmons:1997}. 
 
Because of the large variety of models and their parameters, ATLAS 
performed a study~\cite{atlasphystdr,atlasphysnote99038} 
of the sensitivity to a ``generic'' narrow resonance decaying to $t \bar t$. 
Events of the single lepton plus jets topology 
$t \bar t \rightarrow WWb\bar b \rightarrow (\ell\nu)(jj)b \bar b$ 
were selected by requiring $\etmiss > 20$ GeV, and the presence of an 
isolated electron or muon with $p_T > 20$ GeV and $|\eta| < 2.5$. In addition, 
it was required that there were between four and ten jets, each with 
$p_T > 20$ GeV and $|\eta | < 3.2$. At least one of the jets was required to 
be tagged as a $b$-jet. After these cuts, the background to the 
$t \bar t$ resonance search 
was dominated by continuum $t \bar t$ production. 
 
The momentum of the neutrino was reconstructed, as described previously, 
by setting $M_\nu$ = 0, assigning $E_T^\nu$ = $\etmiss$, and 
calculating $p_z^\nu$ (with a quadratic ambiguity) 
by applying the constraint that $M_{\ell\nu }$ = $M_W$. 
The hadronic $W \rightarrow jj$ 
decay was reconstructed by selecting pairs of jets 
from among those not tagged as $b$-jets. In cases where there were at 
least two $b$-tagged jets, candidates for $t \rightarrow Wb$ were formed by 
combining the $W$ candidates with each $b$-jet. In events 
with only a single $b$-tagged jet, this was assigned as one of the 
$b$-quarks and each of the still unassigned jets  was then considered as 
a candidate for the other $b$-quark. 
 
Among the many different possible jet-parton assignments, the 
combination was chosen that minimised the following $\chi^2$: 
$$\chi^2 = (M_{jjb} - m_t)^2/\sigma ^2(M_{jjb}) + (M_{\ell\nu b} 
 - m_t)^2/\sigma ^2(M_{\ell\nu b}) + (M_{jj} - M_W)^2/\sigma 
 ^2(M_{jj})$$ 
Events were rejected if either $M_{\ell\nu b}$ or $M_{jjb}$ disagreed 
with the known value of $m_t$ by more than 30~GeV. 
 
\begin{figure}[t] 
\begin{center} 
\includegraphics[width=0.6\textwidth,clip]{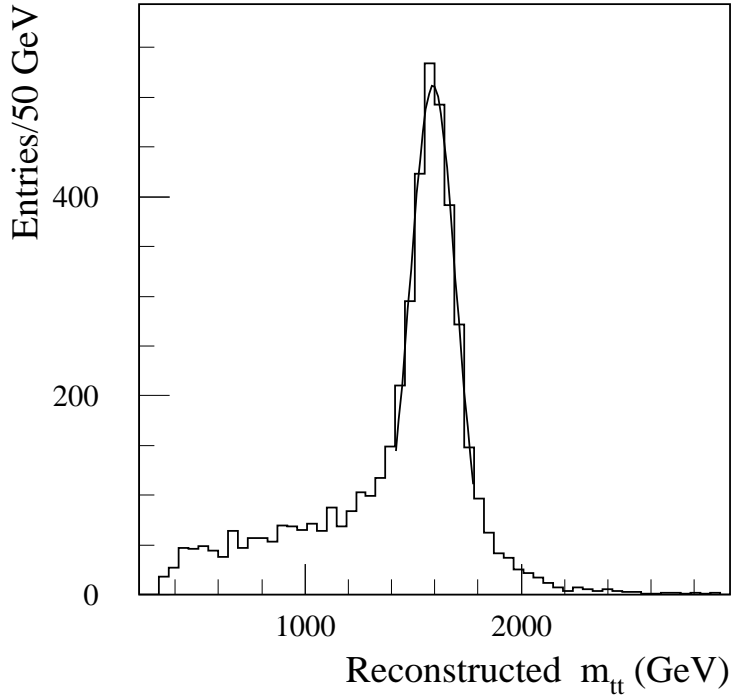} 
\vskip -0.5cm 
\caption{Measured $t\bar t$ invariant mass distribution for 
reconstruction of a narrow resonance of mass 1600 GeV decaying to 
$t\bar t$.} \label{fig:ttbarmass} 
\end{center} 
\end{figure}

For events passing the reconstruction procedure, the measured 
energies were rescaled, according to their resolution, to give the 
correct values of $M_W$ and $m_t$ for the appropriate 
combinations. This procedure improved the resolution of the mass 
reconstruction of the $t \bar t$ pair to $\sigma(M_{t\bar 
t})/M_{t\bar t} \approx 6.6$\%. As an example, 
Fig.~\ref{fig:ttbarmass} shows the reconstructed $M_{t \bar t}$ 
distribution for a narrow resonance of mass 1600 GeV. The width of 
the Gaussian core is well described by the resolution function 
described above. The size of the tails, which are dominated by 
incorrect jet-parton assignments, is such that approximately 65\% 
of the events are contained within $\pm 2\sigma $ of the peak. 
 
The reconstruction efficiency, not including branching ratios, for 
$t\bar t \rightarrow WWb\bar b \rightarrow (\ell\nu )(jj)b \bar b$ 
was about 20\% for a resonance of mass 400 GeV, 
decreasing gradually to about 15\% for $M_{t \bar t}$ = 2 TeV. 
 
\begin{figure} 
\begin{center} 
\includegraphics[width=0.6\textwidth,clip]{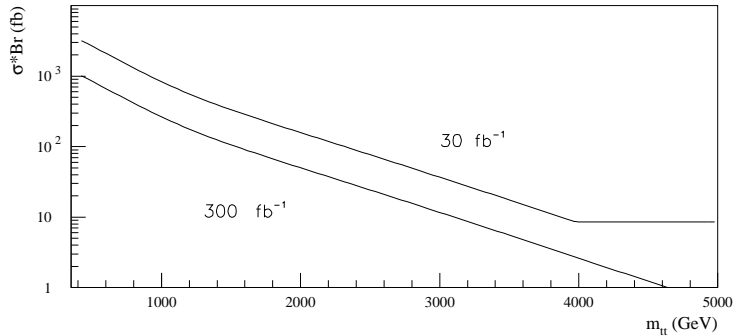} 
\vskip -0.5cm 
\caption{Value of $\sigma \times $ BR required for a 5$\sigma $ 
discovery potential for a narrow resonance decaying to $t\bar t$, 
as a function of $M_{t\bar t}$, and for an integrated luminosity 
of either 30 or 300 fb$^{-1}$.} 
    \label{fig:resnlimit} 
\end{center} 
\end{figure}

For a narrow resonance $X$ decaying to $t \bar t$, Fig.~\ref{fig:resnlimit} 
 shows the required $\sigma \times $ BR($X \rightarrow t\bar t$) for 
discovery of the resonance. The criterion 
used to define the discovery potential was observation within a $\pm 2 \sigma$ 
mass window of a signal above the $t\bar t$ continuum background, where the 
required signal must have a statistical significance of at least 5$\sigma $ 
and must contain at least ten events. Results are shown versus $M_X$ for 
integrated luminosities of 30 fb$^{-1}$ and 300 fb$^{-1}$. For example, with 
30 fb$^{-1}$, a 500 GeV resonance could be discovered provided its 
$\sigma \times $ BR 
is at least 2560 fb. This value decreases to 830 fb for $M_X$ = 1~TeV, 
and to 160 fb for $M_X$ = 2 TeV. The corresponding values for an 
integrated luminosity of 300 fb$^{-1}$ are 835 fb, 265 fb, and 50 fb for 
resonances masses $M_X$ = 500 GeV, 1 TeV, and 2 TeV, respectively. 
 
\begin{figure} 
  \begin{center} 
    \includegraphics[width=0.4\textwidth,clip]{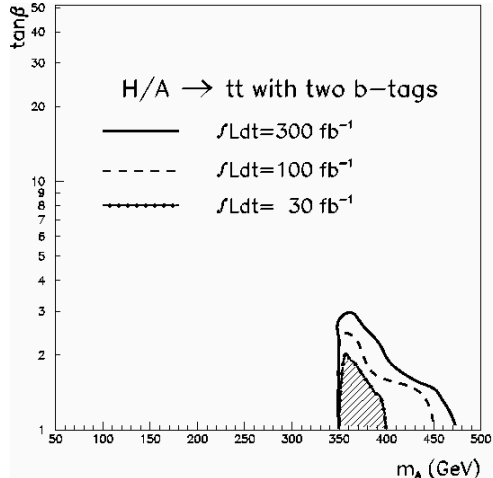} 
\vskip -0.5cm 
    \caption{For various integrated luminosities, 
      5$\sigma $ discovery contours in the MSSM ($M^A$, $\tan \beta$) 
      plane for the channel $H,A \rightarrow t \bar t$.} 
    \label{fig:higgsttbar}  
  \end{center}  
\end{figure}

Once predictions from models exist for the mass, natural width, and 
$\sigma \times $ BR for a specific resonance, the results in 
Fig.~\ref{fig:resnlimit} can be 
used to determine the sensitivity and discovery potential for those 
models. As discussed above, for the case of scalar or pseudo-scalar 
 Higgs resonances, extra care must be taken due to possible interference effects. While such 
effects are small for the case of a narrow resonance, they can be 
significant once the finite widths of heavy resonances are taken 
into account. For example, ATLAS has performed an 
analysis~\cite{atlasphystdr,atlasphysnote99016} of the decays 
$H/A \rightarrow t \bar t$ in MSSM with $\tan \beta = 1.5$ and 
$M_{H,A} > 2 m_t$.  Assuming the $t \bar t$ continuum 
background is well known, a combined $H+A$ signal would be visible 
for Higgs masses in the range of about 370 - 450 GeV. However, the 
interference effects produce an effective suppression of the combined 
$H+A$ production rates of about 30\% for $M_{H,A} = 370$ GeV, increasing 
to 70\% for masses of 450 GeV, essentially eliminating the 
possibility to extract a signal for higher Higgs masses, and 
thereby severely limiting the MSSM parameter space for which 
this channel has discovery potential (see Fig.~\ref{fig:higgsttbar}). 
 
\section{TOP QUARK MASS\protect\footnote{Section coordinators: M.~Beneke, 
    M.L.~Mangano, I.~Efthymiopoulos (ATLAS), P.~Grenier (ATLAS), 
    A.~Kharchilava (CMS).}} 
\label{sec:mass} 
As discussed in Section~\ref{sec:weiglein}  
one of the main motivations for top physics at the LHC 
is an accurate measurement of the top mass. 
Currently the best Tevatron single-experiment results on $\mt$ 
are obtained with the lepton plus jets final states. These 
yield: \mt = 175.9 $\pm$ 4.8~(stat.) $\pm$ 5.3~(syst.) 
(CDF)~\cite{Abe:1998vq} and 173.3 $\pm$ 5.6~(stat.) $\pm$ 5.5~(syst.) 
(D\O)~\cite{Abachi:1997jv}. The systematic errors in both measurements 
are largely dominated by the uncertainty on the jet energy scale 
which amounts to 4.4~GeV and 4~GeV for CDF and D\O , 
respectively. On the other hand, the systematic errors in the di-lepton 
channels are somewhat less, but the statistical errors are 
significantly larger, by a factor of $\gsim$~2, as compared to the 
lepton plus jets final states. Future runs of the Tevatron with 
an about 20-fold increase in statistics promise a measurement of the 
top mass with an accuracy of up to $\sim$~3~GeV~\cite{tev:future}; 
in the lepton plus jets channel the error is dominated by the 
systematics while in the di-lepton channels the limiting factor is  
still the statistics. 
 
Several studies of the accuracy which can be expected  
with the LHC experiments have been performed in the 
past~\cite{top:aachen}. It is interesting to see  
whether one can use the large statistics available after a  
few years of high-luminosity running to push the precision  
further. In particular, it is interesting to study the ultimate  
accuracy achievable at a hadronic collider, and the factors that  
limit this accuracy. 
 
In the following subsections, we begin with general remarks on the top 
quark mass and a very brief review of the present status of the 
theoretical understanding of top quark mass measurement in the 
threshold scan at a future $e^+ e^-$ collider. We then present 
the results of a recent studies of top mass reconstruction at the LHC. 
The techniques used include the study of the lepton plus jets final 
states (inclusive, as well as limited to high-$\pt$ top quarks), 
di-lepton final states (using the di-leptons from the leptonic decay of 
both $W$'s, as well as samples where the isolated $W$ lepton is paired 
with a non-isolated lepton from the decay of the companion $b$ 
hadron).  A very promising analysis using the $J/\psi$~from the $b$ 
hadron decay paired with the lepton from the leptonic decay of the $W$ 
is discussed at the end. The conclusions of these studies indicate 
that an accuracy of 2~GeV should be achievable with the statistics 
available after only 1 year of running at low luminosity. An  
accuracy of 1~GeV accuracy could be achieved after the high luminosity  
phase. 
 
\subsection{General remarks and the top mass measurement in  
$\bf e^+ e^-$ annihilation} 
 
Although one speaks of ``the'' top quark mass, one should keep in  
mind that the concept of quark mass is convention-dependent.  
The top quark pole mass definition is often implicit, but in a  
confining theory it can be useful to choose another convention. This  
is true even for top quarks when one discusses mass measurements  
with an accuracy of order of or below the strong interaction scale.  
Since different mass conventions can differ by 10$\,$GeV  
(see Section~\ref{sec:topprop}), the question arises which mass is  
actually determined to an accuracy of $1$-$2\,$GeV by a particular  
measurement. 
 
The simple answer is that a particular measurement determines those  
mass parameters accurately in terms of which uncalculated higher  
order corrections to the matrix elements of the process are  
small. This in turn may depend on the accuracy one aims at and the  
order to which the process has already been calculated. To clarify  
these statements we briefly discuss the top quark mass measurement  
at a high energy $e^+ e^-$ collider. 
 
``The'' top quark mass can be measured in $e^+ e^-$ collisions by  
recontructing top quark decay products in much the same way as at the  
LHC. In addition, there exists the unique possibility of determining 
the mass in pair production near threshold. This is considered to be  
the most accurate method \cite{Accomando:1998wt} and it appears that an 
uncertainty of $\delta \overline{m}_t\approx 0.15\,$GeV can be  
achieved {\em $\,$for the top quark $\overline{MS}$ mass} with the  
presently available theoretical input 
\cite{Beneke:1999qg}. This is a factor two improvement  
compared to the accuracy that could be achieved with the same  
theoretical input if the cross section were parametrised in terms of  
the top quark pole mass. The fundamental reason for this difference  
is the fact that the concept of a quark pole mass is intriniscally  
ambiguous by an amount of order $\Lambda_{\rm QCD}$ 
\cite{Beneke:1994sw} and this conclusion remains valid even if the  
quark decays on average before hadronisation \cite{Smith:1997xz}.  
In the context of perturbation theory this ambiguity translates into  
sizeable higher order corrections to the matrix elements of a given  
process renormalized in the pole mass scheme. This makes it preferable 
to choose another mass convention if large corrections disappear in  
this way as is the case for the total cross section in $e^+ e^-$  
annihilation, because the total cross section is less affected by  
non-perturbative effects than the pole mass itself. Note, however,  
that despite this preference the position of the threshold is  
closer to twice the pole mass than twice the $\overline{\rm MS}$ mass, 
hence a leading order calculation determines the pole mass more  
naturally. It is possible to introduce intermediate mass  
renormalizations that are better defined than the pole mass and  
yet adequate to physical processes in which top quarks are close to  
mass shell \cite{Beneke:1999qg,Beneke:1998rk}. The  
conclusion that the top quark pole mass is disfavoured is based on  
the existence of such mass redefinitions and the existence of  
accurate theoretical calculations. 
 
The situation with mass determinations at the LHC appears much more  
complicated, since the mass reconstruction is to a large extent an  
experimental procedure based on leading order theoretical 
calculations, which are not sensitive to mass renormalization at all.  
Furthermore the concept of invariant mass of a top quark decay  
system is prone to ``large'' non-perturbative corrections of 
relative order  
$\Lambda_{\rm QCD}/m_t$, because the loss or gain of a soft particle  
changes the invariant mass squared by an amount of order  
$m_t\Lambda_{\rm QCD}$. The parametric magnitude of non-perturbative  
corrections is of the same order of magnitude as for the top quark  
pole mass itself and cannot be decreased by choosing another mass  
renormalization prescription. For this reason, top mass measurements  
based on reconstructing $m_t$ from the invariant mass of  
the decay products of a single top quark should be considered 
as measurements of the top quark pole mass. From the remarks above  
it follows that there is a limitation of principle on the accuracy of  
such measurements. However, under LHC conditions the experimental  
systematic uncertainty discussed later in this section is the 
limiting factor in practice. A potential exception is the measurement of  
$m_t$ in the decay mode $\ell J/\psi X$ discussed at the end of this  
section, since the systematic error is estimated to be below $1\,$GeV  
and since the systematic error is to a large extent theoretical. It  
would be interesting to investigate non-perturbative power corrections 
and principle obstructions to an accurate mass measurement for this  
process. This analysis has however not yet been carried out in any  
detail, comparable to the threshold scan in $e^+ e^-$ annihilation.  
 
\subsection{$\bf m_t$~in the lepton plus jets channel. Inclusive sample} 
\label{sec:masslj} 
The inclusive lepton plus jets channel provides a large and clean 
sample of top quarks for mass reconstruction. Considering only 
electrons and muons, the branching ratio of this channel is 29.6\%. 
Therefore, one can expect 
more than 2 millions events for one year of running at low 
luminosity. 
ATLAS performed an analysis in that channel using events generated 
using \pyth~\cite{Sjostrand:1994yb} and the ATLAS detector fast 
simulation package ATLFAST~\cite{atl:atlfast}. The top mass is determined 
using the hadronic part of the decay, as the invariant mass of the 
three jets coming from the same top: \mt = \mjjb. 
The leptonic top decay is used to tag the event with the presence 
of a high \pt~lepton and large \etmiss. For the background 
processes, the \herw~\cite{Marchesini:1992ch,Mangano:1993kp} generator 
was used for the background process \Wbbbar. 
 
The following background processes have been considered: \bbbar, 
$W + jets$ with \Wlnu, $Z + jets$ with \Zll, $WW$ with one 
\Wlnu~and the other \Wqqbar, $WZ$ with \Wlnu~and 
\Zqqbar, $ZZ$ with one \Zll~and \Zqqbar, and \Wbbbar~ 
with \Wlnu. Events are selected by requiring an isolated lepton 
with \pt$>20$~GeV and \abseta$< 2.5$, \etmiss$> 20$~GeV, and four 
jets with \pt$>40$~GeV and \abseta$<2.5$, of which two of them 
were required to be tagged as $b$-jets. Jets were reconstructed 
using a fixed cone algorithm with \dr = 0.4. Although at 
production level the signal over background is very unfavourable, 
after the selection cuts and for an integrated luminosity of 
10~\infb, 126000 signal events and 1922 background events were 
kept, yielding a value of $S/B  = 65$ (see Table~\ref{tabl:lpjets}). 
\begin{table} 
\begin{center} 
\caption{Efficiencies (in percent) for the inclusive \ttbar~ 
  single lepton plus jets signal and for background processes, as a 
  function of the selection cuts applied. No branching ratios are 
  included in the numbers. The last column gives the equivalent number 
  of events for an integrated luminosity of 10~\infb, and the 
  signal-to-background ratio.} 
 \label{tabl:lpjets} \vspace*{0.1cm} 
\begin{tabular}{|l|cccc|} \hline 
              & { $ \ptl>20$~GeV} & { as before, } & { as 
              before,} & { events,} \\ 
 { Process} & { $ \etmiss>20$~GeV} & { plus $ N_{jet}\geq4$} & 
 { plus $ N_{b-jet}\geq 2$} & { per 10~$ \infb$} \\ \hline 
 \hline 
 { $ \ttbar$ signal} & 64.7 & 21.2  & 5.0 & 126000 \\ \hline 
 { $W+jets$} & 47.9 & 0.1 & 0.002 & 1658 \\ 
 { $Z+jets$} & 15.0 & 0.05 & 0.002 & 232 \\ 
 { $WW$}     & 53.6 & 0.5 & 0.006 & 10 \\ 
 { $WZ$}     & 53.8 & 0.5 & 0.02 & 8 \\ 
 { $ZZ$}     & 2.8  & 0.04 & 0.008 & 14 \\ 
 { Total background} & & & & 1922 \\ \hline 
 {$ S/B $}             & & & & 65 \\ \hline 
\end{tabular} 
\end{center} 
\end{table} 
 
\begin{figure} 
\begin{center} 
\includegraphics[width=0.45\textwidth,clip]{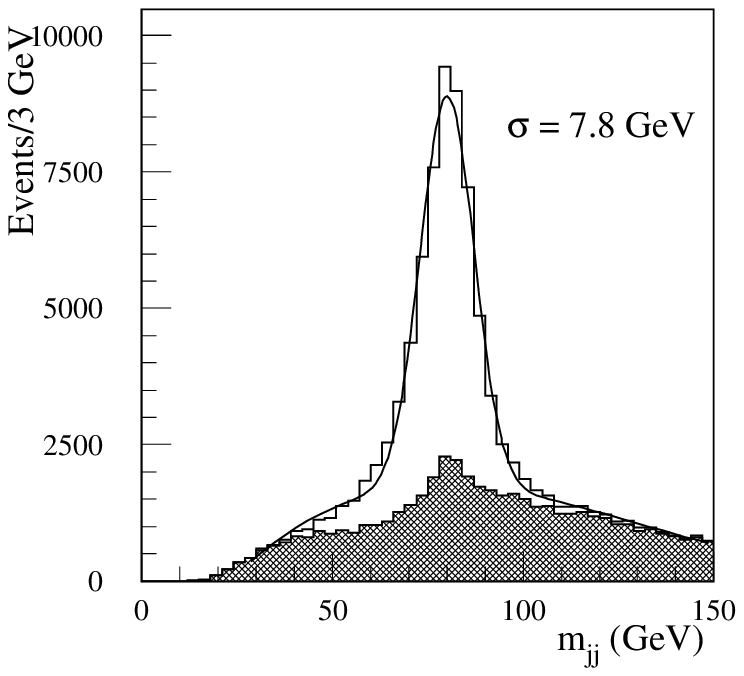} \hfill 
\includegraphics[width=0.45\textwidth,clip]{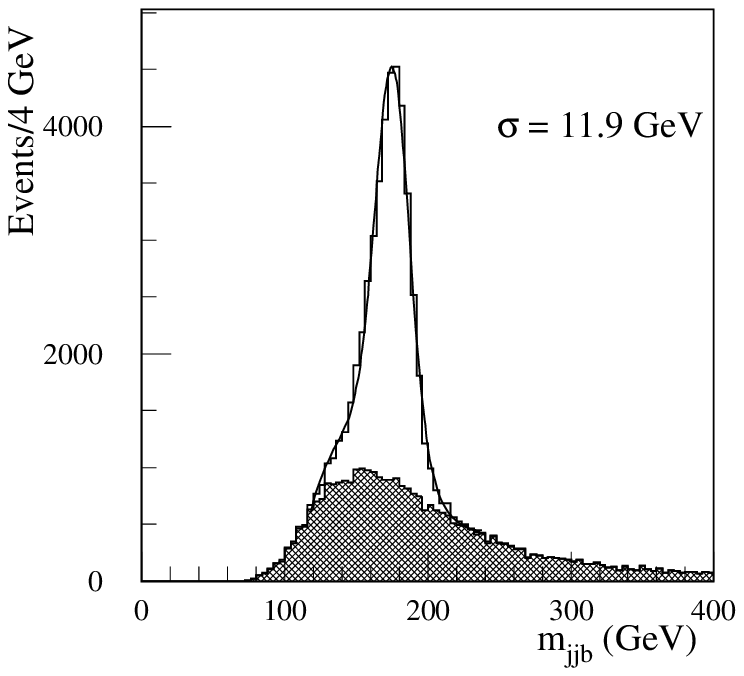} 
\end{center} 
\vskip -1.5cm 
\caption{Left: invariant mass distribution of the selected 
\jj~pairs.  Right: invariant mass distribution of the selected 
\jjb~combination. Both distributions are normalised to an integrated 
luminosity of 10\infb. The shaded area shows the combinatorial background.} 
\label{fig:incl-mtop1} 
\end{figure}

The reconstruction of the decay \Wjj~is first performed. The 
invariant mass \mjj~of all the combinations of jets (with \pt$> 
40$~GeV and \abseta$<2.5$) that were not tagged as $b$-jets is 
computed and the jet pair with an invariant mass closest to \mW~is  
selected as the $W$ candidate. Fig.~\ref{fig:incl-mtop1} represents 
the invariant mass distribution of the selected jet pairs. The 
reconstructed $W$ mass is consistent with the generated value, the 
mass resolution being 7.8~GeV.  Within a window of $\pm20$~GeV around 
the $W$ mass, the purity (P) and the overall efficiency (E) of the $W$ 
reconstruction are respectively P=67\% and E=1.7\%. Additional pair 
association criteria, such as requiring the leading jet to be part of 
the combination, did not improve significantly the purity and have not 
been considered further in the analysis. 
$W$ candidates, retained if $|m_{jj} - M_W|< 20$~GeV, have then 
to be associated with one $b$-tagged jet to reconstruct the decay 
\tWb. 
To reconstruct the right combination, some association criteria have 
been tried, such as choosing the $b$-jet furthest from the isolated 
lepton, the $b$-jet closest to the reconstructed $W$, and choosing the 
\jjb~combination having the highest \pt~for the reconstructed top. 
These various methods gave similar results. Fig.~\ref{fig:incl-mtop1} 
presents the invariant mass distribution of the reconstructed top when 
the \jjb~ combination having the highest \pt~has been used as 
association criteria. No $M_W$~constraint is applied for the light 
quark jets.  For an integrated luminosity of 10\infb, the total number 
of reconstructed top is 32000 events, of which 30000 are within a 
window of $\pm35$~GeV around the generated top mass \mt = 175~GeV. The 
total number of combinatorial events is 34000, of which 14000 are 
within the mass window. The number of background events coming from 
other processes is negligible. The \mjjb~ distribution fitted by a 
Gaussian plus a third order polynomial yields a top mass consistent 
with the generated value of 175~GeV and a top mass resolution of 
11.9~GeV. The resulting statistical uncertainty for an integrated 
luminosity of 10\infb~is \dmt = 0.070~GeV. 
\begin{figure} 
\begin{center} 
\includegraphics[width=0.45\textwidth,clip]{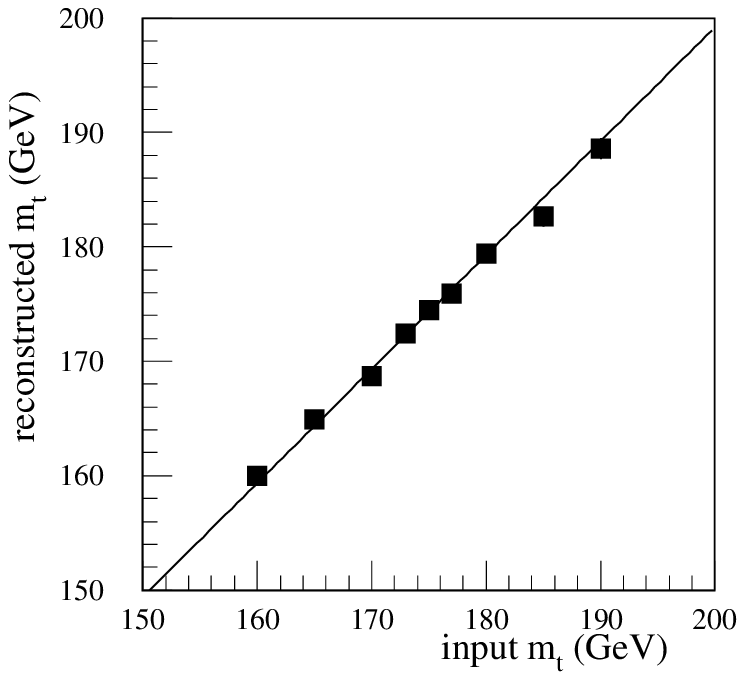} \hfill 
\includegraphics[width=0.45\textwidth,clip]{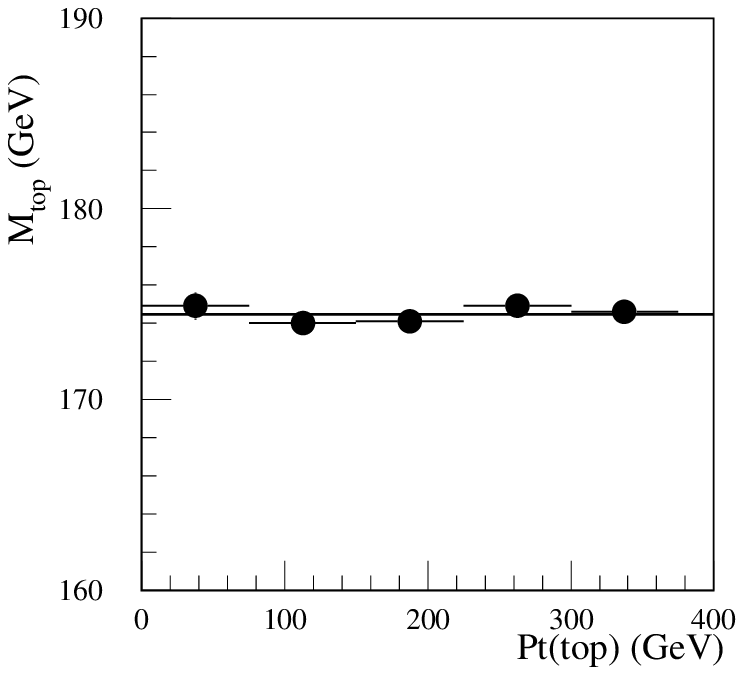} 
 \end{center} 
\vskip -1.5cm 
\caption{Left: dependence of the reconstructed top mass on the generated 
value. Right: dependence of the reconstructed top mass on the transverse 
momentum(\pt) of the reconstructed top.} 
\label{fig:incl-mtop2} 
\end{figure} 
 
The dependence of the top reconstruction algorithm on the top mass 
has been checked using several samples of \ttbar~events 
generated with different values of \mt~ranging from 160 to 
190~GeV. The results, shown in Fig.~\ref{fig:incl-mtop2}, 
demonstrate a linear dependence of the reconstructed top mass on 
the generated value: the data points are fitted to a linear 
function with $\chi^2/ \rm ndf = 6.7/8$. The stability of the mass 
value as a function of the transverse momentum of the 
reconstructed top (\pt (top)) was also checked. As shown in 
Fig.~\ref{fig:incl-mtop2}, no significant \pt (top) dependence 
is observed: the data points are fitted to a constant with 
$\chi^2/ \rm ndf = 6.25/5$. For more details of this analysis, 
see~\cite{atlasphysnote99024}. 
 
\begin{figure} 
    \begin{center} 
    \includegraphics[width=0.45\textwidth,clip]{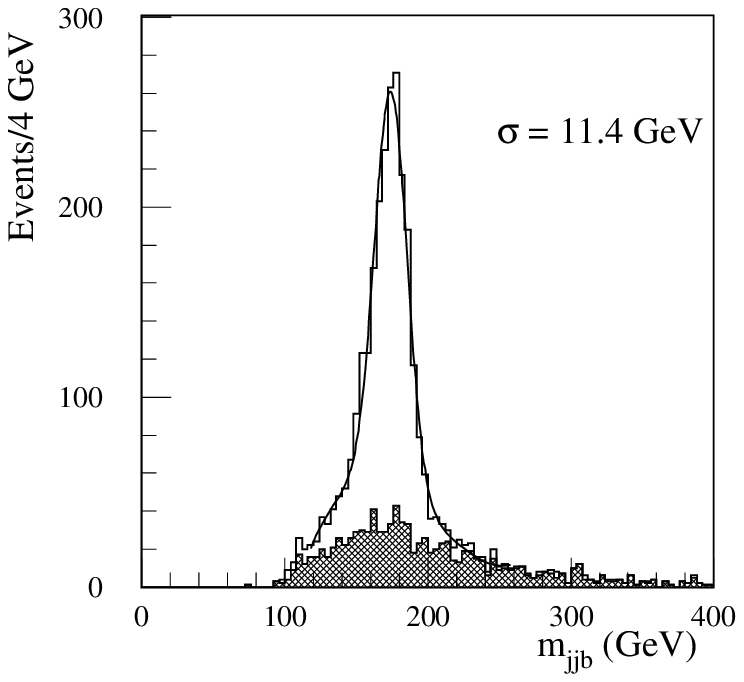} \hfill 
    \includegraphics[width=0.45\textwidth,clip]{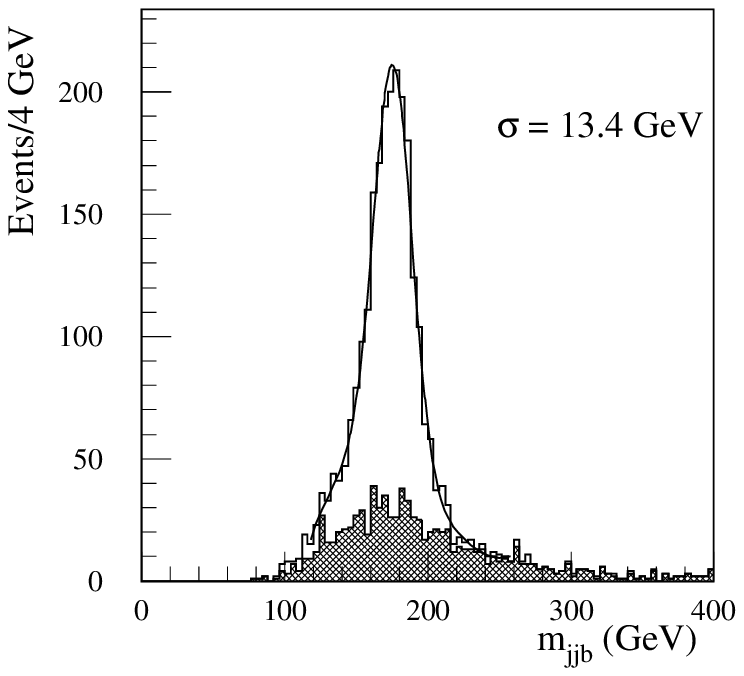} 
    \end{center} 
\vskip -1.5cm 
    \caption{Invariant $jjb$~mass distributions. Left: from fast simulation. 
    Right: from full simulation.} 
    \label{fig:incl-mtop3} 
\end{figure}

The results presented above, obtained with a fast simulation 
package, have been cross-checked with 30000 events passed through 
the ATLAS GEANT-based full simulation package~\cite{atl:DICE}. In 
full simulation, in order to save computing time, events have been 
generated under restrictive conditions at the generator level. The 
comparison is done by using the same generated events which have 
been passed through both the fast and full simulation packages. 
The results, in terms of purity, efficiency and mass resolutions 
show a reasonable 
agreement between fast and full simulation. In addition, as it is 
shown in Fig.~\ref{fig:incl-mtop3}, the shape and amount of the 
combinatorial background for the \mjjb~distributions are in 
good agreement between the two types of simulations. 
 
It has to be noted that for this analysis as well as for the other 
top mass reconstruction studies performed within \ATLAS, the 
jets were calibrated using the ratio \pt(parton)/\pt(jet) obtained 
from Monte Carlo samples of di-jet events or  $H\rightarrow 
b\bar{b}$ with $m_H=100$~GeV. In that aspect this calibration does 
not include all possible detector effects and corrections. More 
details can be found in Chapter~20 of~\cite{atlasphystdr} and 
in Appendix~\ref{app:btagjet}. 
 
\subsection{$\bf m_t$~in the lepton plus jets channel. High $\bf p_T$~sample} 
 
An interesting possibility at the LHC, thanks to the large \ttbar~ 
production rate, is the use of special sub-samples, such as events  
where the top and anti-top quarks have high \pt. 
In this case, they are produced back-to-back in the lab-frame, and the 
daughters from the two top decays will appear in distinct 
``hemispheres'' of the detector. This topology would greatly 
reduce the combinatorial background as well as the backgrounds 
from other processes. Furthermore, the higher average energy of 
the jets to be reconstructed should reduce the sensitivity to 
systematic effects due to the jet energy calibration and to 
effects of gluon radiation. However, in this case a competing 
effect appears which can limit the resulting precision: as the top 
\pt~increases, the jet overlapping probability increases as well, 
which again affects the jet calibration. \ATLAS~performed 
a preliminary study of this possibility using two different 
reconstruction methods: 
\begin{itemize} 
\item in the first one an analysis similar to the inclusive case is 
  done, with \mt~being reconstructed from the three jets  
  in the one hemisphere (\mt=\mjjb); 
\item in the second one, \mt~is reconstructed summing up the energies 
in the calorimeter towers in a large cone around the top direction. 
\end{itemize} 
In the following paragraphs, highlights of these analyses are discussed. 
 
\subsubsection{Jet Analysis} 
 
\begin{figure} 
\begin{center} 
\includegraphics[width=0.45\textwidth,clip]{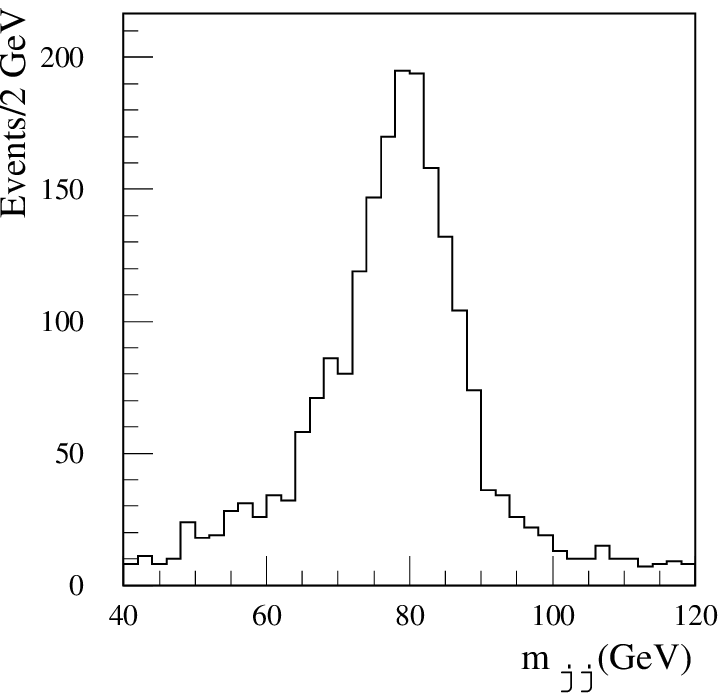} \hfill 
\includegraphics[width=0.45\textwidth,clip]{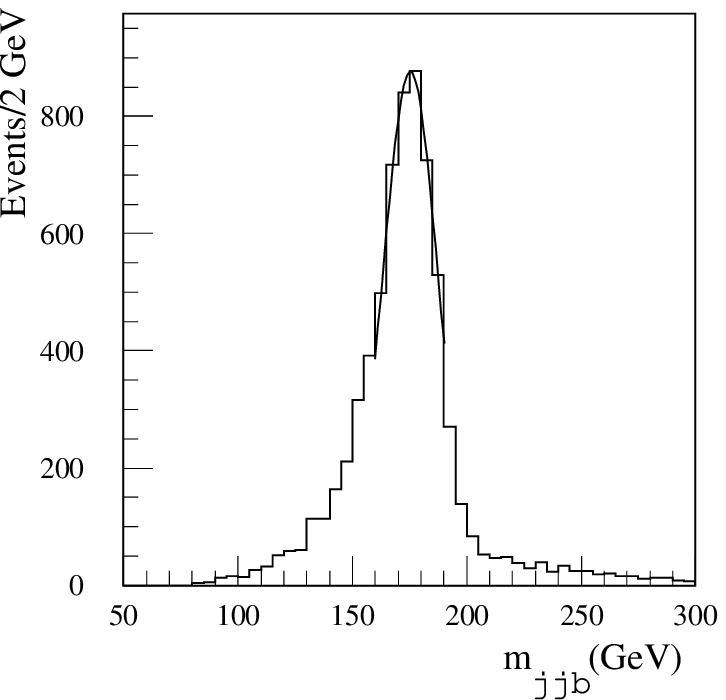} 
\end{center} 
\vskip -1cm 
\caption{Left: invariant mass distribution of the selected di-jet 
combinations for the high \pt(top) sample. Right: invariant mass 
distribution of the accepted combinations for the high \pt(top) 
sample. Both distributions are normalised to an integrated luminosity 
of 10\infb.} 
\label{fig:Fig18-8} 
\end{figure}

High \pt~\ttbar~events were generated using 
\pyth~5.7~\cite{Sjostrand:1994yb} with a \pt~cut on the hard scattering 
process above 200~GeV. The expected cross-section in this case is 
about 120~pb, or about 14.5\% of the total \ttbar~production 
cross-section.  The selection cuts required the presence of an 
isolated lepton with \pt$>30$~GeV and \abseta$<2.5$, and 
\etmiss$>30$~GeV. The total transverse energy of the event was 
required to be greater than 450~GeV. Jets were reconstructed using a 
cone algorithm with radius \dr=0.4. The plane perpendicular to the 
direction of the isolated lepton was used to divide the detector into 
two hemispheres. Considering only jets with \pt$>40$~GeV and 
\abseta$<2.5$, the cuts required one $b$-tagged jet in the same 
hemisphere as the lepton, and three jets, one of which was $b$-tagged, 
in the opposite hemisphere. Di-jet candidates for the \Wjj~decay were 
selected among the non-$b$-tagged jets in the hemisphere opposite to 
the lepton. The resultant \mjj~ invariant mass distribution is shown 
in Fig.~\ref{fig:Fig18-8}~(left). Fitting the six bins around the peak 
of the mass distribution with a Gaussian, yielded a $W$ mass 
consistent with the generated value, and a \mjj~resolution of 7~GeV, 
in good agreement with that obtained for the inclusive sample. Di-jets 
with 40~GeV$<$\mjj$<$120~GeV were then combined with the $b$-tagged 
jet from the hemisphere opposite to the lepton to form 
\tjjb~candidates. Finally, the high \pt(top) requirement was imposed 
by requiring \pt(\jjb)$>250$~GeV. With these cuts, the overall signal 
efficiency was 1.7\%, and the background from sources other than 
\ttbar~was reduced to a negligible level. The invariant mass 
distribution of the accepted \jjb~combinations is shown in 
Fig.~\ref{fig:Fig18-8}~(right). Fitting the six bins around the peak 
of the mass distribution with a Gaussian, yielded a top mass 
consistent with the generated value of 175~GeV, and a \mjjb~ mass 
resolution of 11.8~GeV. For an integrated luminosity of 10~\infb, a 
sample of 6300 events would be collected in \ATLAS, leading to a 
statistical error of \dmt(stat.) = $\pm$0.25~GeV, which remains well 
below the systematic uncertainty. As in the case of the inclusive 
sample, no strong \pt~dependence was observed and the reconstructed 
mass depends linearly on the Monte Carlo input value. 

\subsubsection{Using a large calorimeter cluster} 
\label{mtop_ue} For sufficiently high \pt(top) values, the jets 
from the top decay are close to each other with a large 
possibility of overlap. In such a case it might be possible to 
reconstruct the top mass by collecting all the energy deposited in 
the calorimeter in a large cone around the top quark direction. 
Such a technique has the potential to reduce the systematic 
errors, since it is less sensitive to the calibration of jets and 
to the intrinsic complexities of effects due to leakage outside 
the smaller cones, energy sharing between jets, etc. Some results 
from a preliminary investigation of the potential of this 
technique are discussed here. More details of the analysis can be 
found in~\cite{atlasphystdr,atl:iefthy99}. 
 
\begin{figure} 
  \begin{center} 
    \includegraphics[width=0.45\textwidth,clip]{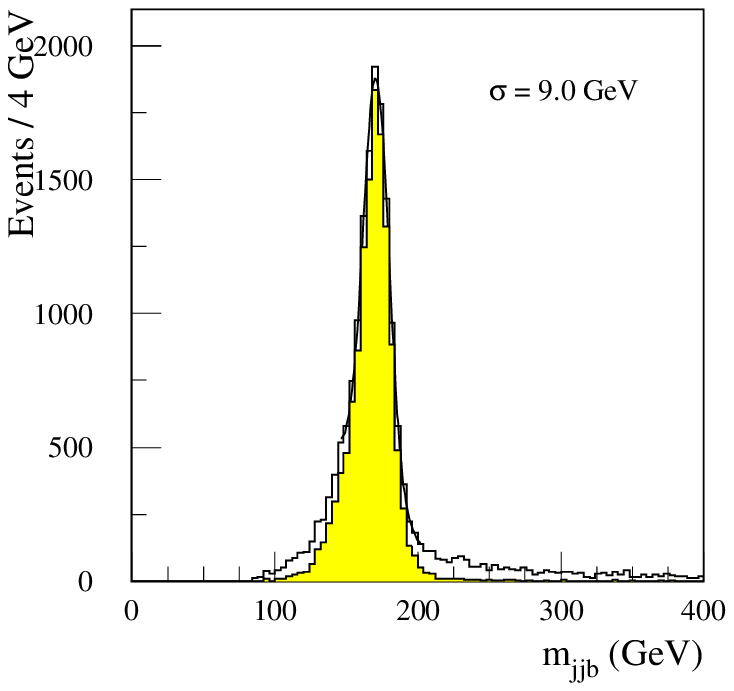} \hfill 
    \includegraphics[width=0.45\textwidth,clip]{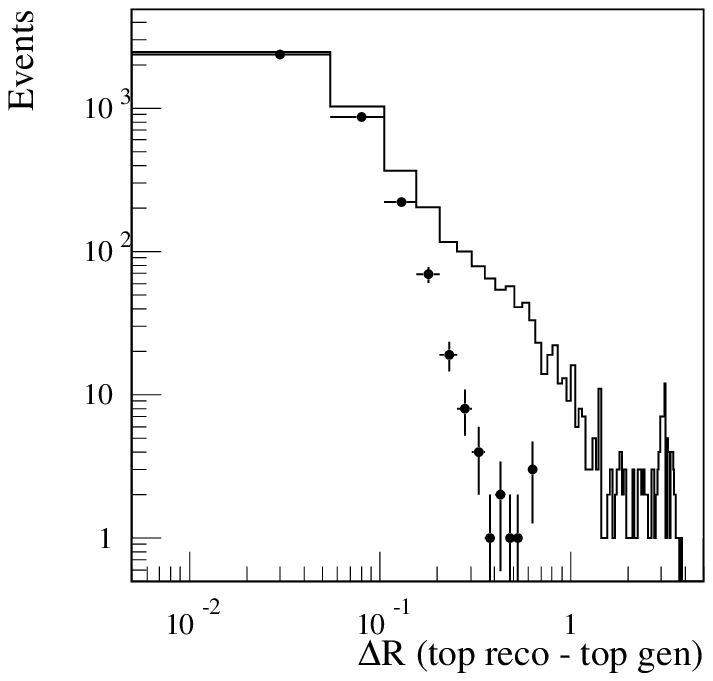} 
    \end{center} 
\vskip -1cm 
    \caption{Left: invariant mass distribution 
      of the selected \jjb~combination, using \dr = 0.2 cones for the 
      high \pt (top) sample, normalised to an integrated luminosity of 
      10\infb. The shaded area corresponds to the combinations with 
      the correct jet-parton assignments.  Right: distance \dr~ 
      between the reconstructed and the parton level direction of the 
      top quark. The dots correspond to the correct 
      \jjb~combinations.} 
    \label{fig:Fig18-11} 
\end{figure}

Similar event selection criteria as in the previous case were 
used: an isolated lepton with \pt$>20$~GeV and \abseta$<2.5$, 
$\etmiss>20$~GeV, one $b$-tagged jet 
(with \dr=0.4 and \pt$>20$~GeV) in the lepton hemisphere, and 
at least 3 jets in the hemisphere opposite to the lepton (\dr=0.2, 
\pt$>20$~GeV) with one of them $b$-tagged. For the accepted 
events, the two highest \pt~non-$b$-tagged jets were 
combined with the highest \pt~$b$-jet candidate in the 
hemisphere opposite to the lepton to form candidates for the 
\jjb~hadronic top decay. The selected \jjb~combination 
was required to have \pt$>150$~GeV and \abseta$<2.5$. With these 
selection criteria, about 13000 events would be expected in the 
mass window from 145 to 200~GeV, with a purity of 90\%, for an 
integrated luminosity of 10\infb. The reconstructed invariant mass 
of the \jjb~combination is shown in 
Fig.~\ref{fig:Fig18-11}~(left). The direction of the top quark 
was then determined from the jet momenta. 
Figure~\ref{fig:Fig18-11}~(right) shows the distance \dr~in 
$(\eta,\phi)$ space between the reconstructed 
and the true top direction at the parton level, 
demonstrating good agreement. 
%
   
\begin{figure} 
  \begin{center} 
    \includegraphics[width=0.45\textwidth,clip]{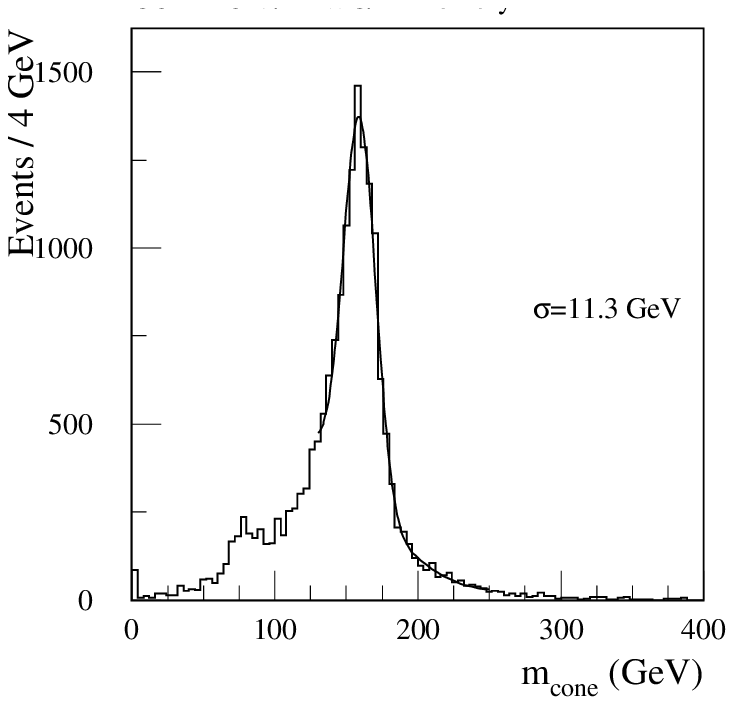} \hfill 
    \includegraphics[width=0.45\textwidth,clip]{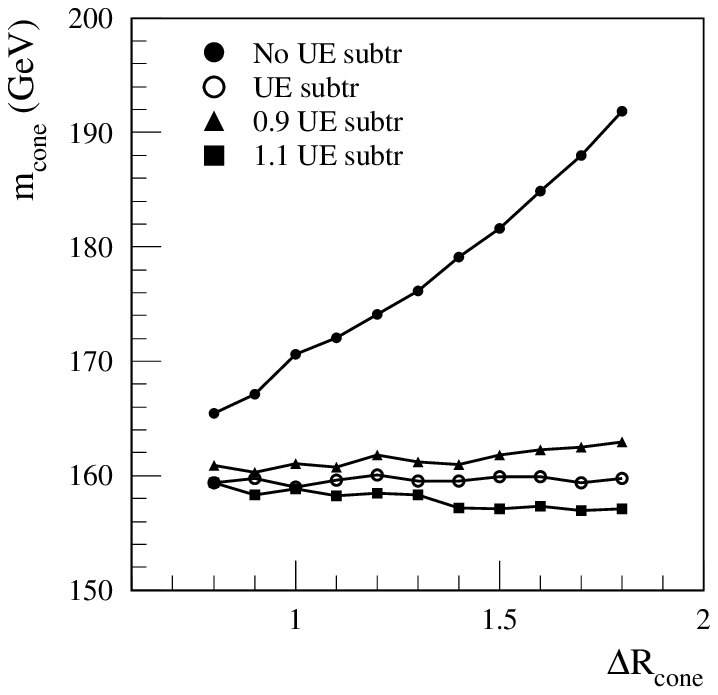} 
  \end{center} 
\vskip -1.5cm 
  \caption{Left: reconstructed \tjjb~mass spectrum obtained using 
    cells in a single cone of size \dr = 1.3, normalised to an 
    integrated luminosity of 10 \infb. Right: the fitted top mass 
    using cells in a single cone, before and after the underlying 
    event (UE) subtraction and as a function of the cone size.} 
  \label{fig:Fig18-13} 
\end{figure}

A large cone of radius \dr~was then drawn around the top quark 
direction, and the top mass was determined by adding the energies of 
all calorimeter ``towers'' within the cone. A calorimeter tower has a 
size of \detadphi = 0.1$\times$0.1, combining the information of both 
the EM and hadronic calorimeters. The invariant mass spectrum is 
shown in Fig.~\ref{fig:Fig18-13}~(left) for a cone size \dr= 1.3, and 
exhibits a clean peak at the top quark mass. The fitted value of the 
reconstructed top mass is shown in Fig.~\ref{fig:Fig18-13}~(right), 
where it displays a strong dependence on the cone size. If initial 
(ISR) and final (FSR) state radiation 
in \pyth~are turned off, the fitted mass remains constant 
(to within 2\%), independently of cone size. 
 
The large dependence of the reconstructed top mass on the cone 
size can be attributed to the underlying event (UE) contribution. 
A method was developed to evaluate and subtract the underlying event 
contribution using the calorimeter towers not associated 
with the products of the top quark decay. The UE contribution was 
calculated as the average \et~deposited per calorimeter 
tower, averaged over those towers which were far away from the 
reconstructed jets of the event. As expected, the average 
\et~per calorimeter tower increases as more activity is 
added, especially in the case of ISR. However, only a rather small 
dependence is observed on the radius \dr~used to isolate the 
towers associated with the hard scattering process. The resulting 
value of the reconstructed mass (\mcone), with and without UE 
subtraction, is also shown in Fig.~\ref{fig:Fig18-13}~(right) as a 
function of the cone radius. As can be seen, after the UE 
subtraction, the reconstructed top mass is independent of the cone 
size used. As a cross-check, the mean \et~per cell 
subtracted was varied by $\pm$10\% and the top mass recalculated 
in each case. As shown superimposed on 
Fig.~\ref{fig:Fig18-13}~(right), these ``miscalibrations'' lead 
to a re-emergence of a dependence of \mt~on the cone size. 
While the prescription for the UE subtraction does lead to a top 
mass which is independent of the cone size, it should be noted 
that the reconstructed mass is about 15~GeV (or 8.6\%) below the 
nominal value, \mt = 175~GeV, implying that a rather large 
correction is needed. 
 
To investigate if this correction can be extracted from the data 
without relying on Monte Carlo simulations, the same procedure was 
applied to a sample of $W+$~jet events generated with a range of 
\pt~comparable to that of the top sample. The $W$ was forced 
to decay hadronically into jets. The UE contribution was estimated 
with the same algorithm as described above. The results agreed 
within 1\% with the values determined for the high \pt(top) 
sample. As in the case of the top events, the reconstructed $W$ 
mass after UE subtraction is independent of the cone size. The 
average value of \mjj~after the UE subtraction is about 
8.5~GeV (or 10.6\%) below the nominal value of \mW. The fractional 
error on \mjj, as measured with the $W$+jet sample, was used as a 
correction factor to \mcone~in the high \pt(top) sample. For 
a cone of radius \dr = 1.3, the top mass after UE subtraction 
increases from 159.9~GeV to 176.0~GeV after rescaling. The 
rescaled values of \mcone~are about 1\% higher than the 
generated top mass. This over-correction of \mt~using the 
value of \mW~measured with the same method, is mainly due to ISR 
contributions. If ISR is switched off, the rescaling procedure 
works to better than 1\%.  
 
\subsection{Systematic uncertainties on the measurement of $\bf m_t$~in 
the single lepton plus jets channel}\label{sec:mtsyst} 
\begin{figure} \centerline{ 
    \includegraphics[width=0.7\textwidth,clip]{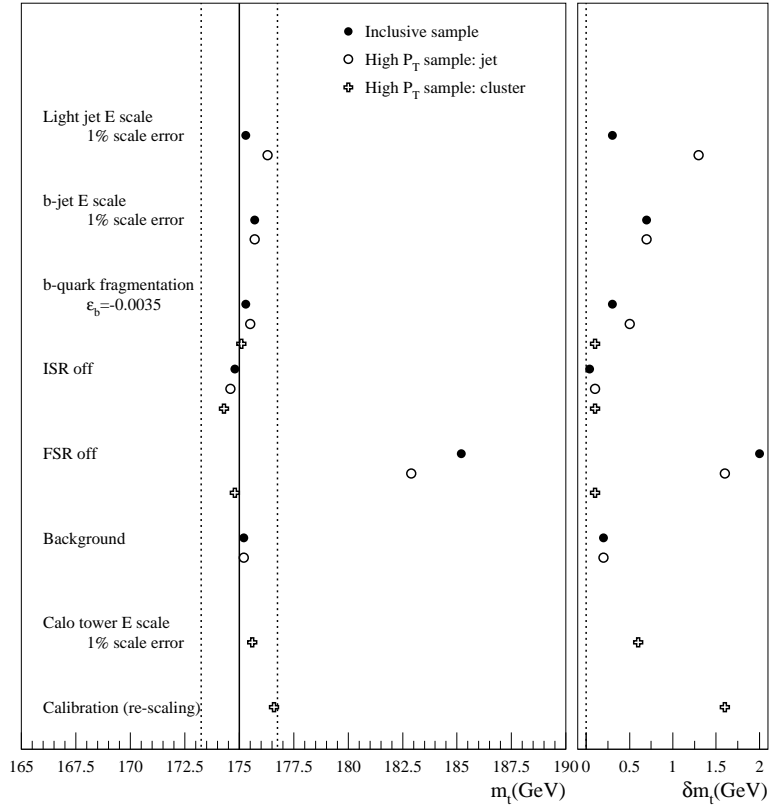}} 
\vskip -0.8cm 
    \caption{Summary of systematic errors in the \mt~measurement. 
    Left: the observed mass shifts for different effects studied. 
    The dashed lines indicates a 1\% mass window.  
    Right: the quoted error in the 
    \mt~measurement.} 
    \label{fig:mtsyst} 
\end{figure}

For the analyses presented above within \ATLAS, a number of sources 
of systematic error have been studied using samples of events generated with 
\pyth~and simulated mainly with the fast detector package ATLFAST, but also 
using a relatively large number of fully simulated events in order to 
cross-check some of the results. The results of these studies are summarised 
in Fig.~\ref{fig:mtsyst} and discussed below. 
 
\underline{{\it Jet energy scale:}} The measurement of \mt~via  
reconstruction of \tjjb~relies on a precise knowledge of 
the energy calibration for both light quark jets and $b$-jets. The 
jet energy scale depends on a variety of detector and physics 
effects, including non-linearities in the calorimeter response, 
energy lost outside the jet cone (due, for example, to energy 
swept away by the magnetic field or to gluon radiation at large 
angles with respect to the original parton), energy losses due to 
detector effects (cracks, leakage, etc.), and ``noise'' due to the 
underlying event. Preliminary studies done in \ATLAS~ 
indicate that a jet energy scale calibration at the level of 1\% 
for both light quark and $b$-jets would be feasible at the LHC (see 
discussion in the Appendix~\ref{app:btagjet}). In the case of the 
\mt~reconstructed from the invariant mass of the three jets 
(\mjjb) the $b$-jet energy scale enters directly in the 
measurement and therefore it must be calibrated from other 
sources, while the energy of the two light quark jets can be 
calibrated event-by-event using the $W$ mass constraint. This 
would work quite well at least for the inclusive sample, where the 
jets are well separated. In the high \pt~case, energy sharing 
algorithms and corrections for the two jets are needed, and 
therefore in order to be conservative we assume in the following  
that no such an event-by-event correction can be made. To estimate 
the effect of an absolute jet energy scale uncertainty, different 
``miscalibration'' coefficients were applied to the measured jet 
energies. A linear dependence was observed. 
 
\underline{{\it $b$-quark fragmentation:}} The  fraction of 
the original $b$-quark momentum which will appear as visible energy in 
the reconstruction cone of the corresponding $b$-jet depends on the 
fragmentation function of the $b$-quark. This function is usually 
parametrised in \pyth~in terms of one variable, \eb, using the 
Peterson fragmentation function~\cite{Peterson:1983}. To estimate the 
systematic error in \mt, the ``default'' value for 
\eb~(=-0.006) was varied within its experimental uncertainty 
(0.0025)~\cite{ALEPH:1998, lepewwg:1999} and the difference in the 
reconstructed \mt~was taken as the systematic error \dmt. 
 
\underline{{\it Initial and final state radiation:}} The presence 
of ISR or FSR 
can impact the measurement of \mt. To estimate the systematic 
error due to these, data samples were generated where ISR or FSR 
in the \pyth~generator were switched off. In the case of FSR, a 
large mass shift was observed for a jet cone of \dr=0.4. This is 
reduced as expected when a larger cone is used. Clearly this case 
is rather pessimistic since the knowledge in both ISR and FSR is 
typically at the level of  10\%. Therefore as a conservative 
estimate of the resultant systematic errors in \mt, 20\% of the 
mass shifts were used. 
 
An alternative approach uses the measured jet multiplicity to search, 
event-by-event, for the presence of hard gluon radiation.  Following 
the convention for this approach adopted at the 
Tevatron~\cite{Abachi:1997jv,Abe:1998vq}, the mass shift would be 
defined not by comparing events with radiation switched on and events 
with radiation switched off, but by the difference, \Dmt, between the 
value of \mt~determined from events with exactly four jets and that 
determined from events with more than four jets. The systematic error 
due to effects of initial and final radiation would then be considered 
as \dmt = \Dmt/$\sqrt{12}$. Such a calculation would yield systematic 
errors of approximately 0.4-1.1~GeV, smaller than the more 
conservative approach adopted here. 
 
\underline{{\it Background:}} Uncertainties in the size and shape 
of the background, which is dominated by ``wrong combinations'' in 
\ttbar~events, can affect the top mass reconstruction. The 
resultant systematic uncertainty on \mt~was estimated by 
varying the assumptions on the background shape in the fitting 
procedure. Fits of the \mjjb~distribution were performed 
assuming a Gaussian shape for the signal and either a polynomial 
or a threshold function for the background. Varying the background 
function resulted in a systematic error on \mt~of 0.2~GeV. 
The structure of the UE can affect the top mass 
reconstruction. However, as discussed above, it 
is possible to estimate and correct for this effect using data. 
Given the large statistics available at the LHC, it is assumed 
that the residual uncertainty from the underlying event will be 
small compared to the other errors (note that the UE 
denotes here a minimum bias event, since the impact of ISR 
has already been accounted for). 
 
For the particular case of the \mt~reconstructed using a 
large calorimeter cluster, similar procedures 
were adopted to estimate the the systematic errors. It is important 
to notice that, as expected, the use of a large cone substantially 
reduces the effects of FSR and $b$-quark fragmentation, each of 
which gives rise to a systematic error of 0.1~GeV. The uncertainty 
arising from ISR, which can affect the determination of the UE 
subtraction, is about 0.1~GeV as well. However, the main 
uncertainty in this technique comes from the calibration 
procedure. The calibration with the $W+$~jet sample produces a 
value of \mt~which is about 1\% above the generated value. 
Furthermore, the \Wjj~events would suffer from background 
from QCD multi-jet events. On-going studies suggest that one could 
calibrate using \Wjj~decays from the high \pt(top) events 
themselves, selecting those events in which the $b$-tagged jet is 
far away from the other two jets of the $W$ decay and then 
reconstructing the \Wjj~decay using a single cone of size 
\dr = 0.8. Further study is required to reliably estimate the 
potential of this calibration procedure, and therefore a 
conservative systematic uncertainty of 1\% is assigned to it. 
 
\subsection{$\bf m_t$~in the di-lepton channel} 
 
Di-Lepton events can provide a measurement of the top quark mass 
complementary to that obtained from the single lepton plus jets 
mode. The signature of a di-lepton event consists of two isolated 
high \pt~leptons, high \etmiss~$\!$due to the neutrinos, 
and two jets from the $b$-quarks. The measurement 
of \mt~using di-lepton events is not a direct measurement as 
in the previous case but it relies on the relation between the 
kinematical distributions of the top decay products and \mt, 
and on how they can be reproduced by the Monte Carlo simulation. 
About 400000 di-lepton \ttbar~events are expected to be 
produced in a data sample corresponding to an integrated 
luminosity of 10~\infb. Backgrounds arise from Drell-Yan processes 
associated with jets, \Ztt~associated with jets, $WW+$~jets 
and \bbbar~production. 
 
Of the many possible kinematic variables which could be studied, 
\ATLAS~performed a preliminary study using: the mass \mlb~of 
the lepton+$b$-jet system, the energy of the two highest \et~jets,  
and the mass \mll~of the di-lepton system formed with 
both leptons originating from the same top decay (i.e. 
\tlnub~followed by \blnuc). The event selection criteria 
required two opposite-sign leptons within \abseta$<2.5$, with 
\pt$>35$ and 25~GeV respectively, and with \etmiss$> 40$~GeV. Two 
jets with \pt~$> 25$~GeV were required in addition. After 
the selection cuts, 80000 signal events survived, with $S/B$ 
around 10. 
 
\subsubsection{Top mass measurement using \mlb} 
 
In this analysis, the value of \mt~was estimated using the 
expression: 
\begin{equation} 
  \mt^2 = \mw^2+2\dot\avmlb/[1-\avthetalb] 
\label{eq:mllb} 
\end{equation} 
Here, \avmlb~is the squared mean invariant mass of the 
lepton and $b$-jet from the same top decay. The mean value of 
\avthetalb, the angle between the lepton and the $b$-jet in the 
$W$ rest frame, can be regarded as an input parameter to be taken 
from Monte Carlo. To obtain a very clean sample, the two highest 
\pt~jets were required to be tagged as $b$-jets, leaving a 
total of about 15200 signal events per 10~\infb. One cannot 
determine, in general, which lepton should be paired with which 
$b$-jet. The pairing which gave the smaller value of \avmlb~ 
was chosen, and checking the parton-level information showed that 
this criterion selected the correct pairing in 85\% of the cases, 
for a generated top mass of 175~GeV. The mean value \avmlb~ 
was measured for samples generated with different input top masses $m$, 
and then \mt~was calculated from the expression above. 
For an integrated luminosity of 10~\infb, the expected statistical 
uncertainty on \mt~using this method is $\pm0.9$~GeV. 
Major sources of systematics include uncertainty on the $b$-quark 
fragmentation function, which produces a systematic error on 
\mt~of 0.7~GeV if defined as described in 
Section~\ref{sec:mtsyst}. Systematic errors due to the effects of 
FSR and ISR together are about 1~GeV, while those due to varying 
the jet energy scale by 1\% are 0.6~GeV. Further studies are 
required to estimate the uncertainties due to the reliance upon 
the Monte Carlo modelling of the \ttbar~kinematics. 
\begin{figure} 
  \begin{center} 
    \includegraphics[width=0.45\textwidth,clip]{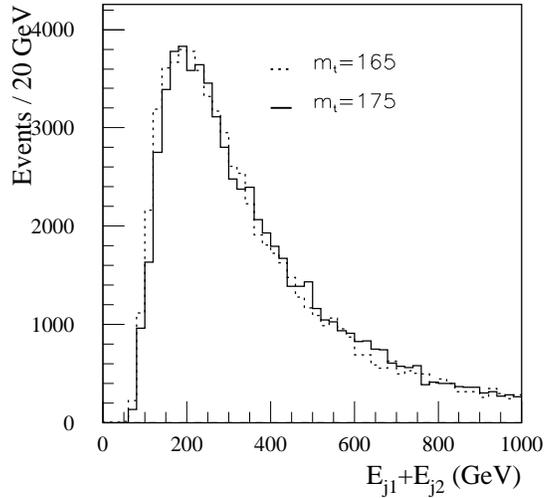} 
  \end{center} 
\vskip -1cm 
  \caption{Template distributions of the total energy of 
    the two leading jets in \ttbar~events for top quark masses 
    of 165 and 175~GeV. The two distributions are normalised to the 
    same area.} 
  \label{fig:Fig18-22} 
\end{figure} 
 
\subsubsection{Top mass measurement using the energy of the two leading jets} 
 
Increased sensitivity could be obtained with a technique which 
utilises not only the mean, but also the shape of the kinematic 
distribution. As an example, a study has been made of the 
sensitivity to \mt~obtained by comparing to ``template'' 
distributions the energy of the two highest-\et~jets. The 
template distributions were made by generating \pyth~samples of 
\ttbar~events with different values of \mt~in the 
range 160-190~GeV, in steps of 5~GeV. 
Figure~\ref{fig:Fig18-22} shows,  
as an example, the templates obtained for $m_t = 165$~GeV 
and 175~GeV. For each possible top mass value $m$, a $\chi^2(m)$ 
was obtained by comparing the kinematical distribution of the 
simulated data with the templates of mass $m$. The best value for 
the mass was the value which, for the ``data'' set, generated with 
\mt = 175~GeV, gave the minimum $\chi^2$. For an integrated 
luminosity of 10~\infb, the expected statistical sensitivity on 
\mt\ corresponds to about $\pm$0.4~GeV. Varying the calorimeter jet 
energy scale by 1\% produced a systematic error on \mt~of 
1.5~GeV. Other sources of systematic error result from the dependence 
of the method on the Monte Carlo modelling of the \ttbar~ 
kinematics, and require further study. As an example, changing the 
choice of the structure functions used in the Monte Carlo 
simulation (for example, from CTEQ2L to CTEQ2M or EHQL1) led to 
differences in the top mass of $\pm$0.7~GeV.  
 
\subsubsection{Top mass measurement using $m_{ll}$ in tri-lepton events} 
 
The invariant mass distribution of the two leptons from the same 
top quark decay (\ie \tlnub~followed by \blnuc) is quite 
sensitive to \mt. It has been shown that the mass distribution of 
lepton pairs from the same top quark decay is much less sensitive 
to the top quark transverse momentum distribution than that of 
lepton pairs from different top quarks~\cite{top:aachen}. Signal 
events are expected to contain two leptons from the decay of the 
$W$ bosons produced directly in the top and anti-top quark decays, 
and one lepton from the $b$-quark decay. In addition to the cuts 
described above, one non-isolated muon with \pt$>15$~GeV was 
required. For an integrated luminosity of 10~\infb, the expected 
signal would be about 7250 events, yielding a statistical 
uncertainty on the measurement of \mt~of approximately 
$\pm$1~GeV. This technique is insensitive to the jet energy scale. 
The dominant uncertainties arise from effects of ISR and FSR and 
from the $b$-quark fragmentation, which sum up to about 1.5~GeV.

\subsection{$\bf m_t$~from $\bf  t\to l+J/\psi+X$~decays} 
\label{sec:tjpsi} 
\begin{figure} \centerline{ 
    \includegraphics[width=0.75\textwidth,clip]{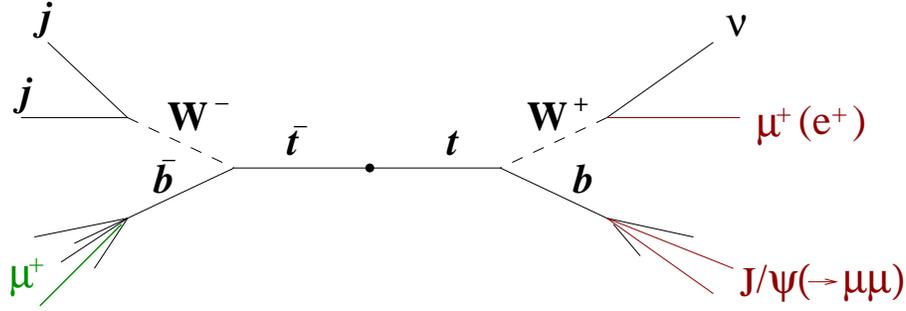}} 
  \caption{Schematics of the top decay to leptonic final states with 
    \jpsi.} 
  \label{fig:avto1} 
\end{figure}

An interesting proposal~\cite{ref:CMSLoI} by CMS, explored in 
detail during the workshop~\cite{Kharchilava:1999yj}, is to take 
advantage of the large top production rates and exploit the 
correlation between the top mass and the invariant mass 
distribution of the system composed of a \jpsi~(from the 
decay of a $b$ hadron) and of the lepton ($\ell=e,\mu$) from the 
associated $W$ decay (see Fig.~\ref{fig:avto1}). 
 
The advantage of using 
a \jpsi~compared to the other studies involving leptons as presented 
above is twofold: first, the large mass of the \jpsi~induces 
a stronger correlation with the top mass (as will be shown later). 
Second, the identification of the \jpsi~provides a 
much cleaner signal. 
%
In order to uniquely determine the top decay topology one can tag 
the charge of the $b$ decaying to \jpsi~by requiring the 
other $b$-jet to contain a muon as well. The overall branching 
ratio is $5.3\times10^{-5}$, taking into account the charge 
conjugate reaction and \Wenu~decays. In spite of this strong 
suppression, we stress that these final states are experimentally 
very clean and can be exploited even at the highest LHC 
luminosities. Furthermore, one can also explore other ways to 
associate the \jpsi~with the corresponding isolated lepton -- for 
example by measuring the jet charge of identified $b$'s. One 
should say that all these methods of top mass determination 
essentially rely on the Monte-Carlo description of its production 
and decay. Nonetheless the model, to a large extent, can be 
verified and tuned to the data. 
 
\subsubsection{Analysis} 
In the following we assume a \ttbar~production 
cross-section of 800~pb for \mt\  = 175~GeV. Events are 
simulated with the \pyth 5.7 \cite{Sjostrand:1994yb} or \herw~5.9 
\cite{Marchesini:1992ch} event generators. Particle momenta are 
smeared according to parameterisations obtained from detailed 
simulation of the CMS detector performance. Four-lepton events are 
selected by requiring an isolated lepton with \pt$>$15~GeV 
and \abseta$<$2.4, and three 
non-isolated, centrally produced muons of \pt$>$4~GeV and 
\abseta$<$2.4, with the invariant mass of the two of them being 
consistent with the \jpsi~mass. These cuts 
significantly reduce the external (non-\ttbar) background, mainly 
\Wbbbar~production,\footnote{\pyth\ results indicate 
that with the above cuts this source of the background can be kept 
at a per cent level.} which can be further reduced by employing, 
in addition, two central jets from another $W$. 
The resulting kinematical acceptance of the selection criteria is 30\%; 
this rather small value is largely due to soft muons from 
\jpsi~and $b$. In one year high luminosity running of LHC, 
corresponding to an integrated luminosity of 100~\infb , and 
assuming trigger plus reconstruction efficiency of 0.8, we expect 
about $10^5 \times 800 \times 5.3\cdot10^{-5} \times 0.3 \times 
0.8 = 1000$ events. 
 
\begin{figure}[hbtp] 
\begin{minipage}{.48\textwidth} 
\vspace*{-1.5cm} 
  \begin{center} 
    \resizebox{7.65cm}{!}{\includegraphics{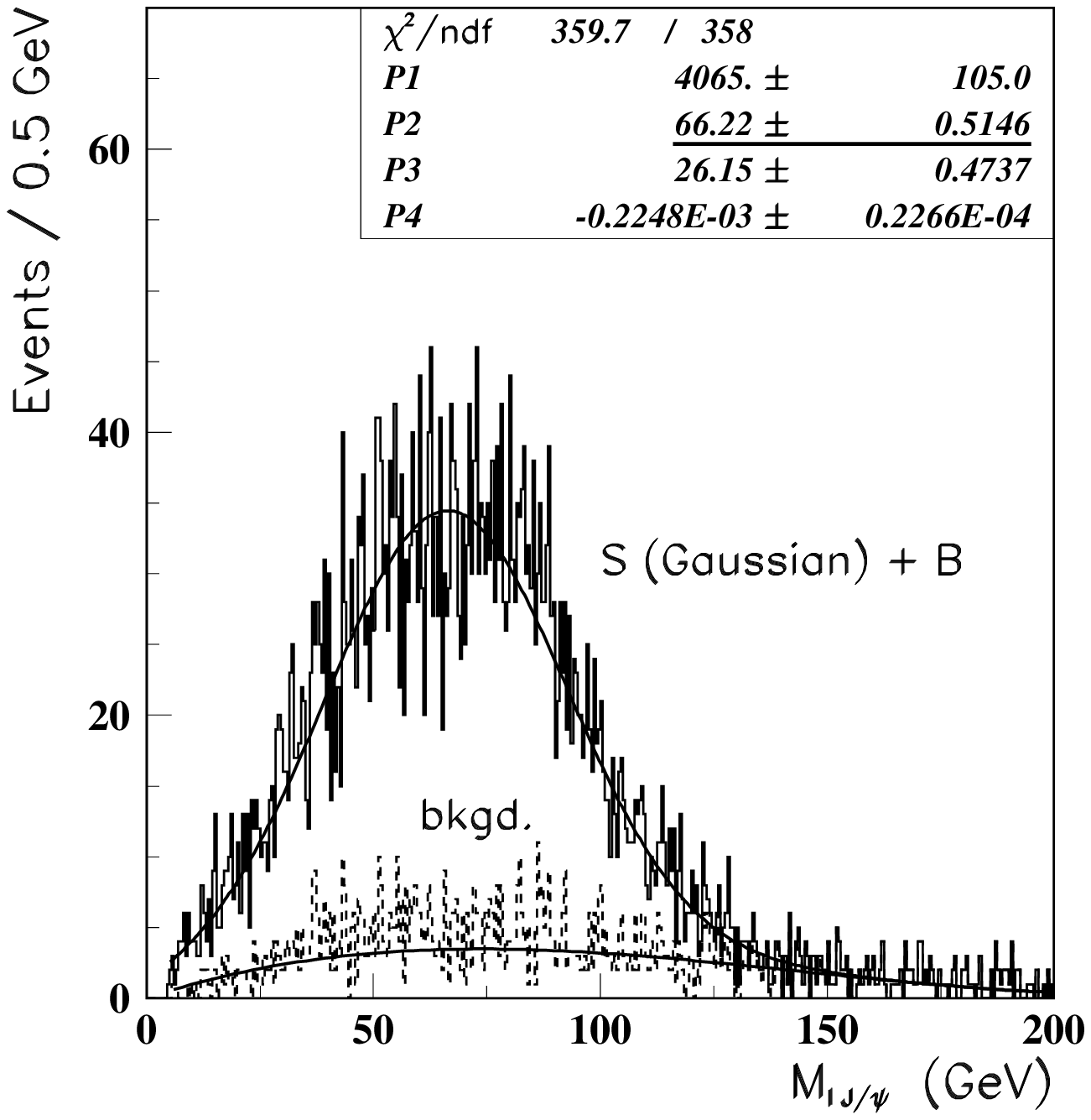}} 
\vspace*{-1.5cm} \caption{Example of the \lj~invariant mass 
spectrum in four-lepton final states. The number of events corresponds 
to four years running at LHC high luminosity.} \label{fig:avto2} 
  \end{center} 
\end{minipage} 
\hfill 
\begin{minipage}{.48\textwidth} 
\vspace*{-1.5cm} 
  \begin{center} 
    \resizebox{7.65cm}{!}{\includegraphics{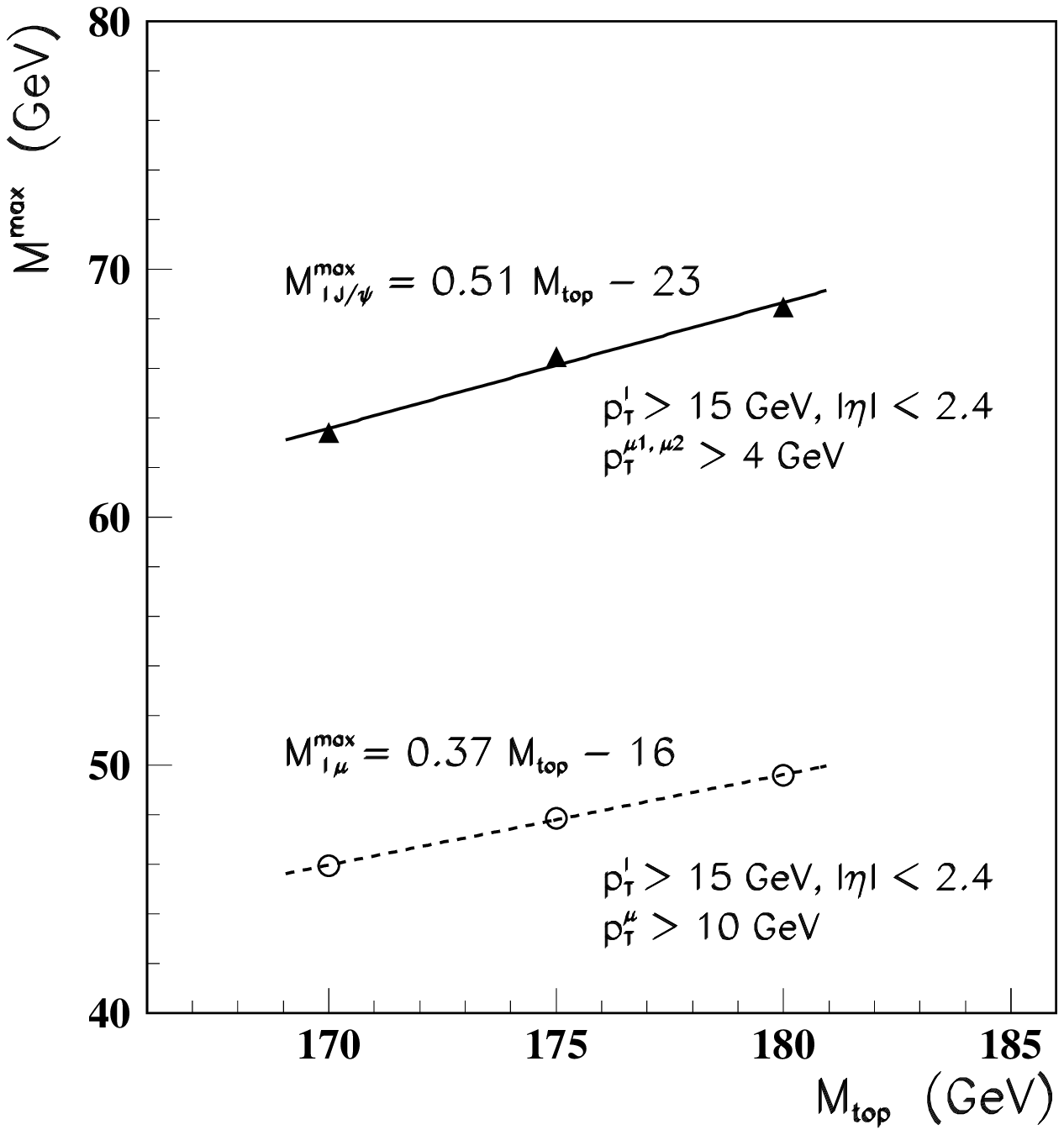}} 
\vspace*{-1.5cm} 
  \caption{Correlation between \mmax~and the top quark mass in isolated 
lepton plus \jpsi\ (solid line) and isolated lepton plus 
$\mu$-in-jet (dashed line) final states.} \label{fig:avto3} 
  \end{center} 
\end{minipage} 
\end{figure}

An example of the \lj~mass distribution with the expected 
background is shown in Fig.~\ref{fig:avto2}. The background is 
internal (from the \ttbar~production) and is due to the 
wrong assignment of the \jpsi~to the corresponding isolated 
lepton. These tagging muons of wrong sign are predominantly 
originating from $B^0/\overline{B}^0$ oscillations, $b \to c \to 
\mu$ transitions, $W(\to c, \ \tau) \to \mu$ decays, $\pi/$K 
decays in flight and amount to $\sim$~30\% of the signal 
combinations. The shape of the signal \lj~events (those with 
the correct sign of the tagging muon) is consistent with a 
Gaussian distribution over the entire mass interval up to its 
kinematical limit of $\sim$~175 GeV. The background shape is 
approximated by a cubic polynomial. The parameters 
of this polynomial are determined with ``data'' made of 
the wrong combinations of \lj~with an admixture of signal. In such 
a way the shape of the background is determined more precisely and 
in situ. Thus, when the signal distribution is fitted, only the 
background normalisation factor is left as a free parameter along 
with the three parameters of a Gaussian. The result of the fit is 
shown in Fig.~\ref{fig:avto2}. We point out that this procedure 
allows to absorb also the remaining external background (if any) 
into the background fit function. 
 
As a measure of the top quark mass we use the mean value (position 
of the maximum of the distribution) of the Gaussian, \mmaxl. In 
four years running at LHC with high luminosity the typical errors 
on this variable, including the uncertainty on the background, are 
about 0.5~GeV. It is composed of $\lsim$~0.5~GeV statistical error 
and $\lsim$~0.15~GeV systematics contribution due to the 
uncertainty on the measurement of the background shape.\footnote{The  
statistical power of the sample can be further 
improved by exploiting full spectrum, rather than its Gaussian 
part.} 
 
The measurement of the \mmaxl~can then be related to the generated top 
quark mass. An example of the correlation between the \mmax~and 
\mt\ is shown in Fig.~\ref{fig:avto3} along with the parameters of a 
linear fit. For comparison, we also show the corresponding dependence 
in a more traditional isolated lepton plus $\mu$-in-jet channel. Not 
surprisingly, the stronger correlation, and thus a better sensitivity 
to the top mass, is expected in the \lj~final states as compared to 
the isolated lepton plus $\mu$-in-jet channel. This is because, in the 
former case, we pickup a heavy object (the \jpsi) which carries a 
larger fraction of the $b$-jet momentum.  The \mmaxl~measurement 
error, statistical and systematic, scales as the inverse slope value 
of the fit, which is a factor of 2 in our case. Hence the statistical 
error on the top mass in this particular example is $\sim 1$~GeV. 
 
It is appropriate to comment on the ways to obtain a larger event 
sample. Encouraging results have been obtained  
in~\cite{jpsiee} to reconstruct the $b\to$\jpsi$\to e^+ e^-$ 
decays for low luminosity runs. The extension of these studies for 
a high luminosity environment is very desirable. 
Another possibility would be to relax the kinematical 
requirements. The choice of $p_T$ cut on soft muons is not 
dictated by the background considerations but by the trigger rates,  
and is set here to 4~GeV rather arbitrarily. For example, the 
di-muon trigger with $\eta$-dependent thresholds which is 
available in CMS for low luminosity runs~\cite{trigmu} allows to 
significantly increase the kinematical acceptance, practically to 
the limit determined by muon penetration up to the muon chambers. 
Therefore, the assessment of the trigger rates at high luminosity with 
lower $p_T$ thresholds and in multi-lepton events clearly deserves 
a dedicated study. 
 
An even larger event sample can be obtained in three lepton final 
states, using instead the jet-charge technique to determine the 
\ttbar\ decay topology instead of the tagging muon. The jet charge is 
defined as a $p_T$-weighted charge of particles collected in a cone 
around the \jpsi~direction. Obviously, this kind of analysis requires 
detailed simulations with full pattern recognition which are under 
way. However, particle level simulations performed with \pyth\ and 
with realistic assumptions on track reconstruction efficiency give 
event samples comparable to the muon-tag performance, with about 10 times 
less integrated luminosity. In any case, through the LHC lifetime, one 
can collect enough events so that the overall top mass measurement 
accuracy would not be hampered by the lack of statistics; it would 
rather be limited by the systematic uncertainties which are tightly 
linked with the Monte-Carlo tools in use, as will be argued in the 
following section. 
 
\subsubsection{Systematics} 
\label{sec:ljpsisyst} 
An essential aspect of the current analysis is to understand 
limitations which would arise from the Monte-Carlo description of 
the top production and decay.  It is important to realize that the 
observable used in this study enjoys two properties: it is Lorentz 
invariant an it does not depend on the detailed structure of the 
jets, but only on the momentum spectrum of the $b$-hadron and of 
the \jpsi~from its decay. 
 
As a result, were it not for distortions of the \lj~mass 
distribution induced by acceptance effects and by the presence of 
an underlying background, the measurement would be entirely 
insensitive with respect to changes in the top production 
dynamics, and in the structure of the underlying event. As a 
result, typical systematics such as those induced by higher-order 
corrections to the production process, or by the ISR 
and by the structure of the minimum bias event, are 
strongly reduced relative to other measurements of \mt. This 
expectation will be shown to be true in the following of this 
section. 
 
The main limitations to an accurate extraction of the top mass 
using this technique are expected to come from: i) the knowledge 
of the fragmentation function of the $b$ hadrons contained in the 
$b$-jet and, ii) the size of the non-perturbative corrections to 
the relation between the top quark mass and the \lj~mass 
distribution. 
The \jpsi~spectrum in the decay of the $b$-hadrons will be 
measured with high accuracy in the next generation of $B$-factory 
experiments. It should be pointed out, however, that the 
composition of $b$-hadrons measured at the $\Upsilon(4S)$ and in 
the top decays will not be the same. In this second case,  
one expects a non-negligible contribution from baryons and from 
$B_s$ states. The size of the relevant corrections to the 
inclusive \jpsi~spectrum in top decays is not known, and,  
although expected to be small, it needs to be studied. 
Additional effects, such as QED corrections to the $W$ leptonic 
decay, $W$ polarisation and spin correlation effects can all be 
controlled and included in the theoretical simulations. 
 
The rest of this section presents the results of a detailed 
study~\cite{Kharchilava:1999yj} of the systematics, mostly based 
on \pyth. 
 
\noindent\underline{{\it Detector resolution:}} Here we have considered 
only Gaussian smearing of particle momenta and the effect on the 
\mmaxl~measurement uncertainty is negligible. A possible 
nonlinearity of the detector response can be well controlled with 
the huge sample of \jpsi, $\Upsilon$ and $Z$ leptonic decays that 
will be available. 
 
\noindent\underline{{\it Background:}} The uncertainty would be mainly due 
to an inaccurate measurement of the background shape and the 
systematics contribution of $\lsim 0.15$~GeV quoted in previous 
section would scale down with increasing statistics. For example, 
already with $\sim 10^4$ events the induced uncertainty is $\lsim 
0.1$~GeV. 
 
 
\noindent\underline{{\it PDF:}} Depending on the relative fraction of 
gluon/quarks versus $x$ in 
  various PDF's the top production kinematics might be different.  No 
  straightforward procedure is available for the moment to evaluate 
  uncertainties due to a particular choice of PDF.  We compared 
  results obtained with the default set {\small CTEQ2L} 
  \cite{Lai:1995bb} and a more recent {\small CTEQ4L} 
  \cite{Lai:1997mg} parameterisations of PDF's.  The observed change in 
  the \mmaxl~value is well within 0.1 GeV. 
 
\noindent\underline{{\it Top \pt~spectrum:}} As shown in 
Section~\ref{sec:topdist}, one does not expect significant 
uncertainties in the prediction of the top $\pt$ spectrum.  However, 
to see an effect we have artificially altered the top $\pt$ spectrum 
by applying a cut at the generator level. We found that even requiring 
all top quarks to have $\pt>100$~GeV gives rise to only a 1$\sigma$ 
change ($\pm 0.7$~GeV) in the fitted value of \mt. 
 
\noindent\underline{{\it Initial state radiation:}} The \mmaxl~value is 
unchanged even switching off completely the ISR. 
 
 
\noindent\underline{{\it Top and $W$ widths:}} Kinematical cuts that are  
usually applied affect the observed 
  Breit-Wigner shape (tails) of decaying particles. Conversely, poor 
  knowledge of the widths may alter the generated $l J/\psi$  
  mass spectrum 
  depending on the cuts.  In our case, only a small change in the 
  \mmaxl~value is seen relative to the zero-width approximation. 
 
\noindent\underline{{\it $W$ polarisation:}} 
  A significant shift is found for the isotropic decays of W when 
  compared to the SM expectation of its $\sim 70$\% longitudinal 
  polarisation. In future runs of the Tevatron the $W$ polarisation will 
  be measured with a $\sim 2$\% accuracy \cite{tev:future}, and at the LHC 
  this would be further improved, so that it should not introduce 
  additional uncertainties in simulations.

\noindent\underline{{\it \ttbar\ spin correlations:}} A ``cross-talk'' 
between $t$ and $\bar{t}$ decay products is 
  possible due to experimental cuts. To examine this effect in detail 
  the $2 \to 6$ matrix elements have been implemented in 
  \pyth~preserving  
  the spin correlations~\cite{serg}. No sizeable difference 
  in the \mmaxl~value is seen compared to the default $2 \to 2$ matrix 
  elements. 
 
\noindent\underline{{\it QED bremsstrahlung:}} 
  Only a small effect is observed when it is switched off. 
  Furthermore, QED radiation is well understood and can be properly 
  simulated. 
 
\noindent\underline{{\it Final State Radiation:}} 
  A large shift of $\sim 7$~GeV is observed when the FSR is switched 
  off.  This is due to the absence of evolution for the $b$ quark, 
  whose fragmentation function will be unphysically hard.  To evaluate 
  the uncertainty we varied the parton virtuality scale $m_{min}$, 
  the invariant mass cut-off below which the showering is terminated. A 
  $\pm 50$\% variation of it around the default (tuned to data) value 
  of 1~GeV induces an uncertainty of $^{+0.1}_{-0.15}$~GeV. 
 
\noindent\underline{{\it $b$ fragmentation, except FSR:}} 
  As a default, in \pyth~we have used the Peterson form for the 
  $b$-quark fragmentation function with $\varepsilon_{b}=0.005$. 
  Variation of this value by $\pm$10\%~\cite{lephfwg} leads to an 
  uncertainty of $^{-0.3}_{+0.25}$~GeV. (The $\pm$10\% 
  uncertainty on $\varepsilon_{b}$ is inferred from LEP/SLD 
  precision of $\sim 1$\% on the average scaled energy of $B$-hadrons.)  
  It should be pointed out that 
  recent accurate measurements of the $b$-quark fragmentation 
  function~\cite{Abe:1999fi} are not well fitted by the Peterson form. 
 
The last two items of this list deserve some additional comments. 
While the separation between the FSR and the non-perturbative 
fragmentation phases seems unnecessary, and liable to lead to an 
overestimate of the uncertainty, it is important to remark that 
our knowledge of the non-perturbative hadronisation comes entirely 
from the production of $b$-hadrons in $Z^0$ decays at LEP and SLC. 
It is important to ensure that the accuracy of both perturbative 
and non-perturbative effects is known, since the perturbative 
evolution of $b$ quarks from $Z^0$ and top decays are not the 
same owing to the different scales involved.  
An agreement between data and Monte Carlo calculations 
for the $b$-hadron fragmentation function at the $Z^0$ does not 
guarantee a correct estimate of the $b$-hadron fragmentation 
function in top decays. 
 
\begin{figure}
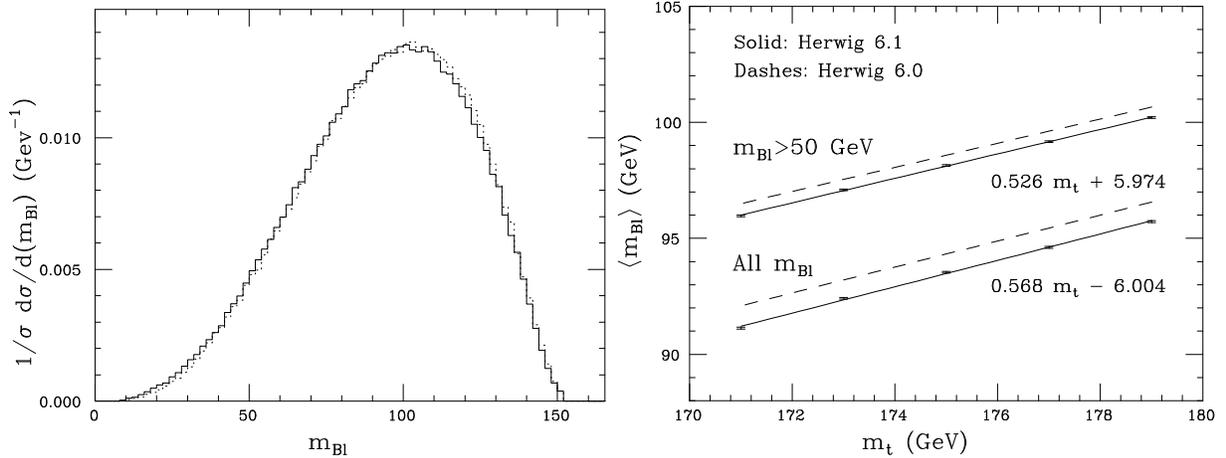
 
  \centerline{ 
    \includegraphics[width=0.5\textwidth,clip]{jpsi_mtop4.eps} \hfil 
    \includegraphics[width=0.5\textwidth,clip]{jpsi_mtop5.eps}} 
\vskip -0.5cm 
  \caption{Left: invariant mass of the $B$-lepton system 
    for $m_t=175$~GeV, according to \herw~6.0 (dotted) and 6.1 (with 
    matrix element corrections, solid). Right: linear fits to the 
    average invariant mass $\langle m_{B\ell}\rangle$  
    as a function of $m_t$.} 
  \label{fig:lhc175norm} 
\end{figure} 
 
To be specific, we shall consider here the effects induced by the 
higher-order matrix element corrections to the radiative top 
decays $t\to bWg$~\cite{Corcella:1998rs}. These effects cannot be 
simulated by a change in the virtuality scale $m_{min}$ as 
explored above in the study based on \pyth, as they have a 
different physical origin. 
The extended phase-space available for gluon emission after 
inclusion of the matrix-element corrections leads to a softening 
of the $b$-quark, and, as a result, of the \lj~spectrum.  For 
simplicity, we study here the invariant mass of the system 
$B\ell$. The resulting invariant mass distributions, for 
\mt$=175$~GeV, with (\herw\ 6.1) and without (\herw\ 6.0) matrix 
element corrections are shown in Fig.~\ref{fig:lhc175norm}. The 
averages of the two distributions, as a function of the top mass, 
are given on the right of the figure, and the difference of the 
averages are given in Table~\ref{tab:mbldif}. Given the slopes of 
the correlation between $\langle m_{B\ell}\rangle $ and \mt, we 
see that the corrections due to inclusion of the exact matrix 
elements are between 1~GeV (for $m_{B\ell}>50$~GeV) and 1.5~GeV (for 
the full sample).   
 
More details of the analysis will be found in~\cite{Corcella00}.  
It is also found there that the 
dependence of $\langle m_{B\ell}\rangle$ on the hadronic center of 
mass energy, or on the partonic initial state producing the 
$\ttbar$ pair, is no larger than 100~MeV. We take this as an 
indication that the effects of non-factorisable non-perturbative 
corrections (such as those induced by the neutralisation of the 
colour of the top quark decay products) are much smaller than the 
1~GeV accuracy goal on the mass. 
 
\begin{table} 
\caption{Negative shift in the average invariant mass $\langle 
  m_{B\ell}\rangle$ after inclusion of matrix element corrections for the 
  top decay in \herw. Left: average over all values of $m_{B\ell}$. Right: 
  average over the sample with $m_{B\ell}> 50$~GeV.} 
\label{tab:mbldif}  
\vspace*{0.1cm} 
\begin{center} 
\begin{tabular}{|l|c|c|}\hline 
$m_t$ & $\langle m_{B\ell}^{6.0}\rangle-\langle m_{B\ell}^{6.1}\rangle$ (all 
$m_{B\ell}$) & $\langle m_{B\ell}^{6.0}\rangle-\langle m_{B\ell}^{6.1}\rangle$ 
($m_{B\ell}>50$~GeV) \\ \hline \hline 
171 GeV & $(0.891\pm 0.038)$ GeV & $(0.479\pm 0.036)$ GeV\\ \hline 
173 GeV & $(0.844\pm 0.038)$ GeV & $(0.479\pm 0.034)$ GeV\\ \hline 
175 GeV & $(0.843\pm 0.039)$ GeV & $(0.510\pm 0.035)$ GeV\\ \hline 
177 GeV & $(0.855\pm 0.039)$ GeV & $(0.466\pm 0.035)$ GeV\\ \hline 
179 GeV & $(0.792\pm 0.040)$ GeV & $(0.427\pm 0.036)$ GeV\\ \hline 
\end{tabular} 
\end{center} 
\end{table}

\begin{figure} 
\begin{center} 
\includegraphics[width=0.6\textwidth,clip]{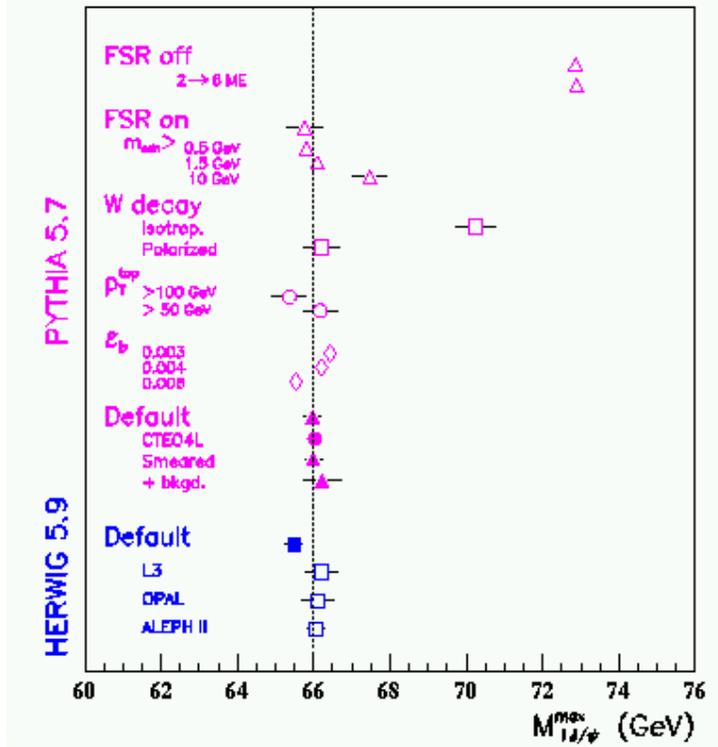} 
\caption{Observed \mt~shifts for the various systematic 
effects studied for the \ljpsi~channel.} \label{fig:avto4} 
\end{center} 
\end{figure}

A summary of these studies is given in Fig.~\ref{fig:avto4}.  One 
sees an impressive stability of the results for reasonable choices of 
parameters. The expected systematic error in the \mmaxl 
determination is $\lsim$~$^{+0.3}_{-0.4}$~GeV which translates 
into a systematic error on the top mass of 
$\delta\mt \lsim$~$^{+0.6}_{-0.8}$~GeV. 
 
In addition to the above studies, we also compared directly the 
results of \herw\ (v5.9) and \pyth.  With \herw~we have tried various 
tunings from LEP experiments as well as its default 
settings~\cite{Marchesini:1992ch}.  They 
all yield comparable results to each other and to \pyth~results, and 
are within $\lsim0.5$~GeV.  This corresponds to a systematic 
uncertainty $\delta\mt~\lsim 1$~GeV.

\subsection{Conclusions for the top mass measurement at the LHC} 
 
The very large samples of top quark events which will be 
accumulated at the LHC lead to a precision measurement of the 
top quark mass. Different statistically independent channels have 
been investigated and from the studies so far a precision of 
better than 2~GeV in each case can be obtained. In particular for 
the lepton plus jets channel where the \mt~is measured 
directly reconstructing the invariant mass of the \mjjb~ 
candidates, such a precision can be achieved within a year or 
running at low luminosity. For the channels involving two or more 
leptons, data from several years have to be combined to limit the 
statistical error in the measurement beyond the expected 
systematic errors. 
 
With the statistical error not being a problem, the emphasis of 
the work was devoted to estimate the systematic error involved in 
each method. For each sample, the contributing systematic errors 
are different, a fact which will allow important cross-checks to be 
made. The results indicate that a total error below 2~GeV 
should be feasible. In the case of the lepton plus jet channel the 
major contribution to the uncertainty  
is identified in the jet energy scale 
(in particular for the $b$-jets) and in the knowledge of FSR. When a 
special sub-sample of high \pt~top events is used and the 
\mt~is reconstructed using a large calorimeter cluster the 
FSR sensitivity is reduced, but further work is required to 
validate it. For the channels using two or more leptons for  
the top decay, the major contribution in the systematic error 
comes from the Monte Carlo and from how well the kinematic observable 
used for the mass measurement is related to the mass of the top 
quark.  
 
In $\ell J/\psi$ final states the top mass can be determined with 
a systematic uncertainty of $\lsim$~1~GeV. These final states are 
experimentally very clean and can be exploited even at highest LHC 
luminosities. The precision would be limited by 
the theoretical uncertainties which is basically reduced to the 
one associated with the $t \to B$ meson transition. This 
method of top mass determination looks very promising, and a final 
definition of its ultimate reach will rely on a better 
understanding of theoretical issues, and on the  
possibility to minimise the model dependence using  
the LHC data themselves. 
 
\section{SINGLE TOP PRODUCTION\protect\footnote{Section coordinators: 
    S.~Willenbrock, D.~O'Neil (ATLAS), J.~Womersley (CMS).}} 
\label{sec:onetop} 
\begin{figure}[tb] 
\begin{center} 
\includegraphics[width=4.5in,clip]{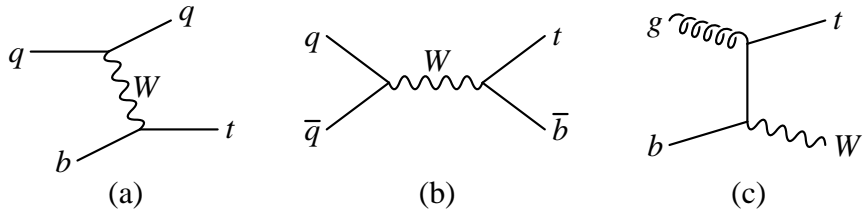} 
\end{center} 
\vskip -0.8cm 
\caption{Feynman diagrams for single-top quark production in 
hadron collisions: (a) $t$-channel process; (b) $s$-channel process; 
(c) associated production (only one of the two diagrams for this process 
is shown).} 
\label{fig:singletop} 
\end{figure}

At the LHC, top quarks are mostly produced in pairs, via the strong process 
$gg \to t\bar t$ (and, to a lesser extent, $q\bar q \to t\bar t$).  However, 
there are a significant number of top quarks that are produced singly, via 
the weak interaction.  There are three separate single-top quark production 
processes of interest at the LHC, which may be characterised by the 
virtuality of the $W$ boson (of four-momentum $q$) in the process: 
\begin{itemize} 
\item $t$-channel: The dominant process involves a space-like 
$W$ boson ($q^2 \le 0$), as shown in Fig.~\ref{fig:singletop}(a) 
\cite{Willenbrock:1986cr}. The virtual 
$W$ boson strikes a $b$ quark in the proton sea, promoting it to a 
top quark. This process is also referred to as $W$-gluon fusion, 
because the $b$ quark ultimately arises from a gluon splitting to 
$b\bar b$. 
\item $s$-channel: If one rotates the $t$-channel diagram 
such that the virtual $W$ boson becomes time-like, as shown in 
Fig.~\ref{fig:singletop}(b), one has another process that produces 
a single top quark \cite{Cortese:1991fw,Stelzer:1995mi}. The 
virtuality of the $W$ boson is $q^2 \ge (m_t+m_b)^2$. 
\item Associated production: A single top quark may also be produced via 
the weak interaction in association with a real $W$ boson ($q^2 = 
M_W^2$), as shown in Fig.~\ref{fig:singletop}(c) 
\cite{Heinson:1997zm,Tait:2000cf}. One of the initial partons is a 
$b$ quark in the proton sea, as in the $t$-channel process. 
\end{itemize} 
 
The total cross sections for these three single-top quark 
production processes are listed in Table~\ref{tab:sigma}, along 
with the cross section for the strong production of top quark 
pairs.  The $t$-channel process has the largest cross section; it 
is nearly one third as large as the cross section for top quark 
pairs.  The $s$-channel process has the smallest cross section, 
more than an order of magnitude less than the $t$-channel process. 
The $Wt$ process has a cross section intermediate between these 
two.  We will argue that all three processes are observable at the 
LHC. The $t$-channel and $s$-channel processes will first be 
observed at the Fermilab Tevatron \cite{Amidei:1996dt}; the $Wt$ 
process will first be seen at the LHC.

There are several reasons for studying the production of single top quarks 
at the LHC: 
\begin{itemize} 
\item The cross sections for single-top quark processes are proportional 
to $|V_{tb}|^2$.  These processes provide the only known way to directly 
measure $V_{tb}$ at hadron colliders. 
\item Single-top quark events are backgrounds to other signals. 
For example, single-top quark events are backgrounds to some signals for the 
Higgs boson \cite{Ladinsky:1991ut}. 
\item Single top quarks are produced with nearly $100\%$ polarisation, 
due to the weak interaction \cite{Heinson:1997zm,Carlson:1993dt,Mahlon:1997pn, 
Mahlon:1999gz}. 
This polarisation serves as a test of 
the $V-A$ structure of the top quark charged-current weak interaction. 
\item New physics may be discernible in single-top quark events.  New physics 
can influence single-top quark production by inducing non-SM weak 
interactions \cite{Carlson:1993dt,Carlson:1994bg,Tait:1997fe, 
Hikasa:1998wx,Boos:1999dd}, 
via loop effects  
\cite{Atwood:1996pd,Simmons:1997ws,Li:1997bh,Li:1997ir,Bar-Shalom:1998si}, 
or by providing new sources of 
single-top quark events 
\cite{Tait:1997fe,Simmons:1997ws,Malkawi:1996dm,Han:1998tp}. 
\end{itemize} 
\begin{table} 
\begin{center} 
\caption[fake]{Total cross sections (pb) for single-top quark 
production and top quark pair production at the LHC, for \mt=175 
$\pm 2$~GeV.  The NLO $t$-channel cross section 
is from~\cite{Stelzer:1997ns}.  The NLO 
$s$-channel cross section is from~\cite{Smith:1996ij}.  The 
cross section for the $Wt$ process is  
from~\cite{Tait:2000cf}; it is leading order, with a subset of the 
NLO corrections included. The uncertainties are 
due to variation of the factorisation and renormalisation scales; 
uncertainty in the parton distribution functions; and uncertainty 
in the top quark mass ($2$~GeV).} \label{tab:sigma} 
\vspace*{0.1cm} 
\begin{tabular}{|l|cccc|} \hline 
  process:    & $t$-channel  & $s$-channel    & $Wt$       & 
  \ttbar 
\\ \hline \hline 
$\sigma$(pb): & $245 \pm 27$ & $10.2 \pm 0.7$ & $51 \pm 9$ & $\sim 800$ 
\\ \hline 
\end{tabular} 
\end{center} 
\end{table} 
 
In the next three subsections we separately 
consider the three single-top quark 
production processes.  The subsection after these discusses the 
polarisation of single top quarks.  In the concluding section, we discuss 
the accuracy with which $V_{tb}$ can be measured in 
single-top quark events at the LHC. 
 
\subsection{t-channel single-top production} 
\label{sec:onetoptch} 
\subsubsection{Theory} 
The largest source of single top quarks at the LHC is via the 
$t$-channel process, shown in Fig.~\ref{fig:singletop}(a) 
\cite{Willenbrock:1986cr, 
Heinson:1997zm,Stelzer:1997ns,Carlson:1993dt,Bordes:1995ki,Stelzer:1998ni, 
Belyaev:1999dn}. A space-like ($q^2 \le 0$) $W$ boson strikes a $b$ 
quark in the proton sea, promoting it to a top quark.  As shown in 
Table~\ref{tab:sigma}, the cross section for this process is about 
one third that of the strong production of top quark pairs.  Thus 
there will be an enormous number of single top quarks produced via 
the $t$-channel process at the LHC. 
 
It is perhaps surprising that the cross section for the weak production of 
a single top quark, of order $\alpha_W^2$, is comparable to that of the 
strong production of top quark pairs, of order $\alpha_s^2$.  There are 
several enhancements to the $t$-channel production of a single top quark that 
are responsible for this: 
\begin{itemize} 
\item The differential cross section for the $t$-channel process is 
proportional to $d\sigma/dq^2 \sim 1/(q^2 - M_W^2)^2$, 
due to the $W$-boson propagator. The total cross section is therefore dominated 
by the region $|q^2| \le M_W^2$, and is proportional to $1/M_W^2$. 
In contrast, the total cross section for the strong production of top quark 
pairs is proportional to $1/s$, where $s \ge 4m_t^2$ is the parton 
center-of-mass energy. 
\item Since only a single top quark is produced, the typical value of the 
parton momentum fraction $x$ is half that of top quark pair production. 
Since parton distribution functions scale roughly like $1/x$ at small values 
of $x$, and there are two parton distribution functions, this leads to an 
enhancement factor of roughly four. 
\end{itemize} 
The fact that the total cross section is dominated by the region 
$|q^2|\le M_W^2$ also has the implication that the final-state light quark 
tends to be emitted at small angles, {\it i.e.}, high rapidities. 
This characteristic feature of the signal proves to be useful when 
isolating it from backgrounds. 
 
The $b$ distribution function in the proton sea arises from the splitting 
of virtual gluons into nearly-collinear $b\bar b$ pairs.  Thus it is 
implicit that there 
is a $\bar b$ in the final state, which accompanies the top quark and 
the light quark.  The final-state $\bar b$ tends to reside at small 
$p_T$, so it is usually unobservable. 
 
\begin{figure}[tb] 
\begin{center} 
\includegraphics[width=3in,clip]{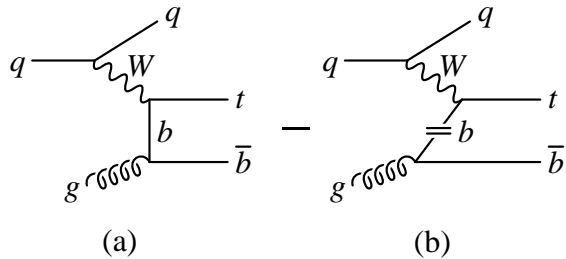} 
\end{center} 
\vskip -0.8cm 
\caption{(a) Initial-gluon correction to single-top quark production 
via the $t$-channel process (the diagram with the $W$ and gluon lines crossed 
is not shown); 
(b) the kinematic region in which the gluon splits to a nearly-collinear 
$b\bar b$ pair (the double line through the $b$ propagator indicates that it 
is nearly on shell) is subtracted from the correction, as it is already 
included at leading order.} 
\label{fig:initial} 
\end{figure}

The total cross section for the $t$-channel production of single 
top quarks has been calculated at NLO 
\cite{Stelzer:1997ns,Bordes:1995ki}; the result is given in 
Table~\ref{tab:sigma}.  A subset of the NLO 
corrections is shown in Fig.~\ref{fig:initial}(a). This correction 
arises from an initial gluon which splits into a $b\bar b$ pair. 
If the $b\bar b$ pair is nearly collinear, then this process 
contributes to the generation of the $b$ distribution function, 
which is already present at leading order; hence, one does not 
include this kinematic region as a contribution to the 
NLO correction.  This is indicated schematically 
in Fig.~\ref{fig:initial}(b). Only the contribution where the 
$b\bar b$ pair is non-collinear is a proper NLO 
correction to the total cross section.\footnote{The formalism for 
separating the nearly-collinear and non-collinear regions, and for 
generating the $b$ distribution function, was developed in 
Refs.~\cite{Barnett:1988jw,Aivazis:1994pi}.} The 
other corrections to this process, due to final-state and virtual 
gluons, as well as corrections associated with the light quark, 
are also included in the cross section given in 
Table~\ref{tab:sigma}. 
 
The central value for the cross section is obtained by setting the 
factorisation scale\footnote{The 
factorisation and renormalisation scales are set equal.} 
of the $b$ distribution function equal to $\mu^2 = -q^2 + m_t^2$. 
The uncertainty in the NLO 
cross section due to the variation of the factorisation 
scale between one half and twice its central value 
is $4\%$.  Due to the similarity with deep-inelastic 
scattering, the factorisation scale of the light 
quark is $\mu^2 = -q^2$, and is not varied \cite{Stelzer:1997ns}. 
 
Since the $\bar b$ tends to reside at low $p_T$, the dominant final state 
is $Wbj$, where the $Wb$ are the decay products of the top quark, and the 
jet is at high rapidity.  However, the $\bar b$ is at $p_T > 20$ GeV in 
roughly $40\%$ of the events, in which case the final state is $Wb\bar bj$. 
From a theoretical perspective, the optimal strategy is to isolate both 
final states and thereby measure the total cross section, which has 
an uncertainty of only $4\%$ from varying the factorisation scale, as 
mentioned above. However, the $Wb\bar bj$ final state 
has a large background from $t\bar t$, and it has not yet been established 
by ATLAS or CMS that this signal can be isolated, although the analysis  
of~\cite{Belyaev:1999dn} gives cause for optimism.  Thus we focus on 
the $Wbj$ final state, demanding that the $\bar b$ have $p_T < p_{Tcut}$. 
For $p_{Tcut} = 20$ GeV,\footnote{The CMS analysis presented below 
uses $p_{Tcut} = 20$ GeV; the ATLAS analysis uses $p_{Tcut} = 15$ GeV.} 
the cross section for this semi-inclusive process is 164~pb, with an 
uncertainty of $10\%$ from varying the factorisation scale 
\cite{Stelzer:1998ni}, about twice the uncertainty of the total cross section. 
Work is in progress to calculate 
the differential cross section $d\sigma/dp_{T\bar b}$ at NLO 
with the goal of reducing this uncertainty \cite{Harris}. 
It would also be desirable to calculate 
the total cross section at next-to-next-to-leading order (NNLO). 
 
Additional theoretical uncertainties stem from the top quark 
mass and the parton distribution functions.  An uncertainty in the 
top quark mass of $2$ GeV yields an uncertainty of only $2\%$ in the 
cross section, which is negligible.  This is due to the fact that the cross 
section scales like $1/M_W^2$ rather than $1/s$.  The uncertainty in the cross 
section due 
to the parton distribution functions is estimated in~\cite{Huston:1998jj} 
to be $10\%$.  That analysis suggests that the uncertainty can be 
reduced below this value.  Combining all uncertainties in quadrature, we 
conclude that the total theoretical uncertainty is presently $15\%$ 
in the $Wbj$ cross section ($11\%$ in the total cross section).  The 
discussion above suggests that this can be significantly reduced with further 
effort. 
 
\subsubsection{Phenomenology} 
Studies of the $t$-channel process have been carried out 
by both ATLAS and CMS. We will first describe the 
CMS study, and then that of ATLAS. 
 
In order to reject the large $t\overline t$ background in this channel, 
it is necessary to impose a cut on jet multiplicity. 
Accurate modelling of jet response and resolution is therefore 
desirable, and so CMS \cite{cmsnote48} used a full GEANT 
calorimeter simulation of the detector. 
The GEANT simulation also allows a more realistic modelling of the 
missing-$p_T$ response of the detector, which is important in 
understanding the mass resolution which can be obtained on the 
reconstructed $t$ quark.  The detailed calorimeter simulation was 
combined with a parameterised $b$-tagging efficiency. 
 
Signal events were generated using \pyth~5.72 \cite{Sjostrand:1994yb}, 
with $m_t = 175$~GeV 
and the CTEQ2L parton distribution functions. Events were preselected at the 
generator level to have one and only one charged lepton 
(with $p_T > 25$~GeV and $|\eta|< 2.5$) and one or two jets 
(generator-level jets were found using the LUCELL 
clustering algorithm, which is part of \pyth). 
Generated events were then passed through the parameterised 
$b$-tagging and the GEANT detector simulation. 
The CMS $b$-tagging performance is taken from a study which used 
a detailed detector simulation combined with existing CDF data on 
impact-parameter resolutions.  The tagging efficiency for 
$p_T > 50$~GeV is typically 50\% for $b$-jets, 10\% for $c$-jets, 
and 1--2\% for light quarks and gluons.  These efficiencies fall 
quite rapidly for lower transverse momenta, and it was assumed no 
tagging could be performed for $p_T < 20$~GeV or $|\eta|> 2.4$. 
The generated 
luminosity corresponded to about 100~pb$^{-1}$ -- only 30 hours 
of running at $10^{33}{\rm cm}^{-2}{\rm s}^{-1}$. 
 
The $t\overline t$ and $WZ$ backgrounds were 
also generated using \pyth~ 5.72. 
The same pre-selections were applied at the generator level. 
The $W+$ jets backgrounds were generated using the 
VECBOS generator \cite{Berends:1991ax}, combined with \herw~ 5.6 
\cite{Marchesini:1992ch} 
to fragment the outgoing partons.\footnote{The version of VECBOS used here, 
and its interface to \herw, were developed for use in CDF~\cite{herprt}, 
and were adapted for CMS by R.~Vidal.} $W+2$ jets and  $W+3$ jets 
processes were generated separately.  Again, events were preselected 
to have a charged lepton with  $p_T > 25$~GeV and $|\eta|< 2.5$, and 
to have a (parton-level) $p_T > 15$~GeV for the final-state jets. 
 
Events were then selected which passed the following requirements: 
\begin{itemize} 
\item One and only one isolated lepton ($\ell = e$ or $\mu$) with 
$p_T > 20$~GeV 
and $|\eta| < 2.5$. This allows the events to pass a reasonable lepton 
trigger. 
\item Missing $p_T > 20$~GeV, and 
transverse mass (of the lepton and missing $p_T$) 
$50< m_T < 100$~GeV. 
These two requirements select $W \to \ell\nu$ candidates. 
\item Exactly two jets with  $p_T > 20$~GeV and $|\eta|< 4$. 
Requiring at least two jets reduces the $W+$ jets background, while 
requiring no more than two jets rejects the 
$t \overline t$ background which naively would produce four jets 
in the final state. 
\item One jet with $p_T > 20$~GeV and $|\eta|< 2.5$, 
the other jet with $p_T > 50$~GeV and $2.5 <|\eta|<4.0$. 
The requirement that the second jet be 
at forward rapidities 
tends to select the desired $t$-channel process. 
\item Leading jet $p_T < 100$~GeV. This helps to reduce the 
$t\overline t$ background. 
\item Exactly one $b$-tagged jet (given the $b$-tagging 
acceptance, this is always the central jet). This requirement 
again reduces $t\overline t$, and 
of course rejects $W+$ jets processes with light-quark or gluon 
jets. 
\item Invariant mass of the two jets in the $80-100$~GeV range. This rejects 
$WZ$ events with $Z \to b\overline b$. 
\end{itemize} 
The single-top signal is then searched for 
in the invariant mass of the $W$ and the $b$-tagged jet (which should peak 
at the top quark mass).  The mass was reconstructed assuming the 
solution for the $W$ kinematics which yields the lower $|p_z^\nu|$. 
(It is possible to use other choices, for example the solution which gives 
the $Wb$ mass closest to $m_t$.  This would result in an apparently 
better top mass resolution but would also severely bias the background 
shape; the statistical significance of the signal would not be improved.) 
 
%
%
\begin{figure}[t] 
\begin{center} 
\includegraphics*[bb=30 140 525 655,height=8cm]{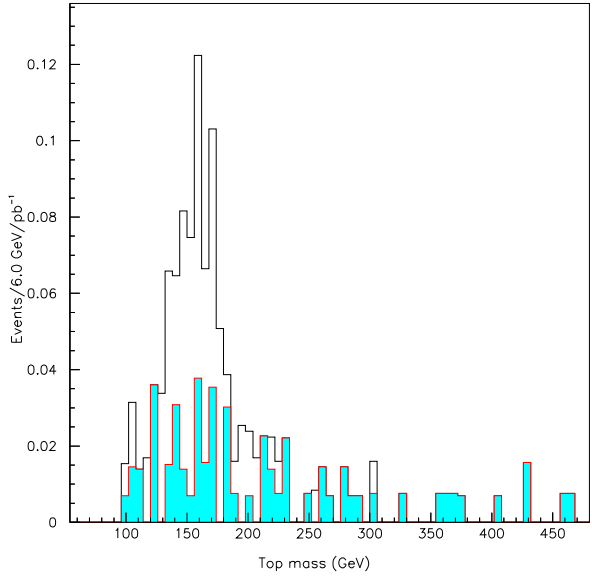} 
\end{center} 
\vskip -1cm 
\caption{Reconstructed top mass 
for signal plus backgrounds (open histogram) and backgrounds only (shaded). 
The backgrounds considered are $t\overline t$, $W+2$ jets and $W+3$ jets. 
The vertical scale is events per 6~GeV mass bin per pb$^{-1}$ 
of luminosity.\label{fig:Wbmass}} 
\end{figure} 
 
Figure~\ref{fig:Wbmass} 
shows the reconstructed mass distribution for signal and 
background combined.  The signal is apparent as an excess over the 
background (the shaded histogram) around 
160~GeV. (Since jet energy scale corrections 
have not been applied to the simulated events, 
the top mass reconstructs to less than its true value.) 
The signal-to-background ratio in a window of $160\pm 20$~GeV is 3.5 
with a clear peak visible in the $Wb$ invariant-mass distribution. 
The number of signal events is 66 in  100~pb$^{-1}$, giving a signal 
efficiency of 1.2\% (after the $W \to \ell\bar\nu$ branching ratio). We 
then find that 
10~fb$^{-1}$ would yield 6600 signal events ($S$) 
and 1900 background ($B$), sufficient for a statistical accuracy 
on the number of signal events of $\sqrt{S+B}/S = 1.4$\%. 
 
The largest background comes from $Wcj$ with the charm jet 
mistagged as a $b$-jet.  It would be worthwhile to develop a 
$b$-tagging algorithm having greater rejection against such mistags, 
even at the cost of some signal efficiency. 
The $Wb\bar b$ background was found to be a small contribution 
to the $W + 2$ jets background at the parton level for the selection cuts 
employed here, and was therefore not explicitly included in the analysis. 
 
The use of the forward jet tag substantially 
improves the signal-to-background ratio, and allows a clear 
reconstructed top-mass peak to be seen. However, it does not 
significantly improve $\sqrt{S+B}/S$ \cite{Stelzer:1998ni}. 
One could therefore 
imagine omitting the forward jet requirement if the systematic uncertainty 
could thereby be reduced. 
 
Compared with earlier studies (for example~\cite{Stelzer:1998ni}), 
this analysis uses more realistic jet and mis\-sing-$p_T$ resolutions, 
and includes initial- and final-state 
gluon radiation.   As a result, the top-mass resolution 
is worsened; but the resolution found here compares well 
with the result of a full simulation of single-top production in CDF. 
 
 
A study of the cross-section measurement for the $t$-channel process was 
also carried out by ATLAS \cite{Pineiro}. Signal events were 
generated using the ONETOP 
parton-level Monte Carlo \cite{Carlson:1995cm} with 
fragmentation, radiation, and underlying event simulated by 
\pyth~5.72. Backgrounds containing top quarks ($t\bar t$ and other 
single-top production) were also generated using ONETOP, while $W+$ jets 
and $Wb\bar b$ backgrounds were generated by \herw~ 5.6.\footnote{The 
$Wb\bar b$ background was generated using the matrix element  
from~\cite{Mangano:1993kp} interfaced to \herw~ 5.6.} These events 
were processed by the ATLAS parameterised detector simulation assuming 
a 60\% $b$-tagging efficiency for $b$-jets, 10\% for $c$-jets, and 1\% for 
light quarks and gluons. The events were then analysed with a view towards 
separating $t$-channel single top from background and measuring its 
cross section. 
 
Event selection criteria were divided into two types: pre-selection 
and selection cuts. The pre-selection criteria were as follows: 
 
\begin{itemize} 
\item at least one isolated lepton with $p_T > 20$ GeV; 
\item at least one $b$-tagged jet with $p_T > 50$ GeV; 
\item at least one other jet with $p_T > 30$ GeV. 
\end{itemize} 
 
These were followed by the selection cuts: 
 
\begin{itemize} 
\item two and only two jets in the event (a jet has $p_T > 15$ GeV); 
\item one jet is a central $b$-tagged jet; 
\item the other jet 
is a forward ($|\eta|>2.5$) untagged jet with $p_T > 50$~GeV. 
\end{itemize} 
 
The application of these cuts, and also the requirement of a 
reconstructed top mass between 150 and 200 GeV, yields the number 
of events shown in Table~\ref{tab:t-channelcuts}. The final signal 
efficiency is 3\% and the signal-to-background ratio is 2.4. This 
implies a statistical precision on the cross-section measurement 
of $\sqrt{S+B}/S=0.9$\% with 10~fb$^{-1}$ of data. Introducing 
other event selection variables (see~\cite{atlasphystdr,Oneil,Ahmedov})  
it is possible to improve 
the signal-to-background ratio to nearly 5, but this does not 
improve the cross-section measurement due to the small remaining 
signal efficiency. 
 
\begin{table}[tb] 
\begin{center} 
\caption{Cumulative effect of cuts on $t$-channel signal and 
  backgrounds. The first four rows of this table refer to cumulative 
  efficiencies of various cuts. The last two rows refer to the number 
  of events for 10~fb$^{-1}$. Only events in which $W\rightarrow$ 
  e$\nu$ or $\mu\nu$ are considered in this table. Uncertainties quoted in 
  this table are due entirely to Monte Carlo statistics. 
\label{tab:t-channelcuts}} \vspace*{0.1cm} 
\begin{tabular}{|c| c| c| c| c|} 
\hline 
cut        &  $t$-channel & $t\bar t$ & $Wb\bar b$ & $W+$ jets\\ 
           & eff(\%) & eff(\%) & eff(\%)&eff(\%) \\ \hline \hline 
 
pre-selection   &18.5  &44.4 & 2.53 & 0.66\\ \hline 
 
njets=2    &12.1  &0.996  &1.55 &0.291 \\ \hline 
 
\begin{tabular}{c} 
fwd jet      \\ 
$|\eta|>2.5$ \\ 
$p_T >$50~GeV 
\end{tabular} 
&4.15    & 0.035  & 0.064  & 0.043 \\ \hline 
 
\begin{tabular}{c} 
M$_{\ell \nu b}$ \\ 
150-200~GeV 
\end{tabular} 
& 3.00 &   0.017 & 0.023 & 0.016 \\ \hline 
\hline 
\begin{tabular}{c} 
events/10~fb$^{-1}$\\ 
(before cuts) 
\end{tabular} 
  &$5.43\times 10^{5}$ & $2.40\times 10^6$ & $6.67\times 10^5$&$4.00\times 
10^{7}$ \\ \hline 
 
\begin{tabular}{c} 
events/10~fb$^{-1}$\\ 
(after cuts) 
\end{tabular} 
&16515$\pm$49 &455$\pm$74 &155$\pm$17  &6339$\pm$265 \\ \hline 
\end{tabular} 
\end{center} 
\end{table} 
 
Both the CMS and ATLAS studies indicate that it will be 
possible to observe $t$-channel single-top production 
with a good signal-to-background ratio and a statistical 
uncertainty in the cross section of less than 2\% with 10~fb$^{-1}$. 
Thus the uncertainty in the extracted value of $V_{tb}$ will 
almost certainly be dominated by systematic uncertainties, as 
discussed in the conclusions. 
 
\subsection{s-channel single-top production} 
\subsubsection{Theory} 
The $s$-channel production of single top quarks is shown in 
Fig.~\ref{fig:singletop}(b) \cite{Cortese:1991fw,Stelzer:1995mi, 
Heinson:1997zm,Smith:1996ij,Stelzer:1998ni,Belyaev:1999dn}. The 
cross section is much less than that of the $t$-channel process 
because it scales like $1/s$ rather than $1/M_W^2$. However, the 
$s$-channel process has the advantage that the quark and antiquark 
distribution functions are relatively well known, so the 
uncertainty from the parton distribution functions is small. 
Furthermore, the parton luminosity can be constrained by measuring 
the Drell-Yan process $q\bar q\to W^* \to \ell\bar \nu$, which has 
the identical initial state 
\cite{Stelzer:1995mi,Dittmar:1997md}.\footnote{The parton 
luminosity can only be constrained, not directly measured, with 
this process.  Since the neutrino longitudinal momentum is 
unknown, the $q^2$ of the virtual $W$ cannot be reconstructed.} 
 
The total cross section for the $s$-channel process has been 
calculated at NLO \cite{Smith:1996ij}; the 
result is given in Table~\ref{tab:sigma}.  The factorisation and 
renormalisation scales are set equal to $\mu^2 = q^2$; varying 
each, independently, between one-half and twice its central value 
yields uncertainties in the cross section of $2\%$ from each 
source.  The uncertainty in the cross section from the parton 
distribution functions is estimated to be $4\%$.  The largest 
single source of uncertainty is the top quark mass; an uncertainty 
of $2$ GeV yields an uncertainty in the cross section of $5\%$. 
The relatively large sensitivity of the cross section to the 
top quark mass is a manifestation of the $1/s$ scaling. Combining 
all theoretical uncertainties in quadrature yields a total 
uncertainty in the cross section of $7\%$.  This is much less than 
the present theoretical uncertainty in the $t$-channel cross 
section. 
 
The Yukawa correction to this process, of order $\alpha_Wm_t^2/M_W^2$, is less 
than one percent \cite{Smith:1996ij}. However, this correction could be 
significant in a two-Higgs-doublet model for low values of $\tan\beta$, 
in which the Yukawa coupling is enhanced \cite{Li:1997bh}. 
 
\subsubsection{Phenomenology} 
In order to evaluate the potential to separate the $s$-channel signal 
from its backgrounds, Monte Carlo events have been processed by a 
fast (parameterised) simulation of an LHC detector. At parton level 
the signal and the $t\bar t$ background were generated by the ONETOP 
Monte Carlo \cite{Carlson:1995cm}. Radiation, showering, and the 
underlying event were added by \pyth~ 5.72 \cite{Sjostrand:1994yb}. 
The $W+$ jets and $Wb\bar b$ 
backgrounds were generated using \herw~ 5.6 
\cite{Marchesini:1992ch}.\footnote{The 
$Wb\bar b$ background was generated using the matrix element  
from~\cite{Mangano:1993kp} interfaced to \herw~ 5.6.} 
Table~\ref{tab:sigma} presents the cross sections assumed for 
the processes containing top quarks. The cross section for the $W+$ jets 
background is normalised to that predicted by the VECBOS Monte 
Carlo \cite{Berends:1991ax} 
and is taken to be 18000~pb.\footnote{This cross section is defined for 
events containing at least two jets, each with $p_T > 15$ GeV 
and $|\eta| < 5$.} 
The $Wb\bar b$ cross section is taken from~\cite{Stelzer:1998ni}  
to be 300~pb. 
 
From a phenomenological standpoint the most important distinction between 
the $s$-channel and $t$-channel sources of 
single top is the presence of a second high-$p_T$ $b$-jet in the 
$s$-channel process. As mentioned previously, in $t$-channel events the 
second $b$-jet tends to be at low $p_T$ and is often not 
seen. 
Therefore, requiring two $b$-jets above 75~GeV $p_T$ will 
eliminate most of the $t$-channel background. Requiring two high-$p_T$ 
$b$-jets in the event also suppresses the $W+$ jets background relative to 
the signal. 
 
In addition to suppressing the $t$-channel background it is also 
necessary, as in other single-top signals, to design cuts to reduce 
the $W+$ jets and $t\bar t$ backgrounds. In order to reduce 
contamination by $W+$ jets events, the reconstructed top mass in each 
event must fall within a window about the known top mass (150-200~GeV), 
and the events must have a total transverse jet 
momentum\footnote{Scalar sum of the transverse momentum of all jets in 
the event.} above 175~GeV. Only events containing exactly two jets 
(both tagged as $b$'s) are kept in order to reduce the $t\bar t$ 
background. 
 
\begin{table} 
\begin{center} 
\caption{Cumulative effect of cuts on $s$-channel signal and 
backgrounds. The first five rows of this table refer to cumulative 
  efficiencies of various cuts. The last two rows refer to the number 
  of events for 30~fb$^{-1}$. Only events in which 
W$\rightarrow$ 
  e$\nu$ or $\mu\nu$ are considered in this table. Uncertainties quoted in 
  this table are due entirely to Monte Carlo statistics. 
\label{tab:wstarcuts}} \vspace*{0.1cm} 
\begin{tabular}{|c| c| c| c| c| c| c|} 
\hline 
cut        &   $s$-channel &  $t$-channel & $Wt$ & $t\bar t$ & 
$Wb\bar b$ & $W+$ jets\\ 
           & eff(\%) & eff(\%) & eff(\%)&eff(\%) &eff(\%) & eff (\%) \\ 
           \hline\hline 
 
pre-selection    &27.0  &18.5  &25.5 &44.4 &2.53 &0.667 \\ \hline 
 
njets=2    &18.4  &12.1  &4.03 &0.996 &1.55 &0.291 \\ \hline 
 
\begin{tabular}{c} 
nbjet=2 \\ 
$p_T>75$ GeV 
\end{tabular} 
  & 2.10 & 0.035 &0.018&0.023 &0.034& 0.0005 \\  \hline 
 
\begin{tabular}{c} 
$\sum^{\mbox{{\scriptsize jets}}}p_T$ \\ 
$>$175~GeV 
\end{tabular} 
& 1.92 & 0.031&0.016&0.021 &0.028& 0.0005\\ \hline 
 
\begin{tabular}{c} 
M$_{\ell \nu b}$ \\ 
150-200~GeV 
\end{tabular} 
&1.36&0.023&0.006&0.012 &0.0097 & 0.00014 \\ \hline 
\hline 
 
\begin{tabular}{c} 
events/30~fb$^{-1}$\\ 
(before cuts) 
\end{tabular} 
&$6.66\times 10^4$  &$1.63\times 10^6$  &$4.5\times 10^5$ 
&$6.9\times 10^6$ & $2.0 \times 10^6$ &$1.2\times 10^8$ \\ \hline 
 
\begin{tabular}{c} 
events/30~fb$^{-1}$\\ 
(after cuts) 
\end{tabular} 
&$908\pm 35$ &$375\pm 13$ &$27\pm 15$ &$853\pm 175$ 
&$194\pm 34$ &$169\pm 76$    \\ \hline 
\end{tabular} 
\end{center} 
\end{table} 
 
Table~\ref{tab:wstarcuts} presents the cumulative effect of all 
cuts on the $s$-channel signal and on the backgrounds. Events from $t$-channel 
single-top production are included in this table as a background to the 
$s$-channel process. 
From this table the predicted signal-to-background ratio for the $s$-channel 
signal is calculated to be 0.56. The results also imply a signal statistical 
significance ($S/\sqrt{B}$) of 23 with an integrated luminosity of 
30~fb$^{-1}$.  The statistical precision on the cross section, 
calculated from $\sqrt{S+B}/S$, is 5.5\% with 30~fb$^{-1}$. 
 
This study indicates that, despite the large anticipated background 
rate, it should be possible to perform a good statistical measurement 
of the $s$-channel single-top cross section.  The accuracy 
with which $V_{tb}$ can be measured is discussed in the conclusions. 
 
\subsection{Associated production} 
\subsubsection{Theory} 
Single top quarks may also be produced in association with a $W$ 
boson, as shown in Fig.~\ref{fig:singletop}(c) 
\cite{Heinson:1997zm,Tait:2000cf,Belyaev:1999dn}. Like the 
$t$-channel process, one of the initial partons is a $b$ quark. 
However, unlike the $t$-channel process, this process scales like 
$1/s$. This, combined with the higher values of $x$ needed to 
produce both a top quark and a $W$ boson, leads to a cross section 
for associated production which is about a factor of five less 
than that of the the $t$-channel process, despite the fact that it 
is of order $\alpha_s\alpha_W$ rather than $\alpha_W^2$. 
 
The total cross section for associated production has been 
calculated at leading order, with a subset of the 
NLO corrections included 
\cite{Tait:2000cf,Belyaev:1999dn}; the result is given in 
Table~\ref{tab:sigma}. This subset is analogous to the 
initial-gluon correction to the $t$-channel process, discussed 
previously.  The other corrections have not yet been 
evaluated.\footnote{The analogous calculation for $Wc$ production 
has been performed in~\cite{Giele:1996kr}.}  The 
initial-gluon correction contains an interesting feature which has 
no analogue in the $t$-channel process.  One of the contributing 
diagrams to the initial-gluon correction ($gg \to Wt\bar b$) 
corresponds to $gg \to t\bar t$, followed by $\bar t\to W\bar b$. 
This should not be considered as a correction to associated 
production, but rather as a background (it is in fact the dominant 
background, as discussed below). Thus, when evaluating the 
initial-gluon correction, it is necessary to subtract the 
contribution in which the $\bar t$ is on shell. This is done 
properly in~\cite{Tait:2000cf}. 
 
The cross section is evaluated with the common factorisation and 
renormalisation scales set equal to $\mu^2 = s$.  The uncertainty in the 
cross section due to varying these scales between one half and twice 
their central value is $15\%$.  This uncertainty would presumably 
be reduced with a full NLO calculation. 
The uncertainty in the cross section 
from the parton distribution functions is estimated to 
be $10\%$ \cite{Huston:1998jj},\footnote{This is the uncertainty in the 
gluon-gluon luminosity at $\sqrt \tau = (m_t+M_W)/\sqrt S \approx 0.02$, 
where $\sqrt S = 14$ TeV.} 
although this could be improved with further study. 
The uncertainty in the cross section due to an uncertainty in the top quark 
mass of $2$ GeV is $4\%$, relatively large due to the $1/s$ scaling 
of the cross section.  Combining all theoretical uncertainties 
in quadrature yields a total uncertainty at present of $18\%$, the largest 
of the three single-top processes. 
 
\subsubsection{Phenomenology} 
The strategy for measuring the cross section for associated production 
($Wt$ mode) is similar to that for the $t$-channel process, as they 
share the same backgrounds. However, the nature of 
associated production makes it relatively easy to separate from $W+$ jets 
and difficult to separate from $t\bar t$ events. This difficulty 
in removing the $t\bar t$ background does not preclude obtaining a 
precise cross-section measurement in this channel, assuming the rate for 
$t\bar t$ can be well measured at the LHC. 
 
Two studies designed to separate signal from background have been 
performed using two different final states. The first is a study by 
ATLAS~\cite{atlasphystdr} which attempts to isolate $Wt$ signal 
events in which one $W$ decays to jets and the other decays to 
leptons. The second study, which is presented  
in~\cite{Tait:2000cf}, attempts to isolate signal events in which both 
$W$'s decay leptonically. 
 
\begin{figure} 
\includegraphics[width=\textwidth]{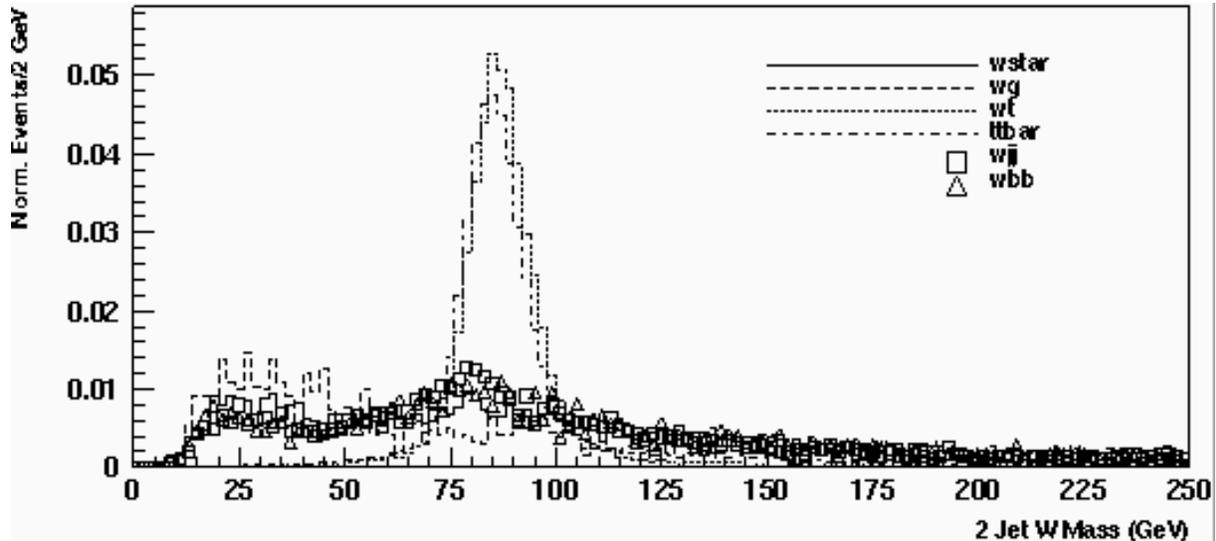} 
\vskip -0.5cm 
\caption[The two-jet invariant-mass distribution.] 
{The normalised two-jet invariant-mass distribution. For each event the 
two-jet combination with mass closest to the $W$ mass is plotted. This 
clearly shows a peak in the distribution for $Wt$ and $t\bar t$ which is 
not present for the other backgrounds. \label{fig:2jetmass}} 
\end{figure}

The first study presented here was done by ATLAS using the same 
event sample described in Section~\ref{sec:onetoptch}. 
Since the presence of a single isolated high-$p_T$ lepton is one 
of the preconditions of this study, the second $W$ must decay to 
two jets to be accepted by the event pre-selection. Therefore 
requiring a two-jet invariant mass within a window around the $W$ 
mass will serve to eliminate most events that do not contain a 
second $W$. The two-jet invariant-mass distribution is shown in 
Fig.~\ref{fig:2jetmass} and clearly demonstrates the presence of a 
sharp peak in the associated-production signal and the $t\bar t$ 
background. This effectively leaves $t\bar t$ as the only 
background to $Wt$ events. 
 
In addition to these special distinguishing features of the $Wt$ signal, 
there are several simple kinematic requirements which can be employed 
to reduce the $t\bar t$ background. By choosing events with exactly three 
jets and with exactly one of them tagged as a $b$-jet, some rejection of the 
$t\bar t$ background is possible. Some further rejection is obtained by 
limiting the selection to events with invariant mass less than 
300~GeV, where the invariant mass of an event is defined as the invariant mass 
obtained by adding the four-vectors of all reconstructed jets and charged 
leptons ($e$ and $\mu$). 
However, even with these cuts the $t\bar t$ background is 
significantly larger than the $Wt$ signal. 
 
Table~\ref{table-wtcuts} presents the cumulative effect of all 
cuts on the $Wt$ signal and on the $t\bar t$ and $W+$ jets backgrounds. The 
$Wb\bar b$ and $t$-channel single-top 
backgrounds are virtually eliminated by the cuts and so are not included 
in the table. 
From this table the predicted signal-to-background ratio for the $Wt$ 
signal is calculated to be 0.24. After three years of running at low 
luminosity (30~fb$^{-1}$), 
this implies a signal statistical significance ($S/\sqrt{B}$) of 25 and 
a statistical error on the $Wt$ cross section ($\sqrt{S+B}/S$) of 4.4\%. 
 
\begin{table} 
\begin{center} 
\caption{Cumulative effect of cuts on $Wt$ signal and 
  backgrounds. Pre-selection cuts are defined in the same way as for 
the ATLAS $t$-channel analysis described earlier in this report. 
The first five rows of this table refer to cumulative 
  efficiencies of various cuts. The last two rows refer to the number 
  of events for 30~fb$^{-1}$. Only events in which 
  W$\rightarrow$ 
  e$\nu$ or $\mu\nu$ are considered in this table. Uncertainties quoted in 
  this table are due entirely to Monte Carlo 
  statistics.\label{table-wtcuts}}\vspace*{0.1cm} 
\begin{tabular}{|c| c| c| c|} 
\hline 
cut        &  $Wt$   & $t\bar t$ & $W+$ jets   \\ 
           & eff(\%)&eff(\%) &eff(\%)     \\ \hline \hline 
 
pre-selection    & 25.5   &44.4    &0.66  \\ \hline 
 
\begin{tabular}{c} 
njets=3    \\ 
$p_T>50$ GeV 
\end{tabular} 
& 3.41   &4.4     &0.030       \\ \hline 
 
\begin{tabular}{c} 
nbjet=1    \\ 
$p_T>50$ GeV 
\end{tabular} 
&3.32        &3.24        &0.028            \\ \hline 
 
\begin{tabular}{c} 
Invariant Mass \\ 
$<300$ GeV 
\end{tabular} 
&0.55 &0.36    &0.00051     \\ \hline 
 
$65<M_{jj}<95$&0.49 &0.14 &0.000085 \\ \hline 
 
  \multicolumn{4}{c}{}\\ \hline 
 
\begin{tabular}{c} 
events/30~fb$^{-1}$ \\ 
(before cuts) 
\end{tabular} 
& 5.3$\times 10^5$ & 7.2$\times 10^{6}$ & 1.2$\times 10^{8}$\\ \hline 
 
\begin{tabular}{c} 
events/30~fb$^{-1}$\\ 
(after cuts) 
\end{tabular} 
& $2608\pm 166$   & $10616\pm 625$  & $102\pm 59$  \\ \hline 
\end{tabular} 
\end{center} 
\end{table} 
 
The second study~\cite{Tait:2000cf} was done at parton level and 
involved the separation of signal from background in the mode in which 
both $W$'s decay to leptons.  This signal contains two high $p_T$ 
leptons and only one jet (the $b$-jet produced from the top decay). In 
this decay channel it was found that, after applying detector 
acceptance cuts, requiring precisely one $b$-tagged jet with $p_T 
> 15$ GeV is enough to yield a signal-to-background ratio of nearly 
unity. Also, the signal efficiency is significantly higher than in the 
ATLAS analysis, allowing more total signal events to pass the cuts 
despite the lower branching ratio for this decay mode.  The 
statistical precision on the cross section measured in this analysis 
is 1.3\% with an integrated luminosity of 30~fb$^{-1}$.  The accuracy with 
which $V_{tb}$ can be extracted is discussed in the conclusions. 
 
\subsection{Polarisation in single-top production} 
\subsubsection{Theory} 
Because single top quarks are produced through the weak interaction, they 
are highly polarised \cite{Heinson:1997zm,Carlson:1993dt,Mahlon:1997pn, 
Mahlon:1999gz,Stelzer:1998ni}. 
In the ultra-relativistic limit, the top quarks are produced in 
helicity eigenstates with helicity $-1/2$ (the top antiquarks have helicity 
$+1/2$), because the $V-A$ structure of the weak interaction selects 
quarks of a definite chirality.  However, if the top quarks are 
not ultra-relativistic, chirality is not the same as helicity. 
Nevertheless, it 
was shown in~\cite{Mahlon:1997pn} that there is a basis in which 
the top quark is $100 \%$ polarised, regardless of its energy.  The 
top quark spin points along the direction of the $d$-type or $\bar d$-type 
quark in the event, 
in the top quark rest frame (the $\bar t$ spin points 
opposite this direction).  In $t$-channel production, 
this is the direction 
of the final-state light quark ($ub \to dt$) or the beam direction 
($\bar d b \to \bar u t$).  In $s$-channel production, 
this is the beam direction ($u\bar d \to t\bar b$).  In associated production 
($gb\to Wt$), 
this is the direction of the $d$ quark (or charged lepton) from the $W$ decay. 
 
We focus our attention on the $t$-channel single-top process for the 
remainder of this section.  The top quark polarisation in the $t$-channel 
process has been calculated at NLO \cite{Mahlon:1999gz}; 
the results below are taken from this study. 
In the case of $t$ production, $80\%$ of the events have the $d$-type 
quark in the final state.  This suggests using the direction of the light-quark 
jet, as observed in the top quark rest frame, to measure the spin. 
This has been dubbed the ``spectator basis'' 
\cite{Mahlon:1997pn}.  The polarisation of the top quark in 
this basis (defined as $P = (N_\uparrow - N_\downarrow)/ 
(N_\uparrow + N_\downarrow)$) is $0.89$.  However, the polarisation 
is increased to nearly $100\%$ when the cuts used in the $t$-channel 
analysis are imposed.  This is because the polarisation is diluted 
by events in which the 
$\bar b$ is produced at high $p_T$; but such events are eliminated 
by the requirement of only two jets. 
 
In the case of $\bar t$ production, $69\%$ of the events have the $d$-type 
quark in the initial state.   This suggests using the beam direction 
to measure the $\bar t$ spin.  However, it turns out that the spectator 
basis again yields the largest polarisation, $P=-0.87$.  This polarisation is 
increased to $P=-0.96$ when cuts are applied.\footnote{With cuts applied, 
the polarisation in the so-called ``$\eta$-beamline basis'' is slightly 
higher, $P=-0.97$.} 
 
Since the top quark decays via the weak interaction, its spin is analysed 
by the angular distribution of its decay products.  The most sensitive 
spin analyser in top decay is the charged lepton, which has a  
(leading order) angular 
distribution with respect to the top quark spin of 
\begin{equation} 
\frac{1}{\Gamma}\frac{d\Gamma}{d\cos\theta_{\ell}} =  
\frac{1}{2}(1+\cos\theta_{\ell}) 
\end{equation} 
in the top quark rest frame \cite{Jezabek:1989ja}.  Hence 
the charged lepton tends to point along the direction of the spectator jet. 
 
\subsubsection{Phenomenology} 
The goal of this analysis is to estimate the sensitivity of ATLAS and 
CMS to the measurement of the polarisation of the top quarks produced 
by the $t$-channel single-top process. The $t$-channel process was chosen 
due to the large statistics available in this channel and the relative 
ease with which it is separated from its backgrounds. The $t$-channel events 
produced by the ONETOP generator and passed through \pyth~ and a 
parameterised detector simulation are analysed to attempt to recover 
the predicted SM top polarisation in the presence of 
background and detector effects. Details of the study are presented  
in~\cite{Pineiro,Oneil}.

The experimental measurement of the polarisation of the top quark is 
essentially a measurement of the angular distribution of its decay 
products in the top quark rest frame. As explained above, 
the most sensitive angle is between the charged lepton from top decay and 
the direction of the spectator jet, in the top quark rest frame. 
In the absence of background or detector effects the angular 
distribution of the charged lepton is given by 
\begin{equation} 
\label{eq:polar_dist} 
f(\cos\theta_{\ell}) = \frac{1}{2} (1 + P\cos\theta_{\ell}) 
\end{equation} 
where $P$ is the polarisation of the sample and can range from $-1$ to 
$1$. 
 
Experimentally, in order to measure the angular distribution of the 
charged lepton in the top quark rest frame, it is necessary to first 
reconstruct the four-momentum of the top quark.  However, 
the reconstruction of the top four-momentum suffers from an 
ambiguity due to the unknown longitudinal momentum of the neutrino produced in 
the top decay. Using the $W$ and top masses as constraints,\footnote{The 
$W$ mass can be used to calculate the neutrino longitudinal momentum to 
within a two-fold ambiguity. Of these two solutions the one which produces the 
best top mass is chosen.} one can reconstruct the top four-momentum, 
but the quality of the reconstruction is degraded by this 
ambiguity. Once the top four-momentum has been reconstructed, one can 
determine the direction of the spectator jet and the charged lepton 
in the top quark rest frame.  The angle between these two directions is 
$\theta_{\ell}$. 
 
In order to extract the value of the top polarisation from the angular 
distribution, reference event samples were created with 100\% 
alignment with the polarisation axis (spin up, $P=+1$) and with 100\% 
anti-alignment with the polarisation  
axis (spin down, $P=-1$). These reference 
distributions were compared to a statistically-independent data set 
with the predicted SM top quark polarisation. This 
comparison was done by minimising 
\begin{equation} 
\label{eq:polar-chi2} 
\chi^{2} = 
\sum_{(\cos\theta)_{i}} 
\frac{(f_{\rm th}(\cos\theta_{\ell})_{i} - f_{\rm d} 
(\cos\theta_{\ell})_{i})^{2}} 
{\sigma_{{\rm th}_{i}}^{2} + \sigma_{{\rm d}_{i}}^{2}} 
\end{equation} 
where the subscript d represents quantities calculated for the data 
distribution and the subscript th refers to the generated reference 
distribution.  The theoretical value 
$f_{\rm th}(\cos\theta_{\ell})$ is 
calculated via 
\begin{equation} 
f_{\rm th}(\cos\theta_{\ell}) = \frac{1}{2} ((1 - P)f_{D}(\cos\theta_{\ell}) + 
(1 + P) f_{U}(\cos\theta_{\ell})) 
\end{equation} 
where $f_{D}$ and $f_{U}$ refer to the value of the 
generated theoretical distribution for the 100\% spin-down and the 
100\% spin-up tops, respectively, and $P$ is the polarisation 
of the top sample. The procedure returns an 
estimate of the top polarisation and an error on that estimate. In this 
way the sensitivity to changes in top polarisation can be quantified. 
 
Moving from the parton-level simulation to a simulation which includes 
both hadronisation and detector effects is certain to complicate the 
measurement of the polarisation of the top quark. In addition, the 
signal could be biased by an event selection 
designed to eliminate background and will be contaminated by residual 
background events. 
 
\begin{figure}[t]
\begin{center} 
\includegraphics[height=0.4\textheight]{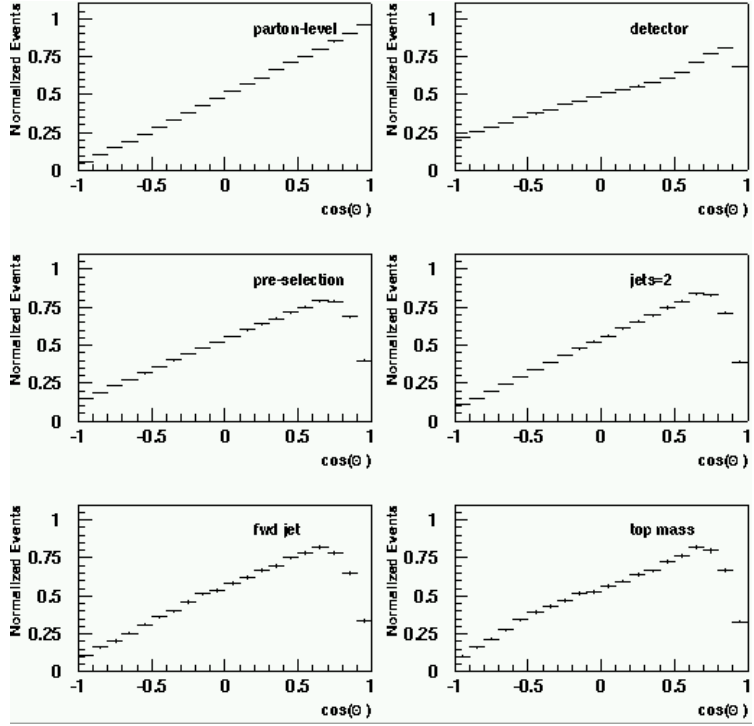} 
\caption[Effect of detector and cuts on signal angular distribution.] 
{Angular distribution of charged lepton in top rest frame for various 
  data samples. The histograms progress from left-to-right, 
  top-to-bottom. The first histogram shows the parton-level 
  distribution. The second histogram is after the simulation of 
  detector and reconstruction effects. The final four histograms 
  illustrate the influence of event selection criteria on the angular 
  distribution. The effects of the cuts are cumulative and are the 
  result of adding pre-selection cuts, a jet-multiplicity requirement, 
  a forward jet tag, and a top mass window, respectively. 
  \label{fig:phen_polar_sig_cuts}} 
\end{center} 
\end{figure}

The first histogram in Fig.~\ref{fig:phen_polar_sig_cuts} shows 
the angular distribution for signal only, at parton-level. The 
second histogram in Fig.~\ref{fig:phen_polar_sig_cuts} shows the 
angular distribution of the charged lepton after detector effects 
have been simulated. In addition to effects associated with 
detector energy smearing, jet and cluster definitions, {\it etc.}, 
this distribution includes the effects of ambiguities in 
reconstructing the top quark due to the absence of information 
about the neutrino longitudinal momentum. It does not, however, 
contain the effects of any event selection in order to separate 
signal from background.  This histogram demonstrates that the 
effect of hadronisation and detector resolution changes the shape 
of the angular distribution but still produces a highly asymmetric 
distribution. 
 
In addition to the effects introduced by the detector resolution, the 
effect of applying the event-selection criteria can be evaluated by 
applying them one at a time and observing the change in shape of this 
distribution. For the purposes of the polarisation analysis the 
event-selection criteria are: 
\begin{itemize} 
 \item{Pre-selection (trigger) cuts as in ATLAS $t$-channel analysis 
described previously;} 
 \item{number of jets = 2;} 
 \item{forward jet ($|\eta|>2.5$) with $p_T >50$ GeV;} 
 \item{reconstructed top mass in the range 150--200~GeV.} 
\end{itemize} 
This set of criteria leads to a signal efficiency of 3.0\%, corresponding 
to more than 16000 events in 10~fb$^{-1}$ of integrated luminosity. 
Fig.~\ref{fig:phen_polar_sig_cuts} demonstrates the effect of applying 
these cuts in a cumulative manner. Again the asymmetry of the $t$-channel 
angular distribution is preserved, though more degradation is clearly 
evident, in particular near $\cos\theta_{\ell} = 1$. The degradation is 
worse at these values of $\cos\theta_{\ell}$ because the leptons from 
these events are emitted in the direction opposite to the top 
boost. This reduces the momentum of the leptons causing more of them 
to fail $p_T$-based selection criteria. 

\begin{figure} 
\begin{center} 
\includegraphics[height=0.4\textheight]{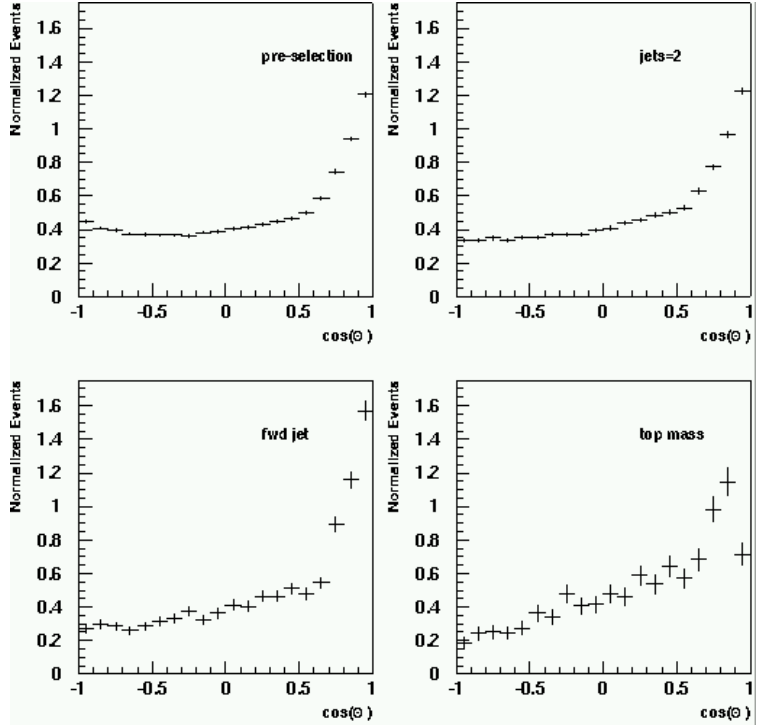} 
\caption[Effect of cuts on Wjj angular distribution.]{The effect of event 
selection cuts on the angular 
distribution of the charged lepton in $Wjj$ events. The effects of the 
cuts are cumulative. The first distribution is the result of applying 
the pre-selection (trigger) cuts only. Further cuts are applied 
cumulatively from left-to-right, 
top-to-bottom.\label{fig:phen_polar_wjj_cuts}} 
\end{center} 
\end{figure}

Since $W+$ jet events dominate the background remaining after cuts, 
they are taken as the only background in this analysis. 
Fig.~\ref{fig:phen_polar_wjj_cuts} shows the cumulative effect of cuts on 
the angular distribution of the charged lepton from $W+$ jets events. A 
peculiar feature of these events is evident in all of these 
distributions. This is the tendency for events to be grouped near 
$\cos\theta_{\ell} = 1$. 
The events which populate this region 
tend to be the highest $p_T$ events. This shows that 
even basic jet and isolated-lepton definitions and pre-selection cuts 
bias the angular distribution of $W+$ jets events. 

\begin{figure} 
\begin{center} 
\includegraphics[width=0.35\textheight]{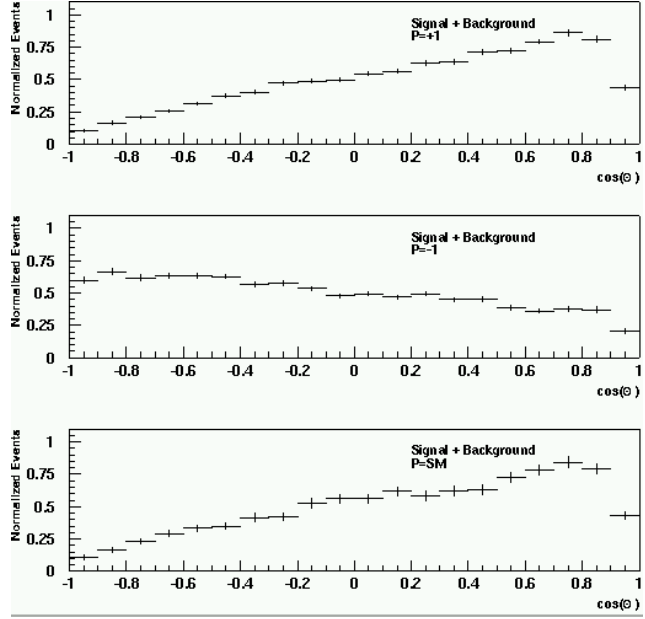} 
\caption[Angular distributions after cuts and addition of 
background.]{The first histogram shows the reference 
distribution for 100\% spin-up top quarks after detector effects 
and event-selection criteria have been applied and the appropriate 
level of background has been mixed in. The second histogram shows the 
reference distribution for 100\% spin-down top quarks. The third 
histogram represents the expected SM distribution for a 
statistically-independent sample of signal and background. 
\label{fig:phen_polar_template}} 
\end{center} 
\end{figure}

When the event-selection criteria described in the previous 
sections are applied, the signal-to-background ratio (treating 
$W+$ jets as the only background) is found to be 2.6. Using the 
methods described earlier it is possible to estimate the 
polarisation of a mixed sample of $t$-channel signal and $W+$ jets 
background. The reference distributions for 100\% spin-down and 
100\% spin-up top quarks mixed with background in a ratio of 2.6 
are shown in Fig.~\ref{fig:phen_polar_template}. Also shown is the 
angular distribution corresponding to a statistically-independent 
data sample with SM polarisation mixed with background 
in the ratio 2.6. 
The $\chi^2$ function presented in~(\ref{eq:polar-chi2}) is 
minimised to obtain an estimate of the polarisation of the top.  To 
estimate the precision for one year of data-taking, the fit 
was done with 3456 signal events and 1345 background events, 
corresponding to 2~fb$^{-1}$ of integrated luminosity 
($\sim$ 1/5 of a year). For this integrated luminosity the 
error on the polarisation measurement is 4.0\%. Then, assuming the 
statistics on the reference distributions, $f_{D}(\cos\theta_{\ell})$ and 
$f_{U}(\cos\theta_{\ell})$, will lead to a negligible source of error, 
this precision improves to 3.5\%. Projecting these results to 
one year of data-taking at low luminosity (10~fb$^{-1}$), 
assuming that the errors scale as the square root of 
the number of events, yields a predicted statistical  
precision of 1.6\% on the 
measurement of the top polarisation. 
 
\subsection{Conclusions on single top production} 
As mentioned in the introduction, single-top quark production is the only 
known way to directly measure $V_{tb}$ at a hadron collider.  In this section 
we estimate the accuracy with which $V_{tb}$ can be extracted at the LHC, 
and discuss what will be required to achieve that accuracy. 
 
There are four sources of uncertainty in the extraction  
of $|V_{tb}|^2$ from 
the single-top 
cross section: theoretical, experimental, statistical, and 
machine luminosity.  As we have seen, the statistical uncertainty with 
30~fb$^{-1}$ of integrated luminosity is less than $2\%$ for both the 
$t$-channel process\footnote{Only 10~fb$^{-1}$ are required to 
achieve this accuracy in the $t$-channel process.} 
and associated production, and is $5.5\%$ for the 
$s$-channel process ($3\%$ with 100~fb$^{-1}$).  It will be a challenge 
to reduce the other sources 
of uncertainty to $5\%$, so we regard the statistical accuracy as 
being sufficient in all three processes. 
 
The traditional uncertainty in the machine luminosity is about $5\%$ 
\cite{Abbott:1999tt}.  It may be possible to reduce the uncertainty below 
this value using Drell-Yan data, but this relies on 
accurate knowledge of the quark distribution functions.  However, the 
process $q\bar q \to W^* \to \ell\bar\nu$ involves the identical combination 
of parton distribution functions as the $s$-channel process, so it can be 
used to almost directly measure the relevant parton luminosity, thereby 
avoiding the need to measure the machine luminosity \cite{Dittmar:1997md}. 
 
The theoretical uncertainty is under the best control in the $s$-channel 
process. The theoretical uncertainty is dominated by 
the uncertainty in the top quark mass; an uncertainty of $2$ GeV 
yields an uncertainty of $5\%$.  This is cut in half if the uncertainty 
in the top mass is reduced to $1$ GeV. The small uncertainty due to 
variation of the factorisation and renormalisation scales can be reduced to 
a negligible amount by calculating the cross section at NNLO 
order, which should be possible in the near future.  The small uncertainty 
from the parton distribution functions can be further reduced 
as described in the previous paragraph; this also obviates the need 
for a measurement of the machine luminosity. 
 
The theoretical uncertainty in the $t$-channel process is presently dominated 
by the factorisation-scale dependence and the parton luminosity.  Although the 
scale dependence of the total cross section is small 
($4\%$), the uncertainty in the semi-inclusive cross section 
($\sigma(p_{T\bar b}) < 20$ GeV) 
is about $10\%$.  This can be reduced 
by calculating the $p_T$ spectrum of the $\bar b$ at NLO. 
It may also prove possible to measure the total cross 
section, although this has yet to be demonstrated.  It is therefore 
plausible that the factorisation-scale dependence will be about $5\%$ 
once the LHC is operating.  It is also likely that the uncertainty from the 
parton distribution functions will be reduced below its present value 
of $10\%$.  The parton luminosity could be directly measured using $Wj$ 
production, which is 
dominated by $gq \to Wq$, and therefore involves the identical combination of 
parton distribution functions as the $t$-channel process. 
Again, this has the desirable feature of 
eliminating the need to measure the machine luminosity. 
 
The theoretical uncertainty in the associated-production cross section 
can be reduced far below its present value of $18\%$.  A full 
NLO calculation should reduce the factorisation-scale 
dependence to roughly $5\%$. It is likely that the uncertainty from 
the parton distribution functions will also be reduced.  Unless it is 
possible to measure the $gg$ luminosity directly, the uncertainty 
from the parton distribution functions will be augmented by the uncertainty 
in the machine luminosity. 
 
As far as experimental systematic uncertainties are concerned, the extraction 
of a signal cross section requires knowledge of the backgrounds 
and of the efficiency and acceptance for the signal. 
These analyses require hard cuts 
on both signal and background, and 
so the processes need to be modelled and understood very well. 
 
For all of these processes, the major backgrounds are $t\overline t$ and 
$W+$ jets. The largest background for the $s$-channel process (where a 
double $b$-tag is employed) and associated production is 
$t\overline t$.  The $t\bar t$ process can be isolated 
in other decay modes and in principle well measured. 
In the $t$-channel process 
the biggest background comes from $Wcj$ with the charm jet 
mistagged as a $b$-jet.  Obviously it would be worthwhile to develop a 
$b$-tagging algorithm having greater rejection against such mistags, 
even at the cost of some signal efficiency, given that the signal rate 
is large.  It may be possible 
to understand the $W+$jets backgrounds by 
comparing with a sample of $Z+$ jets events after applying similar 
selections to those used to select the single-top sample in $W+$ jets. 
The $Z+$ charm rate will be suppressed compared to 
the $W+$ charm rate since the latter is mostly produced from the 
strange sea, which is bigger than the charm sea; nonetheless, 
the cross section, kinematics, jet multiplicities and so on can all be 
compared to our simulations using the $Z+$ jets sample. 
 
The forward jet tag is very effective in enhancing the 
signal-to-background ratio in the $t$-channel process. 
This means that jets need to be found with 
good efficiency up to large rapidities, at least $|\eta|\sim 4$ in the 
calorimeter.  Unfortunately these observations also imply that the 
background estimate is very sensitive to the Monte Carlo predicting 
the correct mix of jet flavours and jet rapidities in the $W+$ jets 
events.  (We note that VECBOS generates very few jets in the tagging 
region, and so far there is no collider data on forward jets in vector-boson 
events which could verify whether this is correct.) 
Of course, effort applied to understanding $W+$ heavy-flavour jets 
backgrounds will pay off in many other searches besides this one, and 
will be a very worthwhile investment.  We also look forward to the 
results of ongoing efforts to improve the Monte Carlo simulation of 
vector-boson plus jet production \cite{Ellis:1998fv}. 
Requiring exactly two jets 
(as was done here to reject the $t\overline t$ background) also means 
that we will be very sensitive to our knowledge of jet 
efficiencies, QCD radiation, {\it etc}. 
The cross-section measurement also requires knowledge of the $b$-tagging 
efficiency.  This should be measurable at the few-percent level 
using control samples of $t\overline t$ events selected with 
kinematic cuts alone. 
 
As mentioned above, the purely statistical uncertainty in the 
cross-section measurement will be less than 5\%, as will most of the 
theoretical uncertainties. 
It will be a considerable challenge to reduce the experimental 
systematic uncertainty to this level. At the present time, the experimental 
systematic uncertainty in the $t\overline t$ cross section at the Tevatron 
(which is a similar challenge in many respects, involving 
jets, $b$-tagging, and background subtraction) is about $19\%$ 
\cite{Abbott:1998nn}. This total is made up of many components 
which are each at the 5\% 
level, so while it will be a lot of work to reduce 
them, there is no obvious ``brick wall'' that would prevent this. 
 
Many of these systematic issues can also be addressed by comparing the 
$t$-channel and $s$-channel single-top processes.  It will be a 
powerful tool to be able to measure $V_{tb}$ in two channels 
which have different dominant backgrounds, different selection 
cuts, and a different balance between theoretical and experimental 
systematic uncertainties. 
 
We are only just now entering the era of precision top physics 
with Run~II at the Tevatron.  Single-top production has not 
yet even been observed.  We will learn a great deal over 
the next few years about how to model top events and their 
backgrounds, and how to understand the systematic uncertainties.  The LHC 
will undoubtedly benefit from all this experience. 
 
If all sources of uncertainty are kept to the $5\%$ level or less, it should 
be possible to measure $|V_{tb}|^2$ to $10\%$ or less.  We therefore regard 
the measurement of $V_{tb}$ with an accuracy of $5\%$ or less as an 
ambitious but attainable goal at the LHC.  We have also seen that a 
measurement of the polarisation of single top quarks produced via the 
$t$-channel process will be possible with a statistical accuracy of $1.6\%$ 
with 10~fb$^{-1}$.  We have not attempted to estimate the systematic 
uncertainty in this measurement. 
 
\section{$\bf t\bar t$ SPIN CORRELATIONS AND CP 
VIOLATION\protect\footnote{Section 
    coordinators: W.~Bernreuther, A.~Brandenburg, V.~Simak (ATLAS),  
    L.~Sonnenschein (CMS).}} \label{TTSPIN} 
\def\rshat{\sqrt{\hat{s}}} 
\def\sp{{\mathbf s}_t} 
\newcommand{\sm}{{\mathbf s_{\bar t}}} 
\newcommand{\kt}{{\hat{\mathbf k}_t}} 
\newcommand{\ph}{\hat{\mathbf p}} 
\newcommand{\eins}{  1\!{\rm l}  } 
\def\vec#1{\mathchoice{\mbox{\boldmath$\mathrm\displaystyle#1$}} 
{\mbox{\boldmath$\mathrm\textstyle#1$}} 
{\mbox{\boldmath$\mathrm\scriptstyle#1$}} 
{\mbox{\boldmath$\mathrm\scriptscriptstyle#1$}}} 
\newcommand{\bm}[1]{\mbox{\boldmath$#1$}} 
\renewcommand{\vec}{\bm} 
\newcommand{\ktb}{\mathbf{\hat{k}}_{\bar t}} 
\newcommand{\lp}{\mathbf{\hat{q}}_+} 
\newcommand{\lm}{\mathbf{\hat{q}}_-} 
\newcommand{\qb}{\mathbf{\hat{q}}_b} 
 
For $t\bar{t}$ production at the LHC quantities associated with the 
spins of the top and antitop quark will be ``good'' observables as 
well. The reason for this is well known. Because of its extremely short 
lifetime $\tau_t$ (see Section~\ref{sec:topprop})  
the top quark decays before it can form hadronic 
bound states.  Thus the information on the spin of the top quark does 
not get diluted.  As the spin-flip time is much larger than $\tau_t$ 
it is, moreover, very unlikely that the top quark changes its 
spin-state by emitting gluon(s) via a chromomagnetic dipole transition 
before it decays. In any case this amplitude is calculable with QCD 
perturbation theory.  Hence by measuring the angular distributions and 
correlations of the decay products of $t$ and $\bar t$ the 
spin-polarisations and the spin-spin correlations that were imprinted 
upon the $t\bar{t}$ sample by the production mechanism can be 
determined and compared with predictions made within the SM or its 
extensions. Therefore these spin phenomena are an additional important 
means to study the fundamental interactions involving the top quark. 
\par 
In this section we are concerned with the production and decay of top-antitop 
pairs. At the LHC the main $t{\bar t} $ production process is gluon-gluon 
fusion, $q{\bar q}$ annihilation being sub-dominant. As the main SM decay mode is  
$t\to W^+ b$ we shall consider here the parton reactions 
\begin{equation} 
gg,  q {\bar q}\rightarrow  t{\bar t} + X \rightarrow  b{\bar b} + 4f + X, 
\label{eq:ttrec} 
\end{equation} 
where $f$ denotes either a quark, a charged lepton or a neutrino. If the final 
state in (\ref{eq:ttrec}) contains two, one,  
or no high $p_T$ charged lepton(s) then we  
call these reactions, as usual, the di-lepton, single lepton, and 
non-leptonic $t\bar{t}$ decay channels, respectively. To lowest order QCD 
the matrix elements for (\ref{eq:ttrec}), including the complete $t\bar{t}$ 
spin correlations and the effects of the finite top and $W$ widths, 
were given in \cite{Kleiss:1988,Barger:1989}. Spin correlation effects 
in  $t\bar{t}$ production in hadron collisions were  
studied within the SM in 
\cite{Barger:1989,Hara:1991yq,Brandenburg:1996,Stelzer:1996gc,Mahlon:1996zn,Mahlon:1997,Chang:1996}. 
\par 
In order to discuss the top  spin-polarisation and correlation 
phenomena that are to be expected at the LHC it is useful to employ the 
narrow-width approximation for the $t$ and $\bar{t}$ quarks. Because 
$\Gamma_t/m_t \ll 1$ one can write, to good approximation, 
the squares of the exact 
Born matrix elements ${\cal M}^{(\lambda)}$, $\lambda=gg,q{\bar q}$, 
in the form 
\begin{equation} 
\mid {\cal M}^{(\lambda)}{\mid}^2 \propto {\rm Tr}\;[\rho R^{(\lambda)}{\bar{\rho}}] 
\equiv\rho_{\alpha'\alpha} 
R^{(\lambda)}_{\alpha\alpha',\beta\beta'}{\bar{\rho}}_{\beta'\beta} . 
\label{eq:trace} 
\end{equation} 
The complete spin information is contained 
in the (unnormalised) spin density matrices $R^{(\lambda)}$ 
for  the production of on-shell $t\bar{t}$ pairs and in the 
density matrices $\rho,{\bar\rho}$ for the decay of polarised $t$ and $\bar t$ 
quarks into the above final states. The trace in (\ref{eq:trace}) is to be taken 
in the $t$ and $\bar t$ spin spaces. 
The decay density matrices will be discussed below. 
The matrix structure of $R^{(\lambda)}$  is 
\be 
R^{(\lambda)}_{\alpha\alpha',\beta\beta'} = 
A^{(\lambda)}\delta_{\alpha\alpha'}\delta_{\beta\beta'} 
+B^{(\lambda)}_{i} (\sigma^i)_{\alpha\alpha'} 
\delta_{\beta\beta'} +{\bar B}^{(\lambda)}_{i} \delta_{\alpha\alpha'} 
(\sigma^i)_{\beta\beta'}  
+\, C^{(\lambda)}_{ij}(\sigma^i)_{\alpha\alpha'} 
(\sigma^j)_{\beta\beta'}  \, , 
\label{eq:Rlam} 
\ee 
where $\sigma^i$ are the Pauli matrices. Using rotational invariance the 
``structure functions''  $B^{(\lambda)}_i,{\bar B}^{(\lambda)}_i$ and 
$C^{(\lambda)}_{ij}$ can be 
further decomposed. A general discussion of the symmetry properties of 
these functions 
is given in \cite{Bernreuther:1994}.  The function 
$A^{(\lambda)}$, which determines the $t\bar{t}$ cross section, is known in 
QCD at NLO~\cite{Nason:1988xz}. Because 
of parity (P) invariance  the vectors   
${\bf B}^{(\lambda)},{\bf\bar B}^{(\lambda)}$ can have,  
within QCD,  only a component normal to the scattering plane. This component, 
which amounts to a normal polarisation of the $t$  quark, 
${\cal P}^{t}_{\bot}$, is induced by the absorptive part of the  
respective scattering amplitude 
and was computed for the above LHC processes to 
order $\alpha^3_s$ \cite{Bernreuther:1996}.   
(${\cal P}^{t}_{\bot}={\cal P}^{\bar t}_{\bot}$  
if CP invariance holds.) The size of the normal polarisation depends on the 
top quark scattering angle  and on the  c.m. energy. 
In the  gluon-gluon fusion process ${\cal P}^{t}_{\bot}$ reaches peak values 
of about 1.5$\%$. 
In $t\bar{t}$ production at the LHC the polarisation of the top 
quark within the partonic scattering plane, which is P-violating, is  
small as well within the SM. Therefore the $t$ and $\bar t$ 
polarisations in the scattering plane are  good observables 
 to search for P-violating 
non-SM interactions in the reactions (\ref{eq:ttrec}) -- see Section 3.4. 
\par 
The $t\bar{t}$ production by the strong interactions leads, on the 
other hand, to a significant correlation between the $t$ and ${\bar 
  t}$ spins.  This correlation is encoded in the functions 
$C^{(\lambda)}_{ij}$. Using P- and charge-conjugation (C) invariance 
they have, in the case of a $t\bar{t}$ final state, the structure 
\begin{equation} 
 C^{(\lambda)}_{ij} = c^{(\lambda)}_1\delta_{ij} + c^{(\lambda)}_2 
{\hat p}_i{\hat p}_j + c^{(\lambda)}_3 
{\hat k}_{ti}{\hat k}_{tj} + c^{(\lambda)}_4 
({\hat k}_{ti}{\hat p}_j + {\hat p}_i{\hat k}_{tj}) , 
\label{eq:cij} 
\end{equation} 
where $\ph$ and $\kt$ are the directions of flight of the 
initial quark or gluon and of the $t$ quark, respectively, in 
the parton c.m. frame. So far the functions $c^{(\lambda)}_r$ are known to 
lowest-order QCD only (see, e.g., \cite{Brandenburg:1996}). 
For a  $t\bar{t} X$ final state a decomposition similar to (\ref{eq:cij}) 
can be made. 
\par 
From (\ref{eq:cij}) one may read off the following set of spin-correlation 
observables \cite{Brandenburg:1996}: 
\begin{equation} 
(\kt\cdot\sp)(\ktb\cdot\sm), 
\label{eq:hbasis} 
\end{equation} 
\begin{equation} 
(\ph\cdot\sp)(\ph\cdot\sm), 
\label{eq:pbasis} 
\end{equation} 
\begin{equation} 
\sp\cdot\sm, 
\label{eq:sbasis} 
\end{equation} 
\begin{equation} 
(\ph\cdot\sp)(\ktb\cdot\sm) + (\ph\cdot\sm)(\kt\cdot\sp), 
\label{eq:phbasis} 
\end{equation} 
where $\sp,\sm$ are the $t$ and $\bar t$ spin operators, respectively.  
The observables 
(\ref{eq:hbasis}), (\ref{eq:pbasis}), and  (\ref{eq:phbasis})  
determine the correlations 
of different $t,\bar t$ spin projections. 
Eq. (\ref{eq:hbasis}) corresponds to a correlation of the $t$ and  
$\bar t$ spins 
in the helicity basis, while (\ref{eq:pbasis}) correlates the spins projected 
along the beam line. We note that the ``beam-line basis'' defined 
 in \cite{Mahlon:1996zn} refers to spin axes being parallel to the 
left- and right-moving beams  in the $t$ and $\bar t$ rest frames,  
respectively. 
The $t\bar{t}$ spin correlation in this basis is a linear combination of 
(\ref{eq:hbasis}), (\ref{eq:pbasis}), 
 and (\ref{eq:phbasis}). 
\par 
A natural question is: what is -- assuming only SM interactions -- 
 the best spin basis 
or, equivalently, the best observable for investigating the $t\bar{t}$ spin  
correlations? 
For quark-antiquark 
annihilation,  which is the dominant production process at the Tevatron,  
it turns out that the spin  
correlation (\ref{eq:pbasis}) \cite{Brandenburg:1996,Chang:1996} 
and the correlation in the beam-line basis 
\cite{Mahlon:1996zn} is stronger  than the correlation in the helicity basis. 
In fact, for ${\bar q}q$ annihilation  
a spin-quantisation axis was constructed in \cite{Mahlon:1997} 
with respect to which the $t$ and $\bar t$ 
spins are 100$\%$ correlated.  
At the LHC the situation is different. For $gg\to t\bar{t}$ at threshold  
conservation 
of total angular momentum dictates that the $t\bar{t}$ is in a $^1\!S_0$  
state. Choosing 
spin axes parallel to the right- and left-moving beams this means that we  
have $t_L{\bar t}_L$ 
and $t_R{\bar t}_R$ states at threshold. On the other hand at very high  
energies helicity conservation 
leads to the dominant production of unlike helicity pairs  $t_R{\bar t}_L$  
and  $t_L{\bar t}_R$. 
One can show that no  
spin quantisation axis exists for $gg\to t\bar{t}$ with respect to which  
the $t$ and $\bar t$ spins are  
100$\%$ correlated.  The helicity basis is a good choice, but one can do  
better. This is  
reflected in the above observables. Computing their expectation values and 
statistical fluctuations one finds  \cite{Brandenburg:1996} that 
(\ref{eq:sbasis}) has a higher statistical significance than the helicity 
correlation (\ref{eq:hbasis}) which in turn is more sensitive  
than (\ref{eq:pbasis}) 
or the correlation in the beam-line basis. 
\par 
The spins of the $t$ and $\bar t$ quarks are to be inferred 
from their P-violating weak decays, i.e., from 
$t\to b W^+\to b\ell^+\nu_{\ell}$ or 
$b q{\bar q}'$  and likewise for $\bar t$ if only SM interactions are relevant. 
As already mentioned and used in previous sections, in this case 
the charged lepton from $W$ decay is the best analyser of the 
top spin. This is seen by considering the decay distribution 
of an ensemble of polarised $t$ quarks decaying into a particle $f$  
(plus anything) with respect to the 
angle between the polarisation vector  ${\mathbf\xi}_t$ of the top quark  
and the 
direction of  flight ${{\bf{\hat q}}_f}$ of the particle $f$ in the $t$ rest  
frame. 
This distribution has the generic form 
\begin{equation} 
\frac{1}{\Gamma}\frac{d\Gamma}{d\cos\theta_f} = \frac{1}{2} 
(1 + {\kappa}_f {\mathbf\xi}_t\cdot{{\bf{\hat q}}_f} ) , 
\label{eq:kappaf} 
\end{equation} 
where the magnitude of the coefficient ${\kappa}_f$ signifies the  
spin-analyser quality of $f$.  
The  SM values for some $f$, collected from 
\cite{Czarnecki:1991,Bernreuther:1992be,Ma:1992ry,Jezabek:1994zv}, 
are given in Table~\ref{tab:kappaf}. 
\begin{table} 
\caption{ 
Correlation coefficient ${\kappa}_f$ for $V-A$ charged current. In the 
last column l.e.j. stands for least energetic jet in the $t$ rest 
frame.} 
\label{tab:kappaf} 
\begin{center} 
\begin{tabular}{c|c|c|c|c|c}\hline 
$f$ & $\ell^+, {\bar d}, {\bar s}$ & $\nu_{\ell}, u, c$ & $b$ & $W^+$ & 
l.e.j. 
from $q{\bar q}'$ \\ \hline 
${\kappa}_f$ & 1 & $-$0.31 & $-$0.41 & 0.41 & 0.51 \\ \hline  
\end{tabular} 
\end{center} 
\end{table} 
The corresponding 
$t$ decay density matrix in the $t$ rest frame is  
read off from (\ref{eq:kappaf}) to be 
$\rho_{\alpha'\alpha} 
= (\eins +{\kappa}_f\,\vec{\sigma} \cdot {{\bf{\hat q}}_f})_{\alpha'\alpha}$. 
The distributions for the decay of polarised antitop quarks are 
obtained  by replacing ${\kappa}_f \to -{\kappa}_f$ 
in (\ref{eq:kappaf}). The order $\alpha_s$ QCD corrections to the decays 
 $t\to b\ell\nu$ and $t\to Wb$ of polarised $t$ quarks 
were computed in \cite{Czarnecki:1991} and \cite{Schmidt:1996}, 
respectively. For $t,\bar t$ polarisation observables these corrections  
are small. 
\par 
From the above table it is clear that the best way to analyse the 
$t\bar t$ spin correlations is 
through angular correlations among the two charged leptons 
$\ell^+\ell'^-$ in the di-lepton final state.  
Using the production and decay density matrices in (\ref{eq:trace}), 
 neglecting 
the 1-loop induced QCD normal polarisation, and integrating over the 
azimuthal angles of the charged leptons 
 one obtains the following normalised double distribution, e.g.  
in the helicity basis 
\begin{equation} 
\frac{1}{\sigma}\frac{d^2\sigma}{d\cos\theta_+ d\cos\theta_-}  = \frac{1 +  
C\kappa_{\ell^+}\kappa_{\ell^-} 
\cos\theta_+ \cos\theta_-}{4}\,\, , 
\label{eq:ddist} 
\end{equation} 
where $\kappa_{\ell^+}\kappa_{\ell^-}=-1$ and $\theta_+(\theta_-)$ is 
the angle between the $t(\bar t)$ direction in the $t\bar t$ 
c.m. frame and the $\ell^+({\ell^-})$ direction of flight in the 
$t(\bar t)$ rest frame. The coefficient $C$, which is the degree of 
the spin correlation in the helicity basis, results from the 
$c^{(\lambda)}_i$ in (\ref{eq:cij}) and it is related to 
\cite{Stelzer:1996gc}: 
\begin{equation} 
C=\frac{N(t_L{\bar t}_L+t_R{\bar t}_R)-N(t_L{\bar t}_R+t_R{\bar t}_L)} 
{N(t_L\bar{t}_L+t_R{\bar t}_R)+N(t_L{\bar t}_R+t_R{\bar t}_L)}\,\, . 
\label{eq:Cdegree} 
\end{equation} 
For partonic final states and to lowest order in $\alpha_s$ 
one gets $C=0.332$ for the LHC. (The number depends somewhat on the 
parton distributions used. Here and below the set CTEQ4L 
\cite{Lai:1997mg} was used.)  The optimum would be to find a spin axis 
with respect to which $|{C}|=1$.  But, as stated above, this is not 
possible for $gg$ fusion.  In addition to (\ref{eq:ddist}), analogous 
correlations among $\ell^+$ from $t$ and jets from $\bar t$ decay (and 
vice-versa) in the single lepton channels, and jet-jet correlations in 
the non-leptonic decay channels should, of course, also be studied. 
While the spin-analysing power is lower in these cases, one gains in 
statistics.  \par From the above example is quite obvious that, for a 
given $t\bar{t}$ decay channel, the $t\bar t$ spin correlation will 
be most visible when the angular correlations among the $t$ and $\bar 
t$ decay products are exhibited in terms of variables defined in the 
$t$ and $\bar t$ rest frames.  An important question is therefore how 
well the 4-momenta of the $t$ and $\bar t$ quarks can be reconstructed 
experimentally?  We briefly discuss the results of a simulation of the 
single lepton and di-lepton channels \cite{Simak:1999} which includes 
hadronisation and detector effects using  
\pyth~\cite{Sjostrand:1994yb} and the ATLFAST~\cite{atl:atlfast} 
software packages.  The transverse momentum of every reconstructed 
object like a jet, a charged lepton, or the missing transverse energy 
of an event has to exceed a certain minimum value $p^{min}_T$.  The 
detector acceptances impose further restrictions on the phase space of 
the detected objects in pseudo-rapidity.  \par In the case of the 
single lepton $t\bar{t}$ decay channels one isolated lepton 
($e^{\pm}$ or $\mu^{\pm}$) is required.  From the missing transverse 
energy of the event and the $W$ mass constraint the longitudinal 
momentum $p_z$ of the neutrino can be determined up to a twofold 
ambiguity.  It turns out that in most cases the lower solution of 
$p_z$ is the correct one. To complete the event topology, four jets 
are demanded.  Two of them have to be identified as $b$-jets coming from 
top 
decay. 
 
The 
two non-tagged jets are often misidentified due to additional activity 
in the detector from initial and final state radiation.  To suppress 
the QCD background the invariant mass of the two jets has to lie in a 
narrow mass window around the known mass of the $W$ boson. After this 
cut the two-jet system is rescaled to the $W$ mass.  Finally there is 
a twofold ambiguity when the $b$-jets are combined with the 
reconstructed $W$ bosons.  The combination which yields the lower 
reconstructed top mass turns out to be the correct one most of the 
time.  \par In the case of the di-lepton decay channels two isolated 
oppositely charged leptons are requested. Moreover two jets have to be 
detected and tagged as $b$-jets.  With the known top and $W$ masses and 
with the missing transverse energy of the event the unknown 3-momenta 
of the neutrino and anti-neutrino can be computed using the kinematic 
constraints of the event. These result in a system of two linear and 
four quadratic equations.  The equations can be solved numerically and 
usually several solutions arise.  Since the experimentally determined 
momenta do not coincide with the corresponding variables at the parton 
level the kinematic constraints have to be relaxed somewhat in order 
to improve the reconstruction efficiency.  The algorithm set up in 
\cite{Simak:1999} was used to solve these equations. The best solution 
can be obtained by computing weights from known distributions. 
Following \cite{Simak:1999} the highest efficiency was obtained using 
the weight given by the product of the energy distributions of 
$\nu_{\ell}$ and ${\bar\nu}_{\ell}$ and the $\cos\theta^*_t$ 
distribution in the $t\bar t$ c.m. frame.  \par
\begin{figure}[t] 
\unitlength 1cm 
\begin{minipage}[t]{7.8cm} 
\begin{picture}(6,7.7) 
\put(-0.8,-0.5){ 
\includegraphics[width=10.1cm]{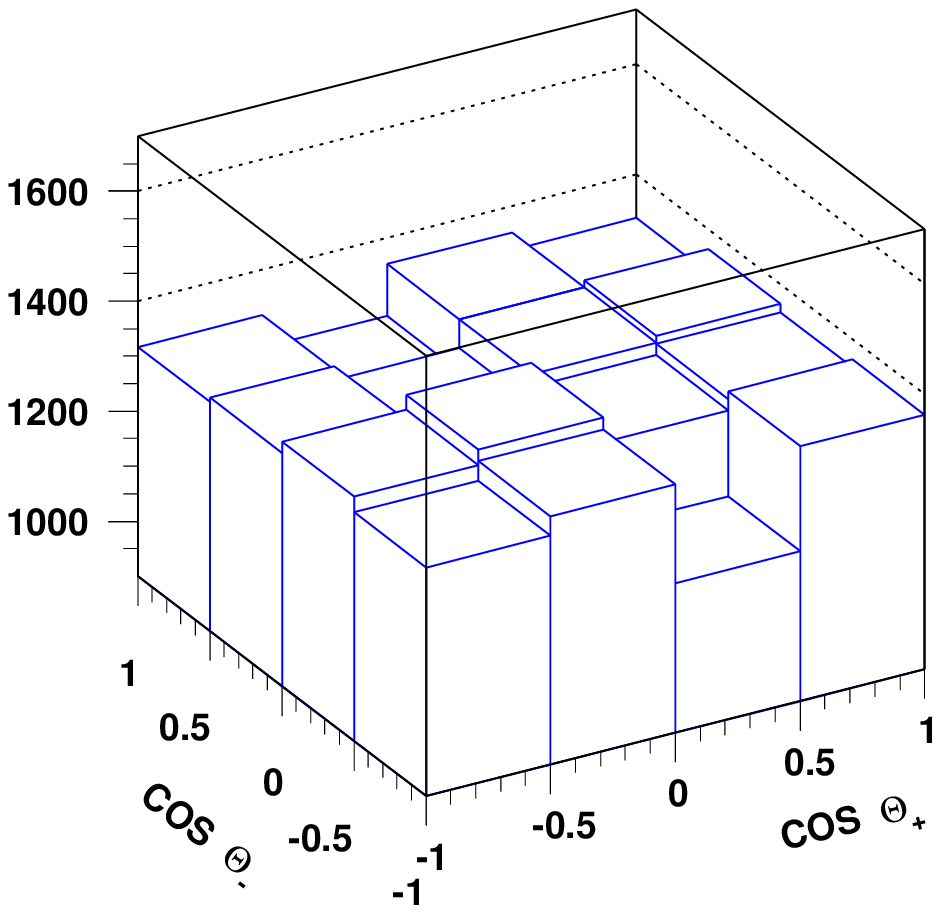} } 
\end{picture} 
\caption{\label{fig:recnospincorr} Joint distribution 
 $d^2N/d\cos\theta_+ d\cos\theta_-$  generated with default \pyth.  
The detector response was simulated with CMSJET.} 
\end{minipage} 
\hfill 
\begin{minipage}[t]{7.8cm} 
\begin{picture}(6,7.7) 
\put(-0.8,-0.5){ 
\includegraphics[width=10.1cm]{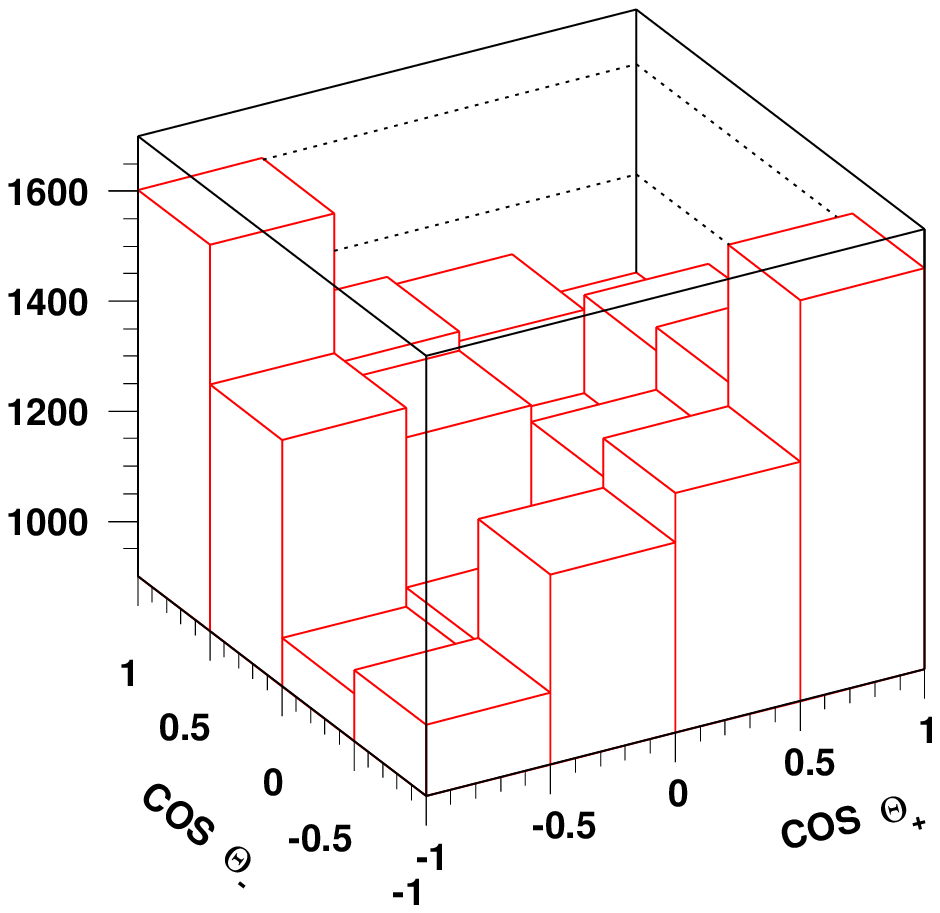} } 
\end{picture} 
\caption{\label{fig:recspincorr} Same distribution as in  
the figure to the left, but 
including the SM $t\bar{t}$ spin correlations. The 
 detector response was simulated with CMSJET.} 
\end{minipage} 
\end{figure}

 For the LHC running at 
low luminosity (${\cal{L}}=10^{33}\,\mbox{cm}^{-2}\mbox{s}^{-1}$), 
about $4\times 10^5$ $t\bar{t}$ events per year are expected in the 
di-lepton decay channels ($\ell = e,\mu)$. A further simulation of 
these channels was performed in order to study the joint distribution 
(\ref{eq:ddist}).  \pyth~5.7 \cite{Sjostrand:1994yb} was used for the 
event generation, CMSJET \cite{Abdullin:1999} for the detector 
response and the algorithm of \cite{Simak:1999} for the reconstruction 
of the $t,\bar t$ momenta. The transverse momenta of the two isolated, 
oppositely charged leptons and of the two jets were required to exceed 
$20\,\mbox{GeV}$. The minimal missing transverse energy of the event 
was chosen to be $40\,\mbox{GeV}$.  A further selection criterion was 
that each jet provides at least two tracks with a significance of the 
transverse impact parameter above $3.0$ to be tagged as $b$-jet.  The 
processes were simulated in two different ways.  First the SM matrix 
elements of \cite{Bernreuther:1998gs} for the reactions 
(\ref{eq:ttrec}), which contain the $t\bar{t}$ spin correlations, 
were implemented into \pyth. For comparison these channels were also 
simulated with the \pyth~ default matrix elements for $gg, q\bar{q}\to 
t\bar{t}$ which do not contain spin correlations. In both simulations 
initial and final state radiation, multiple interactions, and the 
detector response was included.  In Figs.~\ref{fig:recnospincorr}, 
\ref{fig:recspincorr} we have plotted the resulting double 
distributions $d^2N/d\cos\theta_+ d\cos\theta_-$.  They have been 
corrected for the distortions of the phase space due to the cuts. 
A fit to the distribution Fig.~\ref{fig:recspincorr}  
according to (\ref{eq:ddist}) 
yields the correlation coefficient  $C=0.331\pm 0.023$, in  
agreement with the value $C=0.332$ obtained at the parton level without cuts. 
A fit to Fig.~\ref{fig:recnospincorr} leads to 
$C=-0.021\pm 0.022$ consistent with $C=0$. 
Systematic errors, for instance due to background processes, e.g.,  
$Z^*\to \ell^+\ell^-$ accompanied by two $b$-jets, remain to be investigated. 
\par 
From these double distributions one may  form one- or zero-dimensional 
projections, for instance asymmetries as 
considered in \cite{Mahlon:1996zn,Stelzer:1996gc,Chang:1996}. 
Another approach is to study distributions and  expectation values of angular  
correlation observables which would be zero in the absence of 
the $t\bar{t}$ 
spin correlations. 
A suitable set of observables is obtained by transcribing, for instance, 
the spin observables given above  into correlations involving the directions 
of flight of those final state particles that are used to analyse  
the $t$ and $\bar t$ spins. As an example we 
discuss the case of the single lepton channels 
$t\to b q{\bar q}',{\bar t}\to {\bar b}\ell^-{\bar\nu}_{\ell}$. One may choose 
to analyse the $t$ spin by the direction of flight  $\qb^*$ of the $b$-jet in the 
rest frame of the $t$ quark and the $\bar t$ spin  
by the momentum direction $\lm$ of the $\ell^-$ in the laboratory frame. 
The latter is rather conservative in that no reconstruction of the $\bar t$ 
momentum is necessary. 
Then  (\ref{eq:hbasis})-(\ref{eq:sbasis}) are translated into the observables 
\begin{eqnarray} 
{\cal O}_1 &=& (\qb^*\cdot{\ph}_p)(\lm\cdot{\ph}_p),\\ 
{\cal O}_2 &=& (\qb^*\cdot\kt)(\lm\cdot\kt),\\ 
{\cal O}_3 &=& \qb^*\cdot\lm , 
\label{eq:spinob} 
\end{eqnarray} 
where ${\ph}_p$ refers to the beam direction.  
 The pattern of statistical 
sensitivities of the spin observables (\ref{eq:hbasis})-(\ref{eq:sbasis})  
stated above 
is present also in these angular correlations.  
Computing the expectation values $\langle {\cal O}_i\rangle$ 
and the statistical fluctuations $\Delta{\cal O}_i$ 
 and those of the observables for the corresponding 
charge conjugated channels, one gets for the 
statistical significances of these observables  
at the parton level \cite{Brandenburg:1996}: 
${\cal S}_1\approx 0.007 \sqrt{N_{b\ell^-}}$, 
${\cal S}_2\approx 0.025 \sqrt{N_{b\ell^-}}$, and 
${\cal S}_3\approx 0.055 \sqrt{N_{b\ell^-}}$, where $N_{b\ell^-}$  
is the number 
of reconstructed events in the specific single lepton channel. The linear 
combination  
\begin{equation} 
{\cal O}_4={\cal O}_3-{\cal O}_1  
\label{eq:corr4} 
\end{equation} 
has a still higher sensitivity 
than ${\cal O}_1$, namely ${\cal S}_3\approx 0.073 \sqrt{N_{b\ell^-}}$.  
Even with $10^4$ reconstructed $b{\ell}^-$ and ${\bar b}{\ell}^+$ 
events each one would get a 7.3$\sigma$ 
spin-correlation signal with this observable.  
The significance of these observables after the inclusion of  
hadronisation and detector effects remains to be studied. 
\par 
The results of the above simulations are very encouraging for the 
prospect of $t,\bar t$ spin physics.  On the theoretical side the NLO 
QCD corrections to the helicity amplitudes, and to the spin density 
matrices should be computed in order to improve the precision of the 
predictions and simulation tools. 
\par 
If $t\bar{t}$ production and/or decay is affected by non-SM interactions then 
the correlations  
above will be changed. 
One interesting possibility would be the existence of a heavy spin-zero  
resonance $X_0$ 
(for instance a heavy (pseudo)scalar Higgs boson as predicted, e.g.,  
by SUSY models  
or some composite object) that couples strongly to top quarks. 
For a certain range of masses and couplings to $t\bar{t}$ such an object would 
be visible in the $t\bar{t}$ invariant mass spectrum  
\cite{Dicus:1994bm,Bernreuther:1998gs}. 
Suppose one will be fortunate and discover such a resonance at the LHC. 
Then the parity of this state may be inferred from an investigation of  
$t\bar{t}$ 
spin correlations. 
 This is  illustrated by the  following  
example. As already mentioned above,  close to threshold  
gluon-gluon fusion produces  a $t\bar t$ pair  
in a $^1\!S_0$ state.  On the other hand if the pair is produced by the 
$X_0$ resonance, 
$g g\to X_0\to t\bar t$, then  for a scalar (pseudo-scalar) $X_0$ the 
$t\bar t$ pair is in a $^3\!P_0$ ($^1\!S_0$) state 
and has therefore characteristic spin correlations. 
Let us  evaluate, for instance, the observable (\ref{eq:sbasis}). 
Its expectation value at threshold is  
$\langle{\bf s}_{t}\!\cdot\!{\bf s}_{\bar t}\rangle = 1/4$ $(-3/4)$  
if $t\bar t$ is produced by a (pseudo)scalar spin-zero boson,  
ignoring the $gg\to t\bar t$ background. An analysis which includes the 
interference with the QCD  $t\bar t$ amplitude shows characteristic  
differences also away from threshold. By investigating several 
 correlation observables (i.e., employing 
different spin bases) one can pin down the scalar/pseudo-scalar nature 
of such a resonance for a range of $X_0$ masses and couplings to top 
quarks \cite{Bernreuther:1998gs}. 
\par 
Another effect of new physics might be the generation of an anomalously large 
chromomagnetic form factor  $\kappa$ (see Section 7.1) 
in the $t\bar t$ 
production amplitude which would change the spin correlations with respect to  
the SM predictions \cite{Haberl:1996,Cheung:1997}  
(see also \cite{Atwood:1992,Atwood:1995vm}). For the LHC with 
100 fb$^{-1}$ integrated luminosity one obtains from a study of 
asymmetries (that 
were also used in \cite{Cheung:1997}) at the 
parton level a statistical sensitivity of 
$\delta \kappa \simeq 0.02$. 
\par  
The top quark decay modes 
$t\to b\ell^+\nu_{\ell}, b q{\bar q}'$ might also be affected by non-SM  
interactions, 
for instance by right-handed currents or by charged Higgs-boson exchange, and  
this would  alter the angular correlations discussed above as well.  
A Michel-parameter type  analysis of  
the sensitivity to such effects at the LHC remains to be done. 
\par 
The large $t\bar t$ samples to be collected at the LHC offer, in particular, 
an excellent opportunity to search for CP-violating interactions beyond the SM 
in high energy reactions. (The Kobayashi-Maskawa phase induces only tiny 
effects in  $t\bar t$ production and decay.) We mention in passing that 
such interactions are of great interest for  
 attempts to understand the baryon asymmetry 
of the universe. Many proposals and phenomenological studies of CP symmetry  
tests 
in $t\bar{t}$ production and decay at hadron colliders have been made.  
The following general statements apply \cite{Bernreuther:1994}: A P- and  
CP-violating 
interaction affecting $t\bar t$ production  induces   additional terms in  
the production density 
matrices $R^{(\lambda)}$ which generate two types of 
CP-odd spin-momentum correlations, namely 
\begin{equation} 
\kt\cdot(\sp - \sm)  \,\, , 
\label{eq:kabs} 
\end{equation} 
and 
\begin{equation} 
\kt\cdot(\sp \times  \sm)  \,\, , 
\label{eq:kdisp} 
\end{equation} 
and two analogous correlations where $\kt$ is replaced by $\ph$. The 
longitudinal polarisation asymmetry (\ref{eq:kabs}) requires a 
non-zero CP-violating absorptive part in the respective scattering 
amplitude. In analogy to the SM spin correlations above, 
(\ref{eq:kabs}) and (\ref{eq:kdisp}) can also be transcribed into 
angular correlations among the $t$ and $\bar t$ decay products, which 
may serve as basic CP observables (see below).  \par As to the 
modelling of non-SM CP violation two different approaches have been 
pursued.  One is to parameterise the unknown dynamics with form 
factors or, neglecting possible dependences on kinematic variables, 
with couplings representing the strength of effective interactions 
\cite{Atwood:1992,Kane:1992,Ma:1992ry,Yuan:1995,Haberl:1996,Cheung:1997,Larios:1997,Zhou:1998}, 
and compute the effects on suitable observables. This yields estimates 
of the sensitivities to the respective couplings.  For instance if 
$t\bar t$ production is affected by a new CP-violating interaction 
with a characteristic energy scale $\Lambda_{CP}> \rshat$ then this 
interaction may effectively generate a chromoelectric dipole moment 
(CEDM) $d_t$ of the top quark (see Section 7.1).  Assuming $10^7$ 
non-leptonic, $6{\times}10^6$ single lepton, and $10^6$ $t\bar t$ 
di-lepton events, the analysis of \cite{Zhou:1998}, using optimal CP 
observables, comes to the conclusion that a $1\sigma$ sensitivity of  
$\delta (Re\, d_t) \simeq 5{\times}10^{-20} g_s$ cm may be reached at 
the LHC.   
A detector-level study of CP violation in $\ttbar$ decays with 
di-lepton final states  was 
performed in~\cite{Simak:1999b}. 
 
\par Alternatively one may consider specific extensions of 
the SM where new CP-violating interactions involving the top quark 
appear and compute the induced effects in $t\bar t$ production and 
decay, in particular for the reactions (\ref{eq:ttrec}). We mention 
two examples. In supersymmetric extensions of the SM, in particular in 
the minimal one (MSSM), the fermion-sfermion-neutralino interactions 
contain in general CP-violating phases which originate from 
SUSY-breaking terms.  These phases are unrelated to the 
Kobayashi-Maskawa phase. The interaction Lagrangian for the top quark 
coupling to a scalar top ${\tilde t}_{1,2}$ and a gluino $\tilde G$ 
reads in the mass basis 
\begin{equation} 
{\cal L}_{{\tilde G}t\tilde t} = i{\sqrt 2}g_s \sum_{l=1,2} 
(e^{-i\phi_t}{\bar t_{L}}\Gamma_{l}{\tilde G^a} 
T^a {\tilde t}_{l} + e^{+i\phi_t}{\bar t_{R}}\Gamma_{l}'{\tilde G^a} 
T^a {\tilde t_l}) + {\rm h.c.} , 
\label{eq:cpsusy} 
\end{equation} 
where $g_s$ is the QCD coupling.  A priori the phase $\phi_t$ is 
unrelated to the analogous phases in the light quark sector which are 
constrained by the experimental upper bound on the electric dipole 
moment of the neutron.  The CP-violating one-loop contributions of 
(\ref{eq:cpsusy}) to $gg,{\bar q}q\to t\bar{t}$ were computed in 
\cite{Schmidt:1992,Zhou:1998}. A non-zero CP effect requires, apart 
from a non-zero phase $\phi_t$, also non-degeneracy of the masses of 
${\tilde t}_{1,2}$. For fixed phase and ${\tilde t}_1-{\tilde t}_2$ 
mass difference the effect decreases with increasing gluino and scalar 
top masses.  Assuming the same data samples as in the CEDM analysis 
above,~\cite{Zhou:1998} concludes from a computation of optimal CP 
observables that a sensitivity $|\phi_t| \gtrsim 0.1$ can be reached 
at the LHC if the gluino and squark masses do not exceed 400 GeV. 
\par 
Another striking possibility would be CP violation by an extended 
scalar sector manifesting itself through the existence of 
non-degenerate neutral Higgs bosons with undefined CP parity. Higgs 
sector CP violation can occur already in extensions of the SM by an 
extra Higgs doublet (see, for instance 
\cite{Weinberg:1990me}). It may also be sizable in 
the MSSM within a certain parameter range \cite{Pilaftsis:1998}. The 
coupling of such a neutral Higgs boson $\varphi$ with undefined CP 
parity to top quarks reads 
\begin{equation} 
{\cal L}_{Y}=-(\sqrt{2}G_F)^{1/2}  m_t(a_{t} \bar{t}t + 
\tilde{a}_{t} \bar{t}i\gamma_5 t)\,\varphi \,\, ,  
\label{eq:cphiggs} 
\end{equation} 
where $a_t$ and ${\tilde a}_t$ denote the reduced scalar and 
pseudo-scalar Yukawa couplings, respectively (in the SM $a_t=1$ and 
${\tilde a}_t=0$).  The CP-violating effects of (\ref{eq:cphiggs}) on 
$gg,{\bar q}q\to t\bar{t}$ were investigated for light $\varphi$ in 
\cite{Schmidt:1992a} and for $\varphi$ bosons of arbitrary mass in 
\cite{Bernreuther:1993,Bernreuther:1994} (see also 
\cite{Zhou:1998,Bernreuther:1998qv}). The exchange of $\varphi$ bosons 
induces, at the level of the $t\bar t$ states, both types of 
correlations (\ref{eq:kdisp}), (\ref{eq:kabs}) (the CP asymmetry 
$\Delta N_{LR} = [N(t_L{\bar t}_L) - N(t_R{\bar t}_R)]/({\rm all}\, 
t{\bar t})$ considered in \cite{Schmidt:1992a} corresponds to the 
longitudinal polarisation asymmetry $\langle\kt\cdot(\sm - 
\sp)\rangle$).  If the mass of ${\varphi}$ lies in the vicinity or 
above $2m_t$ the $s$-channel $\varphi$-exchange diagram $gg \to\varphi 
\to t\bar{t}$ becomes resonant and is by far the most important 
$\varphi$ contribution. 
\begin{figure}[t] 
\unitlength 1cm 
\begin{minipage}[t]{7.6cm} 
\begin{picture}(5.4,6.93) 
\put(-0.4,-0.8){ 
\includegraphics[width=8.8cm]{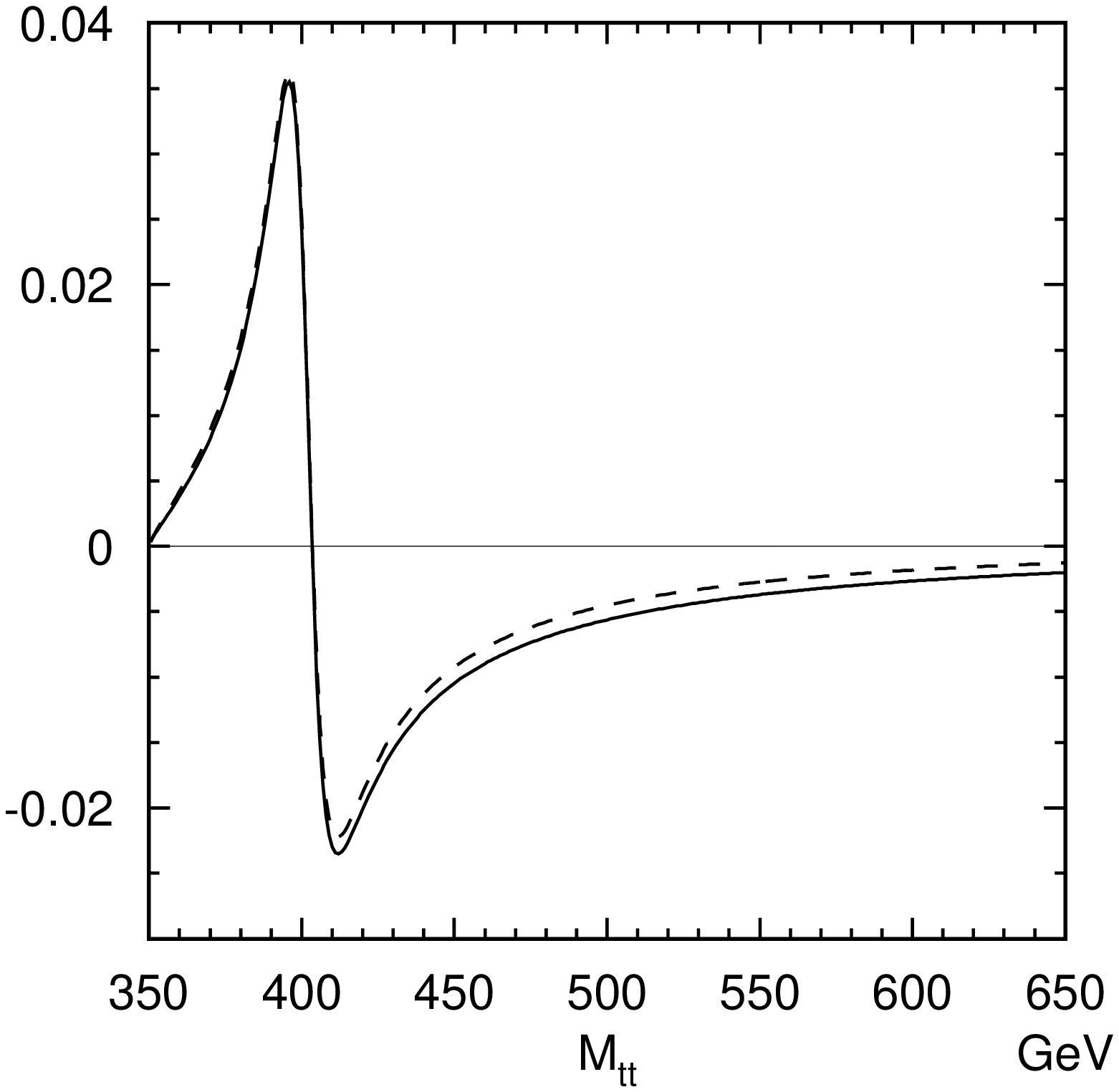} } 
\end{picture} 
\vskip 0.2cm 
\end{minipage} 
\hfill 
\begin{minipage}[t]{7.6cm} 
\begin{picture}(5.4,6.93) 
\put(-0.4,-0.8){ 
\includegraphics[width=8.8cm]{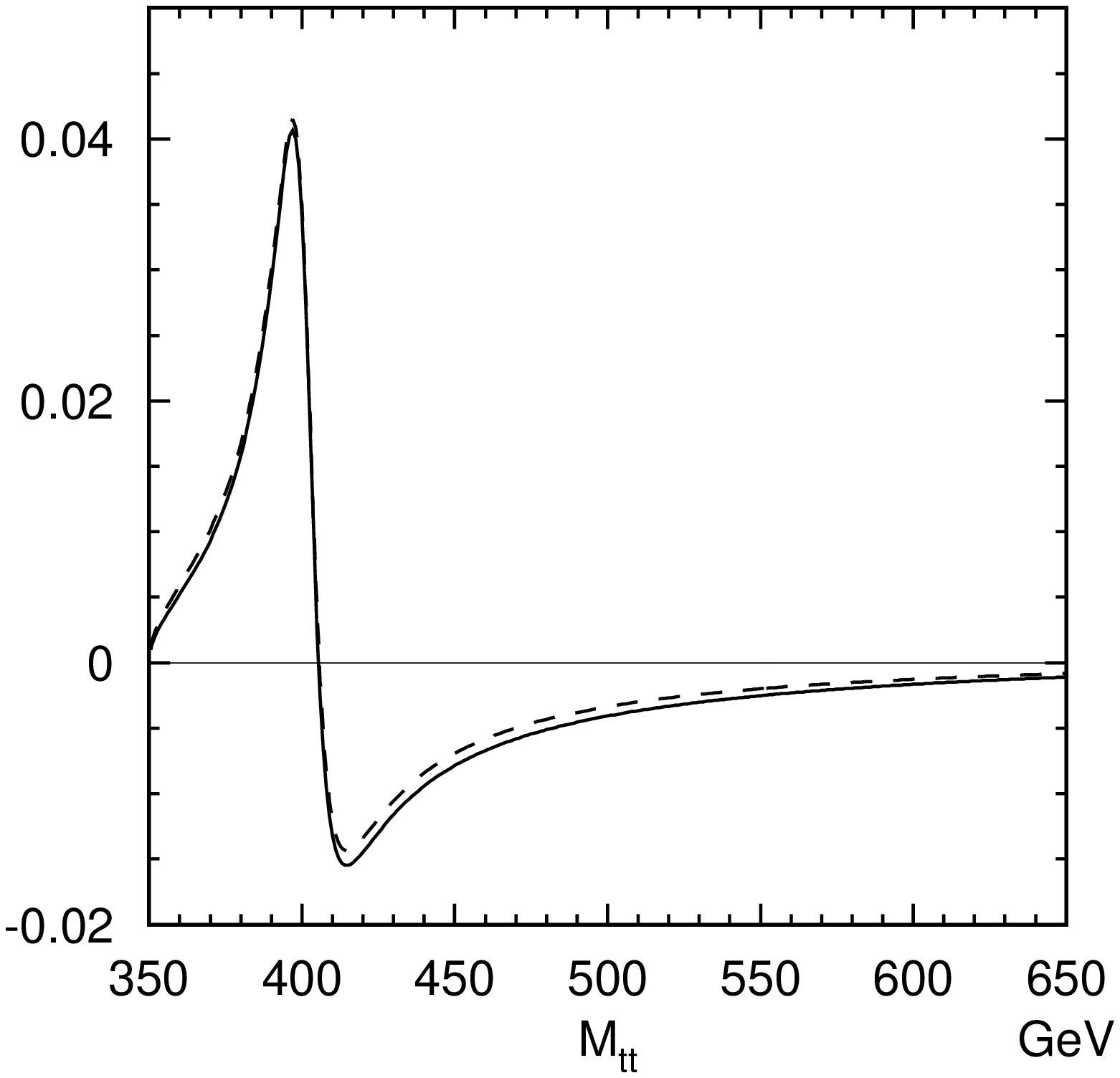} } 
\end{picture} 
\vskip 0.2cm 
\end{minipage}
\vskip0.2cm 
\caption{\label{fig:tspin3}Right: differential expectation  
value of $Q_{1}$ as a function of the $t\bar t$ invariant mass 
at $\sqrt{s}=14$ TeV for 
reduced Yukawa couplings $a_t=1$, ${\tilde a}_t=-1$,  
and a Higgs boson mass $m_{\varphi}= 400$ GeV. The dashed  
line represents the resonant and the solid line the sum of the 
resonant and non-resonant $\varphi$ contributions. 
Left: same as figure to the left, but for the 
observable $Q_2$ \cite{Bernreuther:1998qv}.} 
\end{figure} 
\par  
Simple and highly sensitive observables and asymmetries were 
investigated for the different $t\bar t$ decay channels in 
\cite{Bernreuther:1998qv}.  For the di-lepton channels the following 
transcriptions of (\ref{eq:kabs}) and (\ref{eq:kdisp}) may be used: 
\begin{equation} 
Q_{1}  =  \kt\cdot\lp - \ktb\cdot\lm \,\, , 
\label{eq:qabs} 
\end{equation} 
\begin{equation} 
Q_{2}  =  (\kt - \ktb)\cdot(\lm\times\lp)/2 \,\, , 
\label{eq:qdisp} 
\end{equation} 
where $\kt,\, \ktb$ are here the $t,\, \bar{t}$ momentum directions in 
the $t \bar{t}$ c.m. frame and $\lp$,$\lm$ are the $\ell^+$,\, 
$\ell^-$ momentum directions in the $t$ and $\bar{t}$ quark rest 
frames, respectively.  Note that $Q_1=\cos\theta_+ - \cos\theta_-$ 
where $\theta_{\pm}$ are defined after (\ref{eq:ddist}).  When taking 
expectation values of these observables the channels $\ell^+,\, 
\ell'^-$ with $\ell, \ell' = e, \mu$ are summed over.  The sensitivity 
to the CP-violating product of couplings $\gamma_{CP}\equiv 
-a_t{\tilde a}_t$ of heavy Higgs bosons is significantly increased 
when expectation values of (\ref{eq:qabs}), (\ref{eq:qdisp}) are taken 
with respect to bins of the $t\bar t$ invariant mass $M_{t\bar t}$. 
Two examples of these ``differential expectation values'' are shown in 
Fig.~\ref{fig:tspin3}. In order to estimate the 
measurement errors we have used a sample of di-lepton events, obtained 
from a simulation at the detector level using the same selection 
criteria as in the simulation described above, and determined the 
resulting error on the expectation value of $Q_1$, choosing $M_{t\bar 
  t}$ bins with a width of 10 GeV.  With $2\times 10^5$ reconstructed 
di-lepton events in the whole $M_{t\bar t}$ range we find that the 
error on ${\langle Q_1\rangle}_{M_{t\bar t}}$ is slightly below $1\%$ 
for a bin at, say, $M_{t\bar t}=400$ GeV. In addition one may employ 
the following asymmetries which are experimentally more robust than 
${\langle Q_i\rangle}$: 
\begin{equation} 
A(Q_{i})  =  \frac{N_{\ell\ell}(Q_{i}>0) - N_{\ell\ell}(Q_{i}<0)} 
                    {N_{\ell\ell}} \, , 
\label{eq:cpasym} 
\end{equation} 
where $i=1,2$ and ${N_{\ell\ell}}$ is the number of di-lepton events. 
From an analysis of these observables and asymmetries and analogous 
ones for the single lepton channels at the level of partonic final 
states the conclusion can be drawn \cite{Bernreuther:1998qv} that one 
will be sensitive to $|\gamma_{CP}| \gtrsim 0.1$ at the LHC.  This 
will constitute rather unique CP tests. 
 
\section{TOP QUARK ANOMALOUS INTERACTIONS\protect\footnote{Section   
   coordinators: 
    F.~del~Aguila, S.~Slabospitsky, M.~Cobal (ATLAS), E.~Boos (CMS).}} 
 \label{ANOM}  
In the SM the gauge couplings of the top quark  
are uniquely fixed 
by the gauge principle, the structure of generations and 
the requirement of a lowest dimension interaction Lagrangian.  
Due to the large top mass, top quark physics looks simple 
in this renormalisable and unitary quantum field theory.   
Indeed, 
\begin{itemize} 
\item[$\bullet$] the top quark production cross section is known with a 
 rather good accuracy ($\sim (10 - 15)$~\%), 
\item[$\bullet$] there are no top hadrons (mesons or baryons), 
\item[$\bullet$] the top quark decay is described by pure $(V-A)$ weak  
interactions, 
\item[$\bullet$] only one significant decay channel is present: $t \to b W^+$  
(other decay channels are very suppressed by small mixing angles). 
\end{itemize} 
This simplicity makes the top quark a unique place to search for  
new physics beyond the SM. 
If anomalous top quark couplings exist, they will affect 
top production and decay at high energies, as well as precisely  
measured quantities with virtual top quark contributions.  
 
We do not know which type of new physics will be responsible for a future  
deviation from the SM predictions.  
However, top quark couplings can be parametrized in a model independent way  
by an effective Lagrangian. 
The top quark interactions of dimension 4 can be written (in standard  
notation~\cite{Caso:1998tx}): 
\begin{eqnarray} 
{\mathcal L}_4  & = &  
-  g_s \bar t \gamma^\mu T^a  t G^a_\mu  
-  \frac{g}{\sqrt 2} \sum_{q=d,s,b} \bar t \gamma^\mu (v_{tq}^W - 
 a_{tq}^W \gamma _5) q W^+_\mu \nonumber \\ 
& & -  \frac {2}{3} e \bar t \gamma^\mu  t  A_\mu     
-  \frac{g}{2 \cos \theta _W} \sum_{q=u,c,t} 
\bar t \gamma^\mu (v_{tq}^Z  - a_{tq}^Z \gamma _5) q Z_\mu 
 \label{anomeq:1} 
\end{eqnarray} 
plus the hermitian conjugate operators for the  
flavour changing terms. 
$T^a$ are the Gell-Mann matrices 
satisfying ${\rm Tr}\, (T^a T^b) = \delta^{ab}/2$.  
Gauge invariance fixes the strong and electromagnetic  
interactions in~(\ref{anomeq:1}) and  
hemiticity implies real diagonal couplings $v^Z_{tt}, a^Z_{tt}$,  
whereas the non-diagonal ones  
$v^{W,Z}_{tq}, a^{W,Z}_{tq}$ can be complex in general.  
Within the SM $v^W_{tq}=a^W_{tq}=\frac {V_{tq}}{2}$,  
with $V_{tq}$ the Cabbibo-Kobayashi-Maskawa (CKM) matrix 
elements, $v^Z_{tt}=\frac{1}{2}-\frac {4}{3}\sin^2\theta_W,\   
a^Z_{tt}=\frac {1}{2}$, and the non-diagonal  
$Z$ couplings are equal to zero. 
Typically modifications of the SM couplings can be traced back to   
dimension 6 operators in the effective Lagrangian description valid  
above the EW symmetry breaking  
scale~\cite{Buchmuller:1986jz,Gounaris:1995ce, Whisnant:1997qu}  
(see also~\cite{Peccei:1990kr, Carlson:1994bg,Malkawi:1994tg}). 
Hence, they are in principle  
of the same order as the other dimension 5 and 6 couplings below the  
EW scale. However, in specific models the new couplings in  
Eq.~(\ref{anomeq:1}) can be large~\cite{delAguila:1999v}.   
Moreover, the present experimental  
limits are relatively weak and these couplings can show up in simple  
processes and can be measured with much better precision at the LHC.  
 
The dimension 5 couplings to one on-shell gauge boson, 
after gauge symmetry breaking, have the generic form:~\cite{Hollik:1999vz} : 
\begin{eqnarray} 
{\mathcal L}_5 & = &  
-  g_s \sum _{q=u,c,t} 
\frac {\kappa ^g_{tq}}{\Lambda} \bar t \sigma^{\mu \nu}T^a  
 (f_{tq}^g +i h^g_{tq} \gamma _5) q G^a_{\mu \nu} 
-  \frac{g}{\sqrt 2}  \sum _{q=d,s,b} 
\frac {\kappa ^W_{tq}}{\Lambda} \bar t \sigma^{\mu \nu}  
 (f_{tq}^W +i h^W_{tq}\gamma _5) q W^+_{\mu \nu} 
 \nonumber \\ 
& & \hspace*{-0.5cm}-  e  \sum _{q=u,c,t} 
 \frac {\kappa ^\gamma _{tq}}{\Lambda} \bar t \sigma^{\mu \nu}  
 (f_{tq}^\gamma  +i h^\gamma _{tq}\gamma _5) q A_{\mu \nu} 
-   \frac {g}{2\cos \theta _W}  \sum _{q=u,c,t} 
\frac {\kappa ^Z_{tq}}{\Lambda} \bar t \sigma^{\mu \nu}  
 (f_{tq}^Z +i h^Z_{tq} \gamma _5) q Z_{\mu \nu} 
 \label{anomeq:2} 
\end{eqnarray} 
plus the hermitian conjugate operators for the flavour changing terms. 
$G^a_{\mu \nu}$ is  
$\partial _\mu G^a_\nu - \partial _\nu G^a_\mu $ (see, however, below)  
and similarly for the other gauge bosons.  
We normalise the couplings by taking  
$\Lambda = 1$~TeV. $\kappa$ is real and positive and  
$f,h$ are complex numbers satisfying 
for each term $|f|^2 + |h|^2 = 1$.  
As in the dimension 4 case these dimension 5 terms  
typically result from dimension 6  
operators after the EW breaking.  
They could be large, although they are absent at tree level and  
receive small corrections in  
renormalizable theories. At any rate the LHC will improve  
appreciably their present limits.  
 
There are also dimension 5 terms with two gauge bosons.  
However, the only ones required by the unbroken gauge symmetry   
$SU(3)_C\times U(1)_Q$, and taken into account here,  
are the strong couplings with two gluons and the EW  
couplings with a photon and a $W$ boson.  
They are obtained including also the bilinear  
term $g_sf^{abc} G^b_\mu G^c_\nu $,  
with $f^{abc}$ the $SU(3)_C$ structure constants, in the  
field strength $G^a_{\mu \nu}$ in~(\ref{anomeq:2}) and  
the bilinear  
term $-ie(A_\mu W^+_\nu - A_\nu W^+_\mu)$ in $W^+_{\mu \nu}$,  
respectively.  
We do not consider any other  
dimension 5 term with two gauge bosons for their  
size is not constrained by $SU(3)_C\times U(1)_Q$  
and/or they only affect to top quark processes with more complicated  
final states than those discussed here.  
We will not elaborate on operators of dimension 6, although  
the first $q^2$ corrections to dimension 4 terms could be  
eventually observed at large hadron colliders~\cite{Hikasa:1998wx}.   
In this section    
we are not concerned with the effective top couplings to Higgs  
bosons either.  
 
In what follows we study the LHC potential for measuring or putting bounds  
on the top quark anomalous interactions in~(\ref{anomeq:1}), (\ref{anomeq:2})  
through production processes.  
Results from top quark decays are presented is Section~\ref{RARE} 
The $t\bar{t}$ couplings to gluons are considered first, since they are  
responsible for $\ttbar$ production. 
Secondly we discuss the top quark couplings $\bar tbW$.  
In the SM this coupling is not only responsible for  
almost $100 \%$ of the top decays but it also leads to an  
EW single top production mode, as reviewed in Section~\ref{sec:onetop}. 
Finally we deal with the $t$ flavour changing neutral currents  
(FCNC).  
The $\gamma  t\bar{t} $ and $Z  t\bar t$ vertices have not been  
considered here because 
$e^+e^-$ and $\mu^+\mu^-$ colliders can give a cleaner environment for  
their study. 
 
With the exception of the summary Table~\ref{tab:summary}, we will 
quote limits from the literature without attempting to compare them. 
In Table~\ref{ta:1} we illustrate statistics frequently used and which 
we will refer to in the text when presenting the bounds.  As can be 
observed, the number of signal events, and the limit estimates, vary 
appreciably with the choice of statistics.  We do correct for the 
different normalizations of the couplings used in the literature. 
\begin{table}[htb] 
\caption{Limits on the number of signal events $S$ obtained with  
different statistics. 
$B$ is the number of background events. In the other columns  
we gather $S$ for (1): $99\%$ CL (3 $\sigma$) measurement,  
$\frac{S}{\sqrt {S+B}} \geq 3$;  
(3): $99\%$ CL (3 $\sigma $) limit, $\frac{S}{\sqrt {B}} \geq 3$;  
and (5): $99\%$ CL for the Feldman-Cousins (FC)  
statistics~\cite{Feldman:1998qc}; and similarly   
for (2), (4), and (6), for the 
$95\%$ CL (1.96 $\sigma$), respectively.} 
\begin{center} 
\begin{tabular}{|c|c|c|c|c|c|c|} 
\hline 
$B$ & (1) & (2) & (3) & (4) & (5) & (6) \\ 
\hline \hline 
0 & 9 & 3.84 & 0 & 0 & 4.74 & 3.09 \\ 
5 & 12.57 & 6.71 & 6.71 & 4.38 & 8.75 & 6.26 \\  
10 & 15 & 8.41 & 9.49 & 6.20 & 10.83 & 7.82 \\  
15 & 16.96 & 9.75 & 11.62 & 7.59 & 12.81 & 9.31 \\ 
\hline 
\end{tabular} 
\label{ta:1} 
\end{center} 
\end{table} 

\subsection{Probes of anomalous $\bf gt\bar{t}$ couplings } 
 
The combination $\frac{4m_t}{\Lambda}\kappa_{tt}^gf^g_{tt}$  
(see~(\ref{anomeq:2})) can be identified with the anomalous  
chromomagnetic dipole moment of the top quark, which, as is the case of 
QED, receives one-loop contributions in QCD. Therefore, its natural  
size is of the  
order of $\alpha_s/\pi$. As we observed above, when this  
coupling is non-zero   
a direct $gg\bar tt$ four-point vertex is induced as a result of  
gauge invariance.  
 
On the other hand the combination $\frac{4m_t}{\Lambda}\kappa_{tt}^g 
h^g_{tt}$ can be identified as the anomalous chromoelectric dipole 
moment of the top quark. Within the SM this can arise only beyond two 
loops ~\cite{Donoghue:1978bw}.  On the other hand it can be much 
larger in many models of CP violation such as multi-Higgs-doublet 
models and SUSY~\cite{Soni:1992tn}. Therefore, such a non-vanishing 
coupling would be a strong indication of BSM physics. 
 
Considering the gluonic terms in~(\ref{anomeq:1}), (\ref{anomeq:2}) for the 
process of light quark annihilation into $\bar tt$ one 
obtains~\cite{Atwood:1995vm, Lampe:1997sj} 
%
\be 
 \frac{d \sigma_{q\bar{q}} }{d t} = \frac{2 \pi \alpha_s}{9 \, \hat s^2} 
\left [ 2 - \beta^2 (1 - z^2) 
-  
\frac{8m_t}{\Lambda}\kappa ^g_{tt} (f^g_{tt}+ f^{g *}_{tt})  
+ \frac{32m_t^2}{\Lambda ^2} (\kappa ^{g}_{tt})^2 | f^g_{tt} |^2  
+ \frac {4\hat s}{\Lambda ^2} (\kappa ^{g}_{tt})^2 \beta^2 (1-z^2)  
 \right ], 
\label{anomeq:gtt4} 
\ee 
 $\hat s$ being the incoming parton total energy squared, 
$z$ being the cosine of the scattering angle $\theta^*$ in the cms of the  
incoming partons, and $\beta = \sqrt{1 - 4 m^2_t/\hat s}$.  
 
The squared matrix element for $gg$ annihilation is a more complicated 
expression; we refer to~\cite{Atwood:1995vm, Lee:1997up} for exact 
formulas.  If the (anomalous) couplings are assumed to be functions 
(form-factors) of $q^2$ and then corrected by operators of dimension 
higher than 5, the $gg$ annihilation amplitude would be evaluated at 
different scales (for the $\hat t(\hat u)$ and $\hat s$ channels), and 
an additional violation of the $SU(3)_C$ gauge invariance could be 
made apparent.  For a detailed discussion of this problem see, for 
example,~\cite{Atwood:1995vm} and references therein. 
 
\begin{figure}[t] 
  \centerline{ 
    \includegraphics[width=0.5\textwidth,angle=-90]{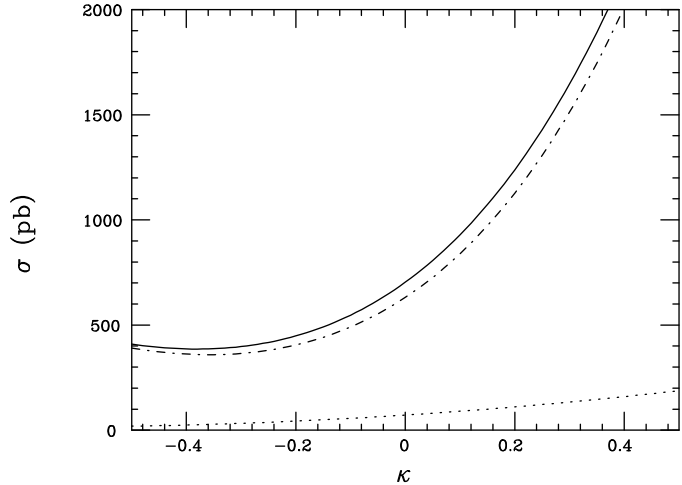}} 
  \vskip -1cm 
  \caption{Cross section for $t\bar t$ production (solid) at the LHC 
    as a function of $\kappa $.  The part of the cross section arising 
    from the $gg(q\bar q)$ annihilation is shown by the dash-dotted 
    (dotted) curve (see~\cite{Rizzo:1996zt} for details).} 
  \label{fig:gtt} 
\end{figure}

The effects associated with $\kappa^g_{tt}f^g_{tt}$  
were examined in~{\cite{Atwood:1995vm, Cheung:1996nt, Rizzo:1996zt}}.   
As shown in~\cite{Hikasa:1998wx} they will be easily distinguishable 
from the effects of $q^2$ corrections to the strong coupling  
due to operators of dimension 6, which are relatively 
straightforward to analyse~{\cite{Whisnant:1997qu}}  
in $\ttbar$ production since  
the effective coupling would be a simple  
rescaling of the strength of the ordinary QCD coupling by an additional 
$q^2$-dependent amount.   
It was shown in~\cite{Rizzo:1996zt} that the high-end tail of the top 
quark $p_T$ and $\mtt$ distributions are the observables most  
sensitive to non-zero values of $\kappa^g_{tt}f^g_{tt}$, with a  
reach for   
$\kappa =\frac{4m_t}{\Lambda}\kappa^g_{tt}f^g_{tt}$ as 
small as $\simeq 0.03$. For these values of $\kappa$,  
only a minor 
change in the total $\ttbar$ rate is expected (see 
Fig.~\ref{fig:gtt}). 
The effect of a non-zero $\kappa^g_{tt}h^g_{tt}$ was analysed, in 
particular, in~\cite{Lee:1997up, Atwood:1992vj, Bernreuther:1992be}. 
It was shown in~\cite{Lee:1997up} that information on 
$\kappa^g_{tt}h^g_{tt}$ could be obtained by studying the following 
correlation observables between $\ell^+ \ell^-$ lepton pairs produced 
in $\ttbar$ in di-lepton decays: 
\begin{eqnarray*} 
T_{33}&=&2({\vec p}_{\bar \ell} 
 -{\vec p}_\ell)_3({\vec p}_{\bar \ell}\times{\vec p}_\ell)_3, \\ 
 A_E&=&E_{\bar \ell}-E_\ell,\qquad 
 Q^\ell_{33}=2({\vec p}_{\bar \ell}+{\vec p}_\ell)_3({\vec p}_{\bar \ell}-
{\vec p}_\ell)_3 
      -\frac{2}{3}({\vec p}^2_{\bar \ell}-{\vec p}^2_\ell). 
\end{eqnarray*} 
Table~\ref{tab:gtt1} 
 shows the 1 $\sigma$ sensitivities of these correlations to 
$Re(d_t)$ and $Im(d_t)$ (where, 
$d_t \equiv g_s \frac{2}{\Lambda}\kappa^g_{tt}h^g_{tt}$). 
Quantitatively, $T_{33}$ and $Q_{33}$ enable us to probe $Re(d_t)$ and 
$Im(d_t)$ of the order of $10^{-17}g_s {\rm cm}$, respectively, and 
$A_E$ allows us to probe $Im(d_t)$  down to the order of 
$10^{-18}g_s {\rm cm}$ (see~\cite{Lee:1997up} for details). 
\begin{table} 
\begin{center} 
\caption{Attainable 1$\sigma$ limits on $Re(d_t)$ and $Im(d_t)$, through 
         $T_{33}$, $A_E$ and $Q_{33}$ for one year of the LHC running 
 at low luminosity (10~fb$^{-1})$~\cite{Lee:1997up}. } 
\vspace*{0.1cm} 
\begin{tabular}{|c|c|} 
\hline 
     Observable & Attainable 1$\sigma$ limits \\ \hline \hline 
       $T_{33}$ & $|Re(d_t)|=0.899\times 10^{-17} g_s{\rm cm}$\\ 
       $A_E$    & $|Im(d_t)|=0.858\times 10^{-18} g_s{\rm cm}$\\ 
       $Q_{33}$ & $|Im(d_t)|=0.205\times 10^{-17} g_s{\rm cm}$\\ 
        \hline 
\end{tabular} 
\label{tab:gtt1} 
\end{center} 
\end{table} 
 
\subsection {Search for anomalous $\bf Wtb$ couplings } 
 
The $Wtb$ vertex structure can be probed and measured using either top 
pair or single top production processes.  The total $\ttbar$ rate 
depends very weakly on the $Wtb$ vertex structure, as top quarks are 
dominantly produced on-shell~\cite{Boos:1997ud}.  However, more 
sensitive observables, like $C$ and $P$ asymmetries, top polarisation 
and spin correlations provide interesting information, as discussed in 
Section~\ref{TTSPIN}~ The single top production rate is directly 
proportional to the square of the $Wtb$ coupling, and therefore it is 
potentially very sensitive to the $Wtb$ structure. In single top 
events the study of the top polarisation properties potentially 
provides a way to probe a $Wtb$ coupling 
structure~\cite{Parke:1994dx}.  The potential to measure anomalous 
$Wtb$ couplings at LHC via single top from the production rate and 
from kinematical distributions has been studied in several 
papers~\cite{Whisnant:1997qu, Larios:1997dc,Boos:1999dd, 
atlasphystdr}. 
 
In the model independent effective Lagrangian  
approach~\cite{Buchmuller:1986jz,Gounaris:1995ce,  
 Whisnant:1997qu}  
there are four independent form factors describing the   
 $W tb$ vertex (see~\cite{Whisnant:1997qu} 
for details). The effective Lagrangian in the 
unitary gauge ~\cite{Kane:1992bg, Boos:1997ud, Boos:1999dd}  
is given in~(\ref{anomeq:1}), (\ref{anomeq:2}). 
As already mentioned the $(V-A)$ coupling in the SM carries the  
CKM matrix element $V_{tb}$ which 
is  very close to unity.  
 The value of a $(V+A)$ coupling is already bounded by the CLEO 
$b \rightarrow s \gamma$ data~\cite{Alam:1995aw, Larios:1999au} at a  level 
\cite{Whisnant:1997qu, Larios:1999au} such that it will be out of  
reach even at the high energy $\gamma e$ 
colliders. Since we are looking for  small deviations from the SM,  
 in the following $v^W_{tb}$ and $a^W_{tb}$  
will be set to $v^W_{tb}=a^W_{tb}=\frac{1}{2}$ 
 and an analysis is presented only for the two 
'magnetic' anomalous couplings  
$F_{L2}=\frac{2M_W}{\Lambda}\kappa^W _{tb}(-f^{W *}_{tb}-ih^{W *}_{tb})$,  
$F_{R2}=\frac{2M_W}{\Lambda}\kappa^W _{tb}(-f^{W *}_{tb}+ih^{W *}_{tb})$.  
Natural values for the couplings $|F_{L(R)2}|$ are in the region of 
$\frac{\sqrt{m_b m_t}}{v} \sim 0.1$ \cite{Peccei:1990kr} and do not exceed the 
unitarity violation bounds for $|F_{L(R)2}| \sim 0.6$ \cite{Gounaris:1995ce}. 
 
\begin{figure}[t] 
\begin{center} 
  \includegraphics[width=0.8\textwidth]{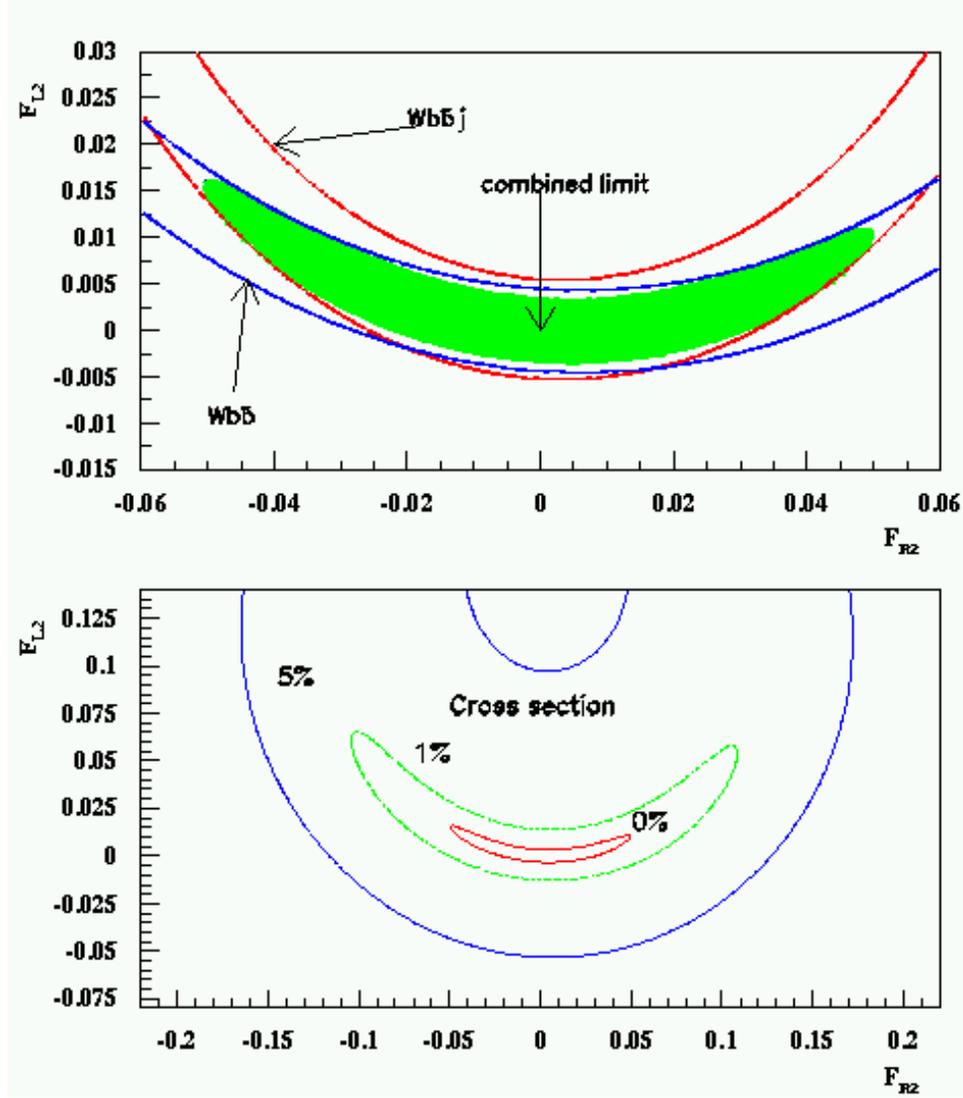} 
\end{center} 
\vskip -0.5cm 
  \caption{\label{fig:anom_coup_LHC}%
    Limits on anomalous couplings after optimised cuts from two 
    processes $ pp \rightarrow b\bar b W$ and $pp \rightarrow b\bar b 
    W+\, {\rm jet}$ (upper plot).  Dependence of the combined limits 
    on the values of systematic uncertainties (lower plot).} 
\label{fig:twb} 
\end{figure}

Calculations of the complete set of diagrams for the two main 
processes $ pp \rightarrow b\bar{b}W$ and $pp \rightarrow 
b\bar{b}W+\,{\rm jet}$ have been performed~\cite{Boos:1999dd} for the 
effective Lagrangian in (\ref{anomeq:1}), (\ref{anomeq:2}), using 
the package CompHEP~\cite{Boos:1994xb}. The calculation includes the 
single-top signal and the irreducible backgrounds.  Appropriate 
observables and optimal cuts to enhance the single-top signal have 
been identified through an analysis of singularities of Feynman 
diagrams and explicit calculations. The known NLO corrections to the 
single top rate~\cite{Smith:1996ij,Stelzer:1997ns} have been included, 
as well as a simple jet energy smearing.  The upper part of 
Fig.~\ref{fig:twb} presents the resulting 2 $\sigma$ exclusion contour 
for an integrated luminosity of $100$~fb$^{-1}$, assuming $e,\mu$ and 
$\tau\to\ell$ decays of the $W$-boson. The combined selection 
efficiency in the kinematical region of interest, including the double 
$b$-tagging, is assumed to be~$50\%$.  Figure~\ref{fig:twb} demonstrates 
that it will be essential to measure both processes $ pp \rightarrow 
b\bar{b}W$ and $pp \rightarrow b\bar{b}W+\,{\rm jet}$ at the LHC. The 
allowed region for each single process is a rather large annuli, but 
the overlapping region is much smaller and allows an improvement of 
the sensitivity on anomalous couplings of an order of magnitude with 
respect to the Tevatron.  Since the production rate is large, even 
after strong cuts, expected statistical errors are rather small, and 
the systematic uncertainties (from luminosity measurements, parton 
distribution functions, QCD scales, $\mt$, \dots) will play an 
important role. As it is not possible to predict them 
accurately before the LHC startup, we 
show here how the results depend on the assumed  
combined systematic uncertainty. Figure~\ref{fig:twb} (lower part) 
shows how the exclusion contours deteriorate when systematic errors 
of~$1\%$ and~$5\%$ are included.  Note that a systematic error of 
10\% at the LHC will diminish the sensitivity significantly and the 
allowed regions will be comparable to those expected at the upgraded 
Tevatron. 
 
The rate of single top production at LHC is different from the rate of 
single anti-top production.  This asymmetry provides an additional 
observable at LHC that is not available at the Tevatron and which 
allows to reduce systematic uncertainties. 
 
The potential of the hadron colliders can be compared to the potential 
of a next generation $e^+e^-$ linear collider~(LC) where the best 
sensitivity could be obtained in high energy $\gamma 
e$-collisions~\cite{Boos:1997ud, Cao:1998at}. The results of this 
comparison are shown in Table~\ref{tb:twb}.  From the table we see 
that the upgraded Tevatron will be able to perform the first direct 
measurements of the structure of the $Wtb$~coupling. The LHC with 
$5\%$ systematic uncertainties will improve the Tevatron limits 
considerably, rivalling with the reach of a high-luminosity (500 
fb$^{-1}$) 500~GeV LC option. The very high energy LC with 500 
fb$^{-1}$ luminosity will eventually improve the LHC limits by a 
factor of three to eight, depending on the coupling under 
consideration. 
 
\begin{table} 
\begin{center} 
  \caption{\label{tb:twb} 
    Uncorrelated limits on anomalous couplings from measurements at 
    different machines.} 
\vspace*{0.1cm} 
    \begin{tabular}{|l|lcl|lcl|}\hline 
        &\multicolumn{3}{|c|}{$F_{L2}$} 
        &\multicolumn{3}{|c|}{$F_{R2}$} \\\hline\hline 
      Tevatron ($\Delta_{{\rm sys.}}\approx10\%$) 
               & $-0.18$ &$\ldots$&$+0.55$ & $-0.24$ &$\ldots$&$+0.25$ \\ 
      LHC ($\Delta_{{\rm sys.}}\approx5\%$) 
               & $-0.052$&$\ldots$&$+0.097$ & $-0.12$ &$\ldots$&$+0.13$ \\ 
      $\gamma e$ ($\sqrt{s_{e^+e^-}}=0.5\,{\rm TeV}$) 
               & $-0.1$ &$\ldots$&$+0.1$ & $-0.1$ &$\ldots$&$+0.1$ \\ 
      $\gamma e$ ($\sqrt{s_{e^+e^-}}=2.0\,{\rm TeV}$) 
               & $-0.008$&$\ldots$&$+0.035$ & $-0.016$&$\ldots$&$+0.016$ \\ 
      \hline 
    \end{tabular} 
\end{center} 
\end{table} 
 
\subsection {FCNC in top quark physics} 
 
In the previous subsections, we analysed top quark anomalous couplings as 
small deviations from the ordinary SM interactions ($g t \bar t$ 
and $tWb$ vertices). Here we consider new 
processes which are absent at tree-level and highly suppressed in the SM, 
namely the FCNC couplings $tVc$ and 
$tVu$ ($V = g, \gamma, Z$).  
The SM predicts very small rates for such 
processes~\cite{Grzadkowski:1991sm} (see Table~\ref{tab:fcnc1}).  The 
top quark plays therefore a unique r\^{o}le compared to the other 
quarks, for which the expected FCNC transitions are much larger: the 
observation of a top quark FCNC interaction would signal the existence 
of new physics. 
As an illustration, Table~\ref{tab:fcnc1} shows predictions for  
the top quark decay branching ratios evaluated in the  
two-Higgs doublet model~\cite{bm:reina},  
the SUSY models~\cite{bm:yang}, and the SM extension  
with exotic (vector-like) quarks~\cite{delAguila:1999v}. 
\begin{table}[htb] 
\begin{center} 
\caption{ Branching ratios for FCNC top quark decays as predicted within the  
SM and in three SM extensions. } 
\vspace {0.4cm} 
\begin{tabular}{|c|c|c|c|c|}   
\hline 
 &SM & two-Higgs~\cite{bm:reina}& SUSY~\cite{bm:yang}  & Exotic 
 quarks~\cite{delAguila:1999v}\\   
\hline 
\hline 
 B$(t \to q  g)$     & $5\times 10^{-11}$ & $ \sim 10^{-5}$ & $\sim 
 10^{-3}$ & $\sim   5\times 10^{-4}$ \\   
 B$(t \to q \gamma)$ & $5\times 10^{-13}$ & $ \sim 10^{-7}$ & $\sim 
 10^{-5}$ & $\sim  10^{-5}$ \\   
 B$(t \to q Z)$      & $\sim 10^{-13}$   & $ \sim 10^{-6}$ & $\sim 
 10^{-4}$ & $\sim  10^{-2}$ \\  
\hline    
\end{tabular} 
\label{tab:fcnc1} 
\end{center} 
\end{table} 
 
In the effective Lagrangian description of~(\ref{anomeq:1}),  
(\ref{anomeq:2}) it is straightforward to calculate the top quark decay 
rates as a function of the top quark FCNC couplings: 
\begin{eqnarray} 
\Gamma({ t \to q g}) \,\,  &=& \left( \frac{\kappa^g_{tq}}{\Lambda} \right)^2 
  \frac{8}{3} \alpha_s m_t^3  \quad \quad , \quad \quad 
\Gamma({  t \to q \gamma}) \,\, = 
\left( \frac{\kappa_{tq}^{\gamma}}{\Lambda}\right)^2 2 \alpha  m_t^3, \\ 
\Gamma({ t \to q Z})_{\gamma} &=& \left(|v^Z_{tq}|^2+|a^Z_{tq}|^2\right)  
 \alpha \, m^3_t 
 \frac{1}{4 M^2_Z \sin^2 2\theta_W }  
 \left ( 1 - \frac{M_Z^2}{m_t^2} \right )^2  
\left ( 1 + 2\frac{M_Z^2}{m_t^2} \right ),  \\ 
\Gamma({ t \to q Z})_{\sigma} &=& \left( \frac{ \kappa^Z_{tq}}{\Lambda} 
\right)^2  
  \alpha \, m^3_t \frac{1}{ \sin^2 2\theta_W }  
 \left ( 1 - \frac{M_Z^2}{m_t^2} \right )^2  
 \left ( 2 + \frac{M_Z^2}{m_t^2} \right ). 
\label{anomeq:br} 
\end{eqnarray} 
For comparison, Table~{\ref{tab:fcnc-2}} collects the rare top decay rates  
normalised to $\kappa^g_{tq} = \kappa^{\gamma}_{tq} =  
|v^Z_{tq}|^2+|a^Z_{tq}|^2 = 
\kappa^Z_{tq} = 1$, and for the SM. We assume  
$m_t = 175$~GeV, { $\Lambda = 1$}~TeV, $\alpha = \frac{1}{128}$, 
$\alpha_s = 0.1$ and sum the decays into $q= u, c$. In this  
'extreme' case with the anomalous couplings equal to one  
the top can decay into a gluon or a $Z$ boson plus a light quark $q=u, c$  
and into the SM mode $bW$ at similar rates. 
\begin{table}[htb] 
\begin{center} 
\caption{Top quark decay widths and corresponding branching ratios  
for the anomalous couplings equal to one and for the SM. In the fourth  
line we gather the values of the corresponding anomalous couplings   
giving the same decay rates as in the SM. }  
\vspace {0.5cm} 
\begin{tabular}{|l|c|c|c|c|c|} \hline  
& \multicolumn{5}{c|}{Top decay mode} \\ 
  & \multicolumn{1}{c}{ $W^+ b$} &  \multicolumn{1}{c}{ $(c+u) g$} & 
  \multicolumn{1}{c}{ $(c+u) \gamma$} &   
\multicolumn{1}{c}{  $(c+u) Z_{\gamma}$} &  
 \multicolumn{1}{c|}{  $(c+u) Z_{\sigma}$} \\ 
 \hline \hline 
{\rm FCNC coupling} 
  && $ 1 $ & $1$ & $1$ & $1$ \\ \hline 
 $\Gamma({\rm GeV})$ & 1.56 & 2.86  & 0.17 & 2.91 & $0.14$ \\ \hline 
 B  & $0.20$ & $0.37$ & $0.022$ & $0.38$ & $0.018$ \\ \hline \hline 
 {\rm FCNC coupling}  
 && $8 \times 10^{-6}$ & $3\times 10^{-6}$ &  
$ 4 \times 10^{-7}$ & \\ \hline 
$\Gamma_{{\rm SM}}({\rm GeV})$ &  
1.56 & $8 \times 10^{-11}$  & $8 \times 10^{-12}$ &  
 $ 2.2\times 10^{-13} $ & \\ \hline 
${\rm B}_{{\rm SM}}$  & 1 & $5\times 10^{-11}$ & $5\times 10^{-13}$ & 
$1.5 \times 10^{-13}$ &\\ \hline 
\end{tabular} 
\label{tab:fcnc-2} 
\end{center} 
\end{table} 

\subsubsection {Current Constraints on FCNC in top quark physics } 
 
Present constraints on top anomalous couplings are derived 
from low-energy data,  
direct searches of  top rare decays,  
deviations from the SM prediction for $t\bar t$ production and  
searches for single top production at LEP2. 
 
\underline { \it Indirect constraints: }  
 The top anomalous couplings are constrained by the experimental 
upper bounds on the induced FCNC couplings between light 
fermions. For example, the $\gamma^{\mu}$ term in the $Ztq$ vertex generates 
 an effective interaction of the form \cite{Han:1995pk} 
\begin{eqnarray} 
 {\cal L}_{eff} = \frac{g}{ \cos \theta_W} a_{ij} \bar f_i \gamma^{\mu}  
\frac{1-\gamma_5}{2} f_j Z_{\mu} \, + \,  
 {\rm h.c.}, 
\end{eqnarray} 
where $f_{i,j}$ are two different light down-type quarks.  
The one-loop estimate of the vertex gives: 
\begin{eqnarray} 
a_{ij} = \frac{1}{16 \pi^2} \frac{m^2_t}{v^2}  
 \left [ V^*_{ti}( v^Z_{tq}+a^Z_{tq}) V_{qj}  
+ V^*_{qi}( v^{Z *}_{tq}+a^{Z *}_{tq}) V_{tj} \right ] \ln 
 \frac{\Lambda^2}{m^2_t}, 
\end{eqnarray} 
where $V_{ij}$ are the CKM matrix elements.   
Then, using the results of~\cite{Han:1995pk} and the experimental 
constraints from~\cite{Caso:1998tx} on $K_L \to \mu^+ \mu^-$,  
the $K_L$-$K_S$ mass difference, $B^0 - \overline{B}^0$ 
mixing, $B \to \ell^+ \ell^-X$ and $b \to s \gamma$,  
one obtains: 
\begin{eqnarray} 
 a_{sd} < 2 \times 10^{-5}, \;\;  
 a_{bd} < 4 \times 10^{-4}, \;\;  
 a_{bs} < 1.4 \times 10^{-3}, 
\end{eqnarray} 
and, taking $v = 250$~GeV, $m_t = 175$~GeV and $\Lambda = 1$~TeV: 
\begin{eqnarray} 
  |v^Z_{tu}+a^Z_{tu}|  < 0.04,\ \    |v^Z_{tc}+a^Z_{tc}|  < 0.11. 
\end{eqnarray} 
$v^Z_{tq}-a^Z_{tq}$ do not contribute to  
$a_{ij}$ for massless external fermions.  
However, both chiralities of the $Ztq$ vertex 
contribute, for instance, to the vacuum polarization tensor  
$\Pi^{\mu \nu}(q^2)$. Thus,  
using the recent value for the $\rho$ parameter,  
$\rho = 0.9998 \pm 0.0008\  (+ 0.0014)$~\cite{Caso:1998tx}, 
the following $2\sigma$ limit is obtained: 
\begin{eqnarray} 
   \sqrt {|v_{tq}^Z|^2 +|a_{tq}^Z|^2 } < 0.15. 
\end{eqnarray} 

\underline { \it CDF results: }   
The CDF collaboration has searched for the decays $t \to 
\gamma c(u)$ and $t \to Z c(u)$ in the reaction $ \bar p p \; \to \; 
\bar t t X$ at $\sqrt{s} = 1.8$~TeV, obtaining the following 
95\%~CL limits~\cite{Abe:1998fz}: 
\be 
 {\rm BR}(t \to c \gamma) + {\rm BR}(t \to u \gamma) < 3.2\%, \quad 
 {\rm BR}(t \to c Z) + {\rm BR}(t \to u Z) < 33\% \; .  
\label{anomeq:cdf-dec2} 
\ee 
These translate into the bounds on the top anomalous couplings 
\begin{eqnarray} 
  \kappa^{\gamma}_{tq} < 0.78,  \quad   
\sqrt {|v^Z_{tq}|^2+|a^Z_{tq}|^2} < 0.73 .   
\label{anomeq:cdf-dec3} 
\end{eqnarray} 

\underline {{\it $t \bar t$} { \it production via 
FCNC:}}   
 Constraints on 
the vertex $gtq$ can be derived form the study of the  
$t \bar t$-pair production 
cross-section.  
Imposing that the $t \bar t$-pair production cross-section, including 
the possible effect of anomalous couplings, should not differ from the 
observed one (assumed in this study to be  
$\sigma_{t\bar t}^{\rm exp}=6.7\pm1.3$~pb~\cite{Abe:1998vq})  
by more than 2~pb, leads to the constraint~\cite{Gouz:1999rk}: 
\begin{eqnarray} 
  \frac{\kappa^g_{tq}}{\Lambda}  \leq 0.47 \; {\rm TeV}^{-1}. 
\end{eqnarray} 

\underline {\it FCNC at LEP~2: }  
Since 1997, LEP2 has run at cms energies in excess of 180~GeV, making 
the production of single top quark kinematically possible through 
the reaction: 
\begin{eqnarray} 
e^+ \; e^- \; \to \; \gamma^{\ast} (Z^{\ast}) \; \to \; \bar{q}. 
\label{anomeq:ee} 
\end{eqnarray} 
Two LEP~experiments \cite{aleph:99,Abreu:1999fw,delphi2:99} have 
presented  the results of their search for this process. A short 
summary of these data is given in Table~\ref{tab:lep}. 
\begin{table} 
\begin{center} 
\caption { Short summary of the LEP 2 results for 
 $e^+e^- \rightarrow t \bar q$. The theoretical value $\sigma_{\rm th}$ 
is evaluated assuming the limit on the corresponding anomalous coupling  
in~(\ref{anomeq:cdf-dec3}). } 
\vspace*{0.1cm} 
\begin{tabular}{|l|c|c|c|c|} 
\hline 
 Collab. & $\sqrt{s}$ (GeV) & ${\cal L}$(pb$^{-1})$ & 
 $\sigma_{\exp}(95\%\,CL)$ & $\sigma_{\rm th}$ \\ \hline \hline 
DELPHI & 183~GeV \cite{Abreu:1999fw} & 47.7 & $< 0.55$~pb & $< 0.15$~pb \\ 
ALEPH  & 189~GeV \cite{aleph:99}   & 174  & $< 0.60$~pb & $< 0.30$~pb \\ 
DELPHI & 189~GeV \cite{delphi2:99} & 158  & $< 0.22$~pb & $< 0.30$~pb 
 \\  
\hline 
\end{tabular} 
\label{tab:lep} 
\end{center} 
\end{table} 
\begin{figure}[t] 
  \centerline{ 
    \includegraphics[width=12cm,clip]{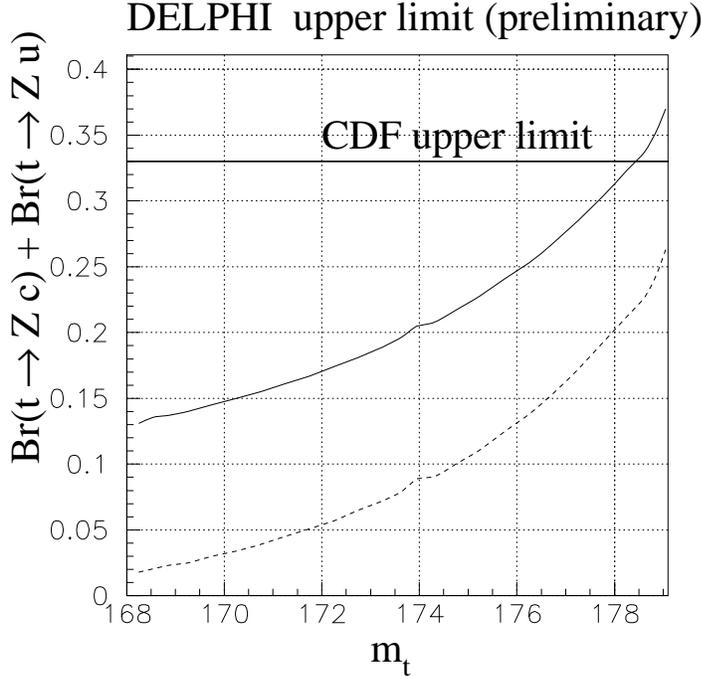}} 
\vskip -1cm 
  \caption{ Upper limit on branching fraction of $t \to Z q$ resulted 
    from LEP 2 data. Dashed curve corresponds to 
    $\kappa^{\gamma}_{tq}=0$, while solid one corresponds to 
    $\kappa^{\gamma}_{tq}<0.78$.  } 
  \label{fig:fig4} 
\end{figure} 
The production cross section is very sensitive to the top quark 
mass, $\sigma_{tq} \sim (1 - \frac{m^2_t}{s})^2$  
(see~\cite{Obraztsov:1998if} for details).  
Therefore, the  upper limit on the 
corresponding branching ratio depends from the exact value of 
$m_t$ as well, as shown in Fig.\ref{fig:fig4}. 
The current constraints on the top quark FCNC processes are 
summarised in Table~\ref{tab:current}. Note that the LEP2 limit 
is slightly better then that  given by CDF~(\ref{anomeq:cdf-dec2}). These 
constraints should further improve once the data from the 
highest-energy runs are analysed. 
\begin{table}[ht] 
\begin{center} 
\caption{ Current constraints on top quark FCNC interactions. } 
\vspace {0.4cm} 
\begin{tabular}{|l|l|cc|} \hline 
$t \to g \, q$ & BR$< 17\%$ & $\kappa^g_{tq} < 0.47 $& 
({\rm  other  FCNC  couplings  zero}) \\ 
$t \to \gamma \, q$ &BR$< 3.2\%$ &$\kappa^{\gamma}_{tq} < 0.78 $& 
 ({\rm  other  FCNC  couplings  zero}) \\ 
$t \to Z \, q$ & BR$< 22\%$ & $  \sqrt {|v_{tq}^Z|^2 +|a_{tq}^Z|^2 }< 0.55 $&  
({\rm  other  FCNC  couplings  zero})  
\\ \hline  
\end{tabular} 
\label{tab:current} 
\end{center} 
\end{table} 

\subsection { Search for  FCNC in top quark production 
processes  } 
 
FCNC interactions of top quarks will be probed 
through anomalous top decays (as discussed in Section~\ref{RARE}), and 
through anomalous production rates or channels, as discussed in the 
remainder of this section. 
 
\subsubsection {Deviations from SM expectations for $t \bar t$ production} 
 
As shown in the previous subsection,  the FCNC $tgq$-vertex  
contributes to $gg\to \ttbar$ transitions, and to  
a possible enhancement of 
the top quark production at large $E_t$ and $M_{t \bar t}$. 
A recent study~\cite{Gouz:1999rk} shows that at 
the LHC the  sensitivity to these couplings is  
equivalent to that found with the data of 
Run~1 at the Tevatron: 
\be 
 \left( \frac{\kappa^g_{tq}}{\Lambda} \right)_{\rm LHC} \simeq \left( 
  \frac{\kappa^g_{tq}}{\Lambda} \right)_{\rm FNAL} \simeq 0.5\,\,{\rm 
 TeV}^{-1}. 
\ee 

\subsubsection {`Direct' top quark production ($2 \to 1$) } 
 
The  `quark--gluon' fusion process~\cite{Hosch:1997gz} 
$g + u(c)  \to  t $ 
is characterised by the 
  largest cross-section for top quark production through FCNC-interactions 
  assuming equal anomalous couplings. 
 At the LHC, using the CTEQ2L structure functions~\cite{Lai:1995bb}, 
 these cross sections for 
$\frac{\kappa^g_{tq}}{\Lambda} = 1$~TeV$^{-1}$ are equal to: 
\be 
  \sigma( u g \to t) \simeq 4 \times 10^4 \; {\rm pb}  \quad,\quad 
  \sigma( \bar u g \to t) \simeq  1\times 10^4 \; {\rm pb} \quad,\quad 
  \sigma( c g \to t) \simeq  6 \times 10^3  \; {\rm pb}. 
\ee 
Note that $\sigma( u g \to t)$ is about 50 times larger  than the SM 
 $t \bar t$ cross section. 
The major source of background to this is the $W+$~jet 
production. The additional background due to single top production, when the 
associated jets are not observed, should not exceed  20\% of the total 
background and was therefore ignored. To reproduce the experimental 
conditions, a Gaussian smearing of the energy of the 
final leptons and quarks was applied (see~\cite{Hosch:1997gz} for details). 
Cuts on the transverse momentum ($p_T > 25$~GeV), pseudo-rapidity 
($|\eta_j| < 2.0$, $|\eta_\ell| < 3.0$), and lepton-jet separation ($\Delta R 
\geq 0.4$) were applied.  A $b$-tagging efficiency of 60\% and a mistagging 
probability of 1\% were assumed. 
 
The criterion ${S}/{\sqrt{S+B}} \geq 3$   
was used to 
determine the minimum values of anomalous couplings.  
The couplings $tgu$ and $tgc$ have been considered separately. The resulting 
constraints on $\kappa^g_{tu}$ and $\kappa^g_{tc}$ are given in 
Table~\ref{tab:sngtop}, which also contains the results of an analysis 
done for the Tevatron.  
\begin{table}[t] 
\begin{center} 
\caption{ Upper bounds on the anomalous couplings  
$\kappa^g_{tu}$ and $\kappa^g_{tc}$ 
from single top production processes. 
The symbols $2 \to 1$ and $2 \to 2$ correspond to the reactions  
quark-gluon fusion, and single top  production,  
respectively~\cite{Hosch:1997gz, Han:1998tp}.} 
\vspace{0.5cm} 
\begin{tabular}{|c|rrr|r|} 
\hline 
& \multicolumn{3}{c|}{Tevatron} & \multicolumn{1}{c|}{LHC}  \\  
& \multicolumn{1}{c}{Run 1} & \multicolumn{1}{c}{Run 2} & 
\multicolumn{1}{c|}{Run 3} & \\ \hline  
$\sqrt{s}$ (TeV) & 1.8 & 2.0 & 2.0 & 14.0 \\  
${\mathcal{L}}({\rm fb}^{-1})$ & 0.1    & 2 & 30 & 10 \\ \hline \hline 
$\kappa^g_{tu}(2\to1)$ & 0.058 & 0.019  & 0.0092 & 0.0033 \\ 
$\kappa^g_{tu}(2\to2)$ & 0.082 & 0.026  & 0.013 & 0.0061 \\ 
$\kappa^g_{tc}(2\to1)$ & 0.22  & 0.062  & 0.030 & 0.0084  \\ 
$\kappa^g_{tc}(2\to2)$ & 0.31  & 0.092  & 0.046 & 0.013 
\\ \hline   
\end{tabular} 
\label{tab:sngtop} 
\end{center} 
\end{table} 
\begin{figure}[t] 
\begin{minipage}{.48\textwidth} 
\vspace*{0cm} 
  \begin{center} 
    \resizebox{6cm}{!}{\includegraphics{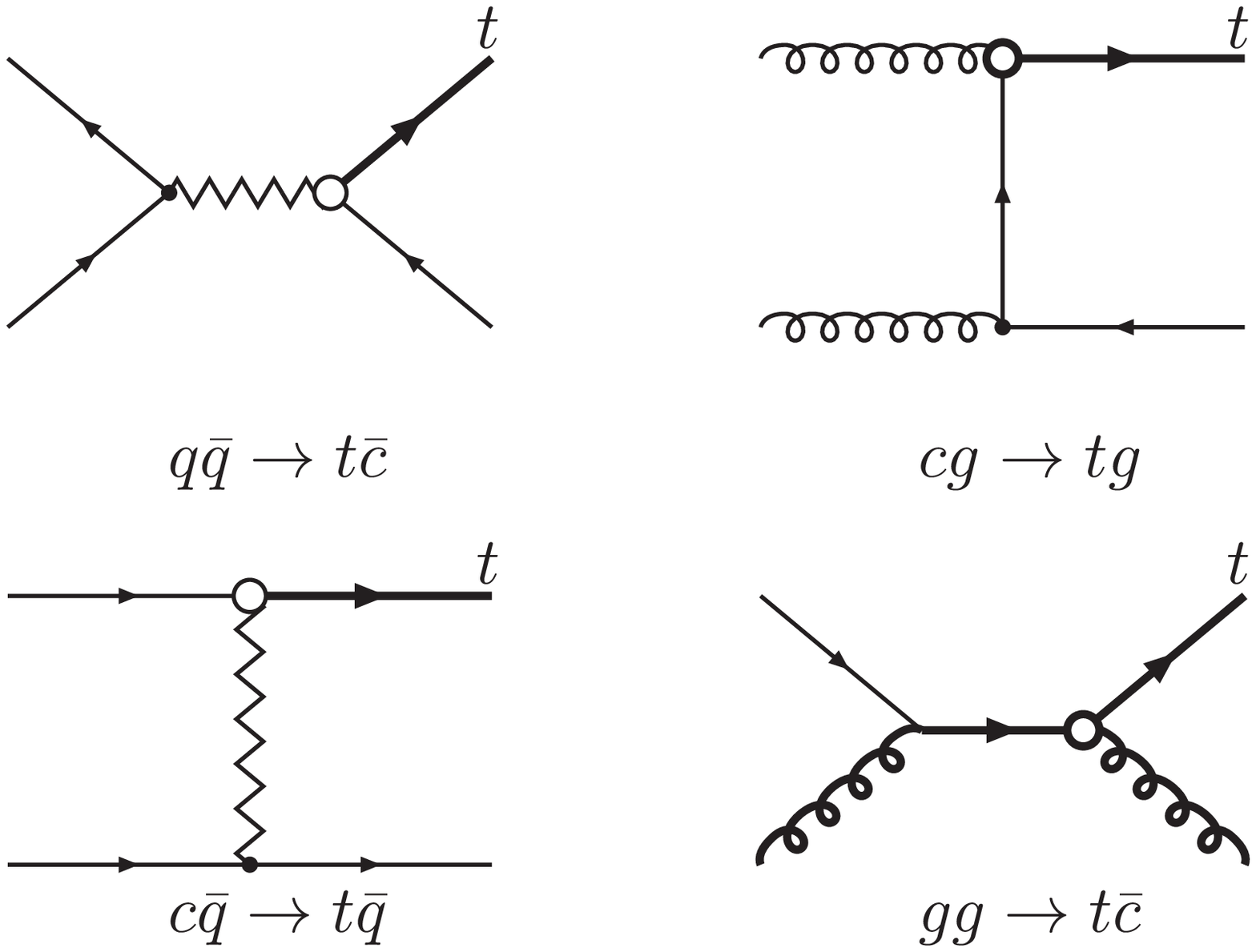}} 
\vspace*{0cm} \caption{ $2 \to 2$ single top quark production.} 
\label{fig:fig11} 
  \end{center} 
\end{minipage} 
\hfill 
\begin{minipage}{.48\textwidth} 
\vspace*{+1.5cm} 
  \begin{center} 
    \resizebox{6cm}{!}{\includegraphics{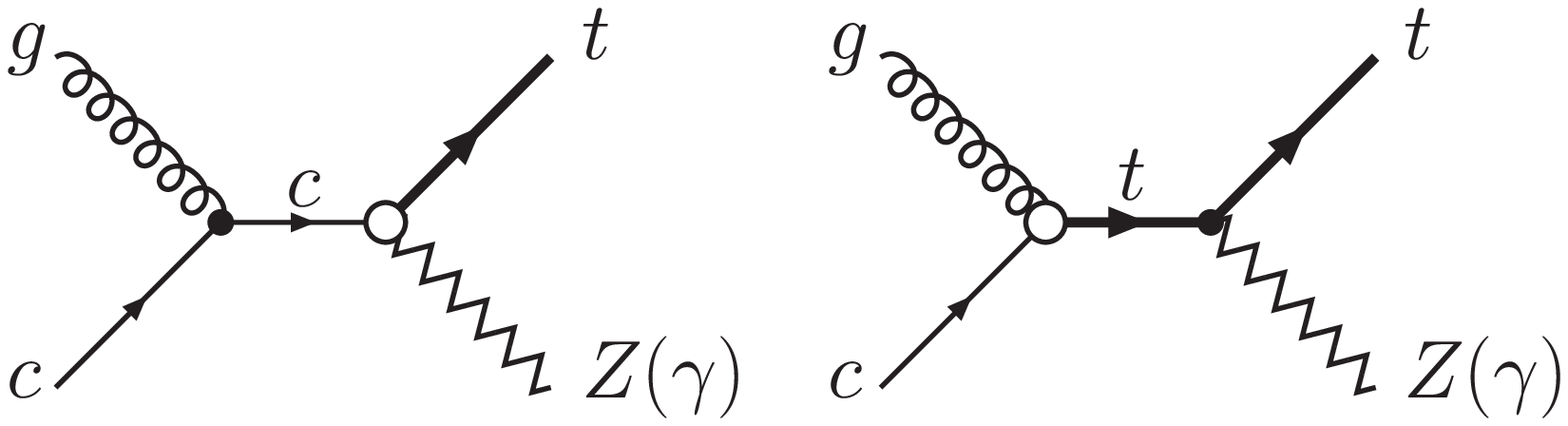}} 
\vspace*{0cm} 
     \caption{ $s$-channel diagrams for $t V$ ($V = Z, \gamma$) 
      production  } 
    \label{fig:fig12} 
  \end{center} 
\end{minipage} 
\end{figure} 

\subsubsection {Single top quark production ($2 \to 2$) } 
 
Single top quark production in $2 \to 2$ processes has been 
studied as well~\cite{Han:1998tp}. There are four different 
subprocesses, which lead to one top quark in the final state 
together with one associated jet (see~Fig.~\ref{fig:fig11} and 
\cite{Han:1998tp} for detailed considerations): 
\begin{eqnarray} 
 q \bar q \to  t \bar q, \quad g g \to t \bar q, \quad 
 q q  \to t q, \quad  q g \to  t g  \label{anomeq:twotwo} 
\end{eqnarray} 
The major background comes from $W+2$ jets and $W+b \bar b$ 
production, as well as from single top production. In addition to the 
cuts and tagging rates used in the above analysis of 'direct' top 
production, additional cuts on the reconstructed top mass (145~GeV~$< 
M_{bW} < 205$~GeV), on $p_{T\,\,b} > 35$~GeV, and on jet-jet and 
lepton-jet separation ($\Delta\,R_{jj} > 1.5$, $\Delta\,R_{lj} > 1.0$) 
were applied here to improve the signal/background separation. 
The corresponding limits on anomalous couplings in the top-gluon 
interaction with $c$ or $u$ quarks are given in 
Table~\ref{tab:sngtop}. 
 
 
\subsubsection {$tZ$ and $t\gamma$ production } 
 
 All the anomalous couplings may contribute to the processes $ q \, g 
\, \to t \, Z(\gamma)$, and were considered 
in~\cite{delAguila:1999ac, delAguila:1999ec}.  The left diagram in 
Fig.~\ref{fig:fig12} corresponds to the $Z(\gamma)tq$ coupling, while 
the right one shows the top-gluon anomalous coupling (the 
corresponding $t$-channel diagrams are not shown).  For all the 
calculations presented here, the MRSA PDF 
set~\cite{Martin:1994kn} with $Q^2 = \hat s$ was used. The resulting 
total cross sections  
for $\kappa _{tq}^\gamma = \sqrt {|v^Z_{tq}|^2+|a^Z_{tq}|^2} =1$ 
are~\cite{delAguila:1999ec}: 
\begin{eqnarray*} 
\begin{array}{lcrclcr} 
 \sigma(u\, g \to \gamma \, t) & = & 73 \,{\rm pb}, & 
 \phantom{leavesomespace} & 
 \sigma(c\, g \to \gamma \, t) & = & 10 \,{\rm pb}, \\ 
 \sigma(u\, g \to Z \, t) & = & 746 \,{\rm pb}, & & 
 \sigma(c\, g \to Z \, t) & = & 114 \,{\rm pb}. 
\end{array} 
\end{eqnarray*} 
 Different background sources ($W+$ jets, $Z+$ jets, $ZW +$ jets, $W b 
\bar b+$ jets, $t \bar t$, and $Wt$ production) were considered. The 
experimental conditions were simulated by a Gaussian smearing of the 
lepton, photon and jet energies (see~\cite{delAguila:1999ec} for 
details).  Cuts on the transverse momenta, $p_T(\ell, \, j, \, \gamma) > 
(15,\, 20,\, 40)$~GeV, on pseudo-rapidities, $|\eta_{j,\ell,\gamma}| < 2.5$, 
and on lepton-jet-photon separation ($\Delta R \geq 0.4$) were 
applied.  A $b$-tagging efficiency of 60\% and a mistagging 
probability of 1\% were assumed. It was found that $b$-tagging 
plays an essential role in tracing the top quark and reducing backgrounds. 
 
It has been shown that the best limits on the top quark FCNC couplings 
can be obtained from the decay channels $Zt \to \ell^+ \ell^- \, \ell 
\nu b$ and $\gamma t \to \gamma \, \ell \nu b$ 
(see~\cite{delAguila:1999ac} and \cite{delAguila:1999ec} for details). 
Upper bounds at 95\% CL are derived using the FC 
statistics~\cite{Feldman:1998qc}.  Table~\ref{tab:tV} collects the 
corresponding limits on eight top anomalous couplings.  Like in 
previous cases the bounds on $u$ and $c$ couplings were obtained under 
the assumption that only one anomalous coupling at a time is non-zero. 
The analysis was done for both Tevatron and LHC but with different 
optimized cuts. 
\begin{table}[ht] 
\begin{center} 
\caption{ Upper bounds on top anomalous couplings  
(see (\ref{anomeq:1},\ref{anomeq:2})) from  $Zt$  
and $\gamma t$ production. We have corrected for  
the different normalizations used  
in~\cite{delAguila:1999ac, delAguila:1999ec}. } 
\vspace{0.5cm} 
\begin{tabular}{|l|rr|rr|} 
\hline  
& \multicolumn{2}{c|}{Tevatron} &   \multicolumn{2}{c|}{LHC}  \\  
& \multicolumn{1}{c}{Run 1} & \multicolumn{1}{c|}{Run 2} & & \\  
\hline 
$\sqrt{s}$ (TeV) & 1.8 & 2.0 & 14.0 & 14.0 \\  
$\mathcal{L} \rm{~(fb}^{-1}\rm{)}$ & 0.1 & 2 & 10 & 100 \\  
\hline \hline 
$\kappa_{tu}^g       $ & 0.31  & 0.057  & 0.0097 & 0.0052 \\ 
$\kappa_{tc}^g       $ & --    & --     & 0.020 & 0.011 \\ 
$\kappa_{tu}^{\gamma}$ & 0.86   & 0.18   & 0.013 & 0.0060  \\ 
$\kappa_{tc}^{\gamma}$ & --    & --     & 0.037 & 0.018  \\    
$\sqrt{|v^Z_{tu}|^2+|a^Z_{tu}|^2}$ & 0.49  & 0.13   & 0.016  & 0.0078  \\ 
$\sqrt{|v^Z_{tc}|^2+|a^Z_{tc}|^2}$ & --    & --     & 0.032  & 0.016  \\ 
$\kappa_{tu}^Z             $ & 1.71  & 0.43   & 0.040  & 0.018  \\ 
$\kappa_{tc}^Z        $ & --    & --     & 0.097  & 0.046  \\ 
\hline   
\end{tabular} 
\label{tab:tV} 
\end{center} 
\end{table} 

\subsubsection { Like-sign  $t t$ ($\bar t \, \bar t$) pair 
  production } 
 
\begin{figure} 
  \centerline{ 
    \includegraphics[width=5cm]{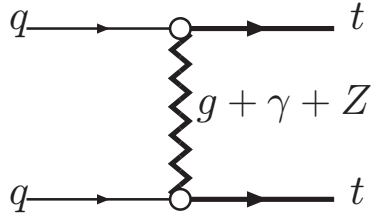}} 
  \caption{ Diagram describing like-sign top quark pair 
    production  } 
  \label{fig:ant:fig13} 
\end{figure}

 Additional evidence for a FCNC $gtq$ coupling can be sought through 
the production of like-sign top pairs (see 
Fig.~\ref{fig:ant:fig13}). 
\begin{eqnarray} 
 p \, p \, \to \, t \, t \, X, \quad 
 p \, p \, \to \, \bar t \, \bar t \, X \label{anomeq:twot} 
\end{eqnarray} 
The ATLAS collaboration performed a detailed investigation of this 
reaction for the case of high luminosity, ${\cal L}_{\rm int} = 
100$~fb$^{-1}$ 
(see~\cite{atlasphystdr} and \cite{Gouz:1999rk} for details). 
 All the three anomalous couplings contribute to this process and the 
kinematics of the $tt$-pair is almost the same as for the conventional 
$t \bar t$-pair production. 
 
An experimentally clean signature of $tt$~$(\bar t \bar t)$ production 
is the production of like-sign high $p_T$ leptons plus 
two hard $b$-jets. The main sources of background 
are $q \bar{q}' \to t \bar t W$ and $qq \to W^{\pm} q' W^{\pm} q'$.  
The expected cross sections 
for the signal (with  
$\kappa_{tq}^g =\kappa_{tq}^\gamma =|v_{tc}^Z|^2+|a_{tc}^Z|^2 = 1)$  
and background processes are equal to: 
\begin{eqnarray} 
\begin{array}{l c c c l c c} 
 \sigma({ tt}) &=& 1920 \;{\rm pb}, & \phantom{leavesomespace} & 
  \sigma({ \bar t \bar t}) &=& 64 \; {\rm pb}, \\ 
 \sigma({ W^+t \bar t}) &=& 0.5 \;{\rm pb}, & & 
  \sigma({ W^- t \bar t}) &=& 0.24 \; 
  {\rm pb}, \nonumber \\ 
\sigma({ W^+ W^+ qq}) &=& 0.5 \;{\rm pb}, & & 
  \sigma({ W^- W^- qq} ) &=& 0.23 \; 
{\rm pb}. \nonumber 
\end{array} 
\end{eqnarray} 
CTEQ2L structure functions~\cite{Lai:1995bb} were used with the 
evolution parameter $Q^2 = m^2_t$ for the signal and $Q^2 = m^2_W$ for 
the background calculations. \pyth~5.7~\cite{Sjostrand:1994yb} was 
used for the fragmentation and all events were passed through the 
ATLFAST detector simulation.  An additional reducible like-sign 
di-lepton background is due to $\ttbar$ events with a $b$ semi-leptonic 
decay. The initial selection required therefore two like-sign {\it 
isolated} leptons with $p_T > 15$~GeV and $|\eta| < 2.5$ as well as at 
least two jets with $p_T > 20$~GeV and $|\eta| < 2.5$.  In order to 
get a better signal/background separation 
 jets with $p_T > 40$~GeV (with at least one tagged as a 
$b$-jet) were required 
(see~\cite{atlasphystdr, Gouz:1999rk} for other cuts). 
The potential reach of this study, using the  
$S/\sqrt{S+B} \geq 3$ criterion, is given in  
Table~\ref{tab:twot}. 
\begin{table} 
\begin{center} 
\caption{ The limits on anomalous couplings from an improved ATLAS 
analysis~\cite{atlasphystdr, Gouz:1999rk} of like-sign top-pair 
production at the LHC for the case of high luminosity, ${\cal L}_{\rm int} = 
100$~fb$^{-1}$. The contribution from the $\sigma^{\mu \nu}$ term 
in the $Ztq$ vertex is ignored. } \vspace*{0.1cm} 
\begin{tabular}{|c|c|c|c|c|c|} \hline 
$\kappa_{tu}^g$ & $\kappa_{tc}^g$ & $\kappa_{tu}^{\gamma}$ & 
$\kappa_{tc}^{\gamma}$ & $\sqrt{|v_{tu}^Z|^2+|a_{tu}^Z|^2} $ & 
$\sqrt{|v_{tc}^Z|^2+|a_{tc}^Z|^2} $ \\ \hline \hline 
 0.078 & 0.25  & 0.14 & 0.32 & 0.27 & 0.85 \\ \hline 
\end{tabular} 
\label{tab:twot} 
\end{center} 
\end{table} 

\begin{table} 
\caption{ Summary of the LHC sensitivity to the top quark anomalous couplings 
$\kappa ^g_{tq}, \kappa ^\gamma_{tq} $ and $\sqrt{|v_{tu}^Z|^2+|a_{tu}^Z|^2}$. 
The resulting constraints are presented in terms of `branching ratio', 
 $\Gamma(t \to q V) / \Gamma_{SM}(=1.56\,\,{\rm GeV})$.  
The results for the Tevatron option are also given (see text for explanation). 
$2\to1$, $2\to2$, $tV$, and $t\,t$ stand for quark-gluon fusion, 
single top production, $t+\gamma(Z)$ production, and like-sign 
top-pair final states, respectively.  The `decay', `ATLAS', and `CMS' 
labels denote the results obtained from the study of top decay 
channels, documented in Section~\ref{RARE}}  
\vspace {0.5cm} \leftskip -.5cm 
\begin{small} 
\begin{center} 
\begin{tabular}{|l|r|r|r|r|r|ll|} \hline 
 & \multicolumn{3}{c}{Tevatron} &  \multicolumn{3}{|c}{LHC} &\\ \hline 
 $\sqrt{s}$(TeV)  & 1.8 & 2 & 2 & 14 & 14 & &\\ 
 ${\cal L}$(fb$^{-1})$ & 0.1 & 2 & 30 & 10 & 100 & & \\ \hline \hline 
$t u g$ &$3.1 \times 10^{-3}$ & $3.3 \times 10^{-4}$ & $7.8 \times 10^{-5}$ 
 &$1.0 \times 10^{-5}$ &$3.2 \times 10^{-6}$ & $2\to1$& \cite{Hosch:1997gz}  \\ 
        &$6.2 \times 10^{-3}$ &$6.2 \times 10^{-4}$ &$1.5 \times 10^{-4}$ 
 &$3.4 \times 10^{-5}$ &$1.1 \times 10^{-5}$ & $2\to2$ &\cite{Han:1998tp}   \\  
        &$1.8 \times 10^{-1}$ &$6.0 \times 10^{-3}$ & --               
  &$1.7 \times 10^{-4}$ &$5.0 \times 10^{-5}$ & $tV$ & 
      \cite{delAguila:1999ac, delAguila:1999ec}              \\ 
        & --                 &$1.9 \times 10^{-2}$ &$2.7 \times 10^{-3}$ 
  & -- & -- & decay & \cite{Han:1996ce}  \\ 
        & -- & -- & -- &$1.5 \times 10^{-2}$ &$5.6 \times 10^{-3}$ 
          & $t \,t$ & \cite{Gouz:1999rk, atlasphystdr} \\ \hline 
 
$t c g$ &$4.4 \times 10^{-2}$ &$3.5 \times 10^{-3}$ &$8.3 \times 10^{-4}$ 
 &$6.5 \times 10^{-5}$ &$2.1 \times 10^{-5}$ & $2\to1$ & \cite{Hosch:1997gz}  \\ 
        &$8.8 \times 10^{-2}$ &$7.8 \times 10^{-3}$ &$2.0 \times 10^{-3}$ 
 &$1.6 \times 10^{-4}$ &$4.9 \times 10^{-5}$ & $2\to 2$& \cite{Han:1998tp}  \\  
    & -- & -- & -- & $7.3 \times 10^{-4}$ &$2.2 \times 10^{-4}$ & $tV$ & 
            \cite{delAguila:1999ac, delAguila:1999ec}              \\ 
 & -- &$1.9 \times 10^{-2}$ &$2.7 \times 10^{-3}$ 
  & -- & -- & decay & \cite{Han:1996ce}  \\ 
        & -- & -- & -- &$1.6 \times 10^{-1}$ &$5.7 \times 10^{-2}$ & 
           $t \, t$     & \cite{Gouz:1999rk, atlasphystdr} \\ \hline 
 
$t u \gamma$ &$7.9 \times 10^{-2}$ &$3.5 \times 10^{-3}$ & --                
 &$1.8\times 10^{-5}$ &$3.9 \times 10^{-6}$ & $tV$ & 
               \cite{delAguila:1999ac, delAguila:1999ec}              \\ 
        & -- & -- & -- &$3.0 \times 10^{-3}$ &$1.1 \times 10^{-3}$ 
     & $t \, t$       &  \cite{Gouz:1999rk, atlasphystdr} \\  
        & -- & -- & -- & $1.9 \times 10^{-4}$ & $4.8 \times 10^{-5}$ 
         & ATLAS          & \cite{atlasphystdr} \\  
        & -- & -- & -- & $8.6 \times 10^{-5}$ & $4.0 \times 10^{-5}$ 
         &CMS         & \cite{Stepanov:2000} \\ \hline 
 
$t c \gamma$ & -- & -- & -- & $1.5 \times 10^{-4}$ & $3.5 \times 10^{-5}$ &  
      $tV$ &   \cite{delAguila:1999ac, delAguila:1999ec}              \\ 
        & -- & -- & -- &$1.7 \times 10^{-2}$ &$5.5 \times 10^{-3}$ 
       & $t \, t$        &  \cite{Gouz:1999rk, atlasphystdr} \\  
        & -- & -- & -- & $1.9 \times 10^{-4}$ & $4.8 \times 10^{-5}$ 
     &ATLAS         & \cite{atlasphystdr} \\  
        & -- & -- & -- & $8.6 \times 10^{-5}$ & $4.0 \times 10^{-5}$ 
         & CMS      & \cite{Stepanov:2000} \\ \hline 
 
$t u Z$ &$4.5 \times 10^{-1}$ &$3.2 \times 10^{-2}$ & --                
 &$4.8 \times 10^{-4}$ &$1.1 \times 10^{-4}$ &  $tV$ & 
               \cite{delAguila:1999ac, delAguila:1999ec}              \\ 
 & -- &$1.1 \times 10^{-2}$ &$5.2 \times 10^{-3}$ 
  & $5.8 \times 10^{-4}$ &$1.9 \times 10^{-4}$ & decay &\cite{Han:1995pk}  \\ 
        & -- & -- & -- &$1.9 \times 10^{-1}$ &$6.8 \times 10^{-2}$ & $t \, t$ 
                 &  \cite{Gouz:1999rk, atlasphystdr} \\  
        & -- & -- & -- & $6.5 \times 10^{-4}$ & $1.0 \times 10^{-4}$ 
                  & ATLAS & \cite{atlasphystdr} \\  
        & -- & -- & -- & $1.4 \times 10^{-3}$ & $1.4 \times 10^{-4}$ 
                  & CMS & \cite{Stepanov:2000} \\ \hline 
 
$t c Z$ & -- & -- & -- & $1.9 \times 10^{-3}$ & $4.8 \times 10^{-4}$ &  
      $tV$ &       \cite{delAguila:1999ac, delAguila:1999ec}              \\ 
 & -- &$1.1 \times 10^{-2}$ &$5.2 \times 10^{-3}$ 
  & $5.8 \times 10^{-4}$ &$1.9 \times 10^{-4}$  & decay & \cite{Han:1995pk}  \\ 
        & -- & -- & -- &$1.9 $ &$6.7 \times 10^{-1}$ 
      & $t \, t$      &   \cite{Gouz:1999rk, atlasphystdr} \\  
        & -- & -- & -- & $6.5 \times 10^{-4}$ & $1.0 \times 10^{-4}$ 
        & ATLAS          & \cite{atlasphystdr} \\  
        & -- & -- & -- & $1.4 \times 10^{-3}$ & $1.4 \times 10^{-4}$ 
          & CMS     & \cite{Stepanov:2000} \\ \hline 
 
\end{tabular} 
\end{center} 
\end{small} 
\label{tab:summary} 
\end{table} 

\subsection{Conclusion on $\bf t q V$ anomalous couplings} 
 
 Table~\ref{tab:summary} presents a short summary of LHC sensitivities 
 to anomalous FCNC couplings of the top quark. For comparison, we present also 
the estimates of the corresponding sensitivities at Tevatron.  
For completeness we anticipate and include here the 
results from rare decays discussed in the next section  
(see also~\cite{Han:1995pk, Han:1996ce}). 
To unify the description of the LHC potential to detect top anomalous 
 couplings from production and decay  
processes, all results in Table~\ref{tab:summary} are expressed in 
 terms of limits on top decay branching ratios: 
$\Gamma(t \to q V) / \Gamma_{\rm SM}\,(=1.56\,\,{\rm GeV})$. 
The results were obtained using $m_t = 
175$~GeV, $\alpha_s = 0.1$, and $\alpha = 1/128$.  
When needed the limits quoted in the table have been rescaled to the different 
 luminosities and to the $S/\sqrt{S+B}\geq 3$ criterion by using a simple 
 linear extrapolation of the available bounds  
(see~\cite{atlasphystdr,Stepanov:2000}  and  Section~\ref{RARE}). 
The limits on the top anomalous couplings from $tV$ production  
in Table~\ref{tab:tV} were   
obtained using the FC prescription~\cite{Feldman:1998qc} and   
have been multiplied by a factor of $\sqrt 2$,  
which roughly relates this prescription with the  
statistical criterion adopted in Table~\ref{tab:summary}  
~\cite{delAguila:1999ac, delAguila:1999ec}. 
 
At present, only few cases (like-sign top-pair production, $t \to 
qZ$ and $t \to q \gamma$ decays, see~\cite{atlasphystdr, Stepanov:2000}) 
were investigated with a more or less realistic detector simulation 
(ATLFAST and CMSJET). Other investigations were done at the parton 
level (the final quarks were considered as jets and a simple smearing 
of lepton, jet and photon energies was applied).  Of course, more 
detailed investigations with a more realistic simulation of the 
detector response may change these results. 
 
The most promising way to measure the anomalous FCNC top-gluon coupling 
seems to be the investigation of single top production processes, as 
the search for $t \to g q$ decays would be overwhelmed by background 
from QCD multi-jet events. At the same time, both top quark production 
and decay would provide comparable limits on top quark anomalous FCNC 
interactions with a photon or a $Z$-boson.  In general, the studies 
shown above indicate that the LHC will improve by a factor of at least 
10 the Tevatron sensitivity to top quark FCNC couplings. 
Of course, the results presented here are not 
complete, since other new kinds of interactions may lead to the 
appearance of unusual properties of the top quark. For example, 
recently proposed theories with large extra-dimensions predict a 
significant modification of $t \bar t$ pair production (see, for 
example,~\cite{Rizzo:1999pc} and references therein). It was found 
that the exchange of spin-2 Kaluza-Klein gravitons leads to a 
modification of the total $t \bar t$ production rate as well as to 
a noticeable deviation in the $p_T$ and $M_{t \bar t}$ distributions 
with respect to the SM predictions. Naturally, we may expect also the 
modifications of spin-spin correlations due to graviton exchange. 
 
It has to be stressed that different types of new 
interactions may affect the same observable quantity. Only a careful 
investigation of different aspects of top quark physics may provide a 
partial separation of these interactions.  
 
 
 
\section{RARE DECAYS OF THE TOP QUARK\protect\footnote{Section coordinators: 
    B.~Mele, J.~Dodd (ATLAS), N.~Stepanov (CMS).}} 
\label{RARE} 
The production of $10^7-10^8$ top quark pairs per year at LHC will 
allow to probe the top couplings to both  known and new particles 
involved in possible top decay channels different from the main 
$\tbw$. Thanks to the large top mass, there are several decays 
that can be considered, even involving the presence of on-shell heavy 
vector bosons or heavy new particles in the final states. 
On a purely statistical basis, one should be able to detect 
a particular decay channel whenever its branching ratio (BR) is larger than 
about $10^{-6}-10^{-7}$. In practice, we will see 
that background problems and systematics will 
lower this potential by a few orders of magnitude, the precise reduction 
being dependent of course on the particular signature considered. 
We will see, that the final detection threshold for each channel will 
not allow  the study of many possible final states predicted in the 
SM, unless new stronger couplings come into play.

\subsection{Standard Model top decays} 
\label{bms:sm} 
In this section, we give an overview of the decay channels of the 
top quark in the framework of the SM. 
In the SM the decay $t \to bW$ is by far the dominant one. The 
corresponding width has been discussed in Section~\ref{sec:topprop}. 
The rates for other decay channels are predicted to be smaller by 
several orders of magnitude in  the SM. 
The second most likely decays are the Cabibbo-Kobajashi-Maskawa 
(CKM) non-diagonal decays \tswm and \tdwm. 
Assuming $|V_{ts}|\simeq 0.04$ and $|V_{td}|\simeq 0.01$,  
respectively~\cite{Caso:1998tx}, 
one gets 
\be 
{\rm BR}(\tsw) \sim 1.6 \times  10^{-3} \; \;{\rm and} \; \; 
{\rm BR}(\tdw) \sim 1 \times 10^{-4} 
\label{bme:tsw} 
\ee 
in the SM with three families. 
From now on, for a generic decay channel $X$, we define 
\be 
{\rm BR}(t \to X) = \frac{\Gamma(t \to X)}{\Gt_{\rm SM}}. 
\ee 
The two-body tree-level decay channels are the only ones that the LHC 
could detect in the framework of the SM. With the exception 
of higher-order QED and QCD radiative decays, the next 
less rare processes have rates no larger than $10^{-6}$. 
\begin{figure*}[thbp] 
\begin{center} 
\vspace*{-2.cm} 
\mbox{\epsfxsize=16cm\epsfysize=18.5cm\epsffile{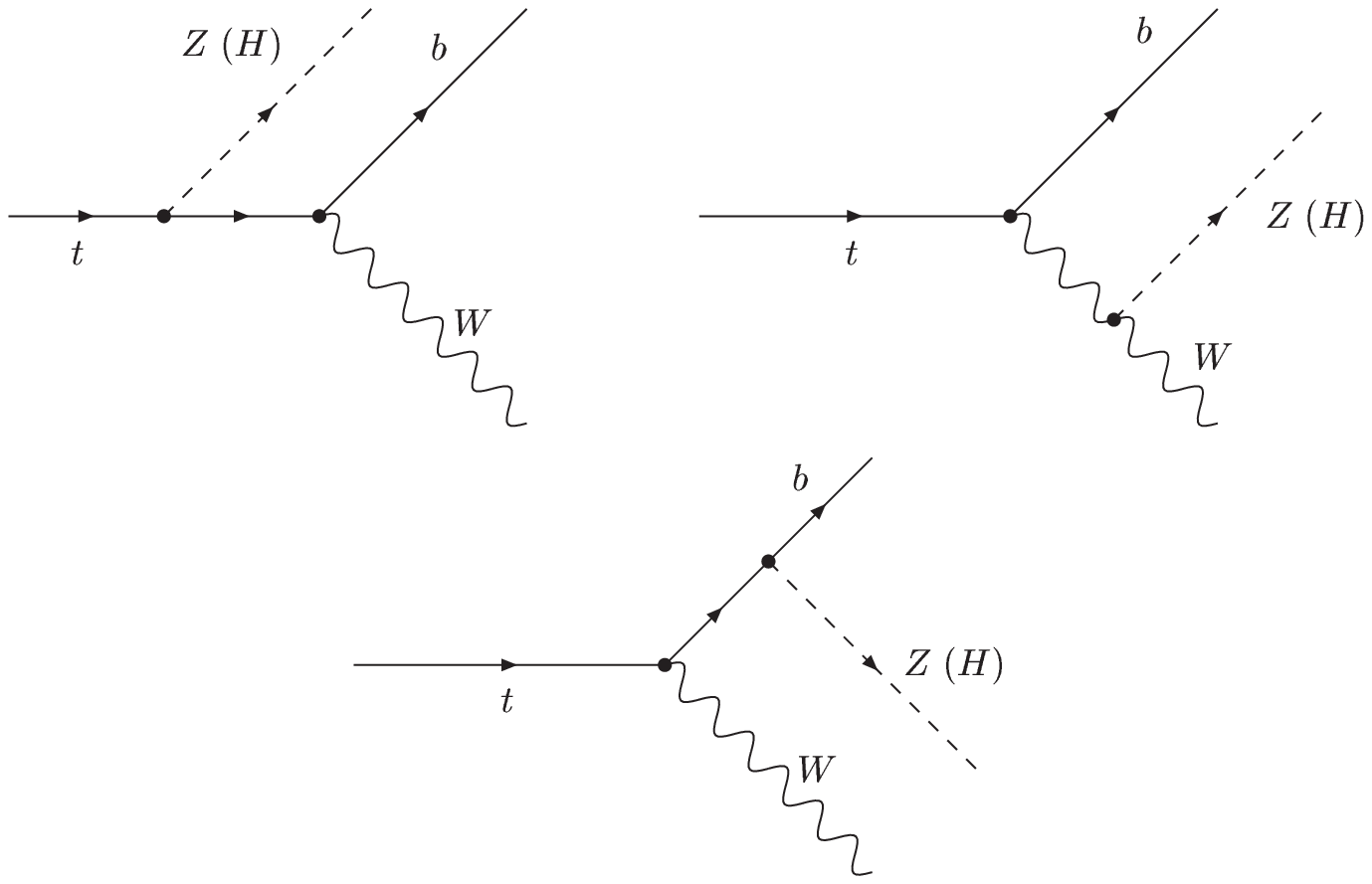}} 
\vspace*{-10.2cm} 
\caption{  Feynman graphs for the decay \tbwzt (\tbwhm). 
 } 
\label{bmf:fig1} 
\end{center} 
\end{figure*} 

At tree level, the decay \tbwzt (Fig.~\ref{bmf:fig1}) has some 
peculiar features, since the process occurs near the kinematical 
threshold ($\mt \sim \mw + 
\mz +\mb$)~\cite{Decker:1993wz,Mahlon:1995us,Jenkins:1997zd,bm:conti}.  
This fact makes the $W$ and $Z$ 
finite-width effects crucial in the theoretical prediction of the 
corresponding width~\cite{Mahlon:1995us}.  
Because the $W$ and $Z$ are unstable and not observed  
directly, more than one definition of the $t\to bWZ$ branching ratio  
is possible. If defined according to  
\be 
\tilde\Gamma(\tbwz)\equiv \frac{\Gamma(t \to b \mu \nu_{\mu} \nu_e 
\bar{\nu}_e)} 
   {{\rm BR}(W \to \mu \nu_{\mu}) {\rm BR}(Z \to \nu_e \bar{\nu}_e)}, 
\label{bme:eq_mahlon_nu} 
\ee 
including a consistent treatment of $W$ and $Z$ width effects,  
the branching ratio is to a very good approximation given by the  
double resonant set of diagrams (shown in Fig.~\ref{bmf:fig1}), since  
the background to the neutrino decay of the $Z$ is negligible.  
One obtains \cite{bm:conti}, for $\mt=175$~GeV, 
\be 
{\rm BR}(\tbwz)=BR_{res}(\tbwz)= 2.1 \times 10 ^{-6}. 
\label{bme:res} 
\ee 
However, the signature $b \mu \nu_{\mu} \nu_e \bar{\nu}_e$ 
is not practical from an experimental point of view. 
In~\cite{Mahlon:1995us}, a first estimate of 
${\rm BR}(\tbwz)$ was given on the basis of the definition  
\be 
\Gamma(\tbwz)\equiv \frac{\Gamma(t \to b \mu \nu_{\mu} e^+e^-)} 
{{\rm BR}(W \to 
  \mu \nu_{\mu}) {\rm BR}(Z \to e^+e^-)} 
\label{bme:eq_mahlon}, 
\ee 
which involves experimentally well-observable decays, but includes  
contributions to the numerator from $t\to bW\gamma$ decays (with  
$\gamma\to e^+e^-$) and other ``background'' diagrams.  
The estimate for the corresponding branching ratio is 
\be 
BR_{cut}(\tbwz)\simeq 6 \times 10^{-7}, 
\label{bme:eq_m} 
\ee 
for $\mt=175$~GeV, 
assuming a minimum cut of $0.8\mz$ on the $e^+e^-$-pair invariant mass. 
This cut tries to cope with the contribution of background graphs where 
the $e^+e^-$ pair comes not from a $Z$ boson but from a photon. 
 
 
If the Higgs boson is light enough, one could also have the decay 
\tbwht (Fig.~\ref{bmf:fig1}), although the present limits on $\mh$ 
strongly suppress its rate. 
For $\mh \gta 100$~GeV, one gets~\cite{Mahlon:1995us}: 
\be 
{\rm BR}(\tbwh) \lta 7 \times 10 ^{-8}. 
\ee 
Finally, the decay \tcwwt is very much suppressed by a GIM factor 
$\frac{\mbb}{\mww}$ in the amplitude. 
One then gets~\cite{Jenkins:1997zd}: 
\be 
{\rm BR}(\tcww) \sim  10 ^{-13}. 
\label{bme:tcww} 
\ee 
One can also consider the radiative three-body decays 
\tbwgt and \tbwfm. These 
channels suffer from infrared divergences and the evaluation of their rate 
requires a full detector simulation, including for instance the effects 
of the detector resolution and the jet isolation algorithm. 
In an idealised situation where the rate is computed in the $t$ rest frame 
with a minimum cut of 10 GeV on the gluon or photon energies, 
one finds~\cite{bm:mahl}: 
\be 
{\rm BR}(\tbwg) \simeq  0.3 \quad , \quad 
{\rm BR}(\tbwf) \simeq  3.5 \times 10 ^{-3}. 
\ee 
 
The FCNC decays \tcgm, \tcft and \tczt occur at one loop, 
and are also GIM suppressed by a factor $\frac{\mbb}{\mww}$ in the amplitude. 
Hence, the corresponding rates are very small~\cite{Eilam:1991zc}: 
\be 
{\rm BR}(\tcg) \simeq  5 \times 10^{-11}  \quad , \quad 
{\rm BR}(\tcf) \simeq  5 \times 10^{-13}  \quad , \quad 
{\rm BR}(\tcz) \simeq  1.3 \times 10^{-13} 
\label{bme:fcnc} 
\ee 
For a light Higgs boson, one can consider also the FCNC decay \tcht. 
A previous evaluation of its rates~\cite{Eilam:1991zc} has now been corrected. 
For $\mh \simeq 100 \;(160)$~GeV, one gets~\cite{bm:soddu}: 
\be 
{\rm BR}(\tch) \simeq 0.9 \times 10^{-13} \;\; (4 \times 10^{-15}). 
\ee 
 
To conclude the discussion of rare SM decays of the top quark, we 
point out here the existence of some studies on {\it semi-exclusive} 
$t$-quark decays where the interaction of 
quarks among the $t$ decay products may lead to final states with one 
hadron (meson) recoiling against a jet. In~\cite{Handoko:1999iu} 
decays with an $\Upsilon$ meson in the final state and  
decays of the top through an 
off-shell $W$ with virtual mass $M_{W^*}$ near to some resonance $M$, 
like $\pi^+, \, \rho^+, \, K^+, \, D_s^+$, were considered. 
An estimate for the latter case is 
\begin{eqnarray} 
 \Gamma(t \to b \, M) \approx \frac{G_F^2 \, m^3_t}{144\pi}f^2_M 
|V_{q q'}|^2.  \label{eq:semdec4} 
\end{eqnarray} 
The typical values of the corresponding branching ratios are too small to 
be measured: 
\be 
{\rm BR}(t \to b \pi) \sim 4 \cdot  10^{-8} \quad , \quad {\rm BR}(t 
\to b D_s) \sim 2 \cdot 10^{-7}. \ee  
 
In Table~\ref{bmt:sm} we summarize the expected decay 
rates for the main top decay channels in the SM. 
\begin{table}[th] 
\begin{center} 
\caption{ Branching ratios for the main SM top decay channels. } 
\label{bmt:sm}\vspace*{0.1cm} { 
\begin{tabular}{|l|c||l|c|} \hline 
 channel  & $BR_{\rm SM}$ &  channel  & $BR_{\rm SM}$  \\ \hline  \hline 
  $b W$   & $1$          & 
  $s W$    & $1.6 \cdot 10^{-3}$ \\ \hline 
  $d W$    & $\sim 10^{-4}$     & 
  $b W g$  &$ 0.3$ {\small($E_{g} > 10$~GeV)} \\ \hline 
  $b W \gamma$  & $ 3.5\cdot 10^{-3}$ {\small $(E_{\gamma} > 10$~GeV)} & 
 $b WZ$ & $2\cdot 10^{-6}$ \\ \hline 
  $c W^+ W^-$ & $\sim 10^{-13}$  & 
  $b W^+ H$ & $< 10^{-7}$ \\ \hline 
 $q g$  & $5\cdot 10^{-11}$ & 
 $q \gamma$ & $5\cdot 10^{-13}$ \\ \hline 
 $q Z$ & $1.3\cdot 10^{-13}$ & 
  $c H$ &  $< 10^{-13}$ \\ \hline 
\end{tabular} 
} 
\end{center} 
\end{table} 
 
\subsection{Beyond the Standard Model decays} 
 
The fact that a measurement of the top width is not available and that  
the branching ratio ${\rm BR}(\tbw)$ is a model dependent quantity   
makes the present 
experimental constraints on the top decays beyond the SM quite weak. 
Hence, the possibility 
of $t$ decays into new massive states with branching fraction of 
order ${\rm BR}(\tbw)$ is not excluded. 
Apart from the production of new final states with large branching fractions, 
we will see that  new physics 
could also give rise to a considerable increase in the rates of 
many decay channels that in the SM framework are below the threshold of 
observability at the LHC. 
 
\subsubsection{$4^{th}$ fermion family} 
Extending the SM with a $4^{th}$ fermion family can alter considerably a few 
$t$ decay channels. First of all, when adding a $4^{th}$ 
family to the CKM matrix 
the present constraints on the $|V_{tq}|$ 
elements are considerably relaxed.  In particular, 
$|V_{ts}|$ and   $|V_{td}|$ can grow up 
to about 0.5 and 0.1, respectively~\cite{Caso:1998tx}. 
Correspondingly, assuming $|V_{tb}|\sim 1$ for the sake of normalisation, 
one can have up to : 
\be 
{\rm BR}_4(\tsw) \sim 0.25 \; \; {\rm and} \; \; {\rm BR}_4(\tdw) \sim 0.01, 
\ee 
to be confronted with the SM expectations in~(\ref{bme:tsw}). 
 
The presence of a $4^{th}$ fermion family could also show up in the 
$t$ direct decay into a heavy $b'$ quark with a relatively small 
mass ($m_{b'} \sim 100$~GeV)~\cite{Atwood:1997iz}. This channel would 
contribute to the \tcwwt decay, 
with a rate: 
\begin{eqnarray} 
 {\rm BR}(t \to W^+ b'(\to W^- c)) \sim 10^{-3}\;(10^{-7}) 
\quad {\rm at} \quad m_{b'} = 100 \;(300) \;\;{\rm GeV}, 
\label{bme:tcwwq} 
\end{eqnarray} 
to be confronted with the SM prediction in~(\ref{bme:tcww}). 
\subsubsection{Two Higgs Doublet  models (2HDM's)} 
The possibility that the EW symmetry breaking involves more 
than one Higgs doublet is well motivated theoretically. 
In particular, three classes of two Higgs doublet models have 
been examined in connection with rare top decays, called model I, II 
and III. The first two are characterised by an {\it ad hoc} 
discrete symmetry which forbids tree-level FCNC~\cite{bm:glashow}, 
that are strongly 
constrained in the lightest quark sector. In model I  and model II, 
the up-type quarks and down-type quarks couple to the same scalar 
doublet and to two different doublets, respectively 
(the Higgs sector of the  MSSM is an example of model II). 
In model III~\cite{bm:hall,bm:hou}, the above discrete symmetry 
is dropped and tree-level FCNC are allowed. In particular, 
a tree-level coupling $tcH$ is predicted with a coupling constant 
$\lta \sqrt{\mt\mc}/v$ (where $v$ is the Higgs vacuum 
expectation value). 
 
Since enlarging the Higgs sector automatically implies the presence of 
charged Higgs bosons in the spectrum, 
one major prediction of these new frameworks is  the decay $t\to bH^+$, 
possibly with rates competitive with ${\rm BR}(\tbw)$ for  
$m_{H^+}\lta 170$ GeV. 
In the MSSM, one expects ${\rm BR}(t\to bH^+)\sim 1$, both  
at small and large values 
of $\tan\beta$. 
 The interaction Lagrangian describing the 
$H^{+} t\,\bar{b}$-vertex 
in the MSSM  is~\cite{Barger:1987nn}: 
\begin{equation} 
{\cal L}_{Htb}={g\over\sqrt{2}M_W}\,H^+\, 
\left [ \bar{t}\, (m_t\cot\beta\,P_L 
+ m_b\tan\beta\,P_R)\,b+ \bar{\nu}\,( m_\ell \tan\beta\,P_R)\,\ell\right ] \, + \, 
{\rm h.c.}\,, 
\label{eq:LtbH} 
\end{equation} 
where $P_{L,R}=1/2(1\mp\gamma_5)$ are the chiral projector operators. 
 
At tree level the corresponding decay widths of 
$t \to b H^+$, $H^+ \to \tau \nu$, and 
$H^{+} \to t \bar{b}$ (or, analogously, of $H^{+} \to c \bar{s}$) 
 are equal to~\cite{Barger:1987nn} 
\begin{eqnarray} 
 \Gamma(t \to b H^+ ) &=& \frac{g^2} {64 \pi M_W^2 m_t} |V_{tb}|^2 
\lambda^{1/2}\left(1, \frac{m_b^2}{m_t^2}, \frac{m_H^2}{m_t^2} \right ) 
 \times \nonumber \\ 
 &&\left[(m_t^2 + m_b^2 -m_H^2)(m_t^2 \cot^2\beta + m_b^2 \tan^2\beta) 
 + 4 m_t^2 m_b^2 \right ], \label{eq:chig3} \\ 
 \Gamma(H^+ \to \tau^+ \nu) &=& \frac{g^2 m_H}{32 \pi M_W^2} 
m_{\tau}^2 \tan^2\beta, \label{eq:chig1} \\ 
 \Gamma(H^+ \to t \bar b) &=& \frac{3g^2 }{32 \pi M_W^2 m_H} |V_{tb}|^2 
\lambda^{1/2}\left(1, \frac{m_b^2}{m_H^2}, \frac{m_t^2}{m_H^2} \right ) 
 \times \nonumber \\ 
 &&\left[(m_H^2 -m_t^2 -m_b^2)(m_t^2 \cot^2\beta + m_b^2 \tan^2\beta) 
 - 4 m_t^2 m_b^2 \right ], \label{eq:chig2} 
\end{eqnarray} 
where $\lambda(a,b,c) = a^2 + b^2 + c^2 - 2(ab +ac + bc)$, 
and $m_H \equiv m_{H^+}$. 
 
Consequently, if $m_H < m_t - m_b$, one  expects 
$H^+ \to \tau^+ \nu$ (favoured for large $\tan\beta$) and/or 
$H^+ \to c \bar{s}$ (favoured for small $\tan\beta$) to be the 
dominant decays. 
Hence, for $\tan\beta >1$ and 
 $H^+\to \tau^+ \nu$  dominant,  one  can look  for the channel 
$t\to bH^+$ by studying a possible excess in the $\tau$ lepton 
signature from the $t$ pair production~\cite{bm:cdf-tau}. 
On the other hand, 
if $\tan\beta < 2$ and $m_H > 130$~GeV, the 
large mass (or coupling) of the $t$-quark causes 
${\rm BR}(H^+ \to t^{*} \bar b \to W^+ b \bar b)$ to exceed 
${\rm BR}(H^+ \to c \bar s)$ (Fig.~\ref{fig:chig}, see~\cite{Ma:1998up} 
for details). 
\begin{center} 
\vskip-0.4cm
  \begin{figure}[th] 
   \centerline{ 
    \includegraphics[width=8cm,clip]{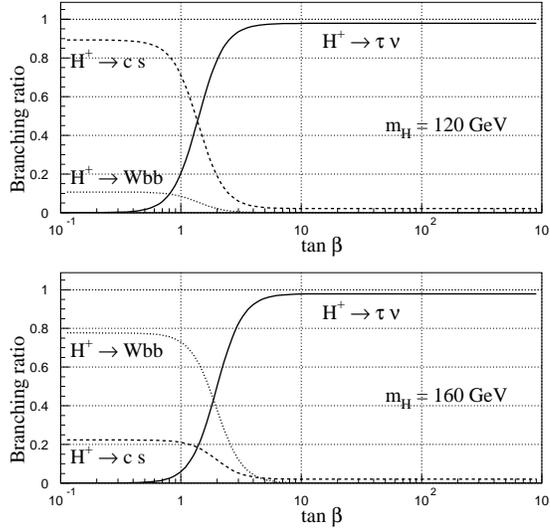}} 
\vskip -0.4cm 
   \caption{Branching fractions for three $H^+$ decay modes for 
two values of $m_{H^+}$ vs. $\tan\beta$.}
\vskip-0.5cm 
   \label{fig:chig} 
  \end{figure} 
\end{center} 
 
As a consequence, new interesting signatures at LHC such as 
leptons plus multi-jet channels  with four $b$-tags, coming from 
the gluon-gluon fusion process $gg\to t\bar b H^-$, 
followed by the $H^-\to \bar t b$ decay, 
have been studied~\cite{bm:stirling}. These processes could 
provide a viable signature over a limited but interesting range of the 
parameter space. 
 
One should recall however that both ${\rm BR}(t\to bH^+)$ 
and  ${\rm BR}(H^+\to W^+b\bar{b})$ are very sensitive 
to higher-order corrections, which are highly model dependent 
\cite{bm:coarasa}. 
 
In model III, the tree-level FCNC decay $t \to ch$ can occur 
with branching ratios up to 10$^{-2}$~\cite{bm:hou}. 
In~\cite{Bar-Shalom:1998sj}, the rate for the channel 
$t \to ch \to cWW (cZZ)$ 
has been studied. Accordingly, 
${\rm BR}(\tcww)$ can be enhanced by several orders of magnitude 
with respect to its SM value. In particular, for an on-shell decay 
with $2\mw\lta m_h \lta \mt$, one can have up to ${\rm BR}(\tcww)\sim 10^{-4}$ 
from this source.  The same process was considered in a wider range of 
models, where the decay \tcwwt can occur not only through 
a scalar exchange but also through a fermion or vector exchange 
\cite{Atwood:1997iz}. In this framework, the fermion exchange too could 
lead to detectable rates for \tcwwm, as in~(\ref{bme:tcwwq}). 
 
In 2HDM's, the prediction for the FCNC decays 
\tcgm, \tcft and \tczt can also be altered. While in models I and II 
the corresponding branching fractions cannot approach the detectability 
threshold~\cite{Eilam:1991zc}, in model III predicts values up to 
${\rm BR}(\tcg) \simeq  10^{-5}$, ${\rm BR}(\tcf) \simeq  10^{-7}$ and 
${\rm BR}(\tcz) \simeq  10^{-6}$~\cite{bm:reina}. 
 
By further extending the 2HDM's Higgs sector and including Higgs triplets, 
one can give rise to a vertex $HWZ$ at tree level in a consistent way 
\cite{bm:diazcruz}. Accordingly, the \tbwzt decay can be mediated 
by a charged Higgs (coupled with $\mt$) that can enhance 
the corresponding branching fraction 
up to ${\rm BR}(\tbwz) \sim 10 ^{-2}$. Large enhancements can also be expected 
in similar models for the channels $t \to sWZ\;$and $t \to dWZ$.

\subsubsection{Minimal Supersymmetric Standard Model (MSSM)} 
Supersymmetry could affect the $t$ decays in different ways. 
(Here, we assume the MSSM framework~\cite{susy}, with 
(or without, when specified) $R$ parity conservation.) 
 
First of all, two-body decays into squarks and gauginos, such as 
$t \to \tilde t_1 \tilde g$, $t \to \tilde b_1 \tilde \chi_1^+$, 
$t \to \tilde t_1 \tilde \chi_1^0$, could have branching ratios 
of order ${\rm BR}(\tbw)$, if allowed by the phase space 
(see, i.e.~\cite{bm:guasch-three} for references). 
QCD corrections to the channel $t \to \tilde t_1 \tilde g$ 
have been computed in~\cite{bm:zhu} and were found to increase 
its width up to values even larger than $\Gamma(\tbw)$. 
Three-body $t$ decays in supersymmetric particles were 
surveyed in~\cite{bm:guasch-three}. 
 
The presence of light top and bottom squarks, charginos and 
neutralinos in the MSSM spectrum could also give rise to a CP 
asymmetry of the order $10^{-3}$ in the partial widths 
for the decays $t \to b W^+\;$ and $\bar t \to \bar b W^-$ 
\cite{Bar-Shalom:1998si,bm:aoki}. 
 
Explicit $R$-parity violating interactions~\cite{bm:roy} could 
provide new flavour-changing $t$ decays, both at tree-level 
(as in the channels $t \to \tilde \tau b$ 
and $t \to \tau b \tilde \chi_1^0\;$~\cite{bm:magro}) 
and at one loop  (as in $t \to c \tilde \nu\;$~\cite{bm:soni}), 
with observable rates. For instance, ${\rm BR}(t \to c \tilde \nu) \sim 
10^{-4}-10^{-3}\;$ in particularly favourable cases. 
 
Another sector where supersymmetric particles could produce 
crucial changes concerns  the one-loop FCNC decays 
\tcgm, \tcfm, \tczt and \tcht, which in the SM are unobservably 
small. In the MSSM with universal soft breaking the situation is 
not much affected, 
while, by relaxing the universality with a large flavour mixing between the 
2$^{nd}$ and 3$^{rd}$ family only, one can reach values such as 
\cite{deDivitiis:1997sh,Li:1994mg}: 
\be 
\mbox{BR}_{\rm MSSM}(\tcg)  \sim    10^{-6}  \quad , \quad 
\mbox{BR}_{\rm MSSM}(\tcf)  \sim    10^{-8}  \quad , \quad 
\mbox{BR}_{\rm MSSM}(\tcz)  \sim    10^{-8}, 
\label{bme:fcncsusy} 
\ee 
which, however, are still not observable. 
The introduction of baryon number violating couplings in broken $R$-parity 
models could on the other hand give large enhancements~\cite{bm:yang}, 
and make some of these channels observable: 
\be 
\mbox{BR}_{R\!\!\!\! /}(\tcg) \sim   10^{-3}  \quad , \quad 
\mbox{BR}_{R\!\!\!\! /}(\tcf) \sim   10^{-5}  \quad , \quad 
\mbox{BR}_{R\!\!\!\! /}(\tcz) \sim   10^{-4}. 
\label{bme:fcncbr} 
\ee 
 
A particularly promising channel is the FCNC decay $t \to c h$ in 
the framework of MSSM, where $h = h^0,H^0,A^0$ is any of the 
supersymmetric neutral Higgs bosons~\cite{bm:sola}. 
By including the 
leading MSSM contributions to these decays (including gluino-mediated FCNC 
couplings), 
one  could approach the detectability 
threshold, especially in the case of the light CP-even Higgs boson, 
for which one can get up to: 
\be 
\mbox{BR}_{\rm MSSM}(t \to c h^0) \sim 10^{-4}. 
\ee 
 
\subsubsection{Anomalous couplings} 
In the framework of the top anomalous couplings described in 
Section~\ref{ANOM}, one can predict large enhancements in different FCNC 
top decay channels. While the \tcgm, \tcfm, and \tczt processes are analysed 
in section~\ref{ANOM}, here we concentrate on the possible 
FCNC contributions to  the top decays into two gauge bosons, 
$t \to q VV$, where $V$ is either a $W$ or a $Z$ and $q=c,u$: 
\be 
 t \to q W^+ Z     \quad , \quad 
 t \to q W^+ W^-   \quad , \quad 
 t \to q Z Z.      \label{eq:tvv} 
\ee 
In the SM, the first two decays occur at 
tree level, while $t \to q Z Z$  proceeds only through loop contributions. 
We will see that within the present experimental limits 
on the top anomalous couplings, the rates for these processes can be large 
with respect to the SM prediction, but are still below the detectability 
threshold at the LHC. 
 
The FCNC contribution to the first channel in (\ref{eq:tvv}), for 
the anomalous coupling $\kappa_z \approx 0.3$, 
has a rate of the the same order of magnitude as 
the SM ${\rm BR}(\tbwz)$~\cite{bm:slab}: 
\begin{eqnarray} 
\mbox{BR}_{\rm FCNC}(t \to c W Z)  \sim 10^{-6}  \approx  
\mbox{BR}_{\rm SM}(t \to bWZ). 
\end{eqnarray} 
 
Top anomalous FCNC interactions with both a photon and a $Z$-boson 
contribute to the second process in eq.(\ref{eq:tvv}). 
Contrary to the SM case this amplitude has no GIM suppression. As a result, 
the corresponding branching ratio can have almost the same value 
as that of the $t \to q W Z$ decay~\cite{bm:slab}: 
\begin{eqnarray} 
  \mbox{BR}_{\rm FCNC}(t \to c W^+ W^-) \sim 10^{-7} \gg 
\mbox{BR}_{\rm SM}(\tcww). 
\end{eqnarray} 
 
For the  $t \to q ZZ$ decay mode, a coupling $\kappa_z \sim 0.3$ 
 gives a branching ratio  much greater than the corresponding SM one 
($\lta 10^{-13}$~\cite{bm:slab}): 
\begin{eqnarray} 
 \mbox{BR}_{\rm FCNC}(t \to q ZZ) \sim 10^{-8} \gg 
\mbox{BR}_{\rm SM}(t \to q ZZ), 
\end{eqnarray} 
but still too small to be detected at LHC. 
 
In summary, the 
observation of any of these decays at LHC would indicate  new physics not 
connected with the top FCNC interactions  (see, 
for example,~\cite{Bar-Shalom:1998sj}). 
 

\subsection{ATLAS studies of (rare) top quark decays and couplings} 
In ATLAS various analyses have been performed on top decays, using 
the \pyth~Monte-Carlo interfaced to a fast detector simulation (ATLFAST). 
In the following, the most relevant results are reported. 
 
\subsubsection{${\rm BR}(t \ra\ bX)$ and measurement of $| V_{tb} |$} 
The SM prediction  ${\rm BR}(t \ra\ \Wpl b) \approx$ 1 can be 
checked by comparing the number of observed (1 or 2) $b$-tags in a \ttbar\ 
sample. The first $b$-tag is used to identify the event as a \ttbar\ event, 
and the second $b$-tag (if seen) is used to determine the 
fraction of top decays producing a $b$ quark. 
Within the three-generation SM, and assuming unitarity of the CKM 
matrix, the ratio of double $b$-tag to single $b$-tag events is given by: 
\begin{equation} 
 {\rm R}_{2b/1b} = {\rm BR}(t \ra\ Wb)/{\rm BR}(t \ra\ Wq) 
                 = | V_{tb} | ^{2}/(| V_{tb} | ^{2} + 
                   | V_{ts} | ^{2} + | V_{td} | ^{2}) 
                 = | V_{tb} | ^{2} 
\end{equation} 
 
The CDF collaboration has used the tagging method in leptonic \ttbar\ events 
to obtain the result  
${\rm R}_{2b/1b} = 0.99 \pm 0.29$~\cite{CDFTart}, 
which translates to a limit of $| V_{tb} |\ >$ 0.76 at the 95\% CL 
assuming three-generation unitarity. If this constraint is relaxed, a lower 
bound of $| V_{tb} |\ >$ 0.048 at the 95\% CL is obtained, implying only 
that $| V_{tb} |$ is much larger than either $| V_{ts} |$ or 
$| V_{td} |$. 
 
The LHC will yield a much more precise 
measurement of ${\rm R}_{2b/1b}$. For example, 
\ttbar\ events in the single lepton plus jets mode can be selected by 
requiring an isolated electron or muon with \pT\ $>$ 20 GeV, \ETmiss\ $>$ 
20 GeV, and at least four jets with \pT\ $>$ 20 GeV. Requiring that at least 
one of the jets be tagged as a $b$-jet produces a clean sample of \ttbar\ 
events, with $S/B$ = 18.6, with the remaining background coming mostly from 
$W$+jet events~\cite{atlasphystdr}. Assuming a $b$-tagging efficiency of 60\%, 
a sample of 820 000 single $b$-tagged events would be selected for an 
integrated luminosity of 10 \invfb . Of these, 276 000 would be expected to 
have a second $b$-tag, assuming the SM top quark branching ratios. 
This ATLAS study indicates that the 
statistical precision achievable would correspond to a relative error of 
$\delta {\rm R}_{2b/1b}/{\rm R}_{2b/1b}$ (stat.) = 0.2\% for an integrated 
luminosity of 10 \invfb . The final uncertainty will be dominated by 
systematic errors due to the uncertainty in the $b$-tagging efficiency and 
fake $b$-tag rates, as well as correlations affecting the efficiency for 
$b$-tagging two different jets in the same event. Further study is needed 
to estimate the size of these systematic uncertainties. 
 
\subsubsection{${\rm BR}(t \ra\ WX)$} 
\label{sec:ttoWX} 
The measurement of the ratio of di-lepton to single lepton events in a \ttbar\ 
sample can be used to determine ${\rm BR}(t \ra\ WX)$. In this case, the first 
lepton tags the \ttbar\ event, and the presence of a second lepton is used 
to determine the fraction of top quark decays producing an isolated lepton, 
which can be then be related to the presence of a $W$ (or other leptonically 
decaying states) in the decay. The SM predicts that 
${\rm R}_{2l/1l}$ = ${\rm BR}(\Wlnu) \approx$ 2/9 where 
$\ell = (e,\mu)$. Deviations from this prediction could be caused by new physics, 
for example, the existence of a charged Higgs boson. 
The 
dominant $H^{+}$ decays in such instances are usually considered to be 
$H^{+} \ra\ \tau\nu$ or $H^{+} \ra\ c \bar{s}$. In either case, the number of 
isolated electrons and muons produced in top decay would be reduced, and 
${\rm R}_{2l/1l}$ would be less than the SM prediction. 
 
A study performed by ATLAS~\cite{atlasphystdr} shows that 
with an integrated luminosity of 10 \invfb , a clean sample of about 443 000 
\ttbar\ events in the single lepton plus jets mode could be selected by 
requiring an isolated electron or muon with \pT\ $>$ 20 GeV, \ETmiss\ $>$ 
20 GeV, and at least two $b$-tagged jets with \pT\ $>$ 20 GeV. To determine 
${\rm R}_{2l/1l}$, one then measures how many of these events have a second 
isolated electron or muon, again with \pT\ $>$ 20 GeV, and of the opposite 
sign to the first lepton. Assuming the SM, one would expect a selected sample 
of about 46 000 di-lepton events with these cuts. Given these numbers, the 
statistical precision achievable would correspond to a relative error of 
$\delta{\rm R}_{2l/1l}/{\rm R}_{2l/1l}$ (stat.) = 0.5\% for an integrated 
luminosity of 10 \invfb . Further study is required to estimate the 
systematic uncertainty on ${\rm R}_{2l/1l}$ due to the lepton identification 
and fake rates. 

\subsubsection{Radiative Decays: $t \ra\ WbZ, t \ra\ WbH$} 
The `radiative' top decay $t \ra\ WbZ$ has been suggested~\cite{Mahlon:1995us} 
as a sensitive probe of the top quark mass, since the measured value of $m_{t}$ 
is close to the threshold for this decay. For the top mass of (173 $\pm$ 5.2) 
GeV~\cite{Caso:1998tx}, 
the SM prediction, based on the $Z\to ee$ signature and a cut $m_{ee}>0.8M_Z$ 
(see Section \ref{bms:sm}), 
is BR$_{cut}(t \ra\ WbZ)$ = $(5.4^{+4.7}_{-2.0}) 
\times 10^{-7}$~\cite{Mahlon:1995us}. Thus, within the current uncertainty 
$\delta m_{t} \approx$ 5 GeV, the predicted branching ratio 
varies by approximately a factor of three.  
A measurement of ${\rm BR}(t \ra\ WbZ)$ 
could therefore provide a strong constraint on the value of $m_{t}$. Similar 
arguments have been made for the decay $t \ra\ WbH$, assuming a relatively 
light SM Higgs boson. 
 
ATLAS has studied the experimental sensitivity to the decay $t \ra\ WbZ$ 
\cite{atlasphystdr,atlraretdk}, with the $Z$ being reconstructed via the 
leptonic decay $Z \ra ll$ $(\ell = e,\mu)$, and the $W$ through the hadronic decay 
$W \ra\ jj$. The efficiency for exclusively reconstructing $t \ra\ WbZ$ 
is very low, due to the soft \pT\ spectrum of the $b$-jet in the $t \ra\ WbZ$ 
decay. Instead, a semi-inclusive technique was used, where a $WZ$ pair close 
to threshold was searched for as evidence of the $t \ra\ WbZ$ decay. Since 
the $t \ra\ WbZ$ decay is so close to threshold, the resolution on $m_{WZ}$ 
is not significantly degraded with respect to the exclusive measurement. The 
selection of $Z \ra\ ll$ candidates required an opposite-sign, same-flavor 
lepton pair, each lepton having \pT\ $>$ 30 GeV and $|\eta |<$ 2.5. 
The clean $Z \ra\ ll$ signal allows a wide di-lepton mass window to be taken 
(60 GeV $<$ \mll\ $<$ 100 GeV) in order to have very high efficiency. 
Candidates for $W \ra\ jj$ decay were formed by requiring at least two jets, 
each having \pT\ $>$ 30 GeV and $|\eta |<$ 2.5, and satisfying 70 GeV 
$< mjj <$ 90 GeV. The $lljj$ invariant mass resolution was $\sigma[m_{WZ}]$ 
= 7.2 $\pm$ 0.4 GeV, and the signal efficiency was 4.3\%. 
 
The dominant backgrounds come 
from processes with a $Z$ boson in the final state, primarily $Z$+jet 
production, and to a much lesser extent from $W$Z and \ttbar\ production. 
In order to reduce the $Z$+jet background, an additional cut requiring a 
third lepton with \pT\ $>$ 30 GeV was made. For the signal process 
$\ttbar\ \ra\ (WbZ)(Wb)$, this cut selects events in which the $W$ from the 
other top decays leptonically. 
After this selection, and with a cut on $m_{WZ}$ of $\pm$10 GeV around the 
top mass, the total expected background was reduced to $\approx$ 1.5 events 
(mostly from $WZ$ production) per 10 \invfb . Requiring at least five 
events for signal observation leads to a branching ratio sensitivity of order 
10$^{-3}$. Since the background has been reduced essentially to zero, the 
sensitivity should improve approximately linearly with integrated luminosity. 
However, even with a factor of ten improvement for an integrated luminosity 
of 100 \invfb , the sensitivity would still lie far above the SM expectation 
of order 10$^{-7}-10^{-6}$. 
 
Given this result, observation of the decay $t \ra\ WbH$ does not look 
possible. The current LEP limit on $m_{H}$ implies that the Higgs is 
sufficiently heavy that, in the most optimistic scenario that the Higgs mass 
is just above the current limit, ${\rm BR}(t \ra\ WbH) \lta $ ${\rm BR}(t \ra\ WbZ)$. 
As $m_{H}$ increases further, ${\rm BR}(t \ra\ WbH)$ drops quickly. Assuming $m_{H} 
\approx m_{Z}$, one would have to search for $t \ra\ WbH$ using the dominant 
decay $H \ra\ b \bar{b}$. The final state suffers much more from background 
than in the case of $t \ra\ WbZ$, where the clean \Zll\ signature is a key 
element in suppressing background. Although ${\rm BR}(H \ra\ b \bar{b})$ in this 
$m_{H}$ range is much larger than {\rm BR}(\Zll\ ), the large increase in 
background will more than compensate for the increased signal acceptance, 
and so one expects the sensitivity to ${\rm BR}(t \ra\ WbH)$ to be worse than for 
${\rm BR}(t \ra\ WbZ)$. The decay $t \ra\ WbH$ has therefore not been studied in 
further detail. 
\subsubsection{$t \ra\ H^{+}b$} 
\label{sec:ttoHb} 
 
Limits on the mass of the charged Higgs have been obtained from a number of 
experiments. 
An indirect limit obtained from world averages of the $\tau$ branching 
ratios excludes at 90\% CL any charged Higgs with $m_{H^{+}} 
< 1.5 \tan \beta$ GeV~\cite{Stahl:1997gu}, where $\tan \beta$ is the 
ratio of the vacuum expectation values of the two Higgs doublets. CLEO 
indirectly excludes $m_{H^{+}} < 244$ GeV for $\tan \beta >$ 50 at 
95\% CL, assuming a two-Higgs-doublet extension to the  
SM~\cite{Alam:1995aw}, while the LEP experiments directly exclude 
$m_{H^{+}} < 59.5$ GeV/\csq\ at 95\% CL 
\cite{Barate:1999rn}. 
Searches at the Tevatron have extended the region of excluded 
$[ m_{H^{\pm}},\tan \beta ]$ parameter space, particularly at small and large 
$\tan \beta$, and set a limit on the branching ratio ${\rm BR}(t \ra\ H^{+}b) <$ 0.45 
at 95\% CL~\cite{Abe:1997rk}. 
Run 2 at the Tevatron will be sensitive to branching fractions 
${\rm BR}(t \ra\ H^{+}b) >$ 11\%~\cite{TeV2000}. 
 
ATLAS has performed an analysis of the experimental sensitivity to the 
$t \ra\ H^{+}b$ decay, followed by $H^{+} \ra\ \tau\nu$, in the 
context of the MSSM~\cite{atlasphystdr,atlH+totaunu}. Since the 
relevant $t \ra\ H^{+}b$ branching ratio is proportional to 
$(m_{t}^{2} \cot^2\beta + m_{b}^{2} \tan^2\beta)$  
(see~(\ref{eq:chig3})), for a given value of $m_{H^{\pm}}$ the branching 
ratio for such decays is large at small and at large $\tan \beta$, but 
has a pronounced minimum at $\tan \beta \sim \sqrt{m_{t}/m_{b}} \sim$ 
7.5. The exact position of this minimum and its depth is sensitive to 
QCD corrections to the running $b$-quark mass. 
 
In the ATLAS analysis, an isolated high-\pT\ lepton with 
\modeta\ $<$ 2.5 is required to trigger the experiment, which in 
signal events originates from the semi-leptonic decay of the second top 
quark. One identified hadronic tau is then required, and at least three jets 
with \pT\ $>$ 20 GeV and \modeta\ $<$ 2.5, of which two are required to be 
tagged as $b$-jets. This reduces the potentially large backgrounds from 
$W$+jet and $b\bar b$ production to a level well below the \ttbar\ signal 
itself. These cuts enhance the $\tau$-lepton signal from $H^{\pm}$ decays 
with respect to that from $W$ decay, and select mostly single-prong 
$\tau$-decays. After the selection cuts and the $\tau$ identification 
criteria are applied, $t \ra\ H^{+}b$ decays appear as final states 
with an excess of events with one isolated $\tau$-lepton compared to those 
with an additional isolated electron or muon. 
 
A signal from charged Higgs-boson production in \ttbar\ decays would be 
observed for all values of $m_{H^{\pm}}$ below $m_{t} - 20$ GeV over most 
of the $\tan \beta$ range. For moderate values of $\tan \beta$, for which 
the expected signal rates are lowest, the accessible values of $m_{H^{\pm}}$ 
are lower than this value by 20 GeV. The limit on the sensitivity to 
${\rm BR}(t \ra\ H^{+}b)$ is dominated by systematic uncertainties, arising mainly 
from imperfect knowledge of the $\tau$-lepton efficiency and of the number 
of fake $\tau$-leptons present in the final sample. These uncertainties are 
estimated to limit the achievable sensitivity to ${\rm BR}(t \ra\ H^{+}b)$ = 3\%. 
 
For charged Higgs masses below 150 GeV and for low values of $\tan \beta$, the 
$H^{\pm} \ra\ cs$ and $H^{\pm} \ra\ cb$ decay  
modes are not negligible. In the same 
mass range, the three-body off-shell decays $H^{\pm} \ra\ hW^{\ast}$, 
$H^{\pm} \ra\ AW^{\ast}$ and $H^{\pm} \ra\ bt^{\ast} \ra\ bbW$ also have 
sizeable branching ratios. 
When the phase-space increases, for 150 GeV $< m_{H^{\pm}} <$ 180 GeV, both 
the $bbW$ and the $hW^{\ast}$ mode could be enhanced with respect to the 
$\tau \nu$ mode. Decays into the lightest chargino $\tilde{\chi}_{1}^{\pm}$ 
and neutralino $\tilde{\chi}_{1}^{0}$ or decays into sleptons would dominate 
whenever kinematically allowed. For large values of $\tan \beta$ the 
importance of these SUSY decay modes would be reduced. However, for values as 
large as $\tan \beta$ = 50, the decay $H^{\pm} \ra\ \tilde{\tau}\tilde{\nu}$ 
would be enhanced, provided it is kinematically allowed and would lead to 
$\tau$'s in the final state. Their transverse momentum spectrum is, however, 
expected to be softer than that of $\tau$'s from the direct 
$H^{\pm} \ra\ \tau\nu$ decays. 
 
The $H^{\pm} \ra\ cs$ decay mode has been considered as a complementary one 
to the $H^{\pm} \ra\ \tau\nu$ channel by ATLAS for low values of $\tan \beta$. 
In the ATLAS analysis, one isolated high \pT\ lepton with \modeta\ 
$<$ 2.5 is required to trigger the experiment, which in signal events 
originates from the semi-leptonic decay of the second top quark. Two $b$-tagged 
jets with \pT\ $>$ 15 GeV and \modeta\ $<$ 2.5 are also required, with no 
additional $b$-jet. Finally, at least two non-$b$ central jets with \modeta\ 
$<$ 2.0 are required for the $H^{\pm} \ra\ cs$ reconstruction, and no 
additional jets above 15 GeV in this central region. Evidence for $H^{\pm}$ is 
searched for in the two-jet mass distribution. 
The mass peak from an $H^{\pm}$ decay can be reconstructed with a resolution 
of $\sigma = (5 - 8)$GeV if the mass of the $H^{\pm}$ is in the range between 
110 and 130 GeV. In this mass range, the peak sits on the tail of the 
reconstructed $W \ra\ jj$ distribution from \ttbar\ background events which 
decay via a $Wb$ instead of a $H^{\pm}b$. In the mass range 110 $< H^{\pm} <$ 
130 GeV, the $H^{\pm}$ peak can be separated from the dominant $W \ra\ jj$ 
background, with $S/B \approx$ 4-5\% and $S/\sqrt{B} \sim$ 5. 
This channel is complementary to the $H^{\pm} \ra\ \tau\nu$ channel for low 
$\tan \beta$ values. Whereas the $H^{\pm} \ra\ \tau\nu$ channel allows only 
the observation of an excess of events, it is possible to reconstruct a mass 
peak in the $H^{\pm} \ra\ cs$ decay mode. 
 
The $H^{\pm} \ra\ hW^{\ast}$, $H^{\pm} \ra\ AW^{\ast}$ and 
$H^{\pm} \ra\ bt^{\ast} \ra\ bbW$ have not been studied so far by ATLAS. With 
the expected $b$-tagging efficiency, these multi-jet decay modes are very 
interesting for a more detailed investigation. 
 
 
\subsubsection{$t \rightarrow Zq $ decay} 
The sensitivity to the FCNC decay $t \rightarrow Zq $ (with $q = u, c$) has 
been 
analyzed~\cite{chikovani} by searching for a signal in the channel $t\bar t 
 \rightarrow (Wb)(Zq)$, with the boson being reconstructed via the leptonic 
 decay $Z \rightarrow ll $. The selection cuts required a pair of isolated, 
opposite sign, same flavor leptons (electrons or muons), each with $p_T$ 
 $>$ 20~GeV and $|\eta|$ $<$ 2.5 and with $|m_{ll} - m_Z|$ $<$ 6~GeV. 
The dominant backgrounds come from  $Z+jet$ and $WZ$ production. 
 Not only cuts were applied on the $Zq$ final state, but 
also on the $Wb$ decay of the other 
top quark in the event, to further reduce the background. 
 Two different possible decay chains have been 
considered: the first (``leptonic mode'') where the $W$ decays leptonically $W 
 \rightarrow \ell \nu$, and the second (``hadronic mode'') with  
$W \rightarrow jj$. 
The hadronic $W$ decay signature has a much larger branching fraction, but 
suffers from larger backgrounds. 
The search in the leptonic mode required, in addition to the leptons from 
 the Z boson decay, a further lepton with $p_T$ $>$ 20 GeV and 
 $|\eta|$ $<$ 2.5, $E^{miss}_T$ $>$ 30~GeV, and at least two jets with 
 $p_T$ $>$ 50~GeV and $|\eta|$ $<$ 2.5. Exactly one of the high 
 $p_T$ jets was required to be tagged as a $b$-jet. 
The invariant mass spectrum of each $Zq$ combination was then 
formed from the $Z \rightarrow ll$ candidates taken with each of the non 
$b$-tagged jets. The $Zq$ invariant mass resolution was 10.1~GeV. Combinations 
were accepted if $m_{Zq}$ agreed with the known top mass within $\pm$ 24 GeV. 
  Assuming an 
  integrated luminosity of 100 fb$^{-1}$, 6.1 signal events survive the cuts 
   with 7 background events. 
  A value of 
  ${\rm BR}(t \ra Zq$) as low as $2\cdot 10^{-4}$ could be discovered at the 
  5$\sigma$ level. 
 
The search in the hadronic mode required, in addition to the $Z \rightarrow 
 ll$ candidate, at least four jets with $p_T$ $>$ 50~GeV and $|\eta|$ $<$ 2.5. 
 One of the jets was required to be tagged as a $b$-jet. 
 To further reduce the background, the decay $t \rightarrow jjb$ was 
 first reconstructed. A pair of jets, among those not tagged as a $b$-jet, 
  was considered a W candidate if $|m_{jj} - M_W|$ $<$ 16~GeV. $W$ candidates 
  were then combined with the $b$-jet, and considered as a top candidate if 
  $|m_{jjb} - m_t|$ $<$ 8~GeV. For those events with an accepted 
  $t\rightarrow jjb$ candidate, the invariant mass of the $Z$ candidate with 
  the remaining unassigned high $p_T$ jets was calculated to look for a signal 
   from $t \rightarrow Zq$ decays.  Combinations were accepted in case 
   $|m_{Zq} - m_t|$ $<$ 24~GeV. 
   Assuming an integrated luminosity of 100 fb$^{-1}$, one would get 0.4 
   signal 
    events, with 2 background events.

 
\subsubsection{ $t\rightarrow  \gamma q$ decay} 
The FCNC decay $t \rightarrow \gamma q$ (with $q = u, c$) 
can be searched for  as a peak in the $M_{\gamma \, j}$ spectrum in the region 
of $\mt$. 
The requirement of a high $p_T$ isolated photon candidate in 
  $t \bar t \rightarrow (Wb)(\gamma q)$ events is not sufficient to reduce the 
  QCD multi-jet background 
  to a manageable level. Therefore, the $t \rightarrow Wb$ decay of the 
  other top (anti-) 
  quark in the event was reconstructed using the leptonic 
  $W \rightarrow \ell \nu$ decay mode, and looking for the 
  $ t \bar t \rightarrow (Wb)( \gamma q) \rightarrow (\ell \nu b)(\gamma q)$ 
  final state. 
For the event selection, 
the ATLAS collaboration~\cite{atlasphystdr, atlraretdk} 
required the presence of an isolated photon with 
 $p_T$ $>$ 40~GeV and $|\eta|$ $<$ 2.5, an isolated electron or muon 
 with $p_T$ $>$ 20~GeV  and 
 $|\eta|$ $<$ 2.5, and E$_T^{miss}$ $>$ 20~GeV. Exactly 2 jets with 
 $p_T$ $>$ 20~GeV  were required, 
 in order to reduce the $t \bar t$ background. At least one of the jets 
 was required to be 
 tagged as a $b$-jet with $p_T$ $>$ 30~GeV and $|\eta|<$ 2.5. 
The $t \rightarrow \ell \nu b$ candidate was first reconstructed. 
The combination was  accepted as 
 a top quark candidate if $m_{\ell \nu b}$ agreed with $m_t$ within 
 $\pm$20 GeV. For these 
 events the $t \rightarrow \gamma q$ decay was sought by combining the 
 isolated photon with an additional hard jet with $p_T$ $>$ 40~GeV 
 and $|\eta|$ $<$ 2.5. 
The invariant mass of the $\gamma j$ system was required to agree 
with the known value 
of $m_t$ within $\pm$20 GeV. The $m_{\gamma j}$ resolution with 
the cuts described above was 
7.7 GeV, and the signal efficiency (not counting branching 
ratios) was 3.3\%, including a $b$-tagging efficiency of 60\%. 
The background (155 events for an integrated luminosity of 100 fb$^{-1}$) 
 is dominated by events with a real $W \rightarrow \ell \nu$ decay and either 
a real or a fake photon. These processes include $t \bar t$, single top 
production, 
 $W+jets$ and $Wbb$ production. 
 The corresponding 5$\sigma$ discovery limit is 
\be 
{\rm BR}(t \rightarrow \gamma q) =   1.0 \times 10^{-4}. \label{eq:qgamma_a} 
\ee 
 
 
\subsubsection{$t \rightarrow gq$ decay} 
 
The search for a FCNC $tgq$ coupling (with $q = u,c$) through the decay 
$t \rightarrow gq$ was analyzed in~\cite{Han:1996ce} for the 
Tevatron. However, as can be seen from Table~\ref{tab:summary} 
in Section~\ref{ANOM}, the sensitivity for such a coupling 
turns our to be much larger in the $t$ production processes 
than in the decay $t \rightarrow gq$, whose signal will be 
overwhelmed by the QCD background. 
We refer the reader to Section~\ref{ANOM} for a detailed discussion of this 
point. 
 
 
\subsection{CMS studies of FCNC top quark decays and 
$\bf t \to\ H^{+}b$} 
 
The CMS sensitivity to $t \rightarrow \gamma (Z) (u,c)$ decays was studied 
recently (see~\cite{Stepanov:2000} for details). 
The \pyth~5.7~\cite{Sjostrand:1994yb} generator was used for 
the signal and background simulations and 
the detector response was simulated at the fast MC level 
(CMSJET~\cite{Abdullin:1999}). 
 For the $t \rightarrow \gamma (u,c)$ signal the exact 
  $2 \rightarrow 5$ matrix elements 
 $gg (q \bar{q}) \rightarrow t \bar{t} \rightarrow \gamma u(c) + 
 W^* b (\rightarrow \ell \nu b)$ were calculated and 
 included in \pyth. 
 The $t \rightarrow \gamma (Z) (u,c)$ decays would be seen as 
 peaks in the $M_{\gamma (Z), jet}$ spectrum in the region of $\mt$. 
 To separate the signal from the background one has to exploit 
 the presence of the additional top decaying to the $ \ell \nu b $ in the same 
 event. The signature with the hadronic decay of the additional top was 
 found to be hopeless.

\subsubsection{$t \rightarrow \gamma (u,c)$} 
\noindent 
 
 In order to separate the $(\gamma q) (\ell \nu b)$  final state 
 from the backgrounds several selection criteria were found to be effective. 
 First, the presence of 
  one isolated photon with $E_t \geq 75$ GeV and $| \eta | \geq 2.5$, 
 one isolated lepton ($\mu, e$) 
  with $E_t \geq 15$ GeV and $| \eta |  \geq 2.5$, and at least two jets 
  with $E_t \geq 30$ GeV and $| \eta |  \geq 2.4$ is required. 
 One top quark has to be reconstructed from the photon and jet 
 ($M_{\gamma, jet} \subset \mt \pm 15$ GeV), the corresponding jet 
 is not allowed to be $b$-tagged. 
 On the  contrary, the jet with maximal $E_t$, which is not involved 
 in the $(\gamma, \mbox{jet})$ system has to be $b$-tagged, should have 
 $E_t \geq 50$ GeV and  contribute to another 
 reconstructed top quark ($M_{\ell \nu j} \subset \mt \pm 25$GeV). 
 There must be no additional jets with $E_t \geq 50$ GeV. 
 The $b$-tagging efficiency was assumed to be $60 \%$ for the purity 
 $1 \% (10 \%)$ with respect to the gluon and light quark jets  
 ($c$-quark jets). 
 After this selection, approximately 
 270 background events dominated by the $t \bar{t}$ and $W$ + jets, 
 including $W b \bar{b}$, 
  survive for the integrated luminosity of 100 fb$^{-1}$, 
 while the signal efficiency is $9.1 \%$. The $S/B$ ratio is about 1 for 
  ${\rm BR}(t \rightarrow \gamma (u,c)) = 10^{-4}$ and the 5 $\sigma$ discovery 
 limit is as low as $3.4 \times 10^{-5}$ for 100 fb$^{-1}$. 
 
\subsubsection{$t \rightarrow  Z q$} 
\noindent 
 
The $t \rightarrow  Z q$ signal was searched for in the 
$t \bar{t} \rightarrow (\ell \bar{\ell} q) \ell \nu b$ final state. 
Three isolated leptons with $E_t \geq 15$ GeV and $| \eta | \leq 2.5$, 
and exactly two jets with $E_t \geq 30$ GeV and $| \eta | \leq 2.5$ 
are required. The pair of the opposite-sign same-flavour leptons has to 
be constrained to the $Z$ mass ($M_{\ell \bar{\ell}}  
\subset M_Z \pm 6 $~GeV) and one 
jet, combined with the reconstructed Z, has to form  the top system 
($M_{\ell \bar{\ell} j} \subset \mt \pm 15 $~GeV). This jet is not allowed to 
 be the $b$-jet, 
but the last "free" jet in the event has to be $b$-tagged. 
For the integrated luminosity of 100 fb$^{-1}$ just $\sim 9$ background 
events coming from the $WZ$, $t \bar{t} Z$ and $Z + jets$ processes survive. 
The signal efficiency is about $6.8\%$ which corresponds, however, 
only to $\sim 12$ events for ${\rm BR}(t \rightarrow Z (u,c)) = 10^{-4}$. 
The indication is that one can reduce the background rate to the nearly zero 
level tightening the selection criteria. In particular, requiring in addition 
$E_{t}^{miss} \geq 30$ GeV and a harder jet involved in the top $(\ell 
\bar{\ell} j)$ 
system ($E_t \geq 50$ GeV) one can reduce the background to the level 
of $\sim 0.6$ events still keeping $\sim 3.7 \%$ of the signal ($6.6$ 
events for ${\rm BR}(t \rightarrow Z (u,c)) = 10^{-4}$ and 100 fb$^{-1}$). 
One can conclude that the $t \rightarrow Z (u,c)$ signal should be very clean 
but, due to the low signal event rate, only $ \sim 3 \times 100 $fb$^{-1}$ of 
integrated luminosity would allow one to probe ${\rm BR}(t \rightarrow Z (u,c))$ 
as low as  $10^{-4}$, 
 provided the present background understanding is correct 
and the detector performance will not be deteriorated during the long run. 
The $5 \sigma$ reach for 100 fb$^{-1}$ is $\sim 1.9 \times 10^{-4}$. 
 
\subsubsection{$t \to\ H^{+}b$} 
CMS has investigated the production of the light charged Higgs, 
$m_{H^{\pm}} < m_{t}$, in $t\overline{t}$ events using the decay chain  
$t\overline{t} \ra H^{\pm} b W b \ra (\tau^{\pm}\nu_{\tau}b)+(\ell\nu b)$ 
\cite{hplustt}. The $H^{\pm}\ra \tau\nu$ branching ratio is 
large $\sim$98\% in this mass range for $\tan\beta>$2 and only slightly 
dependent on $\tan\beta$.  The 
 $t \ra H^{\pm}b$ branching ratio is large both 
 at high and at low $\tan\beta$ values and has a minimum of $\sim$0.8\% 
around $\tan\beta \sim$6. Since the Higgs mass cannot be 
reconstructed in this process the signal can be only inferred 
 from the excess of $\tau$ production over what is expected 
from the SM $t \ra Wb$, $W^{\pm} \ra \tau^{\pm}\nu$ decay. 
 
An isolated lepton with $p_t>$ 20 GeV is required to 
identify the top decay and to trigger the event. The $\tau$'s 
are searched starting from calorimeter jets with $E_t >$40 GeV 
within $|\eta|<$2.4. For the $\tau$ identification the tracker information 
 is used, requiring one hard isolated charged hadron with $p_t >$30 GeV 
 within the cone of $\Delta R<$0.1 inside the calorimeter jet. 
The algorithm thus selects the one prong $\tau$ decays. 
 
The main backgrounds are due to the $t\overline{t}$ events with 
 $t\overline{t} \ra WbWb \ra (\tau^{\pm}\nu_{\tau}b)+(\ell\nu b)$ 
and $W$ + jet events with $W \ra \tau\nu$. The $t\overline{t}$ 
background is irreducible,  but can be suppressed by exploiting 
the $\tau$ polarisation effects~\cite{dproy}. Due to 
the $\tau$ polarisation the charged pion from 
 $\tau \ra \pi^{\pm}\nu$ decay has a harder $p_t$ spectrum 
 when coming from $H^{\pm} \ra \tau\nu$ than from $W^{\pm}\ra \tau\nu$. 
 The decay matrix elements with polarisation~\cite{monojet} 
were implemented in \pyth~\cite{Sjostrand:1994yb}. Due to the polarisation, 
 the efficiency of the above $\tau$ selection is significantly 
 better for $H^{\pm} \ra \tau\nu$ ($\sim$19\%) than 
for $W^{\pm} \ra \tau\nu$ ($\sim$6\%). 
 
The events were required to have at least one $b$-jet with 
 $E_t>$ 30 GeV tagged with an impact parameter 
 method~\cite{BTAG}. This $b$-tagging suppresses efficiently, 
by a factor of $\sim$70, the background from $W+$jet events. 
The efficiency for $t\overline{t}$ events is $\sim$35\%. 
The expected 5$\sigma$ discovery range for 10 fb$^{-1}$ in 
the MSSM ($m_A, \tan\beta$) parameter space was found to 
be: $m_A <$ 110 GeV for all $\tan\beta$ values and 
 somewhat extended ($m_A \lta$ 140) for $\tan\beta \lta$ 2. 
 
\def\lta{\;\raisebox{-.5ex}{\rlap{$\sim$}} \raisebox{.5ex}{$<$}\;} 
\def\gta{\;\raisebox{-.5ex}{\rlap{$\sim$}} \raisebox{.5ex}{$>$}\;} 
 
\subsection{Conclusions on rare top decays} 
In the framework of the SM, the top rare decays (that is any channel 
different from $t\to q W$) are definitely 
below the threshold for an experimental analysis at LHC. 
On the other hand, LHC experiments will be able to probe quite a few  
predictions of possible extensions of the SM. 
 
An extended Higgs sector will be looked for through 
the tree-level decay $t \to b H^+$. ATLAS estimates its sensitivity 
to  this channel in the MSSM,  
 through an excess in the tau lepton signal, 
to be around  ${\rm BR}(t \to H^{+}b)$ = 3\% (that is almost 4 times better than  
what expected from Run 2 at the Tevatron). 
This would allow to probe 
all values of $m_{H^{\pm}}$ below $m_{t} - 20$ GeV over most  
of the $\tan \beta$ range. 
For low $\tan \beta$, the complementary decay mode  
$H^{\pm} \to cs$ has been considered. 
In the mass range 110 $< H^{\pm} <$ 
130 GeV, the $H^{\pm}$ peak can be reconstructed 
and separated from the dominant $W \to jj$  
background. 
 
For CMS, using the $\tau$ excess signature,  
the expected 5$\sigma$ discovery range for 10 fb$^{-1}$ in  
the MSSM ($m_A, \tan\beta$) parameter space is $m_A <$ 110 GeV, 
 for all $\tan\beta$ values, and 
 somewhat extended ($m_A \lta$ 140), for $\tan\beta \lta$ 2.   
  
Other interesting signatures like   
$H^{\pm} \to hW^{\ast}$, $H^{\pm} \to AW^{\ast}$ and  
$H^{\pm} \to bt^{\ast} \to bbW$  
are very promising in particular parameter ranges, but have not yet been 
thoroughly investigated. 
 
ATLAS has studied its sensitivity to 
the radiative decay $t \to WbZ$. This has been found to be at most of  
the order 
10$^{-4}$, that is insufficient for the study of a SM signal ($\sim 10^{-6}$),  
but possibly useful for exploring the predictions of some extended Higgs-sector 
model, for which ${\rm BR}(t \to WqZ)\lta 10^{-2}$. 
On the other hand, the radiative Higgs decay  $t \to WbH$ 
seems out of the reach of LHC in any realistic model. 
 
The LHC reach for the FCNC decays  
$t \to qZ$, $t \to q\gamma \; $ and $t \to qg\; $ has also been thoroughly 
investigated. 
Apart from  the $t \to qg$, which is completely overwhelmed by the hadronic 
background, both ATLAS and CMS have a sensitivity of about $2 \times 10^{-4}$ 
to the $t \to qZ$ channel, while the CMS reach for the $t \to q\gamma \; $ 
channel is about $3.4 \times 10^{-5}$, that is slightly better than the ATLAS 
sensitivity ($1.0 \times 10^{-4}$), 
assuming an integrated  
luminosity of 100 fb$^{-1}$. These thresholds could be largely 
sufficient to detect some manifestation of possible FCNC anomalous couplings 
in the top sector. 
 
ATLAS has also investigated its sensitivity to a measurement 
of $| V_{tb} |$ through a determination of the rate ${\rm BR}(t \to bX)$, 
by comparing the number of observed (1 or 2) $b$-tags in a $t\bar t$\   
sample. Within the three-generation SM,  
the ratio of double $b$-tag to single $b$-tag events is  
${\rm R}_{2b/1b} =| V_{tb} | ^{2}$. 
LHC will allow a much more precise determination of ${\rm R}_{2b/1b}$ 
with respect to the Tevatron (where, presently, one gets  
$| V_{tb} |\ >$ 0.76 at the 95\% CL). 
On a purely statistical basis, the expected relative error on  
${\rm R}_{2b/1b}$ is  
$\delta {\rm R}_{2b/1b}/{\rm R}_{2b/1b}$ (stat.) = 0.2\% for an integrated  
luminosity of 10 fb$^{-1}$, that would imply a relative error 
on $| V_{tb} |$ of about 1\permille. On the other hand, 
the final uncertainty will be dominated by  
systematic errors  related to the $b$-tagging.  
Further study is needed  
to estimate the size of these systematic uncertainties. 
 
\section{ASSOCIATED TOP PRODUCTION\protect\footnote{Section coordinators: 
    A.~Belyaev, L.~Reina, M.~Sapinski (ATLAS), V.~Drollinger (CMS).}} 
\label{sec:ttH} 
The associated production of a Higgs boson (both SM-like and MSSM) 
with a top-antitop pair, is  one of the most promising reactions 
to study  both top quark and Higgs boson physics at the LHC. 
 
The $pp\rightarrow t\bar t H$ channel can be used in the difficult 
search for an intermediate mass Higgs 
($m_H\!\simeq\!100\!-\!130$~GeV), as first proposed in 
\cite{Marciano:1991qq}.  In this mass region, the associated top 
production cross section is quite high but still smaller than the 
leading $gg\rightarrow H$ and $qq\rightarrow Hqq$ cross sections by 
two orders and one order of magnitude, respectively. However, since the 
final state $t\bar t H$ signature is extremely distinctive, even such 
a small signal production rate can become relevant, especially if  
identifying the Higgs through its dominant 
$H\rightarrow b\bar b$ decay becomes realistic, as will be discussed 
in the following. 
 
Associated $\ttbar H$ production will furthermore provide the first direct 
determination of the top quark Yukawa coupling, allowing to 
discriminate, for instance, a SM-like Higgs from a more general MSSM 
Higgs. Processes like $gg\rightarrow H$ or $H\rightarrow\gamma\gamma$ 
are also sensitive to the top Yukawa coupling, but only through large 
top loop corrections.  Therefore loop contributions from other sources 
of new physics can pollute the interpretation of the signal as a 
measurement of the top Yukawa coupling. 
 
\begin{figure}[t] 
\leftfig{ 
  \epsfig{file=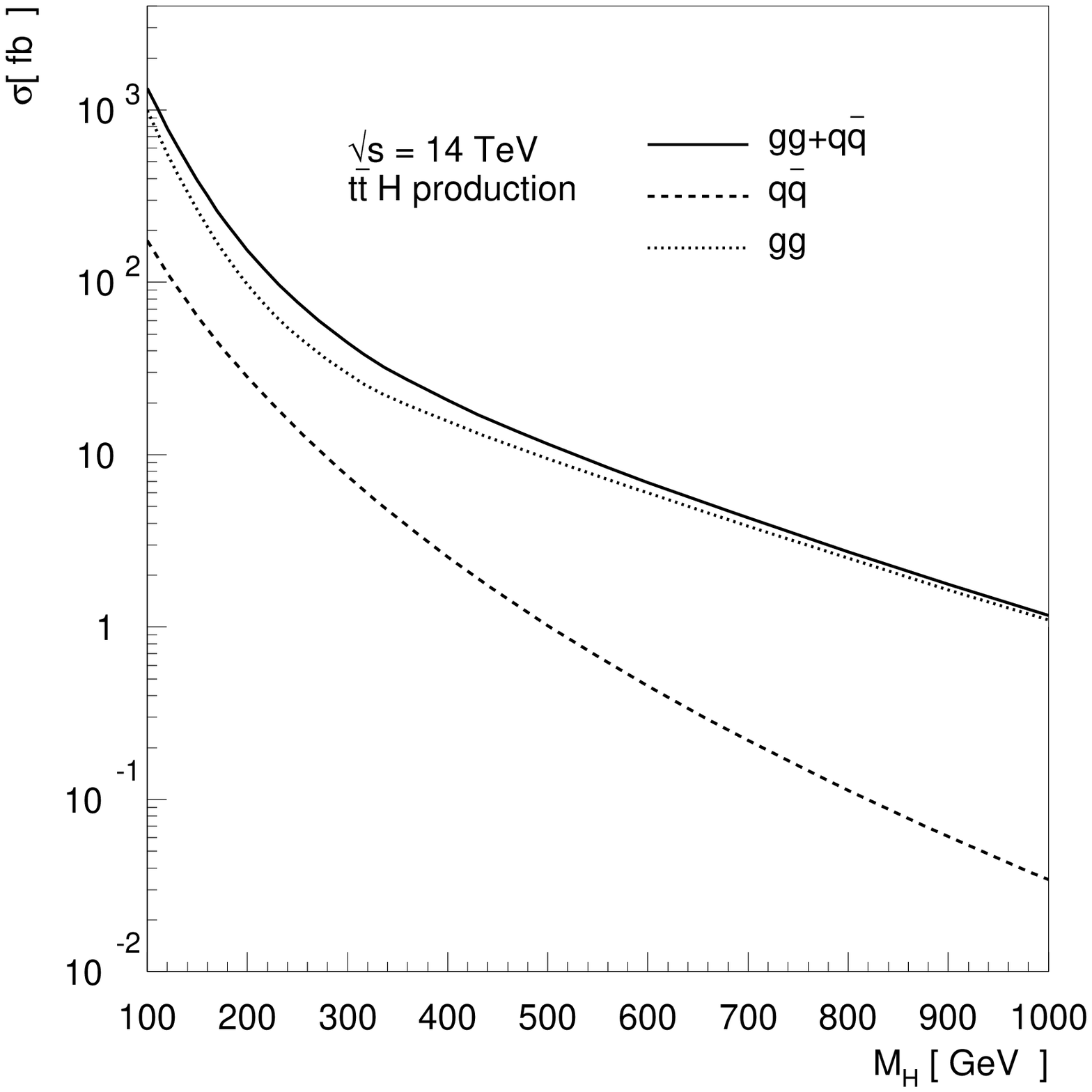,width=0.5\textwidth,height=0.45\textwidth}} 
{\label{fig:cross_sections} \footnotesize  
  Cross section for $t\bar tH$ production at 
  the LHC as a function of the Higgs 
 mass, for $\mu\!=\!m_H$. }%
\vspace*{0.6cm} 
\rightfig{\epsfig{file=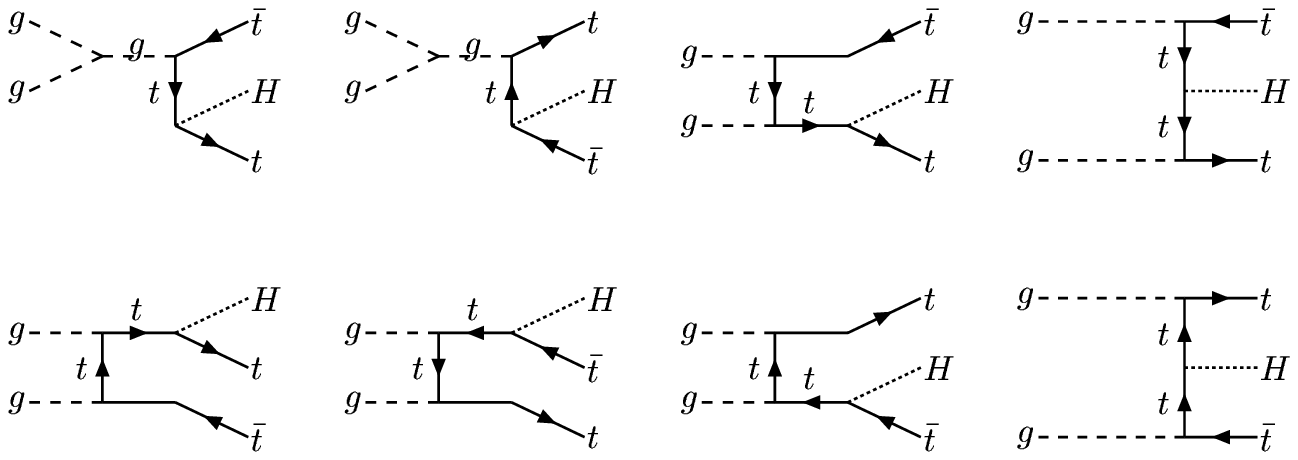,width=0.45\textwidth,height=0.4\textwidth}} 
{\label{fig:tth_diagrams} \footnotesize  
 Diagrams for $gg\to t\bar tH$, the leading parton level process for 
 $pp\rightarrow t\bar t H$.}%
\vspace*{7.7cm} 
\end{figure} 
  
\begin{table}[t] 
 \caption{\label{tab:cross_sections} Leading order cross sections for 
   $t\bar tH$ production at the LHC. The 
   individual parton level channels ($q\bar q\rightarrow t\bar t H$ 
   and $gg\rightarrow t\bar t H$) as well as their sum are given for a 
   few values of the renormalization scale $\mu$.} 
\begin{center} 
\begin{tabular}{ |c|c|c|c|c|c|c|} 
 \hline\hline 
 $m_H$[GeV]&  
 $ \ \ q\bar q$ [fb] &  
 $ \ \ gg$ [fb] &  
 $ \ \ q\bar q$+$gg$ [fb] & $ \ \ q\bar q$+$gg$ [fb] &  
 $ \ \ q\bar q$+$gg$ [fb] &$ \ \ q\bar q$+$gg$ [fb] \\ 
 &$\mu=m_H$&$\mu=m_H$  &$\mu=m_H$ & $\mu=m_t$ & $\mu=2 m_t+m_H$& 
 $\mu=\sqrt{\hat{s}}$ \\ 
 \hline\hline 
 100&   348. &   990.  &  1340.  &  1070.  &   765.   &   685. \\\hline 
 110&   279. &   740.  &  1020.  &   840.  &   596.   &   534.\\\hline 
 120&   227. &   558.  &   785.  &   674.  &   473.   &   422. \\\hline 
 130&   186. &   428.  &   613.  &   542.  &   379.   &   338.\\\hline 
 140&   153. &   334.  &   487.  &   445.  &   308.   &   273.\\\hline 
 150&   128. &   263.  &   391.  &   367.  &   251.   &   224.\\\hline 
 160&   107. &   210.  &   317.  &   306.  &   207.   &   184.\\\hline 
 170&    90.5&   169.  &   260.  &   257.  &   173.   &   152.\\\hline 
 180&    76.8&   139.  &   216.  &   218.  &   145.   &   128.\\\hline 
 190&    65.7&   115.  &   181.  &   187.  &   124.   &   108.\\\hline 
 200&    56.4&    97.1 &   153.  &   162.  &   106.   &    92.4 \\\hline 
 300&    15.0&    29.5 &    44.5 &    55.7 &    33.2  &    28.4 \\\hline 
 400&    5.11&    15.6 &    20.7 &    29.6 &    16.2  &    13.8\\\hline 
 500&    2.04&     9.51&    11.5 &    18.4 &     9.32 &     7.98\\\hline 
 600&   0.909&     6.00&     6.91&    12.1 &     5.73 &     4.93\\\hline 
 700&   0.439&     3.86&     4.29&     8.20&     3.63 &     3.14\\\hline 
 800&   0.226&     2.50&     2.72&     5.62&     2.34 &     2.04\\\hline 
 900&   0.122&     1.65&     1.76&     3.90&     1.54 &     1.35\\\hline 
1000&  0.0684&     1.10&     1.16&     2.73&     1.02 &     0.900\\\hline 
 \hline\hline 
 \end{tabular} 
\end{center} 
 \end{table} 
  In the following we will concentrate on the case of a SM-like Higgs 
 boson, whose top Yukawa coupling ($y_t\!=2^{3/4} G_F^{1/2} m_t$) 
 is enhanced with respect to the corresponding MSSM (scalar Higgs) 
 coupling for $\tan\beta>2$, the region allowed by LEP data. 
 Predictions for the MSSM case can be easily obtained by rescaling 
 both the $t\bar t H$ coupling and any other coupling that appears in 
 the decay of the Higgs boson. 
  
 The cross section for $pp\rightarrow t\bar t H$ at LO in QCD has been 
 known for a long time \cite{Kunszt:1984ri} and has been 
 confirmed independently by many authors. We have recalculated it and 
 found agreement with the literature. Of the two parton level 
 processes ($q\bar q\rightarrow t\bar tH$ and $gg\rightarrow t\bar 
 tH$), $gg\rightarrow t\bar tH$ dominates at the LHC due to the 
 enhanced gluon structure function. The complete gauge invariant set 
 of Feynman diagrams for $gg\rightarrow t\bar{t}H$ is presented in 
 Fig.~\ref{fig:tth_diagrams}.  The corresponding analytical results 
 are too involved to be presented here. The numerical results for 
 $\sqrt{s}\!=\!14$~TeV and a few values of the QCD scale $\mu$ are 
 given in Table~\ref{tab:cross_sections}, and illustrated in 
 Fig.~\ref{fig:cross_sections} as functions of $m_H$, for 
 $\mu\!=\!m_H$.  For consistency, we have used the leading order 
 CTEQ4L PDFs~\cite{Lai:1995bb} as well as the 
 leading order strong coupling constant (for reference, 
 $\alpha_s^{LO}(\mu\!=M_Z)\!=\!0.1317$ for 
 $\Lambda_{QCD}^{(5)}\!=\!0.181$).  The cross section, as expected from 
 a LO calculation, shows a strong scale dependence, as can be see in 
 Table~\ref{tab:cross_sections} , where results for 
 $\mu\!=\!m_H\,,\,m_t$, $m_H+2m_t$ and $\sqrt{\hat s}$ are presented. 
 In comparison with $\mu\!=\!2 m_t+m_H$, for $\mu\!=\!m_H$ we have 
 80-50\% higher cross sections, when 100~GeV $<m_H<$ 200~GeV. Since 
 the choice of the QCD scale at LO is pretty arbitrary, and since we 
 expect NLO QCD corrections to enhance the LO cross section, we decide 
 to use $\mu\!=\!m_H$ in Fig.~\ref{fig:cross_sections} and in the 
 following presentation.  These calculations have been performed 
 independently using the CompHEP software package~\cite{Pukhov:1999gg} 
 and MADGRAPH~\cite{Stelzer:1994ta}+HELAS~\cite{Murayama:1992gi}. 
 
\begin{table}[bht] 
 \caption{\label{tab:ttxx} Leading-order cross sections for various  
  $\ttbar XX$ backgrounds.} 
\begin{center} 
\begin{tabular}{ |c|c|c|c|c|c|} 
 \hline\hline 
 &$t\bar{t}b\bar{b}$  
 &$t\bar{t}\tau\bar{\tau}$ 
 &$t\bar{t}\gamma\gamma$ 
 &$t\bar{t}WW$ 
 &$t\bar{t}ZZ$\\ \hline 
 cuts           & $|\eta_b|<3$        &$|\eta_\tau|<3$      
                & $|\eta_\gamma|<2.5$ &          &    \\ 
                & $E_T^b>15$ GeV      &$E_T^\tau>15$ GeV 
                &$E_T^\gamma>15$ GeV  &   &  \\ 
                & $m_{bb}>90$ GeV&                  & 
       $M_{\gamma\gamma}>90$ GeV &    &    \\\hline 
 $\sigma$ [fb]\ \ $q\bar q$  & 41.2          &2.9               
            & 2.73                    &0.50&1.11\\ 
        $gg$      &  846.         &15.7              
            & 1.82                    &1.52&0.567\\ 
        $q\bar q$+$gg$   &  887.         &18.6              
            & 4.55                    &2.53&1.68 \\\hline 
 \hline\hline 
 \end{tabular} 
\end{center} 
 \end{table} 
  
 The NLO QCD corrections are expected to enhance the cross section, 
 but their complete evaluation is still missing at the moment. 
 Associated top production is in fact the only Higgs production mode 
 for which the exact NLO QCD corrections have not been calculated yet. 
 The task is very demanding, since it requires the evaluation of 
 several one loop five-point functions for the virtual corrections and 
 the integration over a four-particle final state (three of which 
 massive) for the real corrections. 
  
 For large $m_H$, the cross section for $t\bar tH$ has been calculated 
 including a complete resummation of potentially large logarithms, of 
 order $\ln(m_H/m_t)$, to all orders in the strong coupling 
 \cite{Dicus:1989cx}. These effects can almost double the cross 
 section for $m_H\!=\!1$ TeV.

 For an intermediate mass Higgs, the $K$ factor 
 ($\sigma_{NLO}/\sigma_{LO}$) has been estimated in the Effective 
 Higgs Approximation (EHA) \cite{Dawson:1998im}. The EHA neglects 
 terms of $O(m_H/\sqrt{s})$ and higher and works extremely well for 
 $e^+e^-\rightarrow t\bar tH$ already at $\sqrt{s}\!=\!1$~TeV. 
 However, it is a much poorer approximation in the $pp\rightarrow 
 t\bar t H$ case, since it does not include the $t\!$~-channel 
 emission of a Higgs boson for $gg\rightarrow t\bar tH$. Indicatively, 
 at $\sqrt{s}\!=\!14$~TeV, for a SM-like Higgs boson with 
 $m_H\!\simeq\!100\!-\!130$~GeV, the EHA gives $K\!\simeq\!1.2-1.5$, 
 with some uncertainty due to scale and PDF 
 dependence. 
\begin{figure}[t] 
\vskip -0.3cm 
\includegraphics[width=0.9\textwidth,clip]{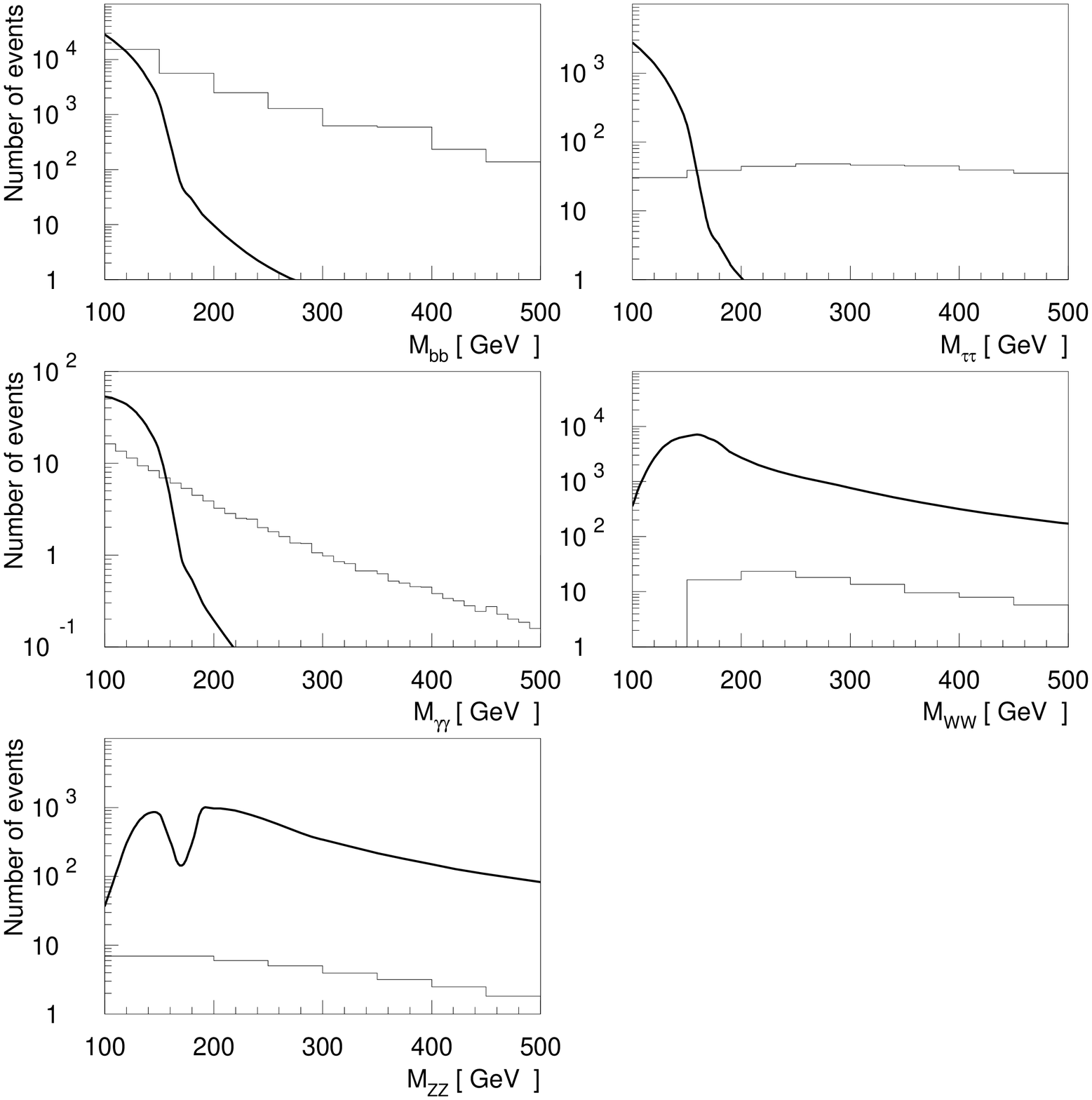} 
\vskip -0.2cm 
\caption{\label{fig:ttxx} 
  Number of events for $t\bar tH$ signal (solid line) and background 
  $t\bar{t}b\bar{b}$, $t\bar{t}\gamma\gamma$, $t\bar{t}WW$, 
  $t\bar{t}ZZ$ signatures (histogram), as a function of the 
  corresponding invariant masses $M_{XX}$, assuming 30~fb$^{-1}$ of 
  integrated luminosity at $\sqrt{s}\!=\!14$~TeV. The bin size is 
  10~GeV for the $M_{\gamma\gamma}$ distribution and 50~GeV for the 
  others.} 
\end{figure} 
Only the complete knowledge of the NLO level of QCD corrections will 
allow to reduce the strong scale and PDF 
dependence of the LO and EHA cross sections.  For the following 
analysis we choose to use the pure LO cross section with no K-factor, 
both due to the uncertainty of the result and for consistency with the 
corresponding background cross sections. However, one should point out 
that, due to the choice of a quite low QCD scale ($\mu\!=\!m_H$), a 
sort of \emph{effective} K-factor has been automatically included in 
our analysis. 
  
In the following subsection we present the analysis and results from the 
ATLAS collaboration as well as a discussion of the main backgrounds. 
The analysis mainly focuses on the search and study of 
an intermediate mass Higgs boson.  To introduce the study,  
it is useful to discuss and qualitatively understand the 
size of the possible irreducible backgrounds in the 
$100\!<\!m_H\!<\!140$~GeV mass region. 
 
Given the relatively small number of events that will be available, 
one should try to consider all possible decay channels of the Higgs 
boson in the intermediate mass region: $H\rightarrow b\bar 
b\,,\,\tau\bar \tau\,,\,\gamma\gamma\,,\,WW$ and $ZZ$.  The 
corresponding irreducible backgrounds are: 1) $t\bar{t}b\bar{b}$, \ 2) 
$t\bar{t}\tau\bar{\tau}$, \ 3) $t\bar{t}\gamma\gamma$, \ 4) 
$t\bar{t}WW$, and \ 5) $t\bar{t}ZZ$.  The number of events expected 
from signal and background signatures for 1)-5) are presented in 
Fig.~\ref{fig:ttxx}.  This figure shows the number of signal and 
background events in each bin of the corresponding invariant mass: 
$M_{bb}$, $M_{\tau\tau}$, $M_{\gamma\gamma}$, $M_{WW}$ or $M_{ZZ}$. 
They are obtained multiplying the $t\bar tH$ cross section by the 
respective Higgs boson branching ratios.  In order to take into 
account finite mass resolution effects, we have chosen 10~GeV bins for 
the $M_{\gamma\gamma}$ distribution and 50~GeV for the others. The 
presented numbers correspond to 30~fb$^{-1}$ of integrated luminosity. 
The corresponding total cross sections are given in 
Table~\ref{tab:ttxx}. 
 
Cross sections for backgrounds 1)-3) were calculated with the 
kinematic cuts shown in Table~\ref{tab:ttxx}, while for 
processes 4) and 5) no cuts were applied.  We have used CTEQ4L PDF 
 and $\mu^2=M_{XX}$, where $XX$ is $b\bar{b}$, 
$\tau^+\tau^-$, $\gamma\gamma$, $W^+W^-$ or $ZZ$ depending on the channel. 
One can see that the $t\bar tb\bar b$ signature has the highest signal 
(and background) event rate. It has been the object of the study of 
the ATLAS collaboration and will be discussed in the next section. 
The $t\bar{t}\gamma\gamma$ channel has also been the subject  
of~\cite{Dubinin:1997rq} where signal as well as reducible and 
irreducible backgrounds have been studied in details at the parton 
level.  However, one can see that other signatures could also be 
interesting and helpful in searching for the Higgs boson and measuring 
the $t\bar tH$ Yukawa coupling, and should be taken into account in 
future studies.

\subsection{$\rm t\bar tH$ : Analyses and Results} 
\label{sec:atlas} 
The ATLAS collaboration has studied several channels in which the 
discovery of a SM-like Higgs boson would be possible and obtained 
a quite complete Higgs discovery potential \cite{atlasphystdr}. One of the 
most important channels for discovery of a low mass Higgs boson 
($100\!-\!130$~GeV) is the $t\bar tH,\,H\rightarrow b\bar b$ channel, in 
which it is possible to obtain quite large signal significance  
\cite{Erichter:1999ac} and 
also to measure the top-Higgs Yukawa coupling. 
 
The final state of this channel consists of two $W$ bosons and four 
$b-\!\!$~jets: two from the decay of the top quarks, and two from the 
decay of the Higgs boson. In order to trigger 
signal events, one $W$ boson is required to decay leptonically. 
The second $W$ boson is 
reconstructed from the decay to a $q'\bar q$ pair. This channel could be 
also investigated with both $W$ bosons decaying leptonically. 
However, for this signature the total branching ratio is much smaller 
and, in addition, it is more difficult to reconstruct two neutrino 
momenta from the measured missing energy. 
 
In the analysis both top quarks are fully reconstructed, and this reduces 
most of the $W$+jets background.  
The reconstruction is done using strategies similar to those discussed 
in Section~\ref{sec:prodmeas} for the kinematic studies of $\ttbar$ 
production. The main backgrounds for this process are: 
\begin{itemize} 
\item the irreducible continuum  $t\bar tb\bar b$ background; 
\item the irreducible resonant $t\bar tZ$ background, which is not very 
  important for this channel as it has a very small cross section; 
\item the reducible backgrounds which contain jets misidentified as 
  $b\!\!$~-jets, such as $t\bar tjj$, $Wjjjjjj$, $WWb\bar bjj$, etc. 
\end{itemize}  
 
After the reconstruction of the two top quarks, it has been found that 
the most dangerous background is $t\bar tb\bar b$ (56\% of all $t\bar 
t$+jets background). In Table \ref{tab:tth-atlas1} we give $\sigma 
\times$BR, where BR represents the product of the branching ratios 
for $t\rightarrow Wb, W_1\rightarrow \ell\nu, W_2\rightarrow q_1\bar 
q_2$, and $H\rightarrow b\bar b$. We also give the number of events 
expected after the reconstruction procedure for 3 years of low luminosity 
operation ($b$-tagging efficiency $\epsilon_b=60\%$; 
probability to mistag $c$-jet as $b$-jet $\epsilon_c=10\%$; 
probability to mistag any other jet as $b$-jet $\epsilon_j=1\%$; 
$p_T^{jet} > 15$~GeV; lepton identification efficiency 
$\epsilon_\ell=90\%$; $p_T^{e,\mu}>20$~GeV), and after one year of high 
luminosity operation (for high luminosity the $b$-tagging 
efficiency is degraded to $\epsilon_b=50\%$ ($\epsilon_c$, 
$\epsilon_j$ and $\epsilon_\ell$ remain unchanged), the threshold on jet 
reconstruction is raised to $p_T>30$~GeV and the electron $p_T$ 
threshold is raised to $p_T^e>30$~GeV). Combined results are also 
shown. 
 
\begin{table}[htb] 
\caption{\label{tab:tth-atlas1} 
  Cross sections multiplied by branching ratios and numbers of events 
  after all cuts, including the $\pm 30 \ m_{b\bar b}$ mass window cut, for 
  30~fb$^{-1}$ (low luminosity detector performance), 100~fb$^{-1}$ 
  (high luminosity detector performance) and combined 100~fb$^{-1}$ 
  (30~fb$^{-1}$ with low luminosity and 70~fb$^{-1 }$ with high 
  luminosity detector performance) of integrated luminosity. } 
\begin{center} 
\vskip0.2cm 
\begin{tabular}{|c|c|c|c|c|} 
\hline 
        & $\sigma\times \mbox{BR}$ & \multicolumn{3}{|c|}{nr. of} \\ 
process & (pb) & \multicolumn{3}{|c|}{reconstructed events} \\ 
\cline{3-5} 
        &    &  low lumi & high lumi & combined \\ 
\hline \hline 
$t\bar tH,~ m_H=120$~GeV & 0.16 & 40 & 62 & 83 \\ 
\hline 
$t\bar t+jets$ & 87 & 120 & 242 & 289 \\  
$Wjjjjjj$ & 65200 & 5 & 10 & 12 \\ 
$t\bar tZ$ & 0.02 & 2 & 5 & 6 \\ 
\hline 
total background & - & 127 & 257 & 307 \\ 
\hline 
$S/B$  & - & 0.32 & 0.24 & 0.27 \\ 
$S/\sqrt(B)$ & - & 3.6 & 3.9 & 4.8 \\ 
\hline 
$S_{H\rightarrow b\bar b}/S_{total}$ & - & 59\% & 50\% & - \\ 
\hline 
$\delta y_t/y_t$ (stat.) & - & 16.2\% & 14.4\% & 11.9\% \\ 
\hline 
\end{tabular} 
\end{center} 
\end{table} 
 
\begin{figure}[t] 
\leftfig{ 
\mbox{\includegraphics[width=6.95cm]{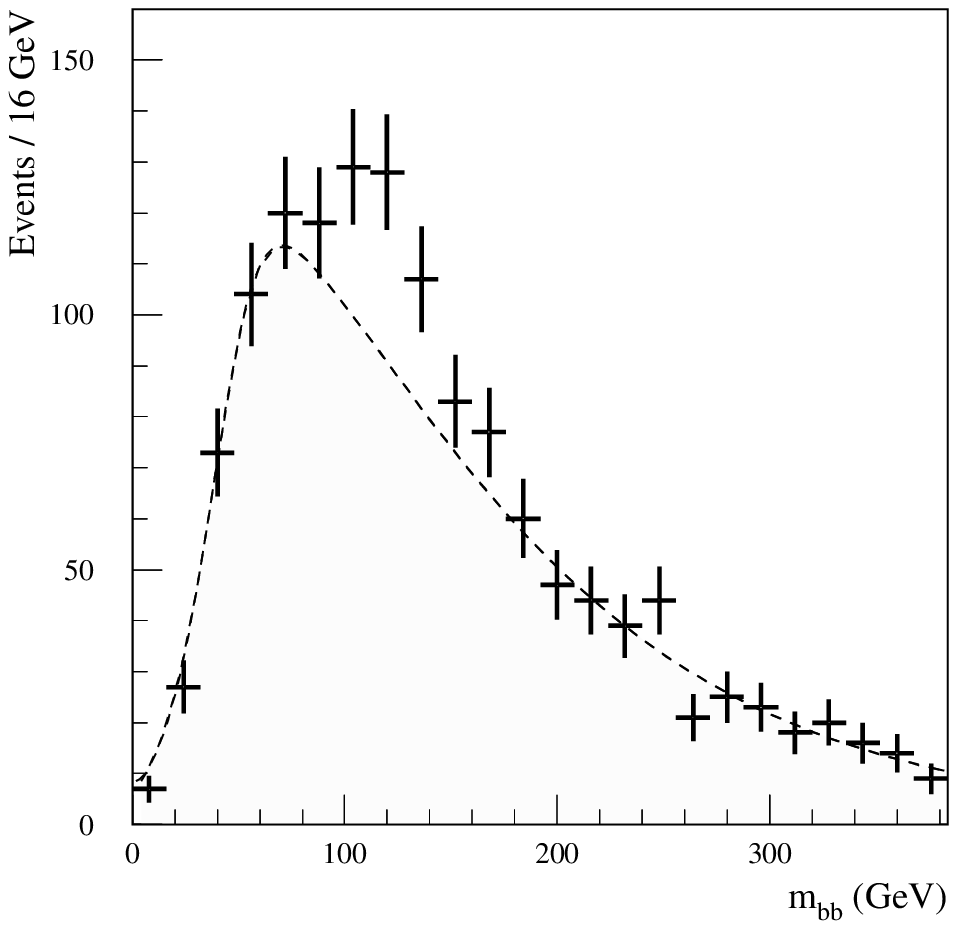}}}
 {\footnotesize Invariant mass distribution of tagged $b$-jet pairs in fully 
reconstructed $t\bar tH$ signal events and background events, obtained 
using the fast simulation of the ATLAS detector, for $m_H\!=\!120$~GeV 
and integrated luminosity of 100~fb$^{-1}$ (30~fb$^{-1}$ at 
low plus 70~fb$^{-1}$ at high luminosity). The points with error bars 
represent the result of a single experiment and the dashed line 
represents the background distribution. 
\label{fig:tth-atlas1}} 
\vspace*{-.27cm} 
\rightfig{ 
\mbox{\includegraphics[width=7.5cm]{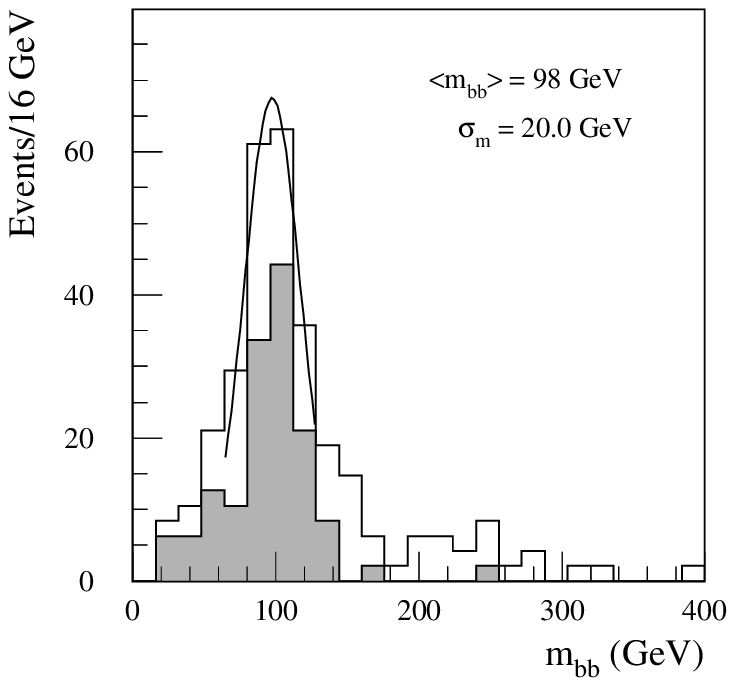}} 
} {\footnotesize Invariant mass distribution of tagged $b$-jet pairs in fully 
reconstructed $t\bar tH$ signal events, obtained using a full (GEANT) 
simulation of the ATLAS detector, for $m_H\!=\!100$~GeV and low 
luminosity performance. The shaded area denotes those events for which 
the jet assignment in the Higgs boson reconstruction is correct. 
\label{fig:tth-atlas2}}%
\vspace{10.3cm} 
\end{figure}

Figure~\ref{fig:tth-atlas1} shows the signal and background shapes for 
$m_H\!=\!120$~GeV and 100~fb$^{-1}$ of integrated luminosity obtained 
with combined detector performance (30~fb$^{-1}$ with low luminosity 
and 70~fb$^{-1 }$ with high luminosity).  On the other hand, 
Fig.\ref{fig:tth-atlas2} illustrates the signal shape for 
$m_H\!=\!100$~GeV, as obtained by using the full (GEANT) simulation 
of the detector. In this figure, the shaded area represents the true 
signal where both $b$-jets come from the Higgs boson, and the solid 
line stands for the signal obtained through the method that we 
described above. The \emph{combinatorial background}, which 
comes from taking at least one $b\!\!$~-jet from a top instead the one 
from the Higgs, is quite large and the signal purity is at the level 
of 60\% for low luminosity. 
 
For the fast simulation the $m_{b\bar b}$ peak mass resolution is 
$\sigma_{m_{b\bar b}}\!=\!19.0$~GeV, while for the full simulation, 
including the influence of electronic noise and the threshold on cell 
energy, a resolution $\sigma_{m_{b\bar b}}\!=\!20.0$~GeV has been 
obtained. 
 
Similar analyses have been performed for the $t\bar tH,\, H\rightarrow 
\gamma\gamma$ channel. Since the signal rate for this channel is very 
small, it will not be useful during the low luminosity period. 
However, thanks to the high purity of the signal, it will be possible 
to obtain between 4 or 5 signal events above 1 event from $t\bar 
t\gamma \gamma$ background per one year of high luminosity operation 
\cite{Eynard:1997ph}. To increase the signal rate, $WH$ and $ZH$ with 
$H\rightarrow \gamma \gamma$ channels have been included into the 
analysis and 14 signal events above 5 background events 
($W\gamma \gamma$, $Z\gamma \gamma$, $t\bar t\gamma \gamma$ and $b\bar 
b\gamma \gamma$) are expected for one year of high luminosity 
operation \cite{atlasphystdr}. 
 
The statistical uncertainty in the determination of the top-Higgs 
Yukawa coupling $y_t$ is given in the last row of 
Table~\ref{tab:tth-atlas1}. These results assume that the theoretical 
uncertainty is small, as we expect to be the case by the time the LHC 
turns on.  Many statistical uncertainties of the direct measurement of 
$y_t$, such as those associated with uncertainties in the integrated 
luminosity and in the $t\bar t$ reconstruction efficiency, could be 
controlled by comparing the $t\bar tH$ rate with the $t\bar t$ rate. 
 
 
To conclude, the $t\bar tH,\,H\rightarrow 
b\bar b$ and $H\rightarrow \gamma \gamma$ channels are very useful for 
Higgs boson discovery as well as for the measurement the of top-Higgs 
Yukawa coupling. 
 
\subsubsection{A closer look at the $t\bar{t}b\bar{b}$ background: CompHEP 
  versus \pyth} 
\label{sec:pithyacomphep}

It is necessary to stress that the correct understanding of the 
$t\bar{t}b\bar{b} $ background is one of the main points of this study. 
One can simulate this background using \pyth, by generating events of 
top pair production and emitting $b\bar b$ pairs from the gluon 
splitting after the initial and final state radiation.  In order to 
understand how good or bad this approximation is, one needs to 
calculate and simulate the complete $t\bar{t}b\bar{b}$ process.  We 
have done this using the CompHEP package~\cite{Pukhov:1999gg}. 
  
In order to compare CompHEP and \pyth~ on the same footing, one should 
take into account the effects of the initial and final state radiation 
in CompHEP. This has been done through a CompHEP-\pyth~  
interface~\cite{Ilyin:comphep-pythia}.  We 
use parton level events generated by CompHEP and link them to \pyth~ 
in order to include initial and final state radiation effects as well 
as hadronization effects. 
 
Table~\ref{tab:comppyt2} presents parton level CompHEP and \pyth~ 
cross sections including branching ratios of the $W$-boson decay, for 
the same choice of structure function (CTEQ4L~\cite{Lai:1995bb}) and 
QCD scale ($\mu^2=m_t^2+ p_T^2$(average)). We can see a good 
agreement for the total cross sections between the exact calculation 
and the gluon splitting approximation. 
 
\begin{table}[htb] 
 \caption{\label{tab:comppyt2} Results for the $t\bar{t}b\bar{b}$ background, 
   assuming an integrated luminosity ${\cal L}_{\rm int} = 30$ fb$^{-1}$: 
   CompHEP (ISR and FSR included) versus \pyth~ (default).} 
\begin{center} 
\begin{tabular}{ |l|c|c|c|} 
\hline 
 Selection & CompHEP & \pyth & CompHEP / \pyth \\\hline 
\hline 
 4 $b$-quarks with                     & 92000 events         & 87600 events         & 1.05 \\ 
 $p_T > $ 15 $\mbox{GeV}/c$ ; $|\eta| < $ 2.5 & $\sigma = $ 3.1 pb & $\sigma = $ 2.9 pb &       \\\hline 
 $\Delta$R($b$,$b$) $>$ 0.5            & 54000 events         & 48900 events         & 1.10  \\ 
 $b$-quarks not from top decay         & 59\% of prev. Step   & 56\% of prev. Step   &       \\\hline 
 \end{tabular} 
\end{center} 
 \end{table} 
 
\begin{figure}[t] 
\centerline{\includegraphics[width=0.9\textwidth,clip]{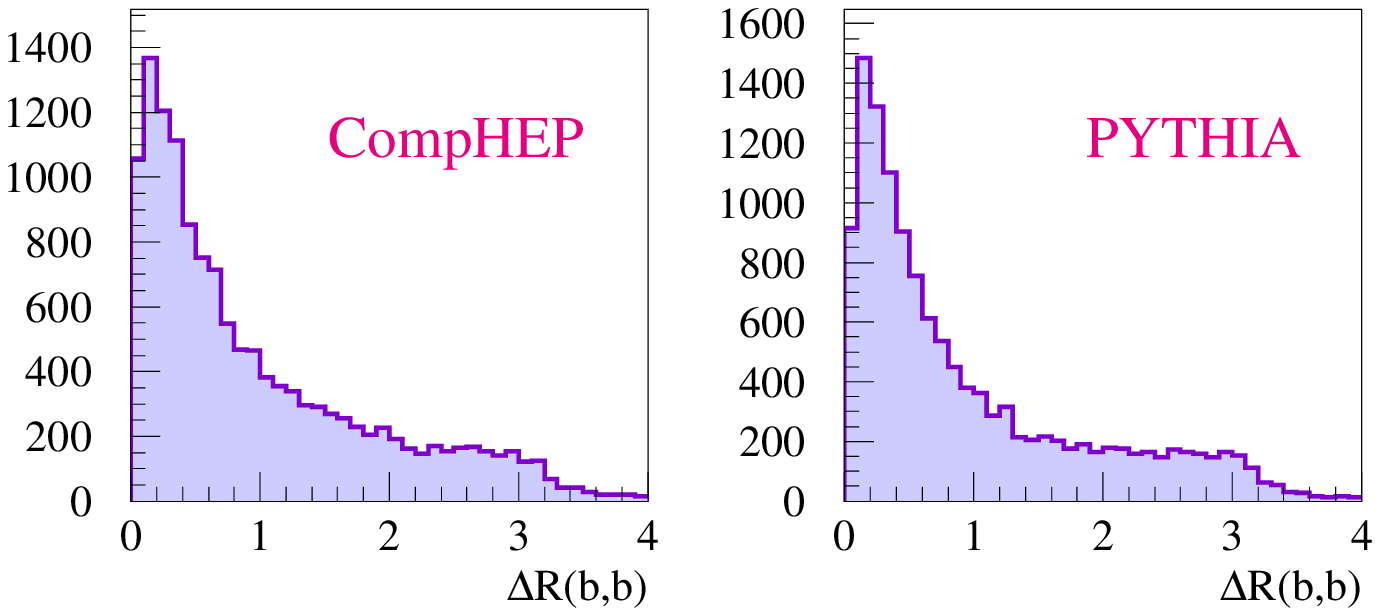}} 
\vskip -1cm 
\caption{\label{fig:compare} Results for $t\bar{t}b\bar{b}$ 
  background, assuming an integrated luminosity $L_{\rm int} = $ 4.4 
  fb$^{-1}$: CompHEP (ISR and FSR included) versus \pyth~ (default).} 
\leftfig{ 
  \epsfig{file=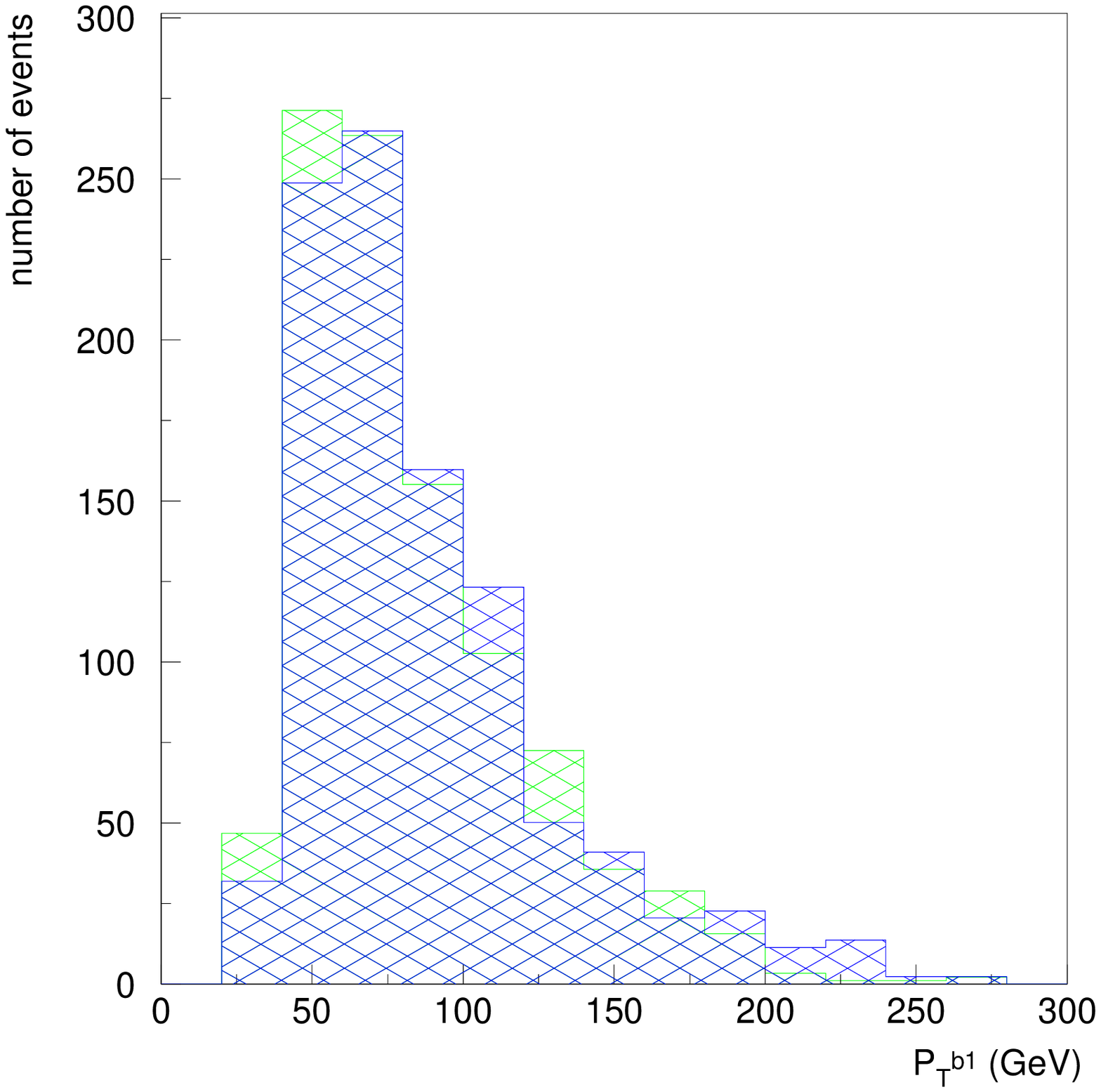,width=0.45\textwidth,height=0.45\textwidth}} 
{\label{fig:compareptb1} \footnotesize  
  Transverse momentum distribution of the most 
  energetic $b$-jet from $t\bar{t}b\bar{b}$ processes at the LHC: 
  CompHEP (dark-grey histogram) versus \pyth~ (light-grey) }%
\vspace*{-0.2cm} 
\rightfig{\epsfig{file=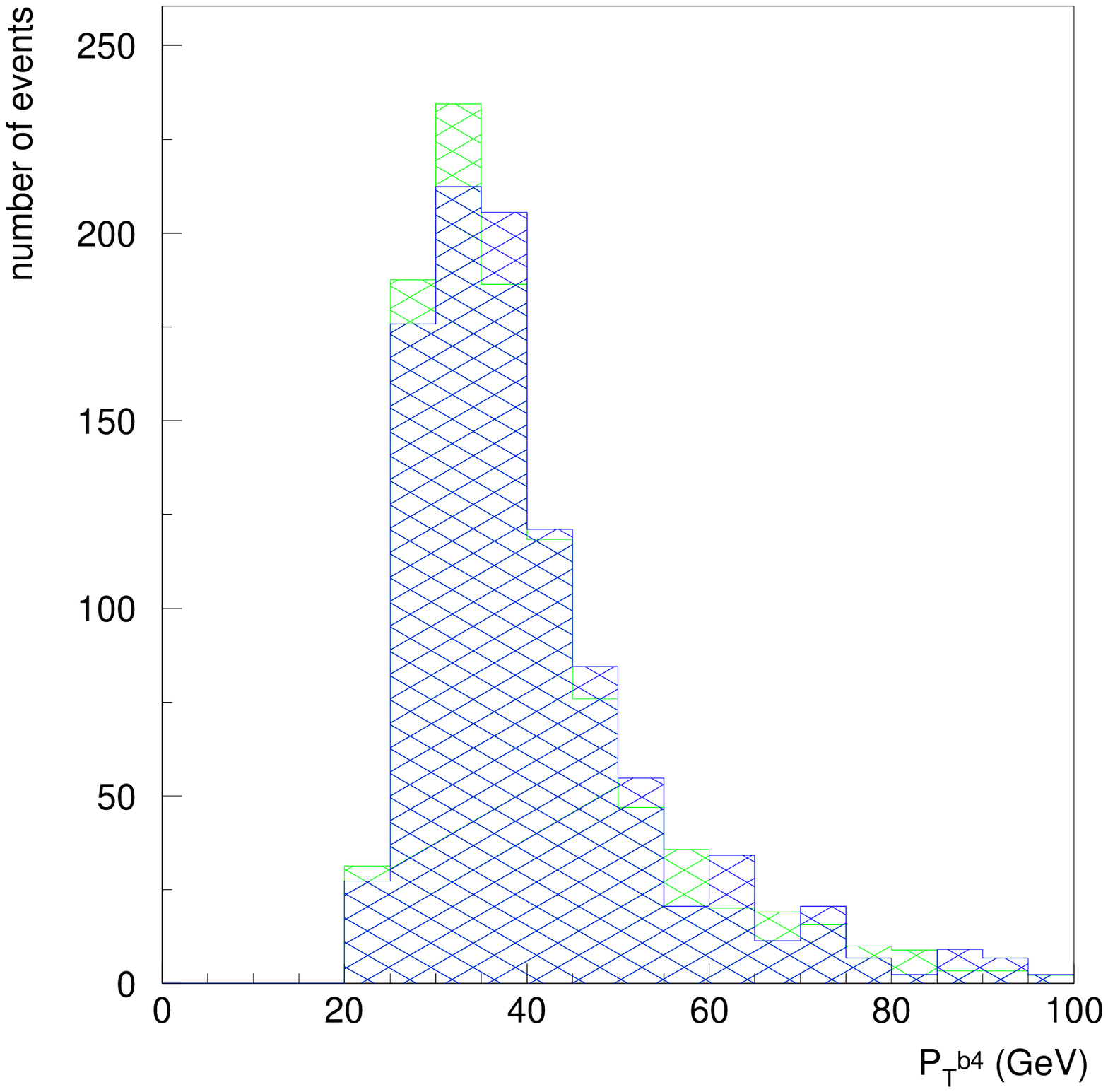,width=0.45\textwidth,height=0.45\textwidth}} 
{\label{fig:compareptb4}  \footnotesize  
 Transverse momentum distribution of the least 
  energetic $b$-jet from $t\bar{t}b\bar{b}$ processes at the LHC: 
CompHEP (dark-grey histogram) versus \pyth~ (light-grey)} %
\vspace*{8.8cm} 
\end{figure} 
 
In Fig.~\ref{fig:compare} we present the distribution of $b$-jet 
separation in $t\bar t b\bar b$ events. One can see a quite good 
agreement between CompHEP and \pyth.  Figures~\ref{fig:compareptb1} 
and ~\ref{fig:compareptb4} compare the transverse momentum 
distributions of the most energetic $b$-jet and of the least energetic 
$b$-jet in $t\bar t b\bar b$ production, as reproduced using \pyth\ and 
CompHEP respectively.  These distributions also confirm that \pyth~ 
describes well the $t\bar{t}b\bar{b}$ background.

\subsection{Summary and conclusions for $\bf t\bar t H$ production} 
The associated production of a Higgs boson with a top-antitop pair is 
important for the discovery of an intermediate mass Higgs boson 
($m_H\!\simeq\!100\!-\!130$~GeV) and provides a direct determination of 
the top-Higgs Yukawa coupling. From studies of the couplings and of 
the CP-parity of the Higgs boson \cite{Gunion:1998hm} it will be 
possible to discriminate, for instance, a SM-like Higgs 
boson from a generic MSSM one. 
 
The ATLAS analysis has focused on the $t\bar tH\,,\,H\rightarrow b\bar 
b$ channel for the low luminosity run of the LHC (30\,fb$^{-1}$ of 
integrated luminosity). The results presented in Section 
\ref{sec:atlas} are very encouraging and indicate that a signal 
significance of 3.6 as well as a precision of 16\% in the determination 
of the Yukawa coupling can be reached (for $m_H\!=\!120$ GeV). 
Better results can be obtained from the high luminosity run of the LHC 
(100\,fb$^{-1}$ of integrated luminosity), when also the high purity 
$t\bar tH,H\rightarrow\gamma\gamma$ channel is available.  
 

\appendix 
 
\section{APPENDIX: 
b-TAGGING AND JET E-SCALE CALIBRATION IN TOP  
EVENTS\protect\footnote{Section coordinator: I.~Efthymiopoulos}}  
\label{app:btagjet} 
For the reconstruction of the top events and in particular for the 
precision measurement of the top mass two important aspects in the 
detector performance have to be considered: 
\begin{itemize} 
\item the $b$-quark jet tagging capabilities and efficiency in top 
events, and 
\item the jet energy scale calibration for the light quark jets 
but in particular for the $b$-jets. 
\end{itemize} 
In both experiments \ATLAS~ and \CMS~ several studies 
have been made on these, highlights of which are presented here. 
From the preliminary results available so far, there is confidence 
that the numbers used or implied in the analyses presented in this report 
are realistic. Needless to say  that these are preliminary 
results and several detailed studies need to be performed with the 
final detector simulations and the first LHC data. 
 
\subsection{b-jet tagging in the top events} 
\begin{figure} 
\begin{center} 
\centerline{ 
\includegraphics[width=0.5\textwidth,clip]{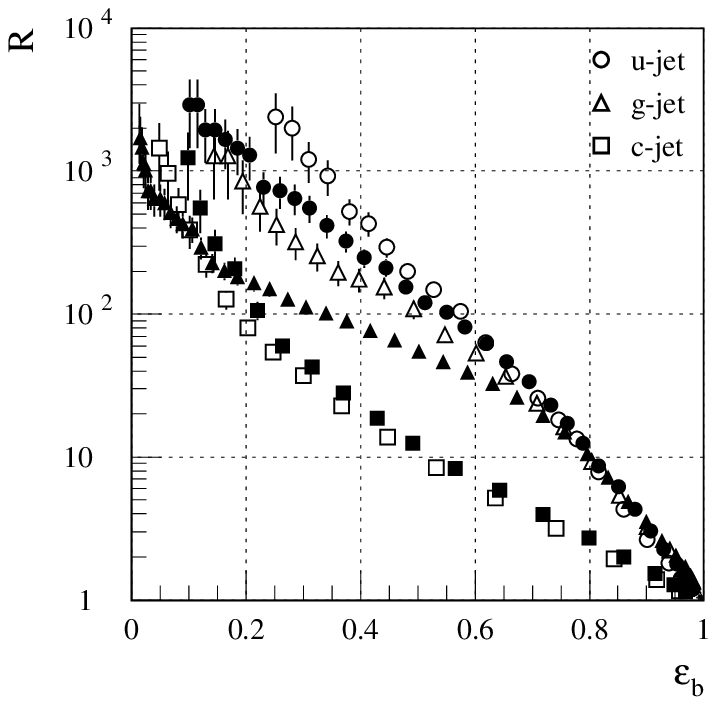} \hfill 
\includegraphics[width=0.5\textwidth,clip]{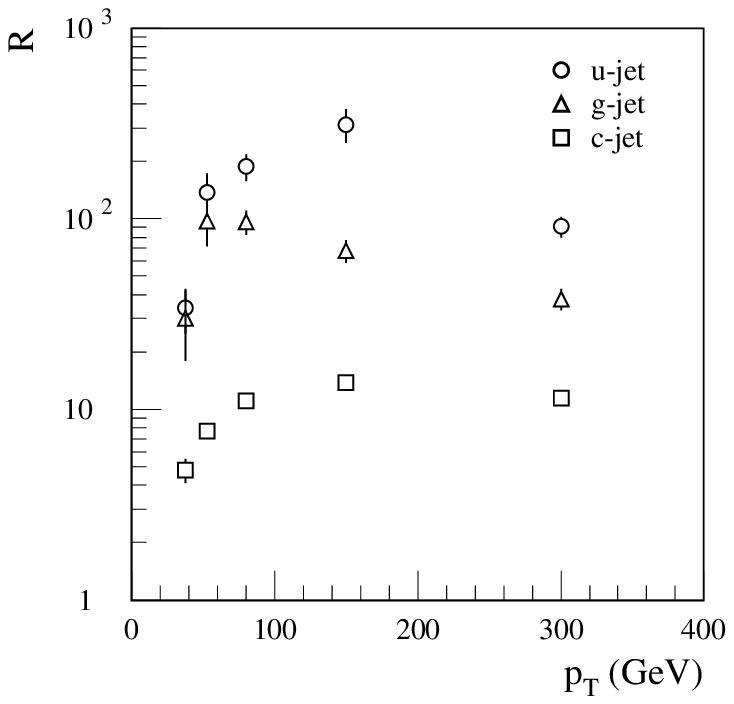} 
} \caption{Left: jet rejection factors for the vertex $b$-tagging 
method, with high luminosity pile-up. Open symbols: $m_H=100$~GeV, 
full symbols: $m_H=400$~GeV. Right: background rejections as a 
function of the \pt~ for $\epsilon_b=50$\% with high 
luminosity pile-up included. From~\cite{atlasphystdr}, Chapter 
10.} \label{fig:btag} 
\end{center} 
\end{figure}

\ATLAS~has done extensive studies for the $b$-tagging 
performance using jets from the decay of 100 and 400~GeV Higgs bosons  
(\cite{atlasphystdr}, Chapter 10). In 
Fig.~\ref{fig:btag} the rejection factors for the light quark jets 
versus the $b$ tagging efficiency and the jet \pt~ are shown. 
 
 Typically in the \ATLAS~analyses discussed here, 
 and in particular for the 
fast simulation studies, an overall $b$-jet tagging efficiency of 
60\% (50\%) for low (high) luminosity of LHC is used. The 
mis-tagging inefficiencies for the $c$-jets (or other light quark 
jets) were 10\% (1\%) for the \pt~range interesting for the 
top physics. Although most of the studies were done with events 
from the Higgs decays, the results were verified with the top 
events themselves and no significant differences were found. 
 
\subsection{Absolute jet energy scale calibration} 
Determining the absolute jet energy scale at LHC will be a rather 
complex issue because it is subject to both physics (initial-final 
state radiation, fragmentation, underlying event, jet algorithm 
etc.) and detector (calorimeter response over a wide range of 
energies and over the full acceptance of the detector, 
non-linearities at high energies, $e/h$ ratio etc.) effects. All 
these have to be understood at the level of a fraction of a 
percent in terms of systematic uncertainties as required for the 
precision measurements of the top mass. 
 
\begin{figure} 
\begin{center} 
\centerline{ 
\includegraphics[width=0.5\textwidth]{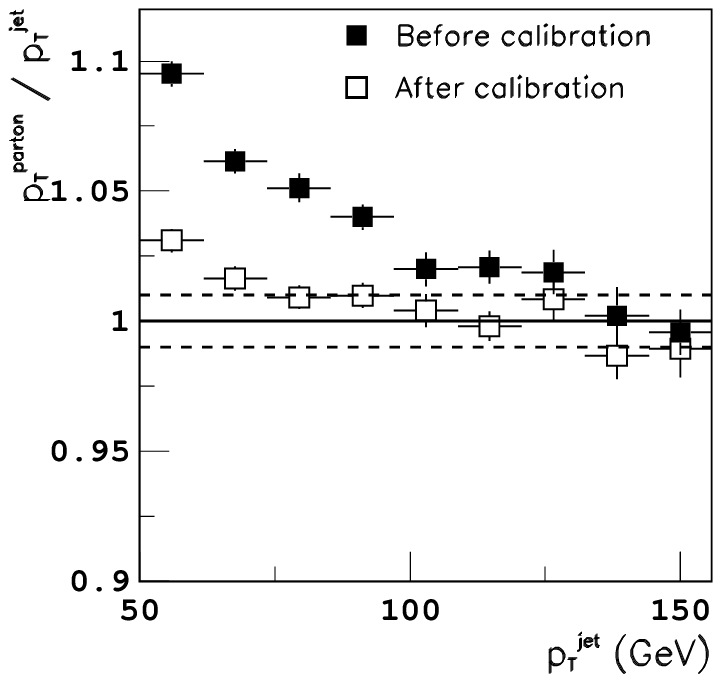} \hfil 
\includegraphics[width=0.5\textwidth]{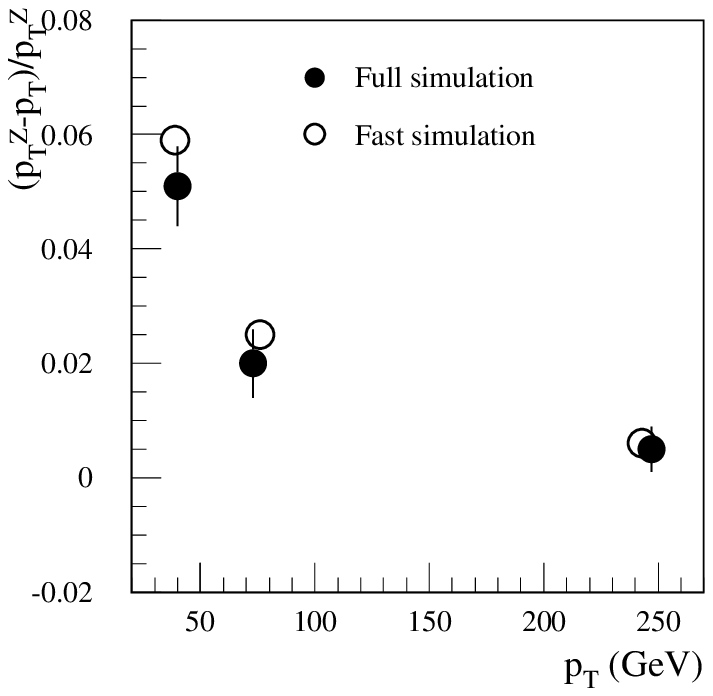}} 
\caption{Left: Ratio of the original parton \pt~to the 
\pt~of the reconstructed jet as a function of the \pt~of the 
jet for the \Wjj~decays reconstructed in include 
\ttbar~ events. The jets were 
reconstruced using a fixed cone jet algorithm with cone size DR$=0.4$, 
(optimized for high luminosity operation of LHC).  
Right: Average fractional imbalance between 
the \pt~ of the $Z$ boson and the \pt~ of the leading 
jet as a function of the \pt~ of the jet for the sample of 
$Z$ + jets events. A cone of DR$=0.7$ is used to collect the jet energy.}  
\label{fig:jscale} 
\end{center} 
\end{figure}

\ATLAS~has done an extensive study of the possible {\em in 
situ} jet scale calibration methods using specific data samples 
available at LHC (\cite{atlasphystdr},  
Chapter 12). In general, good candidate 
event classes at LHC will be: 
\begin{itemize} 
\item reconstruction of \Wjj~ decays within the top events 
themselves~\cite{Abe:1998ev} to obtain the light quark jet calibration and, 
\item events containing a $Z$ boson decaying into leptons balanced with one 
high-\pt~ jet to cross-check the light quark jet calibration 
but in addition to calibrate the $b$-jets and extend the energy 
reach to the TeV range. 
\end{itemize} 
In Fig.~\ref{fig:jscale} the results obtained are shown. As can be 
seen (left plot) for the case of \Wjj~ events, once the jet 
4-vectors are rescaled using the $M_W$ constraint the required 1\% 
uncertainty is reached for jets with \pt$>70$~GeV up to several 
hundred GeV. The lower and upper end of this range will depend on 
how well residual systematic effects can be controlled in the data 
and the Monte Carlo simulation~\cite{ref:escale1}. 
 
The use of the $Z$ + jets sample in LHC is a bit less straightforward 
than at the Tevatron~\cite{ref:escale2} 
due to the ISR radiation which 
produces an additional high-\pt~ jet which degrades the 
quality of the \pt-balance between the $Z$ boson and the leading 
jet.  
In Fig.~\ref{fig:jscale}~(right) the variation of the average 
fractional imbalance between the $p_T$ of the leading jet and the $Z$ boson 
as a function of the $p_T$ of the jet. Rescaling the jet $p_T$ to satisfy  
$p_T$ 
balance with the $Z$ boson and applying tight selection criteria (jet veto 
and difference in azimuth $\delta\phi$ between the reconstructed $Z$ and 
the leading jet) the desired goal of $\pm 1\%$ systematic uncertainty on 
the absolute jet energy scale can be achieved for jets with $p_T>50$~GeV 
and up to the TeV range \cite{MehVic99}.  
 
However, as shown in Fig.~\ref{fig:jscale}~(right), it is 
possible, taking advantage of the large rate and requiring tight 
event selection criteria, to obtain the required precision for jets 
with \pt$>40$~GeV and up to the TeV range. 
 
Clearly more studies are needed, and will be done in the years to 
come, to understand the limitations of the proposed methods and to devise 
possible improvements. 
 
\section{APPENDIX: DIRECT MEASUREMENT OF TOP QUANTUM  
NUMBERS\protect\footnote{Section coordinators: 
 E.L.~Berger, U.~Baur.}} 
\label{app:app1} 
\subsection{Top spin and experimental tests} 
Evidence to date is circumstantial that the top events analysed in 
Tevatron experiments are attributable to a spin-1/2 parent.  The 
evidence comes primarily from consistency of the distribution in 
momentum of the decay products with the pattern expected for the 
weak decay $t \rightarrow b + W$, with $W \rightarrow \ell + \nu$ or 
$W \rightarrow {\rm jets}$, where the top $t$ is assumed to have 
spin-1/2. 
 
It is valuable to ask whether more definitive evidence for spin-1/2 
might be obtained in future experiments at the Tevatron and LHC.  We 
take one look at this question by studying the differential cross 
section $d{\sigma}/d M_{t {\bar t}}$ in the region near production 
threshold \cite{berger:2000xx}.  
Here $M_{t {\bar t}}$ is the invariant mass of the 
$t {\bar t}$ pair.  We contrast the behaviour of $t {\bar t}$ production 
with that expected for production of a pair of spin-0 objects.  We are 
motivated by the fact that in electron-positron annihilation, $e^+ + 
e^- \rightarrow q + {\bar q}$, there is a dramatic difference in 
energy dependence of the cross section in the near-threshold region 
for quark spin assignments of 0 and 1/2. 
 
For definiteness, we compare top quark $t$ and top squark $\tilde t$ 
production since a consistent phenomenology exists for top squark pair 
production, obviating the need to invent a model of scalar quark 
production.  Moreover, top squark decay may well mimic top quark decay. 
Indeed, if the chargino $\tilde \chi^+$ is lighter than the light top 
squark, as is true in many models of supersymmetry breaking, the dominant 
decay of the top squark is 
$\tilde t \rightarrow b + \tilde \chi^+$.  If there are no sfermions 
lighter than the chargino, the chargino decays to a $W$ and the lightest 
neutralino $\tilde \chi^o$.  In another interesting possible decay mode, 
the chargino decays into a lepton 
and slepton, $\tilde \chi^+ \rightarrow \ell^+ \tilde \nu$. 
The upshot is that decays of the top squark 
may be very similar to those of the top quark, but have larger values of 
missing energy and softer momenta of the visible decay products. 
A recent study for Run~II of the Tevatron \cite{bigstop} concluded that 
even with 4 fb${}^{-1}$ of data at the Tevatron, and including the 
LEP limits on chargino masses, these decay modes remain open (though 
constrained) for top squarks with mass close to the top quark mass. 
 
\begin{figure} 
\begin{center} 
\centerline{ 
 \includegraphics[width=0.5\textwidth,clip]{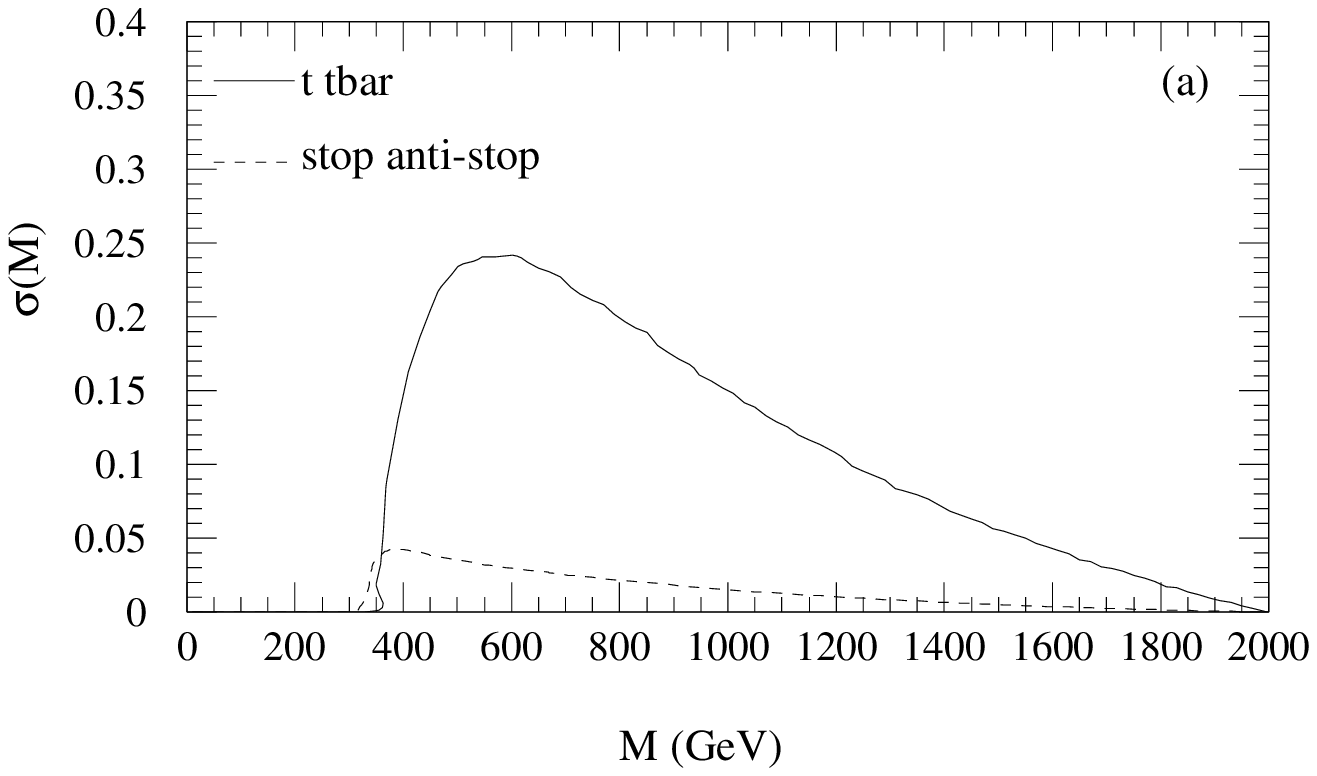} 
 \includegraphics[width=0.5\textwidth,clip]{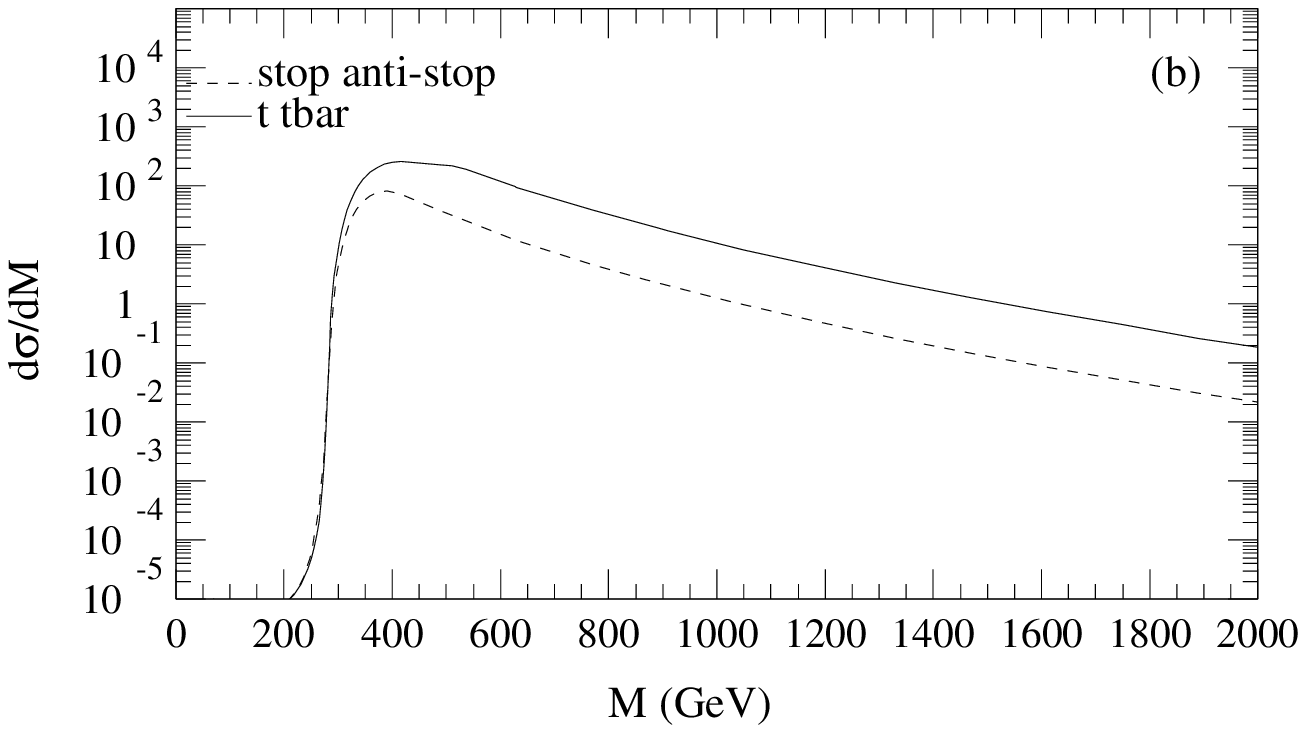}} 
\end{center} 
\vskip -1.5cm 
\caption{(a) Partonic cross sections $\hat \sigma(M)$ as 
          functions of partonic sub-energy $M$ for the $gg$ channel. 
          (b) Hadronic cross sections $d{\sigma}/d M$ in proton-proton 
          collisions at 14 TeV as functions of pair mass. The top 
          quark mass $m_t = 175$ GeV, and the top squark (stop) mass 
          $m_{\tilde t} = 165$ GeV.} 
        \label{fig:1} 
\end{figure} 
 
At the energy of the CERN LHC, production of $t {\bar t}$ pairs 
and of $\tilde{t} {\bar{\tilde t}}$ pairs is dominated by $g g$ 
subprocess, and the threshold behaviours in the two cases do not 
differ as much as they do for the $q {\bar q}$ incident channel. 
In Fig.~\ref{fig:1}(a), we show the partonic cross sections $\hat 
\sigma(\sqrt {\hat s})$ as functions of the partonic sub-energy 
$\sqrt{\hat s}$ for the $g g$ channel.  In Fig.~\ref{fig:1}(b), we 
display the hadronic cross sections for $p p \rightarrow t {\bar 
t} X$ and $p p \rightarrow \tilde{t} \tilde{{\bar t}} X$ at 
proton-proton center-of-mass energy 14 TeV as a function of pair 
mass. We include the relatively small contributions from the $q 
{\bar q}$ initial state.  After convolution with parton densities, 
the shape of the $\tilde{t} {\bar{\tilde t}}$ pair mass 
distribution is remarkably similar to that of the $t {\bar t}$ 
case. 
 
Based on shapes and the normalisation of cross sections, it is 
difficult to exclude the possibility that some fraction (on the order 
of 10\%) of top squarks with mass close to 165 GeV is present in the 
current Tevatron $t {\bar t}$ sample.  The invariant mass distribution 
of the produced objects, $M_{t {\bar t}}$, is quite different at the 
partonic level for the $q \bar q$ initial state (dominant at the Tevatron), 
but much less so for the $g g$ initial state (dominant at the LHC). 
However, after one folds with the parton distribution functions, the 
difference in the $q \bar q$ channel at the Tevatron is reduced to 
such an extent that the $M_{t {\bar t}}$ distribution is not an 
effective means to isolate top squarks from top quarks. 
 
Ironically, the good agreement of the absolute rate for $t {\bar t}$ 
production with theoretical 
expectations~\cite{Berger:1998gz,Bonciani:1998vc} would seem to be the 
best evidence now for the spin-1/2 assignment in the current Tevatron 
sample. 
 
A promising technique to isolate a top squark with mass close to $m_t$ 
would be a detailed study of the momentum distribution of the top 
quark decay products (presumably in the top quark rest frame).  One 
could look for evidence of a chargino resonance in the missing 
transverse energy and charged lepton momentum, or for unusual energy 
or angular distributions of the decay products owing to the different 
decay chains.  One could also look for deviations from the expected 
correlation between angular distributions of decay products and the 
top spin~\cite{Mahlon:1997}. 
        \begin{figure} 
            \epsfxsize = 12 cm   
            \centerline{\epsfbox{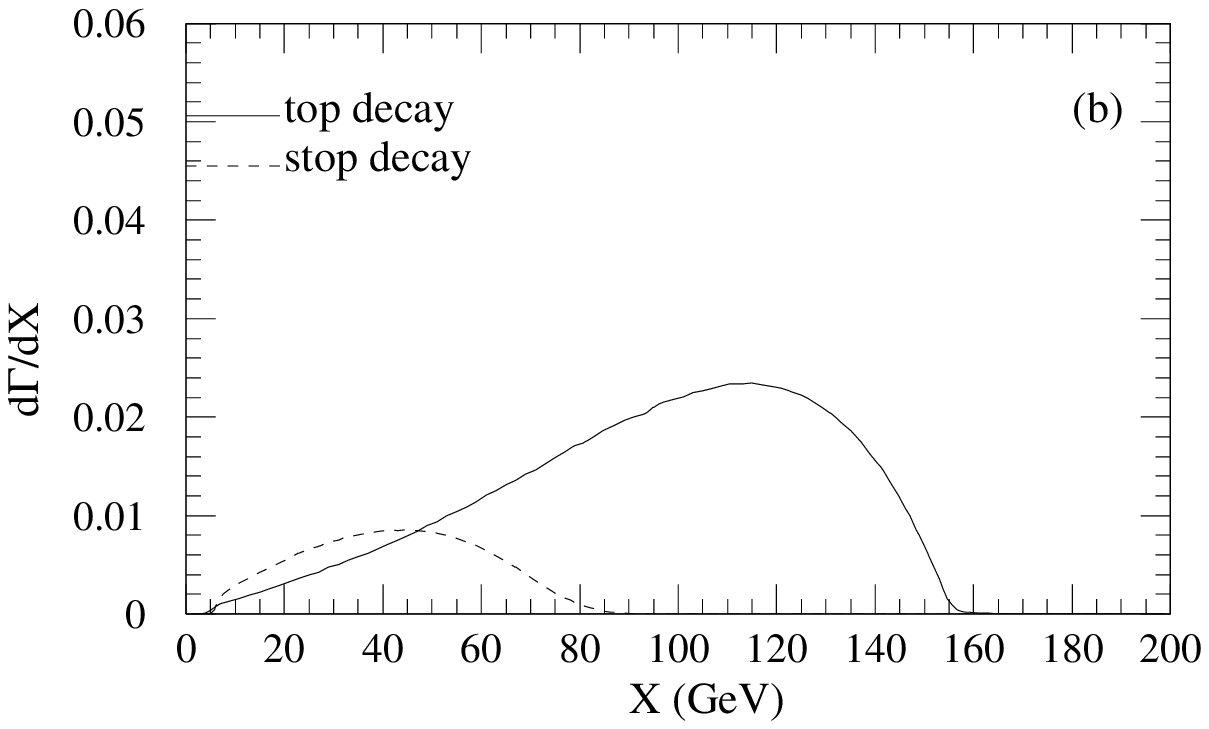}} 
\vskip -1cm 
        \caption{Distribution in the invariant mass of a bottom quark 
          and charged lepton ($X$) for a top quark or top squark 
          decay, with relative size normalized to the cross sections 
          at the LHC.  The top squark decay and sparticle masses are 
          discussed in the text.} 
       \label{fig:2} 
        \end{figure} 
 
As a concrete example of an analysis of this type, in 
Fig.~\ref{fig:2} we present the distribution in the invariant mass 
$X$ of the bottom quark and charged lepton, with $X^2 = (p_b + 
p_{\ell^+})^2$, where the bottom quark and lepton are decay 
products of either a top quark with $m_t = 175$ GeV or a top 
squark $\tilde t \rightarrow \tilde \chi^+ b \rightarrow W^+ 
\tilde \chi^0 b \rightarrow \ell^+ \nu_\ell \tilde \chi^0 b$, with 
$m_{\tilde t} = 165$ GeV, $m_{\tilde \chi^+} = 130$ GeV, 
$m_{\tilde \chi^0} = 40$ GeV, and $m_b = 5$ GeV.  The $X$ 
distribution is a measure of the degree of polarisation of the $W$ 
boson in top quark decay \cite{carlson}, and the figure shows that 
the different dynamics responsible for top squark decay result in 
a very different distribution, peaked at much lower $X$.  The 
areas under the curves are normalised to the inclusive $t \bar t$ 
and $\tilde{t} {\bar{\tilde t}}$ rates at the LHC. 
 
In this simple demonstration potentially important effects are ignored 
such as cuts to extract the $t \bar t$ signal from its backgrounds, 
detector resolution and efficiency, and ambiguities in identifying the 
correct $b$ with the corresponding charged lepton from a single decay. 
Detailed simulations would be required to determine explicitly how 
effective this variable would be in extracting a top squark sample 
from top quark events. Nevertheless, such techniques, combined with 
the large $t \bar t$ samples at the Tevatron Run~II and LHC, should 
prove fruitful in ruling out the possibility of a top squark with mass 
close to the top quark mass, or alternatively, in discovering a top 
squark hidden in the top sample. 
 
\subsection{Direct Measurement of the Top Quark Electric Charge} 
In order to confirm that the electric charge of the top quark is 
indeed $Q_{top}=2/3$, one can either measure the charge of the $b$-jet and 
$W$ boson, or 
attempt to directly measure the top quark electro-magnetic coupling through 
photon radiation in 
\begin{equation} 
pp\to t\bar t\gamma,\qquad pp\to t\bar t,~t\to Wb\gamma. 
\label{eq:top1} 
\end{equation} 
Since the process $pp\to t\bar t\gamma$ is dominated by $gg$ fusion at 
the LHC, one expects that the $t\bar t\gamma$ cross section is 
approximately proportional to $Q_{top}^2$. For radiative top decays the 
situation is more complicated because the photon can also be radiated off 
the $b$-quark or the $W$ line. 
 
The charge of the $b$-jet can most easily be measured by selecting 
events where the $b$-quarks are identified through their semi-leptonic 
decays, $b\to\ell\nu c$ with $\ell=e,\,\mu$. The small semi-leptonic 
branching ratio of the $b$-quark (Br$(b\to\ell\nu c)\approx 10\%$) and 
wrong sign leptons originating from $B-\bar B$ mixing are the main 
problems associated with this method. For a quantitative estimate 
realistic simulations are needed. Nevertheless, we believe that the 
enormous number of top quark events produced at the LHC should make it 
possible to use semi-leptonic $b$-tagging to determine the electric 
charge of the top quark. 
 
In our analysis, we  focus on top charge measurement through the 
photon radiation  processes listed  
in~(\ref{eq:top1}), concentrating on the lepton$+$jets mode, 
\begin{equation} 
pp\to\gamma\ell\nu jjb\bar b. 
\label{eq:top2} 
\end{equation} 
We assume that both $b$-quarks are tagged with a combined efficiency 
of 40\%. Top quark and $W$ decays are treated in the narrow width 
approximation. Decay correlations are ignored. 
To simulate detector response, the following transverse 
momentum, rapidity and separation cuts are imposed: 
\begin{eqnarray} 
p_T(b) >15~{\rm GeV,} & & |y(b)|<2, \\\label{eq:top31} 
p_T(\ell)>20~{\rm GeV,} & & |\eta(\ell)|<2.5,\\\label{eq:top32} 
p_T(j)>20~{\rm GeV,} & & |\eta(j)|<2.5,\\\label{eq:top33} 
p_T(\gamma)>30~{\rm GeV,} & & |\eta(\gamma)|<2.5,\\\label{eq:top34} 
p\llap/_T>20~{\rm GeV,} & & {\rm all~\Delta R's}>0.4. 
\label{eq:top35} 
\end{eqnarray} 
In addition, to suppress contributions from radiative $W$ 
decays, we require that 
\begin{eqnarray} 
m(jj\gamma)>90~{\rm GeV} & {\rm and} & m_T(\ell\gamma;p\llap/_T) 
>90~{\rm GeV}, 
\label{eq:top4} 
\end{eqnarray} 
where $m_T$ is the cluster transverse mass of the $\ell\gamma$ system. 
 
The events passing the cuts listed in~(\ref{eq:top31}) -- 
(\ref{eq:top4}) can then be split into three different subsamples: 
\begin{enumerate} 
\item By selecting events which satisfy 
\begin{eqnarray} 
m(bjj\gamma)>190~{\rm GeV} & {\rm and} & m_T(b\ell\gamma;p\llap/_T) 
>190~{\rm GeV}, 
\label{eq:top5} 
\end{eqnarray} 
radiative top quark decays can be suppressed and an almost pure sample 
of $t\bar t\gamma$ events is obtained (``$t\bar t\gamma$ cuts''). 
 
\item For 
\begin{eqnarray} 
m_T(b_{1,2}\ell\gamma;p\llap/_T) 
<190~{\rm GeV} & {\rm and} & m(b_{2,1}jj\gamma)>190~{\rm GeV}, 
\label{eq:top6} 
\end{eqnarray} 
the process $pp\to t\bar t$, $t\to Wb\gamma$, $W\to\ell\nu$ 
dominates (``$t\to Wb\gamma$, $W\to\ell\nu$ cuts''). 
 
\item Requiring 
\begin{eqnarray} 
m_T(b_{1,2}\ell\gamma;p\llap/_T) 
>190~{\rm GeV} & {\rm and} & 150~{\rm GeV}<m(b_{2,1}jj\gamma)<190~{\rm GeV}, 
\label{eq:top7} 
\end{eqnarray} 
one obtains an event sample where the main contribution originates from 
the process  $pp\to t\bar t$, $t\to Wb\gamma$, $W\to jj$ 
(``$t\to Wb\gamma$, $W\to jj$ cuts''). 
 
\end{enumerate} 
For $m_t=175$~GeV, $Q_{top}=2/3$, 
and $\int\!{\cal L}dt=100~{\rm fb}^{-1}$, one expects about 2400, 11000 
and 9400 events in the  regions of phase space corresponding to the 
three sets of cuts. We have 
not studied any potential background processes. The main background 
should originate 
from $W\gamma + $~jets production and should be manageable in a way 
similar to the  $W+$~jets 
background for regular $t\bar t$ production. 
 
\begin{figure*}[t] 
\centerline{ 
\epsfysize=9.cm 
\epsffile{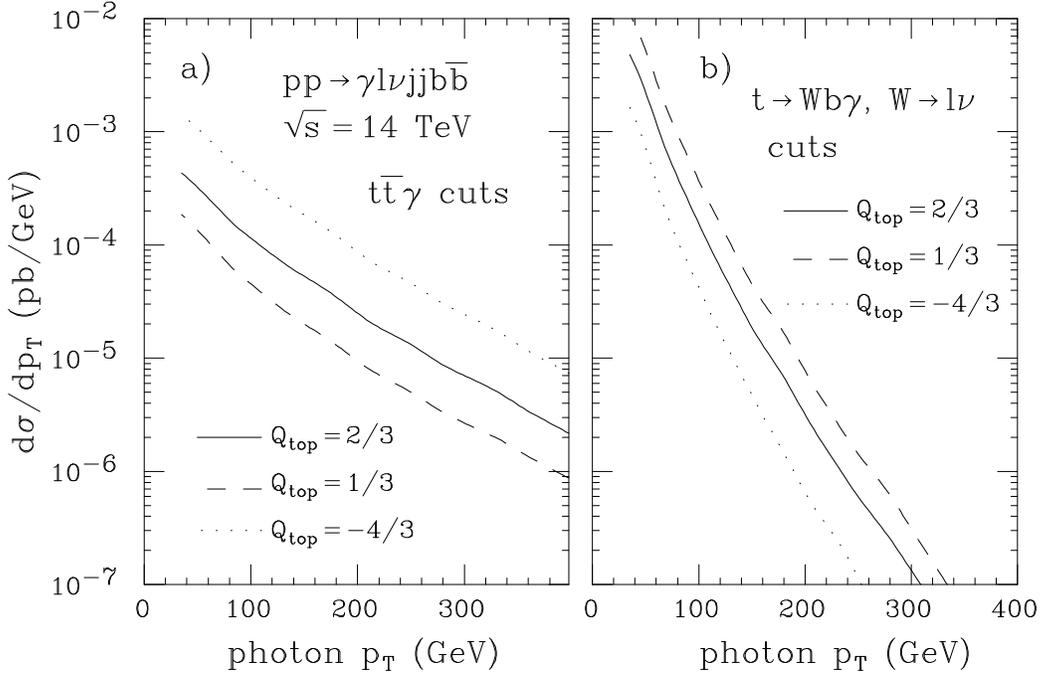}  } 
\caption{The differential cross section for the photon transverse 
momentum in the reaction $pp\to\gamma\ell\nu jjb\bar b$ at the LHC for 
three different ``top'' quark charges.} 
\label{fig:ftop1} 
\end{figure*}

The differential cross section for the photon transverse momentum 
at the LHC is shown in Fig.~\ref{fig:ftop1}. Results are shown for 
$m_t=175$~GeV and three ``top'' quark charges: $Q_{top}=2/3$, 
$Q_{top}=-4/3$, and $Q_{top}=1/3$. For $Q_{top}=-4/3$, the ``top'' 
quark decays into a $W^-$ and a $b$-quark instead of $t\to W^+b$. 
If $Q_{top}=1/3$, the ``$b$''-quark originating from the ``top'' 
decay is a (exotic) charge $-2/3$ quark. In the $t\bar t\gamma$ 
region (Eq.~(\ref{eq:top5}) and  Fig.~\ref{fig:ftop1}a), the 
$pp\to\gamma\ell\nu jjb\bar b$ cross section for a charge $-4/3$ 
($1/3$) ``top'' quark is uniformly a factor~$\approx 3.3$ larger 
($\approx 2.3$ smaller) than that for $Q_{top}=2/3$, reflecting 
the dominance of the $gg\to t\bar t\gamma$ process for which the 
cross section scales with $Q^2_{top}$. On the other hand, for the 
$pp\to t\bar t$, $t\to Wb\gamma$, $W\to\ell\nu$ selection cuts 
(Eq.~(\ref{eq:top6}) and Fig.~\ref{fig:ftop1}b), the cross section 
for $Q_{top}=-4/3$ is a factor~3 to~5 smaller than that for a 
charge $2/3$ top quark, due to destructive interference effects in 
the $t\to Wb\gamma$ matrix element. If $Q_{top}=1/3$, the 
interference is positive, and the cross section is about a 
factor~2 to 2.5 larger than for $Q_{top}=2/3$. The results for the 
$t\to Wb\gamma$, $W\to jj$ selection cuts ((\ref{eq:top7})) 
are similar to those shown in Fig.~\ref{fig:ftop1}b, and are 
therefore not shown here. Note that the photon $p_T$ distribution 
for radiative top decays is much softer than that for $t\bar 
t\gamma$ production. 

From our (simplified) calculation we conclude that the large number of 
double-tagged $\gamma\ell\nu jjb\bar b$ events, 
together with the significant changes in the 
$t\bar t\gamma$ and the $t\bar t$, $t\to Wb\gamma$ cross sections should 
make it possible to accurately determine $Q_{top}$ at the LHC. 
 
\section{APPENDIX: 
4$^{th}$ GENERATION QUARKS\protect\footnote{Section coordinator: M.L.~Mangano}} 
\label{app:4gen} 
\begin{figure}[t]
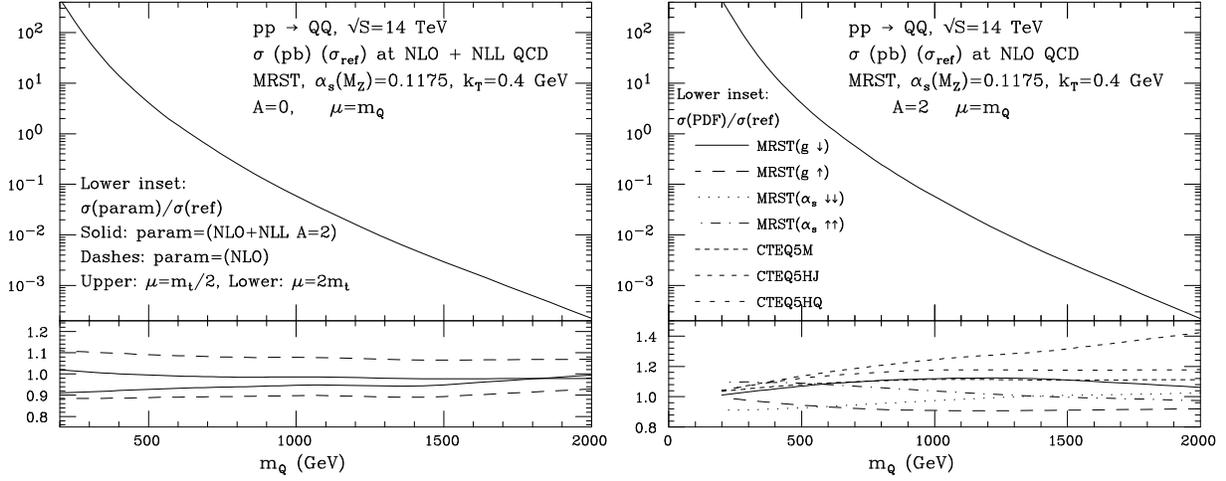
 
   \centerline{ 
    \includegraphics[width=0.5\textwidth,clip]{sighvq_res.eps} \hfil 
    \includegraphics[width=0.5\textwidth,clip]{sighvq.eps}} 
    \caption{Heavy quark total production rates. Left figure: scale 
   dependence at fixed NLO (dashed lines in the lower inset), and at 
   NLO+NLL (solid lines). Right figure: PDF dependence. See the  
   Section~\ref{sec:rates} for details.} 
    \label{fig:sighvq} 
  \vspace*{0.5cm} 
  \end{figure}

\begin{figure}
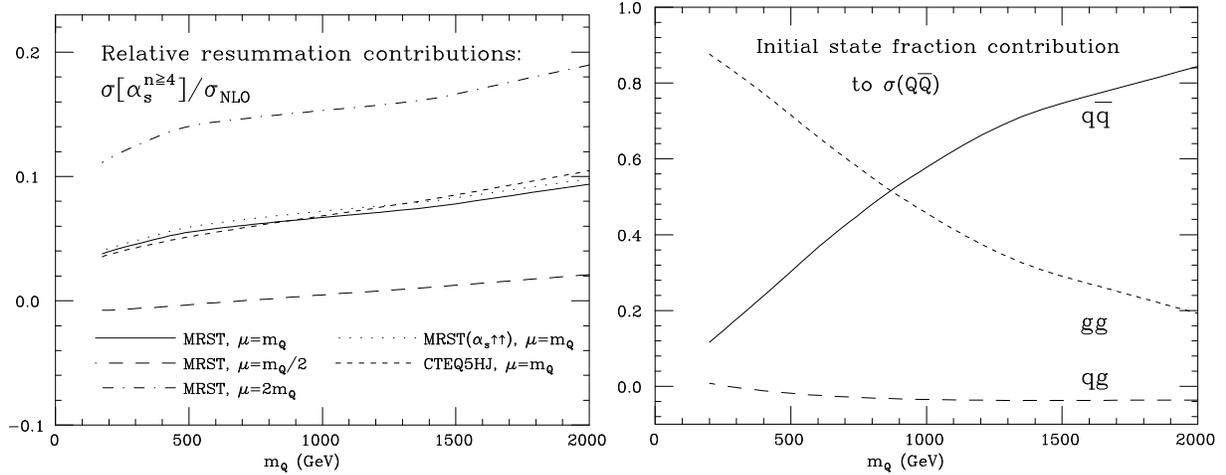
 
   \centerline{ 
    \includegraphics[width=0.5\textwidth,clip]{resfrac.eps} 
    \includegraphics[width=0.5\textwidth,clip]{topisf.eps}} 
    \caption{Heavy quark total production rates. Left figure: 
   fractional contribution induced by resummation contributions of 
   order $\calO(\as^{\geq 4})$. Right figure: initial state 
   composition.} 
    \label{fig:resfrac} 
  \end{figure}

For completeness, we present here results for the total cross section 
 of possible heavy quarks above the top quark mass. The scale and PDF 
 dependences are shown in Fig.~\ref{fig:sighvq}.  The uncertainty due 
 to the choice of scale is comparable to that of the $\ttbar$ cross 
 section, although the effects of the higher order 
 corrections are more and more important at large masses (see 
 Fig.~\ref{fig:resfrac}).  The uncertainty induced by PDF changes 
 becomes very large at large masses, in particular if one considers 
 sets such as CTEQ5HJ which have harder gluons. Notice however that 
 the relative effect due to the resummation corrections depends only 
 very weakly upon the choice of PDF's (cf. Section~\ref{sec:rates}). 
 
\section{APPENDIX: MONTE CARLO TOOLS\protect\footnote{Section coordinators: 
    M.L. Mangano, M.~Seymour.}} 
\label{app:mc} 
\subsection{Parton shower Monte Carlos} 
\label{app:herwig} 
General purpose Monte Carlo event generators like \herw, \pyth~and 
\isajet~ are essential tools for measuring the top quark cross section, 
mass and other production and decay properties.  They are complementary 
to the QCD tools described in Section~\ref{sec:qcdtools} since, 
although they are less reliable for inclusive quantities like the total 
cross section, they provide a fully exclusive description of individual 
events at the hadron level.  These can be analysed in exactly the same 
way as experimental data and can be put through full or fast detector 
simulations to estimate experimental systematics.  In certain kinematic 
regions, particularly the quasi-elastic limit in which accompanying 
radiation is suppressed, they give more reliable QCD predictions than 
the available calculations.  They include approximate treatments of 
higher order perturbative effects, hadronisation, secondary decays and 
underlying events. 
 
The three programs we discuss have the same basic structure, although 
the precise details vary enormously.  Events are generated by starting 
with the hardest (highest momentum scale) interaction, described by 
exact QCD (or EW) matrix elements.  This is usually only done to 
leading order so describes a $2\to2$ scattering process.  The production 
of multi-parton final states is described as the emission of additional 
partons from the incoming and outgoing partons of the hard process. 
This is simulated by a parton shower algorithm in which the partons 
evolve downwards in some energy-like scale according to 
perturbatively-calculable probabilistic distributions.  When the 
evolution scale becomes small the running coupling grows, phase space 
fills with (mostly soft) partons and perturbation theory breaks down. 
At this point a model of the non-perturbative physics is needed: 
the perturbative emission is cutoff by a fixed infrared cutoff and the 
system of partons is confined into hadrons.  Having treated all outgoing 
partons we are left with the remnants of the incoming protons, stripped 
of the partons that participated in the hard process.  These remnants 
can interact with each other, to produce additional soft hadrons in the 
event, known as the underlying event. 
 
Parton shower algorithms are developed by studying the amplitude to emit 
an additional parton into a given process.  This is enhanced in two 
kinematic regions: collinear, where two massless partons are much closer 
to each other than any others or where a massless parton is close to the 
incoming proton direction; and soft, where a gluon is much softer than 
any other parton.  In both cases the enhanced terms are universal, 
allowing a factorisation of emission by a system of partons from the 
process that produced them.  In the collinear case, this factorisation 
works at the level of cross sections, so it is not surprising that a 
probabilistic approach can be set up.  In the soft case however, the 
factorisation theorem is valid at amplitude level and it turns out that 
in any given configuration, many different amplitudes contribute 
equally.  It therefore seems impossible to avoid quantum mechanical 
interference and so to set up the evolution in a probabilistic way.  The 
remarkable result though is that, due to coherence between all the 
coloured partons in an event, the interference is entirely destructive 
outside angular-ordered regions of phase space.  This means that the 
soft emission can be incorporated into a collinear algorithm, simply by 
choosing the emission angle as its evolution variable, as is done in 
\herw.  The most important effects of coherence can be approximately 
incorporated by using some other evolution variable, like virtuality, 
and vetoing non-angular-ordered emission, as is done in \pyth.  If the 
colour-coherence is not treated at all, one obtains the wrong 
energy-dependence of jet properties.  Such models, like \isajet, are 
completely ruled out by $e^+e^-$ annihilation data.  Colour coherence 
effects are also important in determining the initial conditions for the 
parton evolution, resulting in physically-measurable inter-jet 
effects~\cite{Abe:1994nj}, which are also in disagreement 
with \isajet. 
 
Since the top quark decays faster than the typical hadronisation time, 
its width cuts off the parton shower before the infrared cutoff.  Its 
decay then acts as an additional hard process and the resulting bottom 
quark (and two more partons if the W decays hadronically) continue to 
evolve.  Additional coherence effects mean that radiation from the top 
quark is suppressed in the forward direction (the dead cone effect), 
as is radiation in the W direction in the top decay.  These effects 
are again included in \herw, partially included in \pyth~and not 
included in \isajet.  Since the top quark is coloured, the $b$ quark 
in its decay is colour-connected to the rest of the event, meaning 
that its properties are not necessarily the same as in a `standard' 
$b$ production event.  As mentioned in Section~\ref{sec:tjpsi} and as 
discussed in more detail in~\cite{Corcella00}, such non-universal 
effects are small. 
 
Although parton showers are reliable for the bulk of emission, which is 
soft and/or collinear, it is sometimes the rare hard emissions that are 
most important in determining experimental systematics and biases.  Such 
non-soft non-collinear emission should be well described by 
NLO perturbation theory, since it is far from all 
divergences.  However, it is not straightforward to combine the 
advantages of the parton shower and NLO calculation, so it has only been 
done for a few specific cases.  Most notable for hadron-hadron 
collisions are the Drell-Yan process, for which matrix element 
corrections are included in both \herw~and \pyth, and top decay, which 
is included in \herw~and discussed earlier in Section~\ref{sec:ljpsisyst}  
in this report.  The 
corrections to Drell-Yan events are particularly important at high 
transverse momenta, where the uncorrected algorithms predict far too few 
events.  It is likely that implementing corrections to $t\bar{t}$ pair 
production would cure the analogous deficit at high $\ptpair$ seen 
in Fig.~\ref{fig:phicomp}. 
 
Hadronisation models describe the confinement of partons into hadrons. 
Although this process is not well understood from first principles, it 
is severely constrained by the excellent data from LEP, SLD and HERA. 
The string model, used by \pyth, and the cluster model, used by \herw, 
both take account of the colour structure of the perturbative phase of 
evolution, with colour-connected pairs producing 
non-perturbative singlet structures that decay to hadrons.  The biggest 
difference between these models is in how local these colour-singlet 
structures are.  In the string model they stretch from a quark (or 
anti-di-quark) via a series of colour-connected gluons to an antiquark (or 
di-quark).  In the cluster model each gluon decays non-perturbatively to 
a quark-antiquark pair and each resulting quark-antiquark singlet 
(coming one from each of two colour-connected gluons) decays to 
hadrons.  The independent fragmentation model, used by \isajet, on the 
other hand, treats each parton as an independent source of hadrons and 
is strongly ruled out by $e^+e^-$ data, for example on inter-jet effects 
in three-jet events, the so-called string effect.  Of the other two 
models, \pyth~gives the better description of $e^+e^-$ data, but  
\herw~also gives an adequate description, despite having a lot fewer 
adjustable parameters. 
 
Models of the underlying event are not strongly constrained by either 
theoretical understanding or experimental data.  Two extreme models are 
available and the truth is likely to lie 
between them.  In the soft model, used in \herw, the collision of the 
two proton remnants is assumed to be like a minimum bias hadron-hadron 
collision at the same energy.  A simple parametrisation of minimum bias 
data (from UA5~\cite{Alner:1987wb})  
is used with little additional physics input.  In the 
mini-jet model, used in \pyth~and available as an additional package for 
\herw, on the other hand, the remnant-remnant collisions act as a new 
source of perturbative scattering, which ultimately produce the hadrons 
of the underlying event.  To avoid regions of unstable perturbative 
predictions and problems with unitarity, a cutoff must be used, 
$p_{t,min}\sim 1$~GeV.  Presumably for a complete description, some 
soft model should describe the physics below $p_{t,min}$ such that the 
results do not depend critically on its value.  Unfortunately no such 
model exists at present.  Although the two models give rather similar 
predictions for average properties of the underlying event, they give 
very different probabilities for the rare fluctuations that can be most 
important in determining jet uncertainties.  This is an area that needs 
to be improved before LHC running begins. 
 
\subsection{Parton-level Monte Carlos} 
With few exceptions (e.g. 3 or 4-jet final 
states in $e^+e^-$ collisions) multi-jet final states are not 
accurately described by the shower MC's described above. This is 
because emission of several hard and widely separated partons is poorly 
approximated by the shower evolution algorithms, and exact (although 
perhaps limited to the tree level) matrix elements need to be used  
to properly evaluate quantum correlations. 
Parton-level Monte Carlos are event generators for multi-parton final 
states, which incorporate the exact tree-level matrix elements. 
They can be used for parton-level simulations of multi-jet processes, 
under the assumption that each hard parton will be identified with a 
final-state physical jet with momentum equal to the momentum of the 
parent parton. Selection and analysis cuts can be applied directly to 
the partons. In some cases, the partonic final states can be used as a 
starting point for the shower evolution performed using a shower MC 
such as \herw, \pyth, or \isajet. For a discussion of the problems 
involved in ensuring the colour-coherence of the shower evolution when 
dealing with multi-parton final states, see~\cite{Caravaglios:1999yr}. 
 
In the following, we collect some information on the most 
frequently used parton-level MCs used in connection with top quark studies. 
 
\def\vecbos{{\small VECBOS}} 
\subsubsection{\vecbos\protect\footnote{ 
    \vecbos~authors: F.A. Berends, H. Kuijf, B. Tausk and W.T. Giele. 
    Contacts: giele@fnal.gov} } \vecbos~\cite{Berends:1991ax} is a 
Monte Carlo for inclusive production of a $W$-boson plus up to 4 jets 
or a $Z$-boson plus up to 3 jets. \vecbos~is therefore a standard tool 
used in the simulation of backgrounds to $\ttbar$ production. The 
matrix elements are calculated exactly at the tree level, and include 
the spin correlations of the vector boson decay fermions with the rest 
of the event. Various parton density functions are available and 
distributions can be obtained by using the kinematics of the final 
state, available on an event-by-event basis together with the 
corresponding event weight.  The code and its documentation can be 
obtained from: 
\\ 
\hspace*{2cm}{\tt http://www-theory.fnal.gov/people/giele/vecbos.html}\\ 
Documentation on the use of VECBOS within ATLAS can be found  
in~\cite{atlasvecbos}. 
 
\subsubsection{CompHEP\protect\footnote{ 
    CompHEP authors: A.~Pukhov, E.~Boos, M.~Dubinin, V.~Edneral, V.~Ilyin, 
    D.~Kovalenko, A.~Kryukov, V.~Savrin, S.~Shichanin, A.~Semenov. 
    Contacts: pukhov@theory.npi.msu.su, ilyin@theory.npi.msu.su} } 
 
CompHEP~is a package for the calculation of elementary 
particle decay and collision properties in the lowest order of 
perturbation theory (the tree approximation).  The main purpose of 
CompHEP~is to generate automatically transition 
probabilities from a given Lagrangian, followed by the automatic 
evaluation of the phase-space integrals and of arbitrary 
distributions. 
The present version has 4 built-in physical models. Two of them are 
the versions of the Standard Model (SU(3)xSU(2)xU(1)) in the unitary 
and t'Hooft-Feynman gauges. The user can change the models or even 
create new ones. 
 
The symbolic part of CompHEP~ allows the user to perform the following 
operations: 
\begin{enumerate} 
\item to select a process by specifying incoming and outgoing  
      particles for the decays of $1 
\rightarrow 2, \ldots ,1 \rightarrow 5$ types and the collisions of $2 
\rightarrow 2, \ldots , 2 \rightarrow 4$ types,  
\item to generate Feynman diagrams, calculating the analytical expressions  
   for the squared matrix elements,  
 \item to save the algebraic symbolic results and to generate the 
   optimized {\sf Fortran} and {\sf C} codes for the squared matrix 
   elements for further numerical calculations. 
\end{enumerate} 
The numerical part of CompHEP~ allows to convolute the squared matrix 
element with structure functions and beam spectra, to introduce 
various kinematic cuts, to introduce a phase space mapping in order to 
smooth sharp peaks of a squared matrix element, to perform a Monte 
Carlo phase space integration by VEGAS, to generate events and to 
display distributions for various kinematic variables.  Recently, an 
interface with \pyth\ has been created~\cite{Ilyin:comphep-pythia}. 
This allows to perform realistic simulations of the process including 
hadronisation effects as well as the effects of the initial and final 
state radiation. 
 
The CompHEP~ codes and manual are available from  the following Web sites: 
\\ 
\hspace*{2cm}{\tt http://theory.npi.msu.su/\verb|~|comphep}\\ 
\hspace*{2cm}{\tt http://www.ifh.de/\verb|~|pukhov} 
 
\def\ALPHA{{\small ALPHA}} 
\subsubsection{\ALPHA\protect\footnote{ 
    ALPHA~authors: F.~Caravaglios, M.~Moretti. The version for 
    hadronic collisions received additional contributions from 
    M.L.~Mangano and R.~Pittau.  Contact: moretti@fe.infn.it}} \ALPHA\  
is an algorithm introduced in~\cite{Caravaglios:1995cd} for 
the evaluation of arbitrary multi-parton EW matrix elements. This 
algorithm determines the matrix elements from a (numerical) Legendre 
transform of the effective action, using a recursive procedure which 
does not make explicit use of Feynman diagrams.  The algorithm has a 
complexity growing like a power in the number of particles, compared 
to the factorial-like growth that one expects from naive diagram 
counting.  This is a necessary feature of any attempt to evaluate 
matrix elements for processes with large numbers of external 
particles, since the number of Feynman diagrams grows very quickly 
beyond any reasonable value. 
 
An implementation of \ALPHA\ for hadronic collisions was introduced  
in~\cite{Caravaglios:1999yr}, where the algorithm was extended to 
the case of QCD amplitudes (see also~\cite{Draggiotis:1998gr}). 
The main aim of the hadronic version of \ALPHA\ is to allow the QCD 
parton-shower evolution of the multi-parton final state, in a way 
consistent with the colour-coherence properties of the soft gluon 
emission dynamics.  This is achieved by evaluating the QCD 
amplitudes in an appropriate colour basis~\cite{Caravaglios:1999yr}, 
such that the assignement of a specific colour flow configuration 
on an event-by-event basis. The pattern of colour flow defines the 
colour currents required to implement the angular ordering 
prescription which embodies, at the leading order in the $1/N_c$ 
expansion, the quantum coherence properties of soft-gluon radiation, 
as discussed in Appendix~\ref{app:herwig}. A version of the code is 
being completed~\cite{alphaWbb}, which incorporates the evaluation of 
$Wb\bar{b}+n$~jets ($n\le 4$), with all $b$-mass effects included. 
This program will allow a complete evaluation of the $W+$~multijet 
backgrounds to single top and $\ttbar$ production.  The code contains 
3 modules: the first for the generation of parton-level events, with 
the assignement of partonic flavours, helicities and colour flows. The 
second for the unweigthing of the events, and the third for the 
parton-shower evolution of the initial and final states, done using 
the \herw\ MC. The code will soon be available from the URL: 
\\ 
\hspace*{2cm}{\tt http://home.cern.ch/\verb|~|mlm/alpha}

\end{document}